\documentclass[12pt]{report}

\usepackage{latexsym}
\usepackage{amssymb}
\usepackage{amsmath}
\usepackage{alltt}
\usepackage[dvips]{graphicx}
\usepackage{amsmath,amssymb}

\newenvironment{exafont}{\begin{bf}}{\end{bf}}

\newfont{\Bb}{msbm10 scaled\magstep 1}

\newcommand{\bbbone}{{\mathbf 1}}
\newcommand{\scalprod}[2]{\left\langle {#1}, {#2}\right\rangle}
\newcommand{\D}{{\mathcal D}}
\renewcommand{\Re}{Re}
\renewcommand{\Im}{Im}
\newcommand{\fer}[1]{(\ref{#1})}

\newcommand{\ran}{Ran}
\renewcommand{\ker}{Ker}
\newcommand{\repsilonbar}{\,\overline{\!R}_\epsilon}
\newcommand{\repsilon}{R_\epsilon}

\newcommand{\edeltanot}{E_\Delta^0}
\newcommand{\chideltanot}{\chi_\Delta^0}
\newcommand{\cx}{{\mathbf C}}
\renewcommand{\r}{{\mathbf R}}

\newcommand{\supp}{{\rm supp}}

\renewcommand{\H}{{\mathcal H}}
\renewcommand{\S}{{\mathcal S}}
\renewcommand{\O}{{\mathcal O}}
\newcommand{\B}{{\mathcal B}}
\renewcommand{\L}{{\mathcal L}}
\newcommand{\K}{{\mathcal K}}
\renewcommand{\a}{\alpha}
\newcommand{\R}{{\mathcal R}}
\renewcommand{\S}{\Sigma}
\renewcommand{\d}{{\rm d}}
\newcommand{\Pibar}{\overline{\Pi}}
\newcommand{\Pbar}{\overline{P}}
\newcommand{\h}{{\mathbf h}}
\newcommand{\F}{{\mathcal F}}
\newcommand{\U}{{\mathcal U}}
\renewcommand{\u}{{\mathbf u}}
\newcommand{\dom}{{\mathcal D}}
\newcommand{\bfmu}{{\mathbf \mu}}
\newcommand{\unit}{{\mathbf 1}}
\newcommand{\bfQ}{{\mathbf Q}}
\newcommand{\M}{{\mathcal M}}
\newcommand{\N}{{\mathcal N}}
\newcommand{\W}{{\mathcal W}}

\newcommand{\allspace}{{\mathbf R}^3_\pm}

\newcommand{\av}[1]{\left\langle{#1}\right\rangle}

\newcommand{\nr}{{\cal N}_r}
\newcommand{\nl}{{\cal N}_l}
\newcommand{\pp}{{\mathcal P}}
\newcommand{\C}{{\mathcal C}}
\newcommand{\E}{{\mathcal E}}
\newcommand{\ls}{\lambda(s)}
\newcommand{\lsp}{\lambda(s')}

\newcommand{\unitD}{{\overline{D}}}
\newcommand{\unitOmega}{{\overline{\Omega}}}

\newcommand{\cp}{{\mathsf P}}

\newcommand{\id}{{\mathbf 1}}

\newcommand{\tr}{Tr}
\newcommand{\bP}{{\overline{P}}}
\newcommand{\PV}{{\mathcal PV}}
\newcommand{\bn}{{(n)}}

\newcommand{\what}[1]{\widehat{#1}}

\renewcommand{\tilde}[1]{\widetilde{#1}}

\newcommand{\Obl}{{\Omega_{\beta,g}}}
\newcommand{\Obz}{{\Omega_{\beta,0}}}
\newcommand{\obl}{{\omega_{\beta,g}}}

\newcommand{\oblL}{{\omega^\Lambda_{\beta,g}}}
\newcommand{\obzL}{{\omega^\Lambda_{\beta,0}}}
\newcommand{\HlL}{H^\Lambda_g}

\newcommand{\HfL}{H_f^\Lambda}
\newcommand{\hfL}{h_f^\Lambda}
\newcommand{\HzL}{H_0^\Lambda}
\newcommand{\HoL}{H_0^\Lambda}

\begin{document}

\pagestyle{empty}

%{\sf DISS. ETH NO. 16187}

\begin{center}

{\bf NONEQUILIBRIUM QUANTUM STATISTICAL MECHANICS \\
AND THERMODYNAMICS}\footnote{This work is based on the author's doctoral thesis, ETH-Diss 16187.}

\vspace{0.5cm}

Walid K. Abou Salem \footnote{{\it E-mail: walid@itp.phys.ethz.ch.}} \\
Institut f\"ur Theoretische Physik\\
ETH H\"onggerberg\\
8093 Z\"urich, Switzerland

\vspace{0.5cm}

August $22^{nd}$, 2005

\vspace{0.5cm}

{\bf Abstract}
\end{center}

The purpose of this work is to discuss recent progress in deriving
the fundamental laws of thermodynamics ($0^{th},1^{st}$ and
$2^{nd}$-law) from {\it nonequilibrium quantum statistical
mechanics}. Basic thermodynamic notions are clarified and
different {\it reversible} and {\it irreversible} thermodynamic processes are
studied from the point of view of quantum statistical mechanics.
Special emphasis is put on new adiabatic theorems for steady
states {\it close to} and {\it far from} equilibrium, and on
investigating cyclic thermodynamic processes using an extension of
{\it Floquet} theory.

\vspace{12cm}

\pagebreak

%%%%%%%%%%%%%%%%%%%%%%%%%%%%%%%%

\pagestyle{plain}

\pagenumbering{roman} \tableofcontents

\pagebreak

%%%%%%%%%%%%%%%%%%%%%%%%%%%%%%%%%

%%%%%%%%%%%%%%%%%%%%%%%%%%%%%%%%%%%%%%%%%%%%%%%%%%%%%%%%%%%%%%%

\setcounter{page}{1} 
\pagenumbering{arabic}

\chapter{Introduction}

There are numerous examples of many-particle systems in Nature,
such as atoms and molecules in gases, fluids, superfluids, solids
and plasmas, electrons in conductors and semi-conductors, nuclear
matter in neutron stars, and the quark-gluon plasma. It is both fascinating and
intriguing that {\it all} these different systems, when in thermal equilibrium, are
theoretically described by very {\it general} and {\it universal}
physical laws in the thermodynamic limit. The main aim of
thermodynamics is to define appropriate physical quantities, the
so called {\it state quantities}, which characterize the
macroscopic properties of many-particle systems in thermal equilibrium, and to relate
these quantities to each other through {\it universally} valid
equations, independent of the specificity and microscopics of the
physical models. However, the laws of thermodynamics have been
generally viewed as {\it empirical theorems} or axioms. The
problem to derive the $0^{th}$, $1^{st}$, $2^{nd}$ (and $3^{rd}$)
law, from kinetic theory and non-equilibrium statistical mechanics
has been studied since the late $19^{th}$ century, with
contributions by many distinguished theoretical physicists
including Maxwell, Boltzmann, Gibbs and Einstein. In this work, we
discuss some recent results concerning our own attempts to
derive thermodynamics from non-equilibrium quantum statistical
mechanics (NEQSM) and to bring the problem just described closer
to a satisfactory solution. These are necessary steps towards
understanding irreversibility and the emergence of macroscopic
{\it classical} behavior, such as thermodynamics, from more {\it
fundamental} (and {\it time-reversal invariant}) microscopic laws,
such as quantum mechanics.

%%%%%%%%%%%%%%%%%%%%%%%%%%%%%%%%%%%%%%%%%%%%%%%%%%%%%%%%%%%%%%%%%%%%%%%%%%%%%%
%%%% SUMMARY OF RESULTS
%%%%%%%%%%%%%%%%%%%%%%%%%%%%%%%%%%%%%%%%%%%%%%%%%%%%%%%%%%%%%%%%%%%%%%%%%%%%%%

\section{Summary of results}

The main new results in this work are:

\begin{itemize}

\item Extending the positive commutator method together with a suitable Virial
Theorem to prove the property of return to equilibrium (RTE) for a
class of systems composed of a small quantum system coupled to non-relativistic fermionic and bosonic reservoirs, such as a spin impurity interacting with magnons in a magnet or a {\it quantum dot} interacting with electrons in a metal.

\item Studying isothermal processes of a finitely extended, driven quantum system in contact with an infinite heat bath from the point of view of
quantum statistical mechanics. Notions like heat flux, work and
entropy are defined for trajectories of states close to, but
distinct from states of joint thermal equilibrium. In this
context, a theorem characterizing {\it reversible} isothermal
processes as {\it quasi-static} processes ({\it isothermal
theorem}) is proven. Corollaries of the latter concern changes of
entropy and free energy in reversible isothermal processes and the
$0^{th}$ law of thermodynamics. We also specialize to the specific
example of a small system coupled to a fermionic
reservoir, and obtain an explicit estimate on the rate of
convergence to the instantaneous equilibrium states in the
quasi-static limit.

\item Proving a novel adiabatic theorem for generally (non)normal
and unbounded generators of time evolution, and applying this
theorem to the study of adiabatic evolution of states close to
non-equilibrium steady states (NESS).

\item Studying cyclic processes of a finitely extended, periodically driven
quantum system coupled to several reservoirs from the point of
view of quantum statistical mechanics, and proving the convergence
of the state of the system to a time-periodic state by
extending Floquet theory to non-equilibrium quantum statistical
mechanics. Positivity of entropy production and Carnot's
formulation of the second law of thermodynamics follow from the
definite sign of relative entropy and the existence of the large
time limit.

\end{itemize}

Together with extending some known techniques to prove return to equilibrium and clarifying basic notions in thermodynamics based on nonequilibrium quantum statistical mechanics, this work addresses novel questions which are important in understanding the emergence of the laws of thermodynamics. As far as we know, studying reversible isothermal processes, the adiabatic evolution of nonequilibrium steady states, and cyclic thermodynamic processes from the point of view of quantum statistical mechanics have not been attempted before in the literature.

%%%%%%%%%%%%%%%%%%%%%%%%%%%%%%%%%%%%%%%%%%%%%%%%%%%%%%%%%%%%%%%%%%%%%%%%%%%%%%
%%%% METHODS USED
%%%%%%%%%%%%%%%%%%%%%%%%%%%%%%%%%%%%%%%%%%%%%%%%%%%%%%%%%%%%%%%%%%%%%%%%%%%%%%

\section{Methods}

The main methods used in this thesis are the following.

\begin{itemize}

\item Algebraic formulation of quantum statistical mechanics.

\item Spectral analysis techniques: Mourre theory and complex deformation techniques.

\item Spectral approach to RTE using the standard Liouvillain.

\item Spectral approach to NESS using the C-Liouvillian.

\item Spectral approach to cyclic thermodynamic processes using the so called Floquet Liouvillean.

\item Generalization of adiabatic theorems in non-equilibrium
quantum statistical mechanics.

\end{itemize}

%%%%%%%%%%%%%%%%%%%%%%%%%%%%%%%%%%%%%%%%%%%%%%%%%%%%%%%%%%%%%%%%%%%%%%%%%%%%%
%%%% ORGANIZATION
%%%%%%%%%%%%%%%%%%%%%%%%%%%%%%%%%%%%%%%%%%%%%%%%%%%%%%%%%%%%%%%%%%%%%%%%%%%%%

\section{Organization}

The organization of this thesis is as follows. In chapter 2, after
reviewing the basic laws and notions of thermodynamics, we present
an overview of the derivation of these laws from non-equilibrium
quantum statistical mechanics (NEQSM). In chapter 3, we discuss
the mathematical framework, which is the algebraic formulation of
quantum statistical mechanics [BR,Ha]. We review relevant
concepts, such as the standard Liouvillean, KMS states, $C^*$ and
$W^*$-dynamical systems and their perturbations, and we introduce
new notions such as {\it instantaneous} equilibrium states. The presentation in this chapter is meant
to be concrete, but a mathematically more elegant and powerful,
yet more abstract, presentation is given in an appendix to this
chapter. In chapter 4, we list all the model quantum systems we consider
in this thesis as paradigms of thermodynamic systems, together with
the assumptions on these model systems. An appendix to this chapter
discusses some consequences of these assumptions. In chapter 5, we
prove the property of return to equilibrium for a class of quantum
mechanical systems composed of a {\it small} system coupled to a
reservoir of non-relativistic bosons or fermions, by extending the
positive commutator method developed in [M1, M2, FM1, FM2, FMS].
Technical proofs in this chapter are relegated to an appendix. In
chapter 6, we extensively discuss a paradigm of a thermodynamic
system that we consider in the following chapters: a two-level system
coupled to $n$ fermionic reservoirs. We present the method
of complex deformations developed in [JP1,2,3] and
extend it to the study of {\it time-dependent}
Liouvilleans. We also present the C-Liouvillean, and discuss how
it relates to non-equilibrium steady states. In chapter 7, we
study isothermal processes of a small system diathermally coupled
to a single reservoir, and we state and prove the {\it isothermal
theorem}. We define notions like heat flux, work and entropy for
trajectories of states close to, but distinct from states of joint
thermal equilibrium. We also prove a theorem characterizing {\it
reversible} isothermal processes as {\it quasi-static} processes
({\it isothermal theorem}), and we discuss corollaries concerning
the changes of entropy and free energy in reversible isothermal
processes and the $0^{th}$ law of thermodynamics. In chapter 8, we
prove a novel adiabatic theorem for generators of time evolution
which are not necessarily normal or bounded. We also discuss two
applications of this theorem in non-equilibrium quantum
statistical mechanics: an adiabatic theorem for states close to
non-equilibrium steady states, and a concrete example of the
isothermal theorem with an explicit rate of convergence to the
{\it quasi-static limit}. Another important application of this
theorem is an adiabatic theorem for quantum resonances.[A-SF3] In
chapter 9, we investigate cyclic thermodynamic processes from the
point of view of quantum statistical mechanics. We introduce a new
Liouvillean, the Floquet Liouvillean, which generates dynamics on
a suitable Banach space when the perturbation is time periodic,
and we relate the time-periodic state to which the system
converges to a zero-energy resonance of the Floquet Liouvillean.
In principle, we can compute the entropy production per cycle
(which is positive) and the difference between the degree of
efficiency $\eta$ and $\eta^{rev}$ to arbitrary orders in the
coupling, for small enough coupling.
%%%%%%%%%%%%%%%%%%%%%%%%%%%%%%%%%%%%%%%%%%%%%%%%%%%%%%5
%\include{td}

\chapter{Thermodynamics and nonequilibrium quantum statistical mechanics: an overview}

This chapter offers a brief  overview of the progress made towards deriving
the fundamental laws of thermodynamics from non-equilibrium
quantum statistical mechanics. New results which are stated here
without proof are carefully and thoroughly discussed in the
following chapters. Along the way, we clarify certain basic
notions, such as thermodynamic systems, heat baths, different
thermodynamic processes, the meaning of heat energy and the rather
special and {\it important} role of relative entropy. The
discussion at this stage is necessarily {\it heuristic}. However,
various terms and notions appearing in this section are precisely
defined in the following chapters. Before presenting this
overview, we briefly recall some basic thermodynamics.

\section{Thermodynamics}

\subsection{Basic concepts and ideas of thermodynamics}

The purpose of thermodynamics is to describe the average properties of macroscopically extended matter close to a state of thermal equilibrium, with small spatial and temporal variations. (Gravitational effects are usually neglected.) Typical thermodynamic systems are formed of $10^{23}-10^{28}$ particles, and describing the system microscopically by solving the corresponding Hamilton equations or Schr\"odinger equation is an impossible task. Instead, one is interested in describing emergent properties, such as the average energy per particle, using few macroscopically observable quantities, such as the volume $V$ of the system, the total energy $E$, and the magnetization $M.$ These macroscopic quantities, which can be measured simultaneously and with precision, are called {\it thermodynamic observables} or {\it state quantities}. One identifies all microstates having the same thermodynamic observables with one macrostate. 

An {\it isolated system} is a macroscopically large thermodynamic system without any kind of contact or interaction with its {\it environment}. For such a system, the measured values of the thermodynamic observables are stationary, that is, time-independent. It is an observed fact that the state of an isolated system (as time $t\rightarrow\infty$) approaches asymptotically a stationary {\it equilibrium state}, with precise values of thermodynamic observables. (This is the equilibrium postulate for isolated systems, which is part of the $0^{th}$ law of thermodynamics, and which plays an important role in thermodynamics.)  

Thermodynamic systems can be approximately infinite. Physically interesting states of infinitely extended systems are their equilibrium states. Although realistic systems are large but finite, if one waits long enough and one looks at a macroscopically large subset of this system, these systems behave locally as infinite ones. 

Let $N$ be the number of elements in a complete family of independent thermodynamic observables of a thermodynamic system $\S$, which can be measured simultaneously and with precision. Their measured values specify a point $X\in \Gamma^\S$, where $\Gamma^\S\subset {\mathbf R}^N.$ A thermodynamic observable is a real- valued function on $\Gamma^\S,$ and every $X\in \Gamma^\S$ corresponds to an equilibrium state. The space $\Gamma^\S$ of equilibrium states of $\S$ is a connected convex subset of ${\mathbf R}^N.$

One may couple two thermodynamic systems, $1$ and $2$, through local interactions. Initially, when the two systems are uncoupled, the state space is the Cartesian product $\Gamma^{1}\times \Gamma^{2}.$ When one allows interaction between $1$ and $2,$ some symmetries of $1$ and $2$ are broken, and the new family of thermodynamic observables and the space of equilibrium states, $\Gamma^{1\vee 2},$ of the coupled system depends on the type of couplings or contacts between $1$ and $2.$

We now discuss the notion of a {\it thermodynamic process}, which plays a central role in thermodynamics. 

%%%%%%%%%%%%%%%%%%%%%%%%%%%%%%%%%%%%%%%%%%%%%%%%%%%%%%%%%%%%%%%%%%%
%THERMODYNAMIC PROCESSES
%%%%%%%%%%%%%%%%%%%%%%%%%%%%%%%%%%%%%%%%%%%%%%%%%%%%%%%%%%%%%%%%%%%

Let $(X_1, X_2)\in \Gamma^1\times\Gamma^2$ be the initial uncoupled equilibrium state of $1\vee 2$ at time $t_0.$ One or both systems can be approximately infinite. If one turns on a coupling between $1$ and $2$ at time $t_0,$ one is interested in knowing the state of the coupled system at time $t_0+T$, as $T\rightarrow\infty.$ Let $\gamma(t)$ be the microstate of the coupled system. When at least one system, $1$ or $2,$ is finite, the $0^{th}$ law of thermodynamics claims that $\gamma(t)\rightarrow X_{12}\in \Gamma^{1\vee 2},$ as time $t\rightarrow\infty.$ Thermodynamics specifies the mapping $\Gamma^1\times\Gamma^2\ni (X_1,X_2)\rightarrow X_{12}\in \Gamma^{1\vee 2}$ only when the nature of the contact between $1$ and $2$ is known. There are two conceivable processes
\begin{align*}
&(X_1,X_2)\rightarrow X_{12} \; ,\\
&X_{12}\rightarrow (X_1,X_2) \; .
\end{align*}
When $X_{12}\rightarrow (X_1,X_2)$ is not possible, we say that $(X_1,X_2)\rightarrow X_{12}$ is {\it irreversible}.  In contrast, a thermodynamic process $\{ X(t)\}_{t_0\le t <\infty}$ of 
$\S$ is {\it reversible} if $X(t_0)=X_i, \lim_{t\rightarrow\infty}X(t)=X_f,$ and $X(t)\in \Gamma^\S , \forall t>t_0.$ In such processes, infinitesimal changes of variables happen slowly as
compared to some typical relaxation time of the system, and the
real state of the system at time $t$ will be infinitesimally close to the {\it
instantaneous equilibrium state} $X(t)$.

Thermodynamic observables correspond to constants of motion of a system, or equivalently to symmetries of the system. The space of equilibrium states $\Gamma$ consists of the total spectrum of all thermodynamic observables. For each symmetry which is conserved in a thermodynamic process, there corresponds a thermodynamic observable whose value is constant in time. One can hence classify thermodynamic processes from the point of view of symmetries. A contact or coupling will correspond to a perturbation of subsystems which breaks one or more of their symmetries. 

A {\it thermal contact} (diathermal wall) between a thermodynamic system $\S$ and a thermal reservoir $\R$ is an interaction which leaves all symmetries of $\S$ invariant except for time-translation invariance. It leaves all the thermodynamic observables of $\S$ invariant except for its energy. 

Similarly, one can define a thermal contact between two thermodynamic systems $\S_1$ and $\S_2$ as an interaction which preserves all the symmetries of $\S_1$ and $\S_2$ except for time-translation invariance: It leaves all the thermodynamic observables of $\S_1$ and $\S_2$ invariant except for their energies. 

%%%%%%%%%%%%%%%%%%%%%%%%%%%%%%%%%%%%%%%%%%%%%%%%%%%%%%%%%%%%%%%%%%
%FURTHER REMARKS
%%%%%%%%%%%%%%%%%%%%%%%%%%%%%%%%%%%%%%%%%%%%%%%%%%%%%%%%%%%%%%%%%%
%Bring two thermodynamic systems in contact with the same thermal reservoir...
%Another type of coupling is a tunneling junction, which breaks gauge invariance of the first kind.
%Another central concept of thermodynamics is internal energy, which we denote by $U.$ We assume that the difference $U(X_f)-U(X_i)$ can be measured.

%%%%%%%%%%%%%%%%%%%%%%%%%%%%%%%%%%%%%%%%%%%%%%%%%%%%%%%%%%%%%%%%%%%%

\subsection{The laws of thermodynamics}

In this subsection, we recall the fundamental laws of thermodynamics ($0^{th},1^{st},$ and $2^{nd}$ law), which form the axiomatic foundation of thermodynamics.\footnote{We will not discuss the third law of thermodynamics.} We are interested in the physical properties of a thermodynamic system $\S,$ which are described by a finite number $N$ of independent thermodynamic observables, $\xi_1,\cdots, \xi_N.$

\subsubsection{The $0^{th}$-law}

There are several parts to the zeroth law.
\begin{itemize}
\item[(i)] Consider an isolated thermodynamic system, $\S,$ which is approximately infinite. The state of $\S$ converges to an equilibrium state as time $t\rightarrow\infty.$ Each equilibrium state of $\S$ corresponds to a point $X$ in a connected convex subset $\Gamma^\S\subset {\mathbf R}^N.$ 

\item[(ii)] A (weaker) form of the $0^{th}$ law says that there exist, for all practical purposes infinitely, large thermodynamic systems, that return to equilibrium when isolated. One calls such systems thermal reservoirs or heat baths.
\item[(iii)] Two thermal reservoirs $\R_1$ and $\R_2$ are said to be equivalent ($\R_1\sim \R_2$) {\it iff} no energy flows between $\R_1$ and $\R_2$ when a diathermal contact is established between them. Then we say that the two reservoirs $\R_1$ and $\R_2$ are at the same temperature (ie, they are in {\it thermal equilibrium} with each other). Furthermore, given three thermal reservoirs $\R_1,\R_2$ and $\R_3$ such that $\R_1\sim \R_2$ and $\R_2\sim \R_3,$ then $\R_1\sim \R_3$, ie, all reservoirs have the same temperature (transitivity of the property of thermal equilibrium). 

\item[(iv)]  When one brings a finite thermodynamic system $\S$ in thermal contact with a thermal reservoir $\R$ and waits for an infinitely long time, the state of the coupled system will be an equilibrium state at the temperature of the reservoir.
 
\item[(v)] Moreover, if one removes the contact between $\S$ and $\R$ quasi-statically, the final state of $\S$ will be the (Gibbs) equilibrium state at the temperature of the reservoir, while the final state of the reservoir is identical to its initial state.

\end{itemize}

\subsubsection{The $1^{st}$-law}

For each finite thermodynamic system, there exists a thermodynamic observable $U$, the internal energy, which has a definite value in each state in $\Gamma^\S$; ($U$ is unique up to an additive constant). For a thermodynamic process $\gamma$ in which one brings $\S$ in contact with a thermal reservoir $\R$, the {\it heat energy} exchange $\Delta Q(\gamma)$ is a well-defined quantity which depends not only on the initial point $X_i=\partial_i\gamma$ and the final point $X_f=\partial_f\gamma$, but on the whole trajectory $\gamma$. \footnote{If $\Delta Q>0$, heat energy flows from $\R$ to $\S$ (we say $\R$ is hotter than $\S$), and if $\Delta Q<0$ heat flows from $\S$ to $\R$ (we say $\R$ is colder than $\S$).} The difference 
\begin{equation*}
\Delta A(\gamma):= U(X_f)-U(X_i)-\Delta Q(\gamma)\; ,
\end{equation*}
is the {\it work} done on $\S.$

\vspace{0.5cm}

Before stating the $2^{nd}$ law of thermodynamics, we need to introduce the notion of a heat engine. 

A {\it heat engine} is a finite thermodynamic system that works periodically in time and that is brought in contact with at least two inequivalent thermal reservoirs or with its environment. After one cycle, the system returns to its initial state, ie, $\partial_i\gamma=\partial_f\gamma.$ Let $\Delta Q(\gamma)$ be the total heat exchange between the heat engine and the thermal reservoirs in one cycle. Since the internal energy of the heat engine is the same at the beginning and at the end of each cycle, the $1^{st}$ law says that $\Delta Q(\gamma)$ has to be converted into work done by the heat engine on its environment. 

One makes the following (scaling) postulate on heat engines: The size of a heat engine can be enlarged or reduced by a scale factor $\lambda >0.$ Consider a heat engine $\S$ with a space of equilibrium states $\Gamma^\S.$ Then
\begin{equation*}
\Gamma^{\S^\lambda} := \{ X\in {\mathbf R}^N : \lambda^{-1} X\in \Gamma^\S \} \; ,
\end{equation*}
is the space of equilibrium states of the heat engine $\S^\lambda.$ To a cycle $\gamma$ of $\S,$ there corresponds a cycle $\gamma^\lambda$ of $\S^\lambda$ such that 
\begin{equation*}
U(\lambda X)=\lambda U(X) , \; \Delta Q(\gamma^\lambda)=\lambda \Delta Q(\gamma) \; .
\end{equation*}

We are now in a position to state one formulation of the second law of thermodynamics.

\subsubsection{The $2^{nd}$-law}

There does not exist any heat engine that converts {\it all} the heat energy it receives from thermal reservoirs into work done on its environment.

\vspace{0.5cm}

Now consider a heat engine $\S$ connected to two thermal reservoirs $\R_1$ and $\R_2$, such that, in one cycle $\gamma$, it gains a heat energy $\Delta Q_1$ from $\R_1$ and it gives a heat energy $\Delta Q_2$ to $\R_2$. The heat engine performs work if $\Delta Q_1-|\Delta Q_2|=\Delta Q_1+\Delta Q_2>0.$ In this case, the thermal reservoir $\R_1$ is called the hot reservoir, while $\R_2$ is called the cold reservoir. 

It follows from the above formulation of the second law of thermodynamics, that if there exists a heat engine that uses $\R_1$ as a hot reservoir and $\R_2$ as a cold one, then there does not exist any heat engine that uses $\R_2$ as a hot reservoir and $\R_1$ as a cold one. This fact can be used to define an {\it empirical} temperature $\Theta :$ the temperature $\Theta_1$ of $\R_1$ is higher than the temperature $\Theta_2$ of $\R_2$ if there exists a heat engine $\S$ that uses $\R_1$ as a hot reservoir and $\R_2$ as a cold one.

A heat engine is said to be {\it reversible} (or a Carnot machine) if, in a time-reversed cycle, it can work as a heat pump: During a cycle $\gamma^-$, it takes heat $\Delta Q_2$ from $\R_2$ and gives heat $\Delta Q_1$ to $\R_1.$ In this case, the environment must supply the work per cycle $\Delta A=\Delta Q_1-|\Delta Q_2|.$ Reversible heat engines are idealizations of realistic systems.

Define the {\it degree of efficiency} of a heat engine $\S$ as the ratio of the work done per cycle and the heat it gains from the hot reservoir in one cycle,
\begin{equation*}
\eta^\S := \frac{\Delta A}{\Delta Q_1} =\frac{\Delta Q_1 + \Delta Q_2}{\Delta Q_1} = 1 + \frac{\Delta Q_2}{\Delta Q_1} \; .
\end{equation*}
It follows from the second law of thermodynamics that among all heat engines with the same hot and cold reservoirs, the reversible engine has the highest efficiency $\eta^{rev}$. One can use this fact to define an absolute temperature $T$ for a thermal reservoir $\R,$ such that 
\begin{equation*}
\eta^{rev}=\frac{T_1-T_2}{T_1}\; .
\end{equation*}
The fact that $\eta^\S\le \eta^{rev}$ implies that 
\begin{equation*}
\frac{\Delta Q_1}{T_1} + \frac{\Delta Q_2}{T_2}\le 0 \; ,
\end{equation*}
with equality when $\gamma$ is reversible.

This result can be generalized to the case when $\S$ is connected to $n$ thermal reservoirs $\R_1,\cdots ,\R_n$ with temperatures $T_1>\cdots >T_n.$ In this case,
\begin{equation*}
\sum_{i=1}^n \frac{\Delta Q_i}{T_i}\le 0 \; ,
\end{equation*}
with equality when the cyclic process is reversible. Taking the limit $n\rightarrow\infty$ gives
\begin{equation*}
\oint_\gamma \frac{\delta Q}{T} \le 0 \; ,
\end{equation*}
with equality when $\gamma$ is reversible.

Consider $\gamma\in \Gamma^\S$, a reversible cyclic process of $\S,$ and parametrize its curve in $\Gamma^\S$ by (the time) $\tau\in [t_0,\infty).$ We assume that
\begin{equation*}
\dot{\gamma}(\tau):= \lim_{h\searrow 0}\frac{1}{h} (\gamma (\tau+h)-\gamma(\tau )) \; 
\end{equation*}
exists for all $\tau\in [t_0,\infty ).$ 

Denote by $\gamma_t$ the subprocess $\{ \gamma (\tau) \}_{t_0\le \tau \le t}$ from $X_i=\gamma (t_0)$ to $\gamma (t)\in \Gamma^\S.$ From the first law of thermodynamics, $\Delta Q(\gamma_t)$ is a well-defined quantity. For $h>0,$ 
\begin{equation*}
\Delta Q(\gamma_{t+h})-\Delta Q(\gamma_t)=h \cdot K(t)+O(h^2) \; ,
\end{equation*}
where we (assume) $K(t)$ is continuous in $t.$ For every point $X\in\Gamma^\S,$ and each vector $Z\in {\mathbf R}^N,$ there exists a subprocess $\gamma_t$ of a reversible cyclic process $\gamma$ of $\S,$ such that 
\begin{equation*}
\gamma(t)=X \; ; \dot{\gamma}(t) = c Z \; ,
\end{equation*}
where $c\in {\mathbf R}.$ One can use the functional $\Delta Q(\gamma_t)$ over a reversible subprocess $\gamma_t\in \Gamma^\S$ to define a 1-form $\delta Q(\gamma (t))$ with the property that\footnote{One needs to make these arguments mathematically accurate. For further details and references, see for example [LY].} 
\begin{equation*}
\dot{\gamma}(t)\cdot \delta Q(\gamma(t)) = \lim_{h\searrow 0} \frac{1}{h}(\Delta Q(\gamma_{t+h})-\Delta Q(\gamma_t))=K(t) \; .
\end{equation*} The internal energy $U$ of $\S$ is a state function, ie, a function over $\Gamma^\S.$ Denote by $dU$ the 1-form over $\Gamma^\S$ with components equal to the gradients of $U.$ We define the work 1-form to be 
\begin{equation*}
\delta A := dU -\delta Q \; .
\end{equation*}
Let $X_1,\cdots, X_N$ be the coordinates of $\Gamma^\S. $ Then one can write 
\begin{equation*}
\delta A =\sum_{i=1}^N a_i(X) dX_i \; ,
\end{equation*}
where $a_i(X), i=1,\cdots, N,$ are called the work coefficients. They are intensive quantities, meaning that under rescaling, $a_i(\lambda X)=a_i(X), i=1,\cdots, N.$\footnote{Quantities $\xi$ with the property that under rescaling $\xi(\lambda X)=\lambda \xi(X), \lambda > 0,$ are called extensive, such as internal energy $U$ and heat $Q$, while quantities with the property that $\xi(\lambda X)=\xi(X)$ are called intensive, such as the temperature $T$ and work coefficients.}

Using the fact that 
\begin{equation*}
\oint_{\gamma^{rev}} \frac{\delta Q}{T}=0 , \; \forall \; \gamma^{rev}\subset \Gamma^\S, 
\end{equation*}
and the convexity of $\Gamma^\S,$ one can define a state function $S$, the {\it entropy}, over $\Gamma^\S$ such that 
\begin{equation*}
dS=\frac{\delta Q}{T} \; .
\end{equation*}
This gives the following identity 
\begin{equation*}
dU= TdS + \delta A \; ,
\end{equation*}
for reversible changes of state.

Consider $\gamma: X_i\rightarrow X_f$ a thermodynamic process of an isolated system $\S,$ such that $X_{i,f}\in \Gamma^\S.$ It follows from the definition of entropy and the fact that $\oint_\gamma \frac{\delta Q}{T}\le 0$ that 
\begin{equation*}
S(X_f)\ge S(X_i) \; . 
\end{equation*}
Together with the scaling postulate and the connectivity and convexity of $\Gamma^\S,$ one can show that the entropy $S$ is concave. For $\lambda\in(0,1),$
\begin{equation*}
S(\lambda X_1 + (1-\lambda) X_2) \ge \lambda S(X_1) + (1-\lambda) S(X_2) \; .
\end{equation*}

There are several equivalent formulations of the second law of thermodynamics, which we list here.

\begin{itemize}
\item[(i)]{\it Clausius} (1854). When one connects two thermal reservoirs $\R_1$ and $\R_2$ through a thermal contact, heat flows either from $\R_1$ to $\R_2,$ or from $\R_2$ to $\R_1.$ The opposite direction of the flow is not possible.

\item[(ii)]{\it Carnot} (1824). For a heat engine $\S,$ $\eta^\S\le\eta^{rev}.$

\item[(iii)]{\it Caratheodory} (1873-1950). In an arbitrarily small neighborhood of each equilibrium state $X$ of an isolated system $\S,$ there are equilibrium states $X'$ of $\S$ that are not accessible from $X$ via reversible and adiabatic paths (for a mathematically rigorous discussion, see for example [Boy]).

%with the property that, for $\S$ isolated, there is no thermodynamic process $\gamma$ which connects $X$ and $X'.$

\end{itemize}

It follows that for an isolated thermodynamic system, the entropy of the equilibrium state is maximal (extremal principle for entropy).

%%%%%%%%%%%%%%%%%%%%%%%%%%%%%%%%%%%%%%%%%%%%%%%%%%%%%%%%%%%%%%%%%%%%%%%%%%%
\section{Quantum description of thermodynamic systems, heat baths and processes}

We start with a brief review of the quantum theory of
thermodynamic systems and heat baths which are expounded in
chapter 3, and then discuss different types of thermodynamic
processes. This will clarify certain notions that are needed in the ensuing discussion and will help to fix our notation.

A thermodynamic system $\S$ is a quantum mechanical system
confined to a compact region $\Lambda$ of physical space
${\mathbf R}^3$. The pure states of $\S$
are rays in a separable Hilbert space $\H^\S$, and its mixed
states correspond to density matrices $\rho$, which are positive,
self-adjoint operators that have unit trace and that belong to
$\L^1(\H^\S )$, the two sided ideal of trace-class operators in
the space of bounded operators $\B(\H^\S )$ of $\H^\S$. Since
$\rho$ is positive, $k=\rho^{\frac{1}{2}}$ belongs to $\L^2(\H^\S
)=:\K^\S$, the two sided ideal of Hilbert-Schmidt operators in $\B(\H^\S
)$ which is isomorphic to $\H^\S\otimes\H^\S$. The kinematics of
$\S$ is encoded in a $C^*$-algebra $\O^\S$ such that $\O^\S\subseteq \B
(\H^\S)$. The dynamics is generated by a family of ({\it generally time
dependent}) Hamiltonians, $H^\S (t)=H^\S( \lambda_t)$, where
$\lambda_t$ are time-dependent parameters. The time-dependent Hamiltonians are self-adjoint and semi-bounded
operators acting on $\H^\S$. In the Heisenberg picture, the time
evolution of an operator $a\in\O^\S$ is $\alpha_t^\S (a)=U^\S
(t)^* a U^\S (t)$, where $U^\S (t)$ is a unitary operator that
satisfies $\partial_t U^\S (t)=-iH^\S (t) U^\S (t)$ and $U^\S
(t=0)=1$ (in units where $\hbar=1$). In certain cases, $\a_t^\S$ defines a *-automorphism
on $\O^\S$. In the Schr\"odinger picture, the time-evolution of an
element $k\in \K^\S$ is given by $k_t:=\alpha_{-t}^\S(k)=U^\S (t)
k U^\S (t)^*$. The generator of the dynamics on $\K^\S$ is the
Liouvillean or thermal Hamiltonian $\L^\S=ad_{H^\S}$, which can be
shown to be selfadjoint on a dense core of $\K^\S$ (see chapter 3
for an extensive discussion of Liouvilleans). The time evolution
of an element $k\in \K^\S$ is given by $k_t=\tilde{U}^\S (t)k$,
where $\tilde{U}^\S$ satisfies the equation 
$\partial_t \tilde{U}^\S (t)=-i\L^\S (t)\tilde{U}^\S (t)$ and
$\tilde{U}^\S (t=0)=1$. Since $\L^2(\H^\S )$ is a Hilbert space,
one may study the spectrum of $\L^\S (t)$ using the available
methods of spectral theory, even in the thermodynamic limit
$\Lambda\nearrow \allspace$.

%%%%%%%%%%%%%%%%%%%%%%%%%%%%%%%%%%%%%%%%%%%%%%%%%%%%%%%%%%%%%%%%%
% This limit will be in the sense
%of Fisher, essentially meaning that the ratio between the surface
%and the volume of $\S$ goes to zero in the thermodynamic limit
%(see, for example [Ru1]).
%%%%%%%%%%%%%%%%%%%%%%%%%%%%%%%%%%%%%%%%%%%%%%%%%%%%%%%%%%%%%%%%%

A heat bath, or reservoir, $\R$, is the thermodynamic limit
($\Lambda\nearrow\allspace$) of an increasing family
$\R_\Lambda$ of thermodynamic systems. The kinematics of the
reservoir is encoded in the $C^*$-algebra $\O^\R:= \overline{
\bigvee_{\Lambda\nearrow \allspace} \O^{\R_\Lambda}}$,
where $\overline{( \cdot )}$ denotes the norm closure. When $\R$
is isolated, we assume that its state is a {\it KMS}  state on
$\O^\R$ at some inverse temperature $\beta_\R=(k_BT^\R)^{-1}>0$
and chemical potentials $\bfmu=(\mu_1,\ldots,\mu_m)$, which
correspond to conserved charges $\bfQ=(Q_1,\ldots,Q_m)$ affiliated
to $(\O^\R)'$. The expectation value of an operator $a\in
\bigvee_{\Lambda\nearrow \allspace}\O^{\R_\Lambda}$ at equilibrium
is
\begin{equation*}
\omega^\R_{\beta,\bfmu}(a)=\lim_{\Lambda\nearrow\allspace}
Tr(\cp_{R_\Lambda} a) \; ,
\end{equation*}
where $\cp_{\R_\Lambda}=\frac{1}{Z^{\R_\Lambda}_{\beta,\bfmu
}}e^{-\beta[H^{\R_\Lambda}-\bfmu\cdot\bfQ_\Lambda]}$,
$e^{-\beta[H^{\R_\Lambda}-\bfmu\cdot\bfQ_\Lambda]}$ is assumed to
be trace-class, and
$Z^{\R_\Lambda}_{\beta,\bfmu}=Tr(e^{-\beta[H^{\R_\Lambda}-\bfmu\cdot\bfQ_\Lambda]}).$

The dynamics of the coupled system, $\S\vee\R,$ ({\it before taking the
thermodynamic limit}) is generated  by the Hamiltonian $H(t)
\equiv H^{\S\vee\R}(t):= H^\S(t)+H^\R$, where $H^\S(t)=H_0^\S({\bf
\lambda}_t) + g(t)V^{\S\vee\R}$. If $\cp_0$ is the initial state
of the system at time $t=0$, then the true state of the system at
time $t$ is
$\cp_t=U(t)\cp_0U(t)^*=:\alpha^{\S\vee\R}_{-t}(\cp_0)$; it
satisfies the Liouville equation $\dot{\cp}_t=-i[H(t),\cp_t]$. Let
the reference state at time $t$ be $\cp_t^\beta :=
Z_{\beta,\bfmu}(t)^{-1}e^{-\beta[H(t)-\bfmu\cdot\bfQ^{\S\vee\R}]}$.
A key problem in quantum statistical mechanics
is establishing the existence of the thermodynamic limit of the
above quantities. 

\begin{align}
\rho_t(\cdot ) &=\lim_{TD}Tr(\cp_t \cdot ) \; , \; (real \; state) \\
\omega_t^\beta(\cdot ) &=\lim_{TD}Tr (\cp_t^\beta \cdot ) \; , \; (instantaneous \; equilibrium \; state) \\
\rho_t^\S &:= \rho_t |_{\O^\S\otimes\id^\R}\; , \; (restriction \; to \; small \; system)\; 
\end{align}
and the existence of the thermodynamic limit of the dynamics
$\alpha^g_t$, where
``$\lim_{TD}$'' denotes the thermodynamic limit for $\R$. We will not discuss this problem; for results see [GNS, HHW, AWo,AWy,Rob,BR,Ru1].

Equivalently, one may work directly in the thermodynamic limit
using the GNS construction, as explained in chapter 3.

\section{Thermodynamic processes}

We roughly sketch what we mean by different thermodynamic
processes before considering specific ones later. In the example
of a small system $\S$ coupled to reservoirs, the choice
of $\{H^\S (t) \}$ determines the trajectory of states $\{\rho_t^\S\}$
of $\S$, alternatingly in contact with $0,1,2,\cdots,n$ heat baths.
Isothermal processes correspond to diathermal contacts of $\S$ to
a single heat bath (or equivalently heat baths with the same
temperature). Diathermal contacts preserve all extensive quantities except for the internal energy $U^\S$ of $\S$. Circular
processes, in which $\gamma_{t_i}=\gamma_{t_f}$, correspond to the
case when $H^\S(t+t_*)=H^\S(t)$, for $t_*<\infty$. Adiabatic
processes correspond to $\S$ isolated, and reversible processes to
$\rho_t\simeq\omega_t^\beta$.

\section{Return to equilibrium}

We start with the example of an irreversible isothermal process.
We will show in chapter 3 that the KMS
state at inverse temperature $\beta$ of a quantum mechanical
system, assuming that it exists, corresponds to the simple zero eigenvalue of the {\it standard} Liouvillean $\L$,
while the rest of the latter's spectrum is continuous. Hence, the
study of the property of return to equilibrium is equivalent to
the analysis of the spectrum of $\L$. In the following discussion,
consider a small system $\S$ with finitely many degrees of freedom
coupled to a single reservoir $\R$, and work directly in the
thermodynamic limit.

{\it Return to Equilibrium: Under suitable hypotheses, such as
sufficient smallness of the coupling constant $g$, Fermi's golden
rule, appropriate infra-red behavior of the coupling, and
$\int^\infty dt \|(\L_t-\L_\infty)(\L_\infty+i)^{-1}\| < \infty$, one
may prove the property of return to equilibrium  for a class of
systems $\S\vee\R$: for an initial state of $\rho_0$ normal to $\rho_0^\S\otimes\omega_\beta^\R$,
$\lim_{t\rightarrow\infty}\rho_t(a)=\omega_\beta^\infty (a),\;
\forall a\in \O^{\S\vee\R}$, where $\omega^\infty_\beta$ is the
equilibrium state of the coupled system for $t\rightarrow\infty$,
with convergence that may be exponentially fast in time for suitably chosen operators $a$.}

Using methods from spectral theory, the property of
return to equilibrium has been established for a variety of
quantum mechanical systems: complex deformation techniques for the
spin-boson system [JP1,JP2], Feshbach map and RG for a small
system coupled to a thermal reservoir of photons [BFS], and an
extension of Mourre's positive commutator method together with a
suitable Virial Theorem for a small system coupled to a thermal
bath. [M1,FM1] All these results are based on important
insights of [HHW]. (PC methods used in studying return to equilibrium
have been extended to studying thermal ionization of atoms and
molecules in the radiation field. [FM2,FMS])

{\it Part} of the $0^{th}$-law of thermodynamics asserts that when
a reservoir $\R$, with temperature $T^\R$, is locally perturbed,
it always returns to the same unique equilibrium state after a
long time. If a quantum system $\S\vee\R$ possesses the property
of return to equilibrium, the state of the coupled system will
be an equilibrium state at $T^\R$ as
$t\rightarrow\infty$. Moreover, if the contact between $\R$ and
$\S$ is removed, then $\R$ returns to the same equilibrium state
$\omega_\beta^\R$ after a long time. The remaining part of the
$0^{th}$ law will be discussed later.

Proving the property of RTE for simple, yet physically relevant
models, will be discussed extensively in chapter 5; (see also
chapter 6).

\section{Heat energy and entropy of thermodynamic systems}

Some of the observed thermodynamic quantities during a
thermodynamic process $\gamma$ of a system formed of $\S$
coupled to $n$ reservoirs $\R_i,i=1,\cdots,n$, are the internal
energy $U^\S$ of $\S$ and the heat energy $Q^\S$ transferred from
the heat baths $\R_i$ to $\S$. The $1^{st}$-law of thermodynamics
asserts the existence of functionals
\begin{align*}
U^\Sigma \ : &\Gamma^\Sigma \ni \gamma_. \rightarrow
u^\Sigma(\gamma_. ) \\
\Delta Q_{\R_i}^\Sigma \ : &\gamma \rightarrow \Delta
Q_{\R_i}^\Sigma(\gamma )\ ,
\end{align*}
and relates, for an arbitrary processes $\gamma$,
\begin{equation*}
\Delta A^\Sigma (\gamma) := \Delta U^\Sigma - \Delta Q^\Sigma (\gamma) \ ,
\end{equation*}
where $\Delta A^\Sigma$ is the work performed on $\Sigma$ during
$\gamma$.

Indeed, from the quantum theory of thermodynamic systems and heat
baths, the above quantities have explicit definitions. For
simplicity of exposition, let the reservoirs be finite, and take
the thermodynamic limit of well-defined quantities later. The
internal energy of $\S$ is defined by $U^\S(t):=\rho_t(H^\S(t))$,
and the rate of change of heat energy is given by
\begin{align*}
\frac{\delta Q(t)}{dt}
&:=-\sum_{i=1}^n\frac{d}{dt}\rho(H^{\R_i})=-i\sum_{i=1}^n\rho_t([H(t),H^{\R_i}])
\\
&= i\sum_{i=1}^n\rho_t([H^{\R_i},gV^{\S\vee\R_i}(t)]) =:
\sum_{i=1}^n \frac{\delta Q^{\R_i}(t)}{dt} \; ,
\end{align*}
where $\delta (\cdot)$ denotes the {\it imperfect} or {\it inexact}
differential of $(\cdot)$. It follows that
$\dot{U}^\S(t)-\frac{\delta
Q(t)}{dt}=\rho_t(\dot{H}^\S(t))=:\frac{\delta A(t)}{dt}$. The
thermodynamic limit of the above relationship is well defined.
This is nothing but the expression of the $1^{st}$-law of
thermodynamics.

Define the (relative) entropy of $\S$ as
\begin{align*}
S^\S (t) &:= -k_B \lim_{TD} Tr(\cp_t [ \log \cp_t -\sum_i \log \cp^{\R_i} ] ) \\
         &= -k_B \lim_{TD} Tr(\cp_t [ \log \cp_t + \sum_j \{ {\beta}_i (H^{\R_i} - {\mu}_i \cdot Q^{\R_i} )+ \log Z^{\R_i} \} ]) 
\; .
\end{align*}
The usefulness of this definition will become apparent soon. Since
$Tr(B\log B) \ge Tr(B\log A)+Tr(B-A)$ for $A$ and $B$ positive,
self-adjoint, and bounded operators, the relative entropy of $\S$
has a definite sign for all $t\in \mathbf{R}$,\footnote{Proof of a
more general trace inequality will be given in chapter 7.}

\begin{equation*}
S^\S (t)\le 0 \; .
\end{equation*}
An important property of relative entropy is its strong
subadditivity (see [LR]).

Both $Tr \cp_t\log \cp_t$ and $Tr \cp_t \log Z^{\R_i}$ are
time-independent. Moreover, $Tr \cp_t \mu_i Q^{\R_i}$ is
time-independent for diathermal contacts. Therefore, the rate of
change of entropy is
\begin{align*}
\dot{S}^\S (t) &= -\sum_i \frac{1}{T_i}
\frac{d\rho_t(H^{\R_i})}{dt} \\
               &= \sum_i \frac{1}{T_i}
               \frac{\delta Q^{\R_i}(t)}{dt} \; .
\end{align*}

Note that if the rate of entropy production ${\cal
E}=-\lim_{t\rightarrow\infty}\dot{S^\S}(t)$ exists, then ${\cal
E}\ge 0$ due to the fixed sign of $S^\S(t)$. The last statement
implies the second law of thermodynamics in the formulation of Clausius.

\section{Isothermal processes and the isothermal theorem}

The question we want to address in this section is what
characterizes {\it reversible isothermal processes}. Careful
statement of results in this section together with detailed proofs
are given in chapter 7.

Suppose a system $\S\vee\R$ has the property of return to
equilibrium. What happens if this system is slowly perturbed after
return to equilibrium, for example, by {\it quasi-statically}
removing the contact between $\R$ and $\S$?

Again consider the system $\S\vee\R$ directly in the thermodynamic
limit, such that the standard Liouvillean $\L_g^\tau(t):=\L_g(s)$,
where the rescaled time is $s:=\frac{t}{\tau}$. Assume $\L_g(s)$
have a common dense domain for all $s\in I$, where $I\subset
{\mathbf R}$ is a compact interval. Moreover, assume that for all
$s\in I$, $(\L_g(s)+i)^{-1}$ is differentiable in $s$,
$\L_g(s)\frac{d}{ds}(\L_g(s)+i)^{-1}$ is uniformly bounded,
$\sigma_{pp}(\L_g(s))=\{ 0 \}$ and $\sigma(\L_g(s))\backslash \{0\} =
\sigma_c(\L_g(s))$, and that the projection onto the eigenstate
corresponding to the zero eigenvalue of $\L_g(s)$,
$P(s)=|\Omega_{\beta,g}(s)\rangle\langle\Omega_{\beta,g}(s)|$, is
twice differentiable in $s$ for almost all $s\in I$. Note that
$P(s)$ corresponds to the instantaneous equilibrium state, or
reference state, $\omega_\beta^{\tau s}$ at time $t=\tau s$. We
are interested in the quasi-static limit $\tau\rightarrow\infty$.
Physically, this limit corresponds to $\tau\gg\tau_R$, where
$\tau_R$ is the relaxation time.

{\it (Isothermal Theorem): Under these hypotheses, $\rho_{\tau
s}(a)=\omega^\beta_{\tau s}(a) + o(1)$, as
$\tau\rightarrow\infty$, $\forall a\in\O^\S\otimes\O^\R$ and
$\forall s\in I$, where $I \subset{\mathbf R}$ is an arbitrary
compact interval.}

To get a better estimate on the rate of convergence in the
quasi-static limit, we need further information about the spectrum
of the standard Liouvillian $\L_g(s)$. With additional hypotheses
that allow one to apply complex deformation techniques, the rate
of convergence is shown to be $O(\tau^{-1})$ in chapter
8. These hypotheses can be verified in classes of quantum
mechanical systems, such as those for which the property of return
to equilibrium has been established.

We now sketch several consequences of this theorem which will
clarify the notions of heat energy and reversibilty in isothermal
thermodynamic processes and emphasize the role of relative
entropy. Further details can be found in chapter 7 and in [A-SF3].

Without loss of generality, assume that the small system $\S$ is
coupled through diathermal contacts to a single reservoir $\R$,
which will be treated as finite first before taking the
thermodynamic limit. From the discussion of the previous section,
we know that in an isothermal process
\begin{align*}
\dot{U}^\S(t) &= \frac{\delta Q(t)}{dt}+\frac{\delta A(t)}{dt} \; , \\
\dot{S}^\S(t) &= \frac{1}{T^\R} \frac{\delta Q}{dt}(t) \; .
\end{align*}

Consider an isothermal process of $\S\vee\R$ from $t_0=\tau s_0$
till $t_1=\tau s_1$, for $s_0$ and $s_1$ fixed,
$\tau\rightarrow\infty$, and where the initial state
$\omega^\beta$ is independent of $s$ for $s<s_0$. The following
holds.

\begin{itemize}
\item[(i)] {\it Reversible isothermal processes are equivalent to
``quasi-static'' isothermal processes for $\tau \gg \tau_R$.}

This is a standard {\it assumption} in {\it equilibrium
thermodynamics}. Moreover, the definition of entropy in the
latter setting is
\begin{align*}
S^\S_{rev}(t) &=-\lim_{TD}k_B Tr (\cp_t^\beta[\log \cp_t^\beta - \log \cp^\R]) \\
              &= \lim_{TD}[k_B \omega_\beta (\beta H^\S (t)) +k_B \log \frac{Z_\beta (t)}{Z^\R}] \\
              &= \lim_{TD}[\frac{1}{T^\R}(U^\S_{rev}-F^\S)] \; ,
\end{align*}

where $F^\S(t)=-k_B \log \frac{Z_\beta (t)}{Z^\R}$ is the 
free energy of $\S$, and $k_B$ is Boltzmann's constant . Using
the isothermal theorem, one may replace $\omega_{\tau s}^\beta$ by
$\rho_{\tau s}$ up to an error term that vanishes in the
quasi-static limit, and hence, in the thermodynamic limit,

\begin{align*}
T^\R\Delta S^\S_{rev} &= \omega_{\tau s}^\beta (H^\S(s_1)) - \omega_{\tau s}^\beta (H^\S(s_0)) - \int_{s_0}^{s_1}ds \omega_{\tau s}^{\beta}(\dot{H}^\S(s)) \\
                   &= \Delta U^\S -\Delta A + o(1) \\
                   &= \Delta Q + o(1) \; ,
\end{align*}
where we made use of the isothermal theorem in the second step and
the $1^{st}$-law of thermodynamics in the last step. We have just
sketched the proof of the following claim, which asserts the
equivalence of the definition of entropy in equilibrium
statistical mechanics and relative entropy in non-equilibrium
quantum statistical mechanics, in the quasi-static limit.

\item[(ii)]{\it $\Delta S_{rev}^\S = \Delta S^\S + o(1)$ as
$\tau\rightarrow\infty$.}

\item[(iii)] Furthermore, if one slowly removes the contact between $\R$ and
$\S$, the state of $\S$ will be the {\it Gibbs} state at inverse
temperature $\beta_\R$, independently of the diathermal contacts.

\noindent {\it This is part of the $0^{th}$-law of thermodynamics:
If $H^\S(t)\rightarrow H^\S_\infty \in \O^\S$, i.e.
$g(t)\rightarrow 0$ as $t\rightarrow\infty$, then $\rho_{\tau s}$
tends to the {\it Gibbs} state at temperature $T^\R$ for $H^\S_\infty$ as
$\tau\rightarrow\infty$ and $s\rightarrow\infty$.}
\end{itemize}

\section{Clausius' and Carnot's formulation of the $2^{nd}$-law}

As discussed in the previous section, there are several equivalent
formulations of the $2^{nd}$-law of thermodynamics. A standard
consequence of the $2^{nd}$-law is the existence of the absolute temperature $T_\R>0$ of a reservoir $\R$ and the entropy functional
\begin{equation*}
S_{rev}^\S : \Gamma^\Sigma \ni \gamma_{.} \rightarrow
S_{rev}^\S(\gamma_{.}) \ ,
\end{equation*}
where $\Gamma^\Sigma$ are the
equilibrium states of $\Sigma$ sampled in a reversible
thermodynamic process.

When two or more reservoirs, with an initial state $\rho_0$, are
coupled , the convergence to a non-equilibrium steady state (NESS)
$\rho^{NESS}:=w^*-\lim_{t\rightarrow\infty}\rho_0 \circ
\alpha_t$ (or more weakly $w^*-\lim_{T\rightarrow\infty}\frac{1}{T}\int_0^{T}\rho_0 \circ
\alpha_t dt$) has been proven recently in several examples using
different approaches: [FMUe,Ru2,3] use algebraic scattering theory
where one has to establish the existence of scattering
endomorphisms. The latter is based on the work of [He,Rob]. On the
other hand, [JP3] relates the NESS to the zero energy resonance of
the adjoint of the so called C-Liouvillean. In the latter setting
we prove an adiabatic theorem for states close to non-equilibrium
steady states using a novel adiabatic theorem and complex
deformation techniques in chapter 8.

Consider a thermodynamic system $\S$ coupled to heat baths $\R_1,
\cdots, \R_n$, where $n\ge 2$. We have shown that, for diathermal
contacts,

\begin{align*}
& -\infty< S^\S (t) \le 0 \; , \\
& \dot{S}^\S (t)=\sum_i \frac{1}{T_i}\frac{\delta Q^{\R_i}(t)}{dt}
\; .
\end{align*}

{\it (NESS and Clausius) Assume that $H^\S(t)\rightarrow
H^\S_\infty\in \O^\S\otimes\O^\R$, as $t\rightarrow\infty$. If
$\rho_t\rightarrow_{t\rightarrow\infty}\rho^{NESS}$, then

\begin{align*}
& i) \sum_{i=1}^n \frac{\delta Q^{\R_i}}{dt} \rightarrow 0 \\
& ii) \dot{S}^\S (t) \rightarrow - {\cal E} \le 0 \\
&iii)\lim_{t\rightarrow\infty}\sum_{i=1}^n\frac{1}{T_i}\frac{\delta
Q^{\R_i}}{dt} =-{\cal E} \le 0 \; ,
\end{align*}
where ${\cal E}$ is the entropy production rate. A direct
consequence is the following:

i) and iii) imply Clausius' formulation of the $2^{nd}$-law.}

Now consider a cyclic thermodynamic process such that $H^\S (t+
\tau_*)=H^\S (t)$, where the period $\tau_*<\infty$, and assume,
without loss of generality, that the number of reservoirs is
$n=2$. This is an example of a {\it heat engine} or a {\it heat pump}. Let
$\omega_t^{p}:=\lim_{N\rightarrow\infty}\rho_{t+N\tau_*}$. For
small enough coupling between the two reservoirs, one can prove
using a norm-convergent Dyson-Schwinger series that $\omega_t^{p}$
is periodic in time $t$ with period $\tau_*$, and that the change
in entropy per cycle is $-\int_0^{t_*}dt \dot{S}=-\Delta S \ge
0$.[FMUe,FMSUe] We also prove convergence to a time-periodic state
by relating the latter to the zero-energy resonance of the so
called {\it Floquet Liouvillean} in chapter 9. After one period,
\begin{align*}
& \Delta u^\S =0 \; ,\\
& \frac{\Delta Q^{\R_1}}{T_1}+\frac{\Delta Q^{\R_2}}{T_2} = \Delta
S \le 0 \; .
\end{align*}
If the engine does work, $\Delta A = \Delta Q^{\R_1}+\Delta
Q^{\R_2} \ge 0$. Suppose that $T_1 \ge T_2$, then {\it Clausius}
implies that $\Delta Q^{\R_1} \ge 0$. The following is nothing but
{\it Carnot's} formulation of the $2^{nd}$-law of thermodynamics.

{\it (Carnot)Assume that $T_1>T_2$. Then
\begin{align*}
0 \le \eta^\S &:= \frac{\Delta A}{\Delta Q^{\R_1}} = 1 + \frac{\Delta Q^{\R_2}}{\Delta Q^{\R_1}} \\
              & \le 1 - \frac{T_2}{T_1} := \eta^{rev} \; .
\end{align*}}

It is important to note that this result follows {\it only} from
the sign of relative entropy and the existence of time periodic
states in the large time limit. Moreover, the difference
$\eta^{rev}-\eta^\S$ can be explicitly computed in terms of the
entropy production per cycle (see remark in chapter 9 and [FMUe]).

We conclude by mentioning that, recently, transport phenomena have been rigorously investigated from the point of view of quantum statistical mechanics. Transport phenomena
between two reservoirs formed of free fermions at different
temperatures/chemical potentials and coupled through bounded local
interactions has been studied in [FMUe]. Together with showing the
convergence of the coupled system to a NESS using scattering
theory and establishing strict positivity of entropy production,
they show that the Onsager reciprocity relations hold to first
non-trivial order. Furthermore, [JOP] study linear response theory from the point of view of the algebraic formulation of quantum statistical mechanics and prove the Green-Kubo formula and Onsager reciprocity relations for heat fluxes generated by temperature gradients.

%%%%%%%%%%%%%%%%%%%%%%%%%%%%%%%%%%%%%%%%%%%%%%%%%%%%%%%%%%%
%\include{mathframework}

\chapter{Mathematical framework: algebraic
formulation of NEQSM}
%Mathematical framework: algebraic formulation of quantum statistical mechanics

%%%%%%%%%%%%%%%%%%%%%%%%%%%%%%%%%%%%%%%%%%%%%%%%%%%%%%%%5

For the sake of simplicity, we opt for a concrete discussion of
the algebraic formulation of quantum statistical mechanics, while
deferring a mathematically more elegant, yet more abstract,
discussion of it to Appendix 1.[BR, Sa, DJP]

\section{Quantum description of finite thermodynamic systems}

Consider a quantum system confined to a compact region of physical
space. Its pure states are unit rays in a separable Hilbert space
$\H^\S$, and its mixed states are described by density matrices
$\rho$, which are positive trace-class operators such that $Tr
(\rho )=1$. The kinematical algebra of observables is a
$C^*$-subalgebra $\O^\S \subseteq \B(\H^\S)$, where $\B(\H^\S)$ is
the set of bounded operators on $\H^\S$. The dynamics is generated
by a semi-bounded, self-adjoint operator $H^\S$, the Hamiltonian,
such that the time evolution of an operator $A\in \O^\S$ is given
in the Heisenberg picture by
\begin{equation}
\a^t_\S (A) = e^{itH^\S} A e^{-itH^\S} \; , \label{dym1}
\end{equation}
assuming that $\a^t_\S(A)\in\O^\S$ for every $A\in O^\S$.

One may represent the algebra $\O^\S$ on the space of
Hilbert-Schmidt operators, which is a Hilbert space. (In fact, the
latter is a Hilbert algebra.) Consider the two-sided ideal of
trace-class operators $\L^1 (\H^\S)$, and the two-sided ideal of
Hilbert-Schmidt operators $\L^2 (\H^\S)=\K$. (An operator $k\in\K$
if $Tr (k^*k) < \infty$.) $\K$ is a Hilbert space with scalar
product
\begin{align*}
\langle \cdot | \cdot \rangle : \K\times\K &\rightarrow {\mathbf C} \\
(\sigma , k) &\rightarrow \langle \sigma , k \rangle := \tr
(\sigma^* k) \; .
\end{align*}

Given $\rho \in \L^1 (\H^\S)$ and positive, then $k:= \rho^{1/2}\in \K$.
Suppose the system is initially described by a density matrix
$\rho$. The expectation value of an observable $A\in \O^\S$ is
\begin{equation}
\langle A \rangle_\rho := \tr (\rho A) \; .
\end{equation}

Its expectation value after a time $t$ is given by

\begin{align*}
\langle \a^t_\S (A) \rangle_\rho &= \tr (\rho e^{iH^\S t} A e^{-iH^\S t}) \\
 &= \tr (kk^* e^{iH^\S t} A e^{-iH^\S t}) \\
 &= \tr ((e^{-iH^\S t}ke^{iH^\S t})^* A (e^{-iH^\S t}k e^{iH^\S
 t}))\\
 &= \tr (\a^{-t}_\S (k)^* A \a^{-t}_\S (k)) \; ;
\end{align*}
hence, $k_t := \a^{-t}_\S (k) = e^{-i H^\S t} k e^{i H^\S t}$ in
the Schr\"odinger picture. Let $\L^\S:=ad_H^\S$, which is a
selfadjoint operator defined on $\D$, a dense domain in $\K$. (For
example, $\D=span{|\psi_i\rangle\langle\psi_j |}$, where $\{
\psi_i \}_i$ is the orthonormal basis in $\H^\S$ such that $H^\S
|\psi_i\rangle = E_i |\psi_i\rangle$.) It follows that
\begin{equation}
k_t = e^{-i\L^\S t} k \; .
\end{equation}
The operator $\L^\S$ is called the Liouvillean or thermal
Hamiltonian.

A selfadjoint operator $Q$ is said to be affiliated with the
commutant $(\O^\S)'$ if all the spectral projections of $Q$ belong
to $(\O^\S)'$. Typically, a conserved charge of $\S$ corresponds
to a selfadjoint operator $Q$ affiliated to $(\O^\S)'$, such that
all the spectral projections of $Q$ commute with all the spectral
projections of $H^\S$.

According to the Gibbs Ansatz, the equilibrium state of $\S$ at
inverse temperature $\beta>0$ and chemical potential $\mu\in {\mathbf R}$ is described by the density matrix
\begin{equation}
{\mathsf P}^\S_\beta := \frac{e^{-\beta (H^\S-\mu Q)}}{Z^\S_\beta} \;
,
\end{equation}
where $Z^\S_\beta := Tr (e^{-\beta (H^\S-\mu Q)})$ is a normalization
factor.

The expectation value of an operator $A\in \O^\S$ in equilibrium
is given by

\begin{equation}
\omega^\S_\beta (A):= Tr ({\mathsf P}^\S_\beta A) \; . \label{equil1}
\end{equation}
We list some of the main properties of the equilibrium state.

\begin{itemize}

\item [(i)] {\it Time-translation invariance.}
It follows from equations (\ref{equil1}), (\ref{dym1}), and the
cyclicity of the trace that $\omega^\S_\beta (\a_\S^t
(A))=\omega^\S_\beta(A)$.

\item[(ii)] {\it Kubo-Martin-Schwinger (KMS) condition.}
\begin{equation}
\omega^\S_\beta (A \a_\S^t (B)) = \omega^\S_\beta
(\a_\S^{t-i\beta}(B) A)\; , \label{KMS1}
\end{equation}
for $A,B\in \O^\S$. This follows from (\ref{equil1}), cyclicity of
the trace, and the fact that $Q$ is affiliated with $(\O^\S)'$
such that its spectral projections commute with the spectral
projections of $H^\S$,

\begin{align*}
\omega^\S_\beta (A \a_\S^t (B)) &= Tr ({\mathsf P}^\S_\beta A \a_\S^t (B)) \\
   &= Tr (e^{-\beta (H^\S-Q)}A e^{iH^\S t}Be^{-iH^\S t})/Z^\S_\beta \\
   &= Tr (e^{-\beta (H^\S-Q)}e^{iH^\S t +\beta H^\S}B e^{-i H^\S t -\beta H^\S}A)/Z^\S_\beta \\
   &= Tr ({\mathsf P}^\S_\beta \a_\S^{t-i\beta} (B) A)/Z^\S_\beta \\
   &= \omega^\S_\beta (\a_\S^{t-i\beta}(B)A) \; .
\end{align*}

\end{itemize}

There are two representations of the algebra $\O^\S$ on the
Hilbert space of Hilbert-Schmidt operators $\K$. The left
representation $l[ \O^\S ]$ is defined by
\begin{equation}
l[A] k= Ak\; , \label{lrep1}
\end{equation}
for any $A\in O^\S$ and $k\in\K$. A dual representation which
commutes with $l[\O^\S]$ is given by the right representation
\begin{equation}
r[A]k=kA^* \; , \label{rrep1}
\end{equation}
for any $A\in\O^\S$ and $k\in\K$. Note that $r[AB]=r[A]r[B]$ and
$r[zA]=\overline{z}r[A]$, for $A,B\in\O^\S$ and $z\in {\mathbf
C}$.

In fact, one may show that $\K$ is isomorphic to
$\H^\S\otimes\H^\S$. Introduce the isomorphism
\begin{align*}
I_C: \K &\rightarrow \H^\S\otimes\H^\S  \\
     |\psi\rangle\langle \varphi | &\rightarrow \psi\otimes C^\S\varphi \; ,
\end{align*}
where $\psi ,\varphi \in \H^\S$, and $C^\S$ is an antiunitary
involution on $\H^\S$ (i.e., $(C^\S)^2=1$ and $(C^\S\psi
,C^\S\varphi )= (\varphi ,\psi )$, where $(\cdot , \cdot)$ denotes
the scalar product in $\H^\S$). By looking at $I_C l[A] k$ and
$I_C r[A] k$, for $A\in\O^\S$ and $k\in\K$, one may show that
\begin{align}
l[A]=A\otimes {\mathbf 1} \; , \\
r[A]={\mathbf 1}\otimes C^\S AC^\S \;,
\end{align}
on $\H^\S\otimes\H^\S$. Both $l[\O^\S]$ and $r[\O^\S]$ can be
related to each other, but we will defer this discussion to the
more general case of infinite systems.

\section{Quantum description of a thermal reservoir}

In this section, we will discuss the quantum description of an
infinitely extended thermal reservoir or heat bath $\R$. One may
regard the heat bath or reservoir $\R$ as the limit of a sequence
of thermodynamic systems confined to compact regions of physical
space $\{ \Lambda_i \}_{i=1}^\infty$, such that
$\Lambda_i\subseteq \Lambda_j$ for $i<j$ and
$\lim_{i\rightarrow\infty}\Lambda_i = {\mathbf R}_\pm^3$. Denote by
$\O^{\Lambda_i}$ the kinematical algebra of $\Lambda_i$, then we
will assume that $\O^{\Lambda_i}\subseteq \O^{\Lambda_j}$ if
$i<j$. The kinematical algebra of {\it observables} in the
thermodynamic limit is $\O^\R:= \overline{\bigvee _{i\in {\mathbf
N}}\O^{\Lambda_i}}$, where $\overline{( \cdot )}$ denotes the norm
closure.

We make the following assumptions, which need to be verified in
specific physical models (such as those considered in chapter 4;
see also [BR]), regarding the existence of the time evolution and
equilibrium states in the thermodynamic limit. Let $\O^\infty :=
\bigvee_{i\in {\mathbf N}}\O^{\Lambda_i}$.\footnote{There are several ensembles in statistical mechanics: the microcanonical ensemble, where the number of particles and energy are fixed, the canonical ensemble where the number of particles in the system is fixed while the energy varies, and the macrocanonical ensemble where both the number of particles and the energy are allowed to vary. Although different for finite systems, in the thermodynamic limit the three ensembles are equivalent.}

\begin{itemize}

\item[(A1)] {\it Existence of dynamics.}
We assume that
\begin{equation}
n-\lim_{i} \a_{\Lambda_i}^t (A) = : \a_\R^t (A) \; , \label{dyn2}
\end{equation}
exists for all $A\in \O^\infty$, $t\in {\mathbf R}$, and $\{ \a_\R^t
\}_{t\in {\mathbf R}}$ is a one-parameter group of $*$-
automorphisms on $\O^\R$. Note that $\a_\R^t$ need not be norm
continuous, as in the case of bosonic reservoirs where it is only
$\sigma$-weakly continuous (see Appendix 1).

\item[(A2)]{\it Existence of equilibrium states.}
For $A\in \O^\infty$, consider the sequence of equilibrium
expectation values $\omega^{\Lambda_i}_\beta$ at inverse
temperatures $\beta>0$. We assume existence of the limit of a
suitable (sub)sequence $\omega^{\Lambda_i}_\beta (\cdot )$ as
$i\rightarrow\infty$. The limiting equilibrium state $\omega^\R_\beta$ is
$\a_\R^t$-invariant
\begin{equation}
\omega^\R_\beta(\a_\R^t (A))=\omega^\R_\beta (A) \; ,
\label{timetrans1}
\end{equation}
for $A\in \O^\R$ and $t\in {\mathbf R}$. Moreover, it satisfies
the $KMS$ condition, which says that, for $A,B$ in a suitable
(sub)algebra of $\O^\R$,
\begin{equation}
\omega^\R_\beta(A\a_\R^t (B))=\omega^\R_\beta (\a_\R^{t-i\beta}
(B) A) \; . \label{KMS2}
\end{equation}
We will state the $KMS$ condition more carefully.  Let
\begin{align}
F_{AB}(t)&:= \omega^\R_\beta (A\a_\R^t (B)) \; , \\
G_{AB}(t)&:= \omega^\R_\beta (\a_\R^t (B) A) \; ,
\end{align}
where $A\in \O^\R$ and $B\in O^0$. The $KMS$ condition is
equivalent to saying that $F_{AB}(t)$ is the boundary value of the
function $F_{AB}(z)$, which is analytic in the strip $S_\beta :=
\{ z\in {\mathbf C} : 0< \Im z < \beta \}$, bounded and continuous
on its boundary, with
\begin{equation}
\lim_{\eta\nearrow \beta} F_{AB}(t+i\eta)=G_{AB}(t)\; .
\end{equation}
(Equivalently, $G_{AB}(z)$ is analytic in the strip $S_{-\beta}:=
\{ z\in {\mathbf C} : -\beta < \Im z < 0 \}$, bounded and
continuous on its closure, such that $\lim_{\eta\nearrow\beta}
G_{AB}(t-i\eta)=F_{AB}(t)$.) 
Let $\O^0$ be a
dense subalgebra of $\O^\R$ which is invariant under $\a_\R^t$,
\begin{equation}
\O^0 := \{ A_f =\int dt \a_\R^t (A) f(t) : A\in O^\R {\rm and }
\tilde{f}\in C^\infty_0 ({\mathbf R}) \} \; ,
\end{equation}
where $\tilde{f}$ is the Fourier transform of $f$.
It follows from \fer{timetrans1} and
\fer{KMS2} that
\begin{equation}
\omega^\R_\beta (A^* B)=\omega^\R_\beta (\a_\R^{-i\beta
/2}(B)(\a_\R^{-i\beta /2}(A))^*) \; , \label{TKMS}
\end{equation}
for $A,B\in \O^0$.
Using \fer{KMS2} and the Cauchy-Schwartz inequality, one may show
that
\begin{equation}
{\mathcal N}:= \{ A\in \O^\R : \omega^\R_\beta (A^* A) = 0\}
\label{2sidedideal1}
\end{equation}
is a two-sided ideal of $\O^\R$. In particular, if $\O^\R$ is
simple (i.e., if $\O^\R$ does not contain any two-sided ideal
except $\{ 0 \}$ and itself), then $\omega^\R_\beta (A^*A)=0
\Rightarrow A=0$.
\end{itemize}

In order to do computations and to prove theorems in a concrete
setting, it is useful to have a representation of $\O^\R$ on a
certain Hilbert space. This is provided by the $GNS$
(Gel'fand-Naimark-Segal) construction.

Assume there exists a countable subspace $\tilde{\O}\subset\O^\R$
such that, $\forall A\in \O^\R$, there exists a sequence $\{ A_i
\}_{i\in {\mathbf N}}\subset \tilde{\O}$ with the property that
\begin{equation}
\lim_{i\rightarrow\infty} \omega^\R_\beta ((A-A_i)^*(A-A_i)) = 0 \; .
\label{dense1}
\end{equation}
\vspace{1cm}

\noindent {\it GNS.} The $GNS$ construction associates to the data
$(\O^\R, \a_\R^t, \omega^\R_\beta)$ a Hilbert space $\H_\beta$, a
representation $\pi_\beta$ of $\O^\R$ on $\H_\beta$, a vector
$\Omega_\beta\in \H_\beta$, which is cyclic for
$\pi_\beta[\O^\R]$, and a continuous one-parameter group of
unitary operators $\{ e^{-i\L t} \}_{t\in {\mathbf R}}$, where
$\L$ is selfadjoint on $\H_\beta$, such that, for all $A\in
\O^\R$,

\begin{align}
\omega^\R_\beta (A) & = \langle \Omega_\beta | \pi_\beta (A) \Omega_\beta \rangle \; ; \\
\pi_\beta [\a_\R^t (A)] &= e^{i\L t} \pi_\beta [A] e^{-i \L t} \; ; \\
\L\Omega_\beta &= 0 \; .
\end{align}

To construct $\H_\beta$, $\Omega_\beta$, $\pi_\beta$ and $\L$,
let, for all $A\in \O^\R$, $[A]:= A \; mod {\mathcal N}$, where the
two-sided ideal $\mathcal{N}$ has been defined in
\fer{2sidedideal1}. Define on the linear quotient space
$\tilde{\H}:= \O^\R/{\mathcal N}$ the scalar product

\begin{equation}
\langle [A] | [B] \rangle := \omega^\R_\beta (A^*B) \; .
\label{scalar1}
\end{equation}

The Hilbert space $\H_\beta$ is the completion (the closure in the
norm induced by the scalar product defined in \fer{scalar1}) of
$\tilde{\H}$. Note that since $\tilde{\O}$ is countable,
$\H_\beta$ is separable. Let $\Omega_\beta := [1]$, and define
$\pi_\beta : \O^\R\rightarrow \B(\H_\beta)$ by
\begin{equation}
\pi_\beta(A)[B]:=[AB] \; ,
\end{equation}
which extends by continuity to $\H_\beta$. The one-parameter
unitary group on $\H_\beta$ is defined by
\begin{equation}
e^{-i\L t}[A] :=[\a_\R^t(A)] \; ,
\end{equation}
for $A\in \O^\R$. Unitarity follows from the fact that
$\omega^\R_\beta$ is invariant under $\a_\R^t$, and, by Stone's
theorem, the generator of the dynamics is selfadjoint because
$e^{-i\L t}$ is strongly continuous on a separable Hilbert space.

\vspace{1cm}

\noindent {\it Modular operator and modular conjugation.} Assume
that $\Omega_\beta$ is separating for $\pi_\beta (\O^\R)$ (i.e.,
$\pi_\beta (A)\Omega_\beta = 0 \Rightarrow \pi_\beta (A) =0$), and
define the unbounded antilinear operator ${\mathcal S}$ on
$\pi_\beta(\O^0)$ such that
\begin{equation}
{\mathcal S}(\pi_\beta (A) \Omega_\beta) := \pi_\beta (A^*)
\Omega_\beta \; ,
\end{equation}
for $A\in \O^0$. We shall call this operator the modular operator,
which is well-defined since $\Omega_\beta$ is cyclic and
separating for $\pi_\beta(\O^0)$.

Moreover, define the modular conjugation operator $J$ on
$\pi_\beta(\O^0)$ such that
\begin{equation}
J(\pi_\beta (A)\Omega_\beta) := {\mathcal S} \pi_\beta
(\a_\R^{-i\beta/2}(A))\Omega_\beta = \pi_\beta
(\a_\R^{i\beta/2}(A^*))\Omega_\beta \; ,
\end{equation}
for $A\in \O^0$. Some of the properties of the modular conjugation
are the following.

\begin{itemize}

\item[(i)] $J e^{-i\L t} = e^{-i \L t}J$.
Since $\O^0$ is dense in $\O^\R$, and $\Omega_\beta$ is cyclic, it
suffices to show that $J e^{-i \L t}\pi_\beta (A) \Omega_\beta =
e^{-i\L t}J \pi_\beta (A)\Omega_\beta$, for $A\in \O^0$.
\begin{align*}
J e^{-i \L t} \pi_\beta (A)\Omega_\beta
&= J\pi_\beta (\a_\R^t (A))\Omega_\beta \\
&= {\mathcal S}\pi_\beta (\a_\R^{t-i\beta /2}(A))\Omega_\beta \\
&= \pi_\beta (\a^{t+i\beta/2}(A^*)) \Omega_\beta \\
&= e^{-i\L t}\pi_\beta (\a_\R^{i\beta/2}(A^*))\Omega_\beta \\
&= e^{-i \L t}J \pi_\beta (A) \Omega_\beta \; .
\end{align*}

\item[(ii)] $\langle J\pi_\beta (A) \Omega_\beta | J \pi_\beta (B) \Omega_\beta\rangle
= \langle \pi_\beta (B)\Omega_\beta | \pi_\beta
(A)\Omega_\beta\rangle$ , for $A,B\in \O^0$.

%%%%%%%%%%%%%%%%%%%%%%%%%%%%%%%%%%%%%%%
%$$\langle J \pi_\beta ( A ) \Omega_\beta | J \pi_\beta ( B ) \Omega_\beta \ranlge 
% = \langle \a_\R^{\frac{i\beta}{2}} (A^*) \Omega_\beta | \a_\R^{\frac{i\beta}{2}} (B^*) \Omega_\beta \ranlge $$
%$$= \omega^\R_\beta (\a_\R^{-\frac{i\beta}{2}} (A) \a_\R^{\frac{i\beta}{2}} (B^*)) $$
%$$= \omega^\R_\beta (\a_\R^{-i\frac{i\beta}{2}}(B^*)\a_\R^{-i\frac{i\beta}{2}}(A)) $$
%$$= \omega^\R_\beta (B^* A) = \langle \pi_\beta (B) \Omega_\beta | \pi_\beta (A) \Omega_\beta\rangle .$$
%%%%%%%%%%%%%%%%%%%%%%%%%%%%%%%%%%%%%%%

\item[(iii)] The dual representation defined by $\pi_\beta^{\#}=J\pi_\beta J$ commutes with $\pi_\beta$
(i.e., $\pi_\beta^\# (\O^\R)\subset \pi_\beta (\O^\R)'$). It is
enough to look at $\pi_\beta^\# (A) \pi_\beta (B) \pi_\beta (C)
\Omega_\beta$, for $A,B,C\in \O^0$.
\begin{align*}
\pi_\beta^{\#} (A) \pi_\beta (B) \pi_\beta (C) \Omega_\beta &=
J\pi_\beta (A) J \pi_\beta (BC) \Omega_\beta \\
&= J\pi_\beta (A)\pi_\beta (\a_\R^{i\beta/2}(C^*B^*))\Omega_\beta
\\
&=\pi_\beta(\a_\R^{i\beta/2}(\a_\R^{-i\beta/2}(BC)A^*))\Omega_\beta
\\
&= \pi_\beta (BC\a^{i\beta/2}(A^*))\Omega_\beta \\
&=\pi_\beta(B)J\pi_\beta(A)\pi_\beta(\a_\R^{i\beta/2}(C^*))\Omega_\beta
\\
&= \pi_\beta(B) \pi_\beta^\# (A) \pi_\beta (C) \Omega_\beta \; .
\end{align*}

The claim follows by continuity. Note that, with our assumptions,
one may show that $(\pi_\beta (\O^\R))''=(\pi_\beta^\#
(\O^\R ))'$ (see for example [HHW]).
\item[(iv)] Note that
$$J={\mathcal S}e^{\beta \L/2}=e^{-\beta
\L/2}{\mathcal S} .$$ The polar decomposition of ${\mathcal S}$ is
$${\mathcal S}=Je^{-\beta \L/2}=e^{\beta\L/2}J \; ,$$
$|{\mathcal S}|=e^{-\beta \L/2}$ and $J={\mathcal S}|{\mathcal
S}|^{-1}$.
\item[(v)] $$e^{it\L} \pi_\beta^{\#} (A) e^{-it\L}=\pi_\beta^{\#} (\a^t_\R (A)) \; ,$$
where $A\in \O^\R$. This follows from the definition of
$\pi_\beta{^\#}$ and (i).
\item[(vi)] Suppose that $\psi$ is an eigenvector of $\L$ with eigenvalue $\lambda$,
and that $J\psi=\psi$. Then $\lambda =0$.
$$\lambda\psi = \L \psi = \L J\psi =-J\L\psi=-J\lambda\psi=-\overline{\lambda}\psi\; $$
and hence $\lambda+\overline{\lambda}=0$. However, $\lambda\in
{\mathbf R}$ since $\L$ is selfadjoint, and therefore $\lambda=0$.
\item[(vii)] It follows from the definition of $J$ and the fact that $\Omega_\beta=[1]$ that $J\Omega_\beta=\Omega_\beta$.

\end{itemize}

\section{Return to equilibrium}

Part of the zeroth law of thermodynamics is that equilibrium
states of reservoirs are stable under local perturbations, and
that the system returns to equilibrium if the perturbation is such
that it satisfies some conditions (see chapters 5
and 6).

In the thermodynamic limit, the KMS state at inverse temperature
$\beta$ of a quantum mechanical system, assuming that it exists,
corresponds to the simple zero eigenvalue of the Liouvillean $\L$,
while the rest of the latter's spectrum is continuous (see, for
example [JP2]).\footnote{The Liouvillean $\L$ depends on the inverse temperature $\beta$, but we suppress this dependence where there is no confusion.}

\vspace{0.5cm}

\noindent {\it Return to equilibrium} 

(i) Consider a state $\omega$ normal to the equilibrium state $\omega_\beta$ at inverse temperature $\beta$, and suppose that zero is a simple
eigenvalue of the Liouvillian $\L$, such that the spectrum of $\L$
away from zero is continuous. Then the system possesses the
property of return to equilibrium in the ergodic sense,
\begin{equation}
\lim_{T\rightarrow\infty} \frac{1}{T} \int_0^T dt \omega (\a^t
(A))=\omega_\beta (A) \; ,
\end{equation}
for $A\in\O$.

(ii) Moreover, if the spectrum of $\L$ away from zero is
absolutely continuous, then the system possesses the property of
return to equilibrium in the mixing sense

\begin{equation}
\lim_{t\rightarrow\infty} \omega (\a^t (A)) = \omega_\beta (A) \;
,
\end{equation}
for $A\in \O$.

{\it Proof.} To a state $\omega$ normal to $\omega_\beta,$ one associates a density matrix $\rho=\sum_n p_n |\psi_n \rangle\langle \psi_n |,$ such that $\sum_n p_n = 1$ and $\psi_n\in\H_\beta.$
Since any vector $\psi_n\in\H_\beta$ can be
approximated by a sequence of vectors $\psi_n^m=\pi_\beta
(A_n^m)\Omega_\beta$, $A_n^m\in\O^0$, it is enough to prove
\begin{equation}
w-\lim_{T\rightarrow\infty}\frac{1}{T}\int_0^T \langle \pi_\beta
(B)\Omega_\beta | \pi_\beta (\a^t (A)) \pi_\beta (C) \Omega_\beta
\rangle = \omega_\beta (B^*C)\omega_\beta (A) \; ,
\end{equation}
for case (i), and
\begin{equation}
w-\lim_{t\rightarrow\infty}\langle \pi_\beta (B) \Omega_\beta |
\pi_\beta (\a^t (A)) \pi_\beta (C) \Omega_\beta\rangle =
\omega_\beta (B^*C)\omega_\beta (A)\; ,
\end{equation}
for case (ii).

For the first case (i), we use the fact that
\begin{equation}
w-\lim_{T\rightarrow\infty}\frac{1}{T}\int_0^T dt e^{\pm i\L
t}=|\Omega_\beta\rangle\langle \Omega_\beta | \; ,
\end{equation}
and the $KMS$ condition,
\begin{align*}
&\lim_{T\rightarrow\infty}\frac{1}{T}\int_0^T dt \langle \pi_\beta
(B) \Omega_\beta |\pi_\beta (\a^t (A)) \pi_\beta (C) \Omega_\beta
\rangle \\
&= \lim_{T\rightarrow\infty} \frac{1}{T} \int_0^T dt \langle \Omega_\beta | \pi_\beta (B^*) \pi_\beta (\a^t (A)) \pi_\beta (C) |\Omega_\beta \rangle \\
&= \lim_{T\rightarrow\infty}\frac{1}{T}\int_0^T dt \langle\Omega_\beta | \pi_\beta (\a^{-i\beta}(C)) \pi_\beta (B^*) e^{i\L t}\pi_\beta (A) | \Omega_\beta\rangle \\
&= \langle\Omega_\beta | \pi_\beta (\a^{-i\beta}(C))\pi_\beta (B) | \Omega_\beta \rangle \langle\Omega_\beta | \pi_\beta (A) \Omega_\beta \rangle \\
&= \omega_\beta (B^*C)\omega_\beta (A) .
\end{align*}

Case (ii) similarly follows using
\begin{equation}
w-\lim_{t\rightarrow\infty}e^{\pm i\L t}=|\Omega_\beta
\rangle\langle \Omega_\beta | \; ,
\end{equation}
and the $KMS$ condition.

$\Box$

We have reduced proving the property of return to equilibrium to
showing that zero is a simple eigenvalue of the standard
Liouvillian $\L$, while the rest of the spectrum away from zero is
continuous. Since $\H_\beta$ is a Hilbert space, we can make use
of the available spectral methods to prove RTE.

\section{Perturbation of the equilibrium state}

In this section, we discuss perturbation of the KMS equilibrium
state. To simplify the discussion, we will consider bounded
perturbations as in [Ar1]. For the unbounded case, look at the
appendix, where the proof is based on the perturbation of
$W^*$-dynamical systems, as discussed in [DJP].

Suppose the Hamiltonian of the system is perturbed by a local
bounded perturbation $V\in\O$. The perturbed dynamics is given by
the Dyson series expansion
\begin{equation}
\a^t_V (A):= \sum_{n=0}^\infty i^n \int_0^t dt_1 \int_0^{t_1}dt_2
\cdots \int_0^{t_{n-1}} dt_n [\a^{t_n} (V), \cdots
[\a^{t_1}(V),\a^t (A)]\cdots] \; ,
\end{equation}
for $A\in\O$. This defines a one-parameter group $\{ \a^t_V
\}_{t\in {\mathbf R}}$ on $\O$, and $\a^t_V$ can be unitarily
implemented on the GNS representation
\begin{equation}
\pi_\beta[\a^t_V (A)] = e^{it (\L+\pi_\beta (V))} \pi_\beta (A)
e^{-i t(\L + \pi_\beta (V))} \; .
\end{equation}
Note that one may add to $(\L + \pi_\beta (V)$ an element $W\in
\pi_\beta(\O)^\#$. Furthermore,
\begin{align*}
\pi_\beta^\# [\a^t_V (A)] &= J\pi_\beta (\a^t_V (A)) J \\
&= J e^{i t (\L +\pi_\beta (V) + W)} J \pi_\beta^\# (A) J e^{-it(\L +\pi_\beta(V)+W)}J \\
&= e^{it(\L -\pi_\beta^\# (V)-JWJ)}\pi_\beta^\# (A)
e^{-it(\L-\pi_\beta^\# (V)-JWJ)}\; .
\end{align*}
Imposing that the left representation and its dual have the same
generator of dynamics gives
\begin{equation*}
\L_V=\L+\pi_\beta(V)-\pi_\beta^\# (V) + Z \; ,
\end{equation*}
where $Z\in (\pi_\beta (\O)'\cap\pi_\beta^\# (\O)')$. Without loss
of generality, set $Z=0$. The {\it standard} Liouvillian is hence
given by
\begin{equation}
\L_V=\L+\pi_\beta (V)-\pi_\beta^\# (V) \; .
\end{equation}
Note that $J\L_V=-\L_V J$. One may also find an expression for the
perturbed $(\a^t_V, \beta)$-KMS state, noting that $\Omega_\beta$
is in the domain of the unbounded operators $e^{-\beta
(\L+\pi_\beta (V))/2}$ and $e^{\beta (\L+\pi_\beta (V))/2}$.

The expectation value of an operator $A\in \O$ in the perturbed
equilibrium state is
\begin{equation}
\omega_\beta^V (A) := \langle \Omega^V_\beta | \pi_\beta (A)
\Omega_\beta^V\rangle \; ,
\end{equation}
where
\begin{align*}
\Omega_\beta^V &:= (Z_\beta^V)^{-\frac{1}{2}}e^{-\beta (\L +\pi_\beta (V))/2}\Omega_\beta \\
&= (Z_\beta^V)^{-\frac{1}{2}}\sum_{n=0}^\infty
\int_0^{\beta/2}d\tau_1
\int_0^{\tau_1}d\tau_2\cdots\int_0^{\tau_{n-1}}d\tau_n \pi_\beta
(\a^{\tau_n}(V))\cdots\pi_\beta (\a^{\tau_1}(V))\Omega_\beta \; ,
\end{align*}
after Wick rotating ($it\rightarrow\tau$), and $Z_\beta^V :=
\omega_\beta (e^{-\beta (\L + \pi_\beta (V))})$ is a normalization
factor so that $\langle \Omega_\beta^V | \Omega_\beta^V\rangle =
1$. Note that we can also write
\begin{eqnarray}
\Omega_\beta^V=(Z_\beta^V)^{-\frac{1}{2}}e^{\beta (\L-\pi_\beta^\# (V))/2}\Omega_\beta \; ; \\
J\Omega_\beta^V = \Omega_\beta^V \; , \\
\rm{and} \; \L_V\Omega_\beta^V = 0.
\end{eqnarray}

\section{Instantaneous equilibrium states}

Now suppose the perturbation $V=V(t)\in\O$ is time-dependent. We
define the {\it instantaneous} equilibrium states
$\omega_{\beta,t}^V$ by
\begin{equation}
\omega_{\beta,t}^V (\cdot ) := \langle \Omega_\beta^V (t) |
\pi_\beta (\cdot) \Omega_\beta^V (t) \rangle \; ,
\end{equation}
where
\begin{equation}
\Omega_\beta^V (t) := (Z_{\beta,t}^V)^{-\frac{1}{2}} e^{-\beta (\L
+ \pi_\beta (V(t)))/2}\Omega_\beta \; ,
\end{equation}
and
\begin{equation}
Z_{\beta,t}^V := \omega_\beta (e^{-\beta (\L +\pi_\beta (V(t)))})
\; .
\end{equation}
Note that, for each $t\in {\mathbf R}$,
$$ \L_V (t) \Omega_{\beta,t}^V = 0 \; , $$
and
$$J\Omega_{\beta,t}^V = \Omega_{\beta,t}^V .$$

%\end{itemize}

These states will be useful in characterizing {\it reversible}
isothermal processes (see chapter 7).

%%%%%%%%%%%%%%%%%%%%%%%%%%%%%%%%%%%%%%%%%%%%%%%%%%%%%%%%%%%%%%%555

\section{Appendix 1: Basics of operator algebras}

\subsection{Banach algebras}

Consider an associative algebra $\U$ over a field ${\mathcal F}$
(which might be ${\mathbf R}$ or ${\mathbf C}$), with a norm $||
\cdot ||$ such that $\U$ is a Banach space. The algebra $\U$ is
called a Banach algebra if $||AB||\le ||A|| \; ||B||$, for any
$A,B\in\U$. We say $\U$ is unital if it contains the unit element
${\mathbf 1}$. An example of a Banach algebra is the space of
bounded operators on a Banach space with the operator norm.

The spectrum of $A\in\U$ is $\sigma (A):= \{ \lambda\in {\mathbf
C} : A-\lambda {\mathbf 1}$ is not invertible in $\U \}$, and its
spectral radius is $r(A):=\sup \{ |\lambda | :
\lambda\in \sigma(A)\} =
\lim_{n\rightarrow\infty}||A^n||^{\frac{1}{n}}=\inf_{n}||A^n||^{\frac{1}{n}}$.

In order to discuss $C^*$-algebras, we need to introduce the
notion of an adjoint operator. An adjoint operator in an algebra
over ${\mathbf C}$ is an anti-linear map $A \rightarrow A^*$, such
that $(A^*)^*=A$ and $(AB)^*=B^*A^*$, for any $A,B\in\U$. A
$C^*$-algebra $\U$ is a Banach algebra over ${\mathbf C}$ such
that $\| A^* A \| = \| A \|^2$. In particular, $\|A^*\|=\|A\|$. If
$\U$ is a $C^*$-algebra without the unit element, one can extend
the algebraic structure by adjoining the unit element, such that
the norm on the extended algebra $\U_{un}:={\mathbf C1}\oplus\U$
is given by $\| \alpha \unit + A \| := \sup_{B\in\U \; ; \| B \| =
1} \|\alpha B +AB \|$, where $\alpha\in {\mathbf C}$ and $A\in\U$.
One may also show that a $C^*$-algebra $\U$ is isomorphic to an
algebra $\O$ of bounded operators on a complex Hilbert space $\H$,
such that $\O$ is selfadjoint ($\O^*=\O$) and closed in the
operator norm topology. For all practical purposes, we will
consider the latter to be the definition of a $C^*$-algebra.

\subsection{Positive elements}

An element $A$ of a $C^*$-algebra $\O$ is said to be positive
($A\ge 0$) if $A=A^*$ and $\sigma (A) \ge 0$, or equivalently,
$A=B^*B$ for some $B\in\O$.

\subsection{States}

A continuous linear functional $\rho$ on a $C^*$-algebra $\O$ is a
state if $\rho\ge 0$ (ie, for positive $A\in\O$, $\rho (A)\ge 0$),
and $||\rho ||=1$. If $\O$ is unital and $\rho\ge 0$, then $||
\rho ||=1$ iff $\rho(\unit )=1$. A state on a $C^*$-algebra $\O$
has a unique extension to a state on $\O_{un}$. The set $E(\O)$ of
all states on $\O$ is a convex subset of the dual of $\O$. In
particular, if $\O$ is unital, $E(\O)$ is compact in the
$w^*$-topology of the dual.

\subsection{Representations}

A representation of a $C^*$-algebra $\O$ is a pair $(\H,\pi)$,
where $\H$ is a complex Hilbert space and $\pi$ is a morphism of
$\O$ to the $C^*$-algebra $\B(\H)$ of bounded operators on $\H$.
The representation $\pi$ is said to be faithful if, for $A\in\O$,
$\pi(A)=0\Rightarrow A=0$.

A cyclic representation of $\O$ is a triple $(\H,\pi,\Omega)$,
where $\Omega\in\H$ such that $|| \Omega ||=1$ and $\pi(\O
)\Omega$ is dense in $\H$.

\subsection{Groups of automorphisms}

A morphism $g:\O\rightarrow\O$ that has an inverse $g^{-1}$ is
called an automorphism. We say that a state $\rho$ is invariant
under a group of automorphisms $G$ if $\rho (gA)=\rho (A)\forall
g\in G$ and $A\in\O$.

\subsection{GNS construction}

If $(\H,\pi,\Omega)$ is a cyclic representation of a $C^*$-algebra
$\O$, then $A\rightarrow \rho (A):= \langle\Omega |
\pi(A)\Omega\rangle$ defines a state on $\O$. The converse is also
true, and it is known as the GNS construction.

Let $G$ be a group of automorphisms of the $C^*$-algebra $\O$ and
$\rho$ a corresponding $G$-invariant state on $\O$. Then there is
a cyclic representation $(\H_\rho,\pi_\rho,\Omega_\rho)$ of $\O$
such that
\begin{equation}
\pi_\rho (gA) = U_\rho (g) \pi_\rho (A) U_\rho(g)^{-1} \; ,
U_\rho(g)\Omega_\rho=\Omega_\rho \; ,
\end{equation}
for all $g\in G$ and $A\in\O$. The data
$(\H_\rho,\pi_\rho,\Omega_\rho)$ are unique up to unitary
equivalence.

{\it Proof.} Assume $\O$ is unital (adjoin the unit element if
necessary), and let $\N=\{ A\in \O : \rho (A^*A) = 0 \}$, and $[
\cdot ]:\O\rightarrow\O/\N$, the quotient map. Using the
Cauchy-Schwarz inequality, one may show that $\N$ is a two-sided
ideal in $\O$. Define the scalar product over $\O/\N$ by
$$ \langle [A] | [B] \rangle = \rho(A^*B).$$ The Hilbert space $\H_\rho$ is the completion of $\O/\N$ with respect to this scalar product. Moreover,
\begin{align*}
\pi_\rho (A) [B] &:= [AB] \; ; \\
\Omega_\rho &:= [1] \; ;\\
U_\rho (g) [A] &:= [g A] \; ,
\end{align*}
for $g\in G$ and $A,B\in\O$. $\Box$

\subsection{Pure and ergodic states}

Let $E_G$ be the set of states which are $G$-invariant. It is a
convex set. The extremal points of this set are the so-called
$G$-ergodic states. In particular, if $G$ is reduced to the
identity on $\O$, $E_G$ reduces to $E$, and its extremal points
are pure states. Moreover, $\rho$ is ergodic iff the only bounded operators on
$\H_\rho$ commuting with $U_\rho (G)$ are multiples of the identity, ie, $\pi_\rho (\O)\vee U_\rho (G)$ is irreducible.

\subsection{von Neumann algebras}

\subsubsection{Commutant}

Let $\B(\H)$ be the algebra of all bounded operators on a complex
Hilbert space $\H$ and $\mathbf{1}$ the identity operator on $\H$.
The commutant of a set $\U\subset\B(\H)$ is $\U':= \{ A\in\B(\H) :
B\in \U \Rightarrow [A,B]=0\}$. If the commutant $\U'$ consists of
multiples of the identity, then $\U$ is irreducible. Let $\U'' :=
(\U')'$, the double commutant of $\U$.

A selfadjoint algebra $\M$ of $\B(\H)$ is a von Neumann algebra if
it satisfies one of the following {\it equivalent} conditions:
\begin{itemize}
\item[(1)] $\unit\in\M$ and $\M$ is closed in the strong operator topology.
\item[(2)] $\unit\in\M$ and $\M$ is closed in the weak operator topology.
\item[(3)] $\M=\M''$.
\end{itemize}
For all practical purposes, it is useful to think of a von
Neumann algebra $\M$ as a unital $C^*$-algebra on a Hilbert space.
Let $\U$ be a selfadjoint subset of $\B(\H)$. Then $\U'$ is a von
Neumann algebra, and the double commutant $\U''$ is the smallest
von Neumann algebra containing $\U$.

A von Neumann algebra $\M$ is called a factor if
$\M\cap\M'=\lambda\unit$, where $\lambda\in {\mathbf C}$. Factors
are classified into three main types (I,II,III). We will not go
into the discussion of the classification of factors, because
it is beyond this review appendix.

(A von Neumann algebra is called a {\it concrete} $W^*$-algebra.
An equivalent, yet abstract, definition of a $W^*$-algebra is
given by [Sa], whereby a $W^*$-algebra is a $C^*$-algebra with a
predual (see following subsection).)

\subsection{Predual and normal states}

Let $\M$ be a von Neumann algebra on a Hilbert space $\H$. Linear
functionals $\omega$ on $\M$ of the form $A\rightarrow \omega
(A)=\sum_{n}(\psi_n, A \varphi_n)$, where $\sum_n || \psi_n ||^2
<\infty$, $\sum_n || \varphi_n ||^2 <\infty$, form a closed
subspace $\M_*$ of the Banach dual $\M^*$ of $\M$. $\M_*$ is
called the predual of $\M$. The dual of $\M_*$ is $\M$ in the
duality
$$ (A,\omega)\in \M\times\M_* \rightarrow \omega(A)\; .$$

The predual of $\B(\H)$ can be identified with the Banach space
$\L^1(\H)$ of trace-class operators on $\H$ using the duality
$$(A,\rho)\in\B(\H)\times\L^1(\H)\rightarrow Tr(\rho A)\; .$$
States in the predual are called normal: A state $\omega$ on $\M$
is normal iff there is a density matrix $\rho$ (a positive
trace-class operator with unit trace) such that $\omega (\cdot
)=Tr_{\H} (\rho \cdot)$.

\subsubsection{Cyclic and separating vectors} Let $\M$ be a von
Neumann algebra on a Hilbert space $\H$. A vector $\Omega\in\H$ is
cyclic if $\M\Omega$ is dense in $\H$. One says that $\Omega$ is
separating if $A\Omega=0\Rightarrow A=0$. Note that $\Omega$
cyclic for $\M$ is equivalent to $\Omega$ separating for $\M'$.

\vspace{0.5cm}

Suppose that $(\H_\omega, \pi_\omega,\Omega_\omega)$ is the GNS
construction associated with a von Neumann algebra $\M$ and a state $\omega$. Then
$\pi_\omega (\M)$ is again a von Neumann algebra. Moreover, if
$\omega$ is faithful (ie, $\omega (A^*A)=0\Rightarrow A=0$, for
$A\in\M$), then $\Omega_\omega$ is separating for $\pi_\omega
(\M)$. In particular, $\pi_\omega$ is an isomorphism.
$W^*$-algebras are $*$-isomorphism classes of von Neumann
algebras.

\subsection{Tomita-Takesaki theory}

Consider the von Neumann algebra $\M$, with cyclic and separating
vector $\Omega$ on $\M$. Since $\M''=\M$, it follows that $\Omega$
is also cyclic and separating for $\M'$. Define the operators $S$
and $S'$ by
\begin{align*}
S A \Omega &:= A^* \Omega \; , A\in\M \; ; \\
S'B\Omega &:= B^*\Omega \; , B\in\M' \; .
\end{align*}

Both $S$ and $S'$ are closable, and denote their closures by the
same symbols. One can show that there is a unique positive
operator, the modular operator $\Delta$, and a unique anti-linear
operator, the modular conjugation operator $J$, such that
$S=J\Delta^{1/2}$ and $S'=J\Delta^{-1/2}$.

The Tomita-Takesaki theorem says that
$$J\M J=\M'\; , \Delta^{it} \M \Delta^{-it}=\M \; ,$$
for $t\in {\mathbf R}$.

\subsection{Modular automorphism group and the modular condition}

Let $\omega$ be a faithful normal state on the von Neumann algebra
$\M$, and $(\H_\omega,\pi_\omega,\Omega_\omega)$ the corresponding
GNS representation. Moreover, let $\Delta$ be the modular operator
associated with the pair $(\pi_\omega, \Omega_\omega)$. (von
Neumann algebras with a faithful state are called
$\sigma$-finite.) One may show that there is a $\sigma$-weakly
continuous one-parameter group $t\rightarrow\sigma^\omega_t$ of
$*$-automorphisms of $\M$ defined by $$\sigma_t^\omega (A) :=
\pi_\omega^{-1} (\Delta^{it}\pi_\omega (A) \Delta^{-it}).$$ This
is the modular automorphism group associated with $(\M,\omega)$.
The modular condition is
\begin{align*}
\langle \Delta^{1/2}\pi_\omega (A) \Omega_\omega | \Delta^{1/2} \pi_\omega (B)\Omega_\omega \rangle &= \langle J\pi_\omega (A^*)\Omega_\omega | J \pi_\omega (B^*) \Omega_\omega \rangle \\
&= \langle \pi_\omega (B^*)\Omega_\omega | \pi_\omega (A^*)
\Omega_\omega \rangle \; .
\end{align*}

The pair $(\sigma_\omega^t, \M)$ is called a $W^*$-dynamical
system. We will discuss perturbations of $W^*$-dynamical
systems in the following subsection.

\subsection{Standard form of $W^*$-dynamical systems}

A $W^*$-algebra in a standard form is a quadruple
$(\M,\H,J,\H^+)$, where $\H$ is a Hilbert space, $\M\subset\B(\H)$
is a concrete $W^*$-algebra, $J$ is an antiunitary involution on
$\H$, and $\H^+$ is a self-dual cone in $\H$ such that
\begin{itemize}
\item[(1)] $J\M J=\M'$ ;
\item[(2)] $JAJ=A^*$ for $A$ in the center of $\M$;
\item[(3)] $J\psi=\psi$ for $\psi\in\H^+$;
\item[(4)] $AJA\H^+\subset\H^+$ for $A\in\M$.

\end{itemize}

If $\M$ is an abstract $W^*$-algebra, and if $\pi: \M\rightarrow
\B(\H)$ is an injective unital representation and $(\pi(\M), \H
,J, \H^+)$ is a standard form, then we say that $(\pi,\H,J,\H^+)$
is its standard representation.

Suppose $\M$ has a faithful state $\omega$,
$\pi:\M\rightarrow\B(\H)$ the corresponding GNS representation
with cyclic vector $\Omega$, and $\H^+:= \{ \pi(A)J\pi(A)\Omega :
A\in \M \}^{cl}$. Then $\H^+$ is a self-dual cone and
$(\pi,\H,J,\H^+)$ is a standard representation of $\M$. Moreover,
every $W^*$-algebra $\M$ possesses at least one standard
representation. If $(\pi_1,\H_1,J_1,\H_1^+)$ and
$(\pi_2,\H_2,J_2,\H_2^+)$ are two standard representations of
$\M$, then there exists a unique unitary operator
$W:\H_1\rightarrow\H_2$ such that
\begin{align*}
W\pi_1(A)&=\pi_2(A)W \; , \\
W\H_1^+&=\H_2^+ \; ,\\
WJ_1&=J_2W \; .
\end{align*}

(For a proof of these results, see for example [BR].)

From now on we will fix the standard form $(\M,\H,J,\H^+)$. For a
vector $\Omega\in \H$, we associate the corresponding state
$\omega_\Omega$ such that
$$\omega_\Omega (A):= \langle \Omega|A\Omega\rangle \; ,$$
where $A\in\M$. Note that $\omega_\Omega$ is a normal positive
functional on $\M$. The following hold.
\begin{itemize}
\item[(1)] The mapping $\H^+\ni\Omega\rightarrow\omega_\Omega\in\M^+_*$ is a bijection, where $\M^+_*$ is the predual of $\M$ formed of positive linear functionals.
\item[(2)] If $\Psi,\Phi\in\H^+$, then
$$||\Psi-\Phi||^2 \le ||\omega_\Psi - \omega_\Phi || \le || \Psi - \Phi|| \; ||\Psi + \Phi || \; . $$
\item[(3)] If $\Omega\in\H^+$, then $\Omega$ is cyclic $\Leftrightarrow$ $\Omega$ is separating $\Leftrightarrow$ $\omega_\Omega$ is faithful.
\end{itemize}

Suppose that $t\rightarrow\a^t$ is a $W^*$-dynamics on $\M$ and
let $U(\a^t)$ be the standard unitary operator corresponding to
$\a^t$. Then there exists a unique selfadjoint operator $\L$ such
that $U(\a^t)=e^{i\L t}$ and $e^{i\L t}\H^+\subset\H^+$. This
operator is called the {\it standard} Liouvillean of $\a^t$.
Furthermore, $\{ \omega_\Phi : \Phi \in \H^+\cap Ker\L \}=\{
\omega\in \M_*^+ : \omega$ is $\a^t$-invariant $\}$. It follows
that
\begin{itemize}
\item[(1)] $dim \; Ker \L=1 \Leftrightarrow$ there exists one normal $\a^t$-invariant state \; ;
\item[(2)] $dim \; Ker \L=0 \Leftrightarrow$ there are no normal $\a^t$-invariant states.
\end{itemize}

\subsection{Perturbation of $W^*$-dynamical systems}

\subsubsection{Perturbation of the $W^*$-dynamics}
Consider a $W^*$-algebra $\M\subset \B(\H)$ with $W^*$-dynamics
implemented by a selfadjoint operator $\L$, and a Hamiltonian
perturbation $V$, which is a selfadjoint operator affiliated to
$\M$. We will make the following sufficient assumptions to prove
the existence of the perturbed dynamics.
\begin{itemize}

\item[(A1)] $\L+V$ is essentially selfadjoint on $\D (\L)\cap\D(V)$.

\end{itemize}

\vspace{0.5cm}

\noindent {\it Proposition A1.1.} Let $\a^t_V (A):= e^{it (\L+V)}A
e^{-it(\L + V)}$ for $A\in\M$, and suppose (A1) holds. Then
\begin{itemize}
\item[(i)] $\a^t_V$ is a $W^*$-dynamics on $\M$ ;
\item[(ii)] if the perturbation $V$ is bounded, then
\begin{equation*}
\a^t_V (\cdot) = \sum_{n\ge 0} i^n \int_{0\le t_n \le \cdots \le
t_1\le t}[\a^{t_n}(V), [ \cdots
[\a^{t_1}(V),\a^t(\cdot)]\cdots]dt_1 \cdots dt_n \; .
\end{equation*}

\end{itemize}

Since $L$ and $V$ are selfadjoint, and $L+V$ is essentially selfadjoint on $\D(L)\cap\D(V)$, the proof of the above proposition follows from the
Trotter product formula [RS1]
\begin{equation*}
\a^t_V (A) = s-\lim_{n\rightarrow\infty} (e^{it\L/n}e^{itV/n})^n
(A)(e^{-it\L/n}e^{-itV/n})^n\; ,
\end{equation*}
where $A\in\M$. Since $e^{\pm itV/n}\in\M$, $\a^t_V(A)\in\M$.
Therefore, $\a^t_V$ is a $W^*$-dynamics. Claim (ii) is nothing but
a Dyson series expansion when the perturbation is bounded.

\subsubsection{Perturbation of the standard Liouvillean}

Suppose that $(\M,\H,J,\H^+)$ is a standard form of a
$W^*$-algebra, and define the standard Liouvillean as $\L_V:=
\L+V-JVJ$. We will make the following assumption.

\begin{itemize}

\item[(A2)] $\L_V$ is essentially selfadjoint on $\D(\L)\cap\D(V)\cap\D(JVJ)$.

\end{itemize}

\vspace{0.5cm}

\noindent {\it Proposition A1.2.} Suppose that (A1) and (A2) hold.
Then
\begin{itemize}
\item[(i)] $\a^t_V (A)=e^{it\L_V}Ae^{-it\L_V}$;
\item[(ii)] $e^{\pm i \L_V t}\H^+\subset\H^+$.
\end{itemize}
{\it Proof.} Note first that $e^{itJVJ}=Je^{-itV}J\in\M'$. Since
$\D(\L)\cap\D(V)\subset\D(\L+V)$, it follows that
$\D(\L)\cap\D(V)\cap\D(JVJ)\subset\D(\L+V)\cap\D(JVJ)$. Now, (i)
follows from the fact that $\L_V$ is essentially selfadjoint on
$\D(\L+V)\cap\D(JVJ)$ and the Trotter product formula
$e^{it\L_V}=s-\lim_{n\rightarrow
\infty}(e^{it(\L+V)/n}e^{-itJVJ/n})^n$. Moreover, since $e^{itV}$
and $e^{-itJVJ}$ commute, $e^{it(V-JVJ)}=e^{itV}Je^{itV}J$. Hence,
$e^{it (V-JVJ)}\H^+\subset\H^+$. The latter together with
$e^{it\L}\H^+\subset\H^+$ imply (ii). $\Box$

\subsubsection{Relative entropy}

Let $\M$ be a $W^*$-algebra, and $\psi ,\varphi$ be two
functionals in $\M^+_*$ with representation vectors $\Psi,\Phi$
respectively. Before discussing relative entropy, we need to
recall the definition of the relative modular operator. Define the
operator $S_{\Phi, \Psi}$ by $S_{\Phi,\Psi}A\Psi:=A^*\Phi$. The
relative modular operator is
$\Delta_{\Phi,\Psi}=S^*_{\Phi,\Psi}S_{\Phi,\Psi}$.

Denote by $Ent (\psi | \varphi)$ the relative entropy of
$\psi,\varphi$ as defined by Araki in [Ar2]. (We will follow the
sign convention as in [BR].)
\begin{equation}
Ent(\psi | \varphi )=
\begin{cases}
\langle \Psi | \log \Delta_{\Phi,\Psi} \Psi \rangle \; {\rm if} \; S_{\psi}\le S_{\varphi} \\
-\infty \; {\rm otherwise}
\end{cases}.
\end{equation}

\vspace{0.5cm}

\noindent {\it Proposition A1.3.} $$Ent (\psi | \varphi)=\lim_{t\downarrow 0} t^{-1} (\| \Delta_{\Phi,\Psi}^{t/2}\Psi \|^2 - \| \Psi \|^2)$$.

{\it Proof.} We will only sketch the main steps of the proof. The claim follows from the spectral theorem, the monotone
convergence theorem, and the fact that $\log x = \lim_{t\downarrow
0}(x^t-1)$, monotonically on the intervals $x\in
[0,1],[1,\infty)$. 
$\Box$

Let $\M_1$ and $\M_2$ be two $W^*$-algebras. A map $\gamma :
\M_1\rightarrow\M_2$ is a Schwartz map iff $\gamma ({\mathbf
1})=1$ and $\gamma (A^*A)\ge \gamma(A^*)\gamma (A)$.

\pagebreak

\noindent {\it Theorem A1.3 (Uhlmann's monotinicity theorem).}

Let $\psi_i,\varphi_i$ be normal states on $\M_i, i=1,2$, and
$\gamma :\M_1\rightarrow\M_2$ a Schwartz map such that
$\psi_2\circ\gamma=\psi_1$ and $\varphi_2\circ\gamma=\varphi_1$.
Then

$Ent(\psi_2|\varphi_2)\le Ent(\psi_1 | \varphi_1)$.(see
[Uh,Do,Ar2])

{\it Proof.} We prove this result in a standard concrete setting
(which is equivalent to the abstract setting). Let $(\M_i, \H_i,
J_i, \H_i^+)$ be the standard form of $\M_i, i=1,2$,
$\gamma:\M_1\rightarrow\M_2$ a Schwartz map, and
$\psi_i\in\M_{i,*}^+, i=1,2$ with corresponding representation
vectors $\Psi_i$ such that $\psi_2\gamma=\psi_1$. Furthermore, let
$\D_1=\M_1\Psi + (\M_1\Psi)^{\perp}$, and $T:\D_1\rightarrow\H_2$
a linear map defined by
$$T(A\Psi_1+\Theta_1):= \gamma (A) \Psi_2 \; ,$$
for $A\in\M_1,\Theta_1\in (\M_1\Psi_1)^\perp$. Since $\gamma
({\mathbf 1})=1$ and $T\Psi_1=\Psi_2$, $T$ is well defined and
extends to a contraction from $\H_1$ to $\H_2$. The claim of the
theorem follows from the interpolation estimate for the relative
modular operator
$$\| \Delta_{\Phi,\Psi}^{t/2}\Psi_2 \| \le \| \Delta_{\Phi,\Psi_1}^{t/2}\Psi_1 \| \; , 0\le t \le 1 , $$
together with
$$Ent(\psi | \varphi) = \lim_{t\downarrow 0} (\| \Delta_{\Phi,\Psi}^{t/2} \Psi \| - \| \Psi \|^2 .
$$
$\Box$

A direct corollary of this theorem is the following.
\vspace{0.5cm}

\noindent {\it Corollary A1.4.}

Let $\N\subset\M$ be $W^*$-algebras with common identity element,
$\psi,\varphi\in\M_*^+$, and $|_{\N}$ is the restriction to $\N$.
Then $Ent_{\M}(\psi | \varphi)\le Ent(\psi |_{\N} | \phi |_{\N})$.

\subsubsection{Perturbation of KMS states}

Consider $(\M,\H,J,\H^+)$ a $W^*$-algebra in the standard form,
$\a^t$ a $W^*$-dynamics, and $\omega$ a faithful
$(\a^t,\beta)$-KMS state, with $\beta>0$. In this subsection we
discuss the existence of the perturbed KMS state for a large class
of {\it unbounded} perturbations.

We first prove its existence for bounded perturbations and then
extend the proof to unbounded perturbations.

\vspace{0.5cm}

\noindent {\it Theorem A1.5 (Bounded perturbation).}

Suppose $V$ is a bounded perturbation. Then we have the following:

\begin{itemize}
\item[(i)] $\Omega\in\D(e^{-\beta (\L + V)/2})$.

Let $\Omega_V:= e^{-\beta (\L+V)/2}\Omega$ and $\omega_V(A) :=
\langle \Omega_V | A \Omega_V \rangle$, for $A\in\M$.

\item[(ii)] $\Omega_V\in\H^+$ and is cyclic and separating.

\item[(iii)] $\omega_V$ is an $(\a^t_V,\beta)$-KMS state.

\item[(iv)] The Peierls-Bogoliobov inequality holds
$$e^{-\beta\langle \Omega | V \Omega\rangle/2} \le \| \Omega_V \| .$$

\item[(v)] The Golden-Thompson inequality holds
$$\| \Omega_V \| \le \| e^{-\beta V/2}\Omega \| .$$
\end{itemize}
{\it Sketch of the proof.} Although these results have been
obtained in [Ar1], we follow the proof of [DJP] which can be
extended to unbounded perturbations. The first step is to prove
(i)-(v) for analytic perturbations, which approximate bounded
perturbations, and then extend the proof to bounded perturbations.
Suppose $V$ is analytic, and let $E_V(t):= e^{it (\L+V)}
e^{-it\L}$. The latter has an analytic continuation to an entire
function $z\rightarrow E_V(z)$, and $\Omega\in\D(e^{iz(\L+V)})$
for all $z\in {\mathbf C}$. In particular,
$\Omega_V=E_V(i\beta/2)\Omega$. Furthermore, $E_V(i\beta/2) = E_V
(i\beta/4)\a^{i\beta/4}(E_V(i\beta/4)^*)$, and hence
\begin{align*}
\Omega_V &= E_V (i\beta/4)\a^{i\beta /4}E_V(i\beta/4)^*\Omega \\
         &= E_V(i\beta/4)J E_V(i\beta/4)\Omega \; .
\end{align*}
Therefore, $\Omega_V\in \H^+$ for analytic $V$.

The relation $S_\Omega E_V
(i\beta/2)^*A\Omega=A^*\Omega_V=S_{\Omega_V,\Omega}A\Omega$, for
$A\in\M$, implies that the relative modular operator
\begin{align*}
\Delta_{\Omega_V,\Omega} &= S_{\Omega_V,\Omega}^*S_{\Omega_V,\Omega} \\
                         &= E_V(i\beta/2)\Delta_\Omega E^*_V(i\beta/2) \\
                         &= e^{-\beta (\L+V)} \; .
\end{align*}

Hence, $\log \Delta_{\Omega_V,\Omega}=\log \Delta_\Omega -\beta
V$. Let $\tilde{V}:=V+\beta^{-1}\log \| \Omega_V \|^2$, and
$\Omega_{\tilde{V}}:= \Omega_V / \| \Omega_V \|$. Since
$\log\Delta_{\Omega_{\tilde{V}}, \Omega}=\log\Delta_\Omega -
\beta\tilde{V}$, it follows that $Ent(\omega | \omega_V)=-\beta
\omega(\tilde{V})$. The latter together with the inequality
$Ent(\omega | \omega_V)\le 0$ imply that $e^{-\beta \langle \Omega
| V \Omega \rangle/2}\le \| \Omega_V \|$, which is the
Peierls-Bogoliubov inequality.

We still want to show the Golden-Thompson inequality in the case
of analytic perturbations. Consider $\N$ an abelian von Neumann
subalgebra of $\M$ generated by $V$.
\begin{align*}
\log \| \Omega_V \|^2 &= Ent(\omega_V | \omega ) -\beta \omega_V (V) \\
                      &\le Ent (\omega_V |_\N | \omega |_\N)-\beta \omega_V (V)\; ,
\end{align*}
where the last inequality follows from Uhlmann's monotinicity
theorem. Using the estimate $Ent(\psi|\varphi)+\psi(V)\le \log
(\varphi (e^V))$ (which follows from the fact that $e^V$ commutes with $\Delta_{\Phi, \Psi}$ and $\log x \le x-1$) and the last inequality, it follows that
$$\log \| \Omega_V \|^2 \le \log \omega (e^{-\beta V}) = \log \| e^{-\beta V/2}\Omega \|^2 \; ,$$
and hence $\| \Omega_V \| \le \| e^{-\beta V/2}\Omega \|$.

Now, extend the proof of the above statements to bounded
operators. Suppose $V\in\M$ is selfadjoint, then there is a
sequence of selfadjoint $\a$-analytic elements $V_n$ such that
$V_n\rightarrow V$ strongly as $n\rightarrow\infty$. As a
consequence, $\L+V_n\rightarrow \L+V$ and $\L_{V_n}\rightarrow
\L_{V}$ in the strong resolvent sense, while
$e^{-\beta(\L+V_n)/2}$ converges to $e^{-\beta(\L+V)/2}$ in the
weak sense. $\Box$

The above theorem can be extended to unbounded perturbations, but
we need to make one additional assumption.

\begin{itemize}

\item[(A3)] $\| e^{-\beta V/2}\Omega \| < \infty$.

\end{itemize}

\vspace{0.5cm}

\noindent {\it Theorem A1.6.}

Assume (A1), (A2) and (A3). Then

\begin{itemize}

\item[(i)] $\Omega\in\D (e^{-\beta (\L+V)/2})$.

\item[(ii)] $\Omega_V \in\H^+$ and $\Omega_V$ is cyclic and separating.

\item[(iii)] $\omega_V$ is a $(\a_V,\beta)$-KMS state.

\item[(iv)] The Peierls-Bogoliubov inequality holds
$$e^{-\beta \langle \Omega | V \Omega \rangle/2}\le \| \Omega_V \|.$$

\item[(v)] The Golden-Thompson inequality holds
$$\| \Omega_V \| \le \|e^{-\beta V/2}\Omega \|.$$

\end{itemize}

{\it Proof.} Consider the sequence of bounded operators
$V_n:=\Xi_{[-n,n]}(V)V$, where $\Xi_{[-n,n]}$ is the spectral
projection of $V$ on the interval $[-n,n]$. The proof holds for
$V_n$ as in the bounded case. Moreover, $\L+V_n\rightarrow \L+V$
in the strong resolvent sense, and so does $\L_{V_n}\rightarrow
\L_V$. The proof follows for $V$ by taking the limit
$n\rightarrow\infty$. $\Box$
%%%%%%%%%%%%%%%%%%%%%%%%%%%%%%%%%%%%%%%%%%%%%%%%%%%%%%%%%%%%%%%%%
%\include{models}

\chapter{A {\it Zoo} of models}

In this chapter, we carefully list several physically relevant
models that are paradigms of thermodynamic systems. We study
thermodynamic processes of these models in the following chapters,
so the reader may opt to skip this chapter only to return to it
when needed.

\section{Model $A1$: a small quantum system coupled to a reservoir of non-relativistic bosons}

Consider a quantum mechanical system composed of a small system
$\S$, with a finite dimensional Hilbert space, weakly coupled to a
large infinitely extended reservoir $\R$ of non-relativistic
Bosons.

The Hilbert space of the small system is $\H^\S={\mathbf C}^d$,
and the kinematical algebra of observables is $\O^\S =\M
(\mathbf{C}^{d})$, the matrix algebra on $\mathbf{C}^{d}$. Its
dynamics is generated by the Hamiltonian $H^\S$, such that $H^\S
\phi_i = E_i \phi_i, \phi_i\in\H^\S, i=0,\cdots,d-1$, and $E_0\le
E_1\le \cdots \le E_{d-1}$. In the Heisenberg picture, the
time-evolution of an operator $A\in\O^\S$ is given by
\begin{equation}
\a^\S_t (A):= e^{i H^\S t} A e^{-i H^\S t} \; , t\in {\mathbf R}
\; .
\end{equation}
For inverse temperature $0<\beta <\infty$, the $(\a^\S_t,\beta
)$-KMS state is given by
\begin{equation}
\omega^\S_\beta (\cdot ) := \frac{Tr (e^{-\beta H^\S} \cdot)}{Tr
(e^{-\beta H^\S})} \; ,
\end{equation}
where the trace is taken over $\H^\S$, assuming $e^{-\beta H^\S}$
is trace-class for $\beta >0$.

The large system $\R$ is infinitely extended and is described by a
free non-relativistic bosonic gas. Its state is taken to be the
equilibrium state at inverse temperature $\beta > 0$. Let
\begin{equation}
L_0^2:= L^2({\mathbf R}^3,d^3k)\cap L^2 ({\mathbf R}^3,
|k|^{-2}d^3k)\; ,
\end{equation}
and denote by $\W (L^2_0)$ the Weyl algebra over $L^2_0$ (see
Appendix 2). The latter is the $C^*$-algebra generated by the Weyl
operators $W(f),f\in L_0^2$, which satisfy
\begin{equation}
W(f)W(g)=e^{-\frac{i}{2} \Im (f,g)}W(f+g)=e^{-i \Im
(f,g)}W(g)W(f)\; ,
\end{equation}
and $W(f)^*=W(-f),W(0)=1$, such that the brackets $(\cdot ,
\cdot)$ denote the scalar product on $L^2({\mathbf R}^3,d^3k)$.
The state of the large system is described by the
$(\a^f_t,\beta)$-KMS state $\omega^f_\beta$ on $\W (L^2_0)$. It is
quasi-free and completely determined by the two-point function
\begin{equation}
\omega_\beta^f (a^* (k) a(k'))=\frac{\delta (k-k')}{e^{\beta
\omega (k)}-1}\; ,
\end{equation}
where $a$ and $a^*$ are the annihilation and creation operators satisfying the commutation relations
\begin{align}
&[a^\# (k), a^\# (k')] = 0 \\
&[a(k),a^*(k)]=\delta (k-k') \; ,
\end{align}
$\omega(k)=k^2$ the non-relativistic dispersion relation, and
$\delta$ is the Dirac distribution.

The dynamics of the uncoupled reservoir is given by
\begin{equation}
\a^f_t (W(f))=W(e^{it\omega}f)\; .
\end{equation}
(Note that the latter is not norm continuous (see Appendix 1).)

Let
\begin{equation}
\rho(k):= \frac{1}{e^{\beta\omega (k)}-1}>0 \; ,
\end{equation}
and
\begin{equation}
a(f):=\int d^3k f(k)a(k)\; , a^*(f):= \int d^3k f(k)a^*(k)\; .
\end{equation}

According to [ArWo], the GNS representation of the free bosonic
reservoir on the Hilbert space $\F_+\otimes\F_+$ is given by the
following:
\begin{align}
\Omega_\beta^f &= \Omega\otimes\Omega \\
\pi_\beta^f (a(f))&:= a(\sqrt{1+\rho}f)\otimes {\mathbf
1}+{\mathbf
1}\otimes a^* (\sqrt{\rho}\overline{f}) \; , \\
(\pi_\beta^f)^\# (a(f))&:=a^*(\sqrt{\rho}f)\otimes {\mathbf
1}+{\mathbf 1}\otimes a(\sqrt{1+\rho}\overline{f}) \; ,
\end{align}
where $\Omega$ is the vacuum state in the bosonic Fock space $\F_+$, $\Omega_\beta^f$ is
the vector representation of $\omega_\beta^f$, and
$(\overline{\cdot})$ stands for complex conjugation. One may check
by direct computation that
\begin{equation}
\langle \Omega_\beta^f | \pi_\beta^f (a^*(k)a(k')) \Omega_\beta^f
\rangle = \langle \Omega_\beta^f | (\pi_\beta^f)^\# (a^*(k)a(k'))
\Omega_\beta^f \rangle = \rho
(k)\delta(k-k')=\omega_\beta^f(a^*(k)a(k')) \; .
\end{equation}
The Liouvillean of the bosonic reservoir on $\F_+\otimes\F_+$ is
\begin{equation}
\L^f=\int dk \omega(k) [\pi^f_\beta (a^*(k)a(k))-(\pi_\beta^f)^\#
(a^*(k)a(k))]\; .
\end{equation}
Introduce the map $\tau_\beta:L_0^2\rightarrow L^2({\mathbf
R}\times S^2,du\times d\sigma)$ such that
\begin{equation}
\label{gluedmap}
(\tau_\beta f)(u,\sigma) =
\begin{cases}
\sqrt{\frac{1+\rho(u)}{2}} u^{1/4}
f(u,\sigma), & u\ge 0 \\
\sqrt{\frac{\rho(-u)}{2}}(-u)^{1/4}
e^{i\phi}\overline{f}(-u,\sigma), & u<0 \end{cases}
 ,
\end{equation}
where $u=k^2$ and $f\in L_0^2$ is represented in polar
coordinates. The freedom in choosing the phase $\phi$ will be used
to impose continuity at $u=0$, as we shall see later. Now, using
the isomorphism between $L^2({\mathbf R}^3) \oplus L^2({\mathbf
R}^3)$ and $L^2({\mathbf R}\times S^2)$ (the latter is the so
called {\it glued} Hilbert space), map $\F_+( L_0^2)\otimes\F_+
(L_0^2)\rightarrow \F_+ (L^2({\mathbf R}\times S^2))$, such that
$\Omega_\beta^f$ is mapped to the vacuum state of $\F_+
(L^2({\mathbf R}\times S^2))$ and $\L^f=d\Gamma(u)$, the second
quantization of the operator of multiplication by $u\in {\mathbf
R}$ (see Appendix 2 for the construction of the {\it glued} Hilbert space).

One may specify the interaction between $\S$ and $\R$ in a
representation-independent way in terms of a suitable
$*$-automorphism group $\a^t_g$ on the $C^*$-algebra
$\B(\H^\S)\otimes \W(L_0^2)$, where $g$ is a perturbation
parameter and let $\a^t_0:=\a^\S_t\otimes\a^f_t$, the free dynamics
(see [FM1]). However, for the sake of simplicity, we specify the
interaction directly on the GNS Hilbert space. The
$(\a_0^t,\beta)$-KMS state is
\begin{equation}
\omega_\beta^0=\omega_\beta^\S\otimes\omega_\beta^f\; ,
\end{equation}
on the algebra $\U=\B(\H^\S)\otimes\W(L_0^2)$. The representation
Hilbert space is
\begin{equation}
\H=\H^\S\otimes\H^\S\otimes\F_+ \; ,
\end{equation}
such that $\F_+=\F_+ (L^2({\mathbf R}\times S^2, du\times
d\sigma))$ is the bosonic Fock space over $L^2({\mathbf R}\times
S^2)$. The cyclic vector in the GNS construction representing
$\omega^0_\beta$ in $\H$ is
\begin{equation}
\Omega^0_\beta = \Omega_\beta^\S\otimes\Omega_\beta^f \; ,
\end{equation}
where $\Omega_\beta^f$ is the vacuum vector in $\F_+ ,$ and
\begin{equation}
\Omega_\beta^\S=(Tr (e^{-\beta H^\S}))^{-1/2} \sum_{j=0}^{d-1}
e^{-\beta E_j/2}\varphi_j\otimes\varphi_j\; ,
\end{equation}
with $\varphi_j$ the eigenvector of $H^\S$ corresponding to the
eigenvalue $E_j$.

Let $\varphi(h)=\frac{a^*(h)+a(h)}{\sqrt{2}}$, for $h\in
L^2({\mathbf R}\times S^2)$. The representation map
$\pi_\beta:\B(\H^\S)\otimes\W(L_0^2)\rightarrow\B(\H)$ is given by
the product
\begin{equation}
\pi_\beta:=\pi^\S\otimes\pi_\beta^f \; ,
\end{equation}
with
\begin{align}
\pi^\S (A) &:= A\otimes {\mathbf 1}^\S \; , \\
\pi_\beta^f (W(f)) &:= e^{i\varphi (\tau_\beta f)}\; ,
\end{align}
for $A\in\B(\H^\S)$ and $f\in L_0^2$.

The interacting dynamics is generated by the {\it standard}
Liouvillean, as seen in Chapter 3, which is given by
\begin{equation}
\L_g := \L_0 + g I \; ,
\end{equation}
where $\L_0:=\L^\S+\L^f$, $\L^\S=H^\S\otimes{\mathbf
1}^\S-{\mathbf 1}^\S\otimes H^\S$, $\L^f=d\Gamma (u), u\in
{\mathbf R}$, $g$ is a coupling constant, and the interaction $I$
is given by
\begin{equation}
\label{interaction} 
I=\sum_{\alpha} \{ G_\alpha\otimes{\mathbf
1}^\S\otimes\varphi (\tau_\beta (g_\alpha))-{\mathbf 1}^\S\otimes
C^\S G_\alpha C^\S \otimes \varphi (\tau_\beta (e^{-\beta
u/2}g_\alpha)) \} \; .
\end{equation}
Here $G_\alpha$ are bounded selfadjoint operators on $\H^\S$,
$g_\alpha\in L_0^2$ are the form factors, and $C^\S$ is the
antilinear operator of complex conjugation on $\H^\S$.

The corresponding interacting $W^*$-dynamics is defined by the one
parameter group of $*$-automorphisms (see Appendix 1)
\begin{equation}
\a_g^t (\cdot ) := e^{it\L_g} (\cdot ) e^{-it\L_g} \; ,
\end{equation}
on the von Neumann algebra
\begin{equation}
\M_\beta := \pi_\beta (\B(\H^\S)\otimes
\W(L^2_0))''\subset\B(\H)\; ,
\end{equation}
where $''$ denotes the double-commutant (weak closure). The pair
$(\M_\beta ,\a^t_\beta)$ defines a $W^*$-dynamical system. Let
$N:=d\Gamma(1)$, the number operator. We will often make use of
the following relative bounds (see Appendix 2)
\begin{equation}
\| I(N+1)^{-1/2} \| , \| (N+1)^{-1/2} I \| < C (1+1/\beta)\; ,
\end{equation}
where $C$ is a constant independent of the inverse temperature
$\beta$.

In order to prove the existence of the perturbed dynamics and KMS state
(together with the selfadjointness of $\L_g$ as a consequence of
the Glimm-Jaffe-Nelson theorem; see also Appendix 2) and to apply a suitable Virial theorem together with a
PC estimate to study the spectrum of the standard Liouvillian, we
make the following assumptions.

\begin{itemize}

\item[(A1.1)] {\it Smoothness of the form factors.}

The form factors are given by
$g_\a(u,\sigma)=u^p\tilde{g_\a}(u,\sigma)$, where $p=1/4,3/4,5/4$
or $>11/4$, and $\tilde{g_\a}$ is such that, for fixed $\a$ and
$\sigma$, the map $u\rightarrow \tilde{g_\a}(u,\sigma)$ is $C^3$
on $(0,\infty)$, and
\begin{equation}
\| \partial_u^i\tilde{g}_\a \|_{L^2({\mathbf R}\times S^2)} <
\infty \; ,
\end{equation}
for $i=0,1,2,3$. If $p=1/4,3/4$ or $5/4$, then the limits
\begin{equation}
\partial_u^i \tilde{g}_\a (0,\sigma):= \lim_{u\rightarrow 0^+} \partial_u^i \tilde{g}_\alpha (u,\sigma) \;
\end{equation}
exist for $i=0,1,2$, and there exists $\phi_0\in {\mathbf R}$ such
that
\begin{equation}
e^{-i\phi_0}\partial_u^i \tilde{g}_\a(0,\sigma)\in {\mathbf R},
i=0,1,2\; .
\end{equation}
Furthermore, we assume that for $p=1/4,3/4$,
$\partial_u\tilde{g}_\a(0,\sigma)=0$. As mentioned before, one may
use the freedom in choosing the phase $\phi$ (in the glued Hilbert space) to impose continuity of $g$ at $u=0$. For $p=1/4$, choose $\phi=2\phi_0+\pi$, while for other admissible values of $p$, choose $\phi=2\phi_0$.

%%%%%%%%%%%%%%%%%%%%%%%%%%%%%%%%%%%%%%%%%%%%%%%%%%%%%%%%%%%%%%%%%
%PHYSICAL RELEVANCE OF THE ASSUMPTIONS
%%%%%%%%%%%%%%%%%%%%%%%%%%%%%%%%%%%%%%%%%%%%%%%%%%%%%%%%%%%%%%%%%

The physical relevance of this assumption will be apparent in chapter 5, when we discuss proving the property of return to equilibrium for this model using the positive commutator method and a suitable Virial theorem. The analysis in chapter 5 involves estimating $\|g I (N+1)^{-1/2}\|$, which depends on $\beta$ through the interaction $I$. It turns out that one has to be careful in taking the limit $\beta\rightarrow\infty$ in the infra-red singular regime $p=1/4$. Moreover, we will need to estimate the norm of the
difference between the interacting and the non-interacting KMS
states, and hence to find an upper bound on the expectation value of the
number operator $N$ in the interacting KMS state $\Obl$. In a
suitable infra-red regime $p\ge 1/4$, we expect the
KMS-equilibrium states of the non-interacting and interacting
systems to be close to each other for small enough $g$, which will turn out to be independent of $\beta$ for $p>1/4$, but which will go to zero in the limit $\beta\rightarrow\infty$ for $p=1/4$. Furthermore, we will need to control multiple commutators of $\L_g$ with the dilatation generator $A_f=d\Gamma (i\partial_u)$ (defined in section 5.1) in order to prove a suitable Virial theorem (Theorem 5.5, chapter 5). In particular, the third commutator of the interaction $I$ with $A_f$ needs to be well-defined and relatively $N^{1/2}$-bounded. This is satisfied if
\begin{eqnarray}
\partial_u^j\tau_\beta(g_\alpha) \mbox{\ \ is continuous in $u\in\r$ for $j=0,1,2$,
  and }\label{124A}\\
\partial_u^j\tau_\beta(g_\alpha)\in L^2(\r\times
  S^2) \mbox{\ \ for $j=0,1,2,3$}.\ \ \ \ \ \ \ \label{125A}
\end{eqnarray}
One can verify that \fer{124A}, \fer{125A} follow from (A1.1). Let
$p$ and $\phi_0$ be as in assumption (A1.1). Then, for $p=3/4, 5/4$,
$p>11/4$, we choose $\phi=2\phi_0$, while for $p=1/4$, we choose
$\phi=\pi+2\phi_0$.

%%%%%%%%%%%%%%%%%%%%%%%%%%%%%%%%%%%%%%%%%%%%%%%%%%%%%%%%%%%%%%%%

\item[(A1.2)] {\it Fermi Golden Rule condition.}

Assume that
\begin{equation}
\min_{E_m\ne E_n} \int_{S^2} d\sigma |\sum_\a (\phi_m, G_\a
\phi_n) g_\a (| E_m - E_n|, \sigma )|^2 >0 \; .
\end{equation}

This assumption pertains to the positivity of the level-shift
operator, and it can be verified in specific physical models, such
as a spin impurity interacting with (free) magnons in a magnet
(see Appendix 2). Physically, it guarantees that the probability of
absorption and emission processes of field quanta does not vanish
to second order in perturbation theory.

\end{itemize}

\section{Model $A2$: spin impurity interacting with magnons in a magnet}

In this section we discuss a concrete physical model corresponding
to the earlier model.

Consider a ferromagnet, say a cubic lattice ${\mathbf Z}^3$, with
lattice spacing $a$. Its Hamiltonian is given by
\begin{equation}
H=H_{spin}+H_{wave}=-J \sum_{\langle ij \rangle} \hat{S}_i \hat{S}_j - \sum_i
j(i) \vec{s_0} \hat{S}_i \; ,
\end{equation}
where $i\in {\mathbf Z}^3, J>0$. The first term corresponds to the interaction between
neighbouring spins, and the second term corresponds to the interaction between the spins and a spin impurity. The spin operators satisfy the
commutation relation
\begin{equation}
[\hat{S}_i^\a, \hat{S}_j^\beta]=i\epsilon^{\a\beta\gamma}\delta_{ij}
\hat{S}_i^\gamma \; ,
\end{equation}
where $\a,\beta,\gamma=1,2,3$. One may write the Hamiltonian as
\begin{equation}
H=-J\sum_{\langle ij \rangle} \{ \frac{1}{2} (\hat{S}_i^{+} \hat{S}_j^{-} +
\hat{S}_i^{-}\hat{S}_j^{+}) + \hat{S}_i^3\hat{S}_j^3 \} - \sum_{i}j(i)\{ \frac{1}{2}
(s_0^+ \hat{S}_i^{-}+ s_0^{-} \hat{S}_i^{+}) + s_0^3 \hat{S}_i^3 \} \; ,
\end{equation}
where $\hat{S}_i^+=\hat{S}_i^1+i\hat{S}_i^2$ and $\hat{S}_i^- = \hat{S}_i^1-i\hat{S}_i^2$ are the
raising and lowering operators respectively. Using the
Holstein-Primakoff transformation, one may write the above
expression in terms of bosonic creation and annihilation operators
$a^*,a$:
\begin{align}
\hat{S}_i^{+} &=(2S)^{1/2}(1-\frac{a^*_ia_i}{2s})^{1/2}a_i \; ,\\
\hat{S}_i^{-} &=(2S)^{1/2}a_i^*(1-\frac{a^*_i a_i}{2s})^{1/2} \; ,\\
\hat{S}_i^3 &= S- a_i^*a_i \; ,
\end{align}
where $S$ denotes the spin of the system. Regard $a_i^*a_i$ as a
perturbation of $S$, such that $\hat{S}_i^+\approx (2S)^{1/2}a_i$ and
$\hat{S}_i^-\approx (2S)^{1/2}a_i^*$. Substituting back in the total
Hamiltonian,
\begin{equation}
H\approx -nNJS^2 - s_0^3 j_0 M + 2JSn\sum_{i}a_i^*a_i
-\sum_{\langle ij
\rangle}JS(a_ia^*_j+a_i^*a_j)-\sum_{i}j_i\sqrt{\frac{S}{2}}(s_0^+a_i^*+s_0^-a_i)\;
,
\end{equation}
where $M=\sum_{i}j_i s_0^3(S-a_i^*a_i)$, $n$ the number of nearest
neighbors (which is 6 in case of a cubic lattice), $N$ is the
total number of spins, and $j_0=\sum_i j_i$.

For a cubic lattice, take the Fourier transform of the
annihilation and creation operators $a_i=\frac{1}{\sqrt{N}}\sum_k
e^{-ik\cdot i}b_k$ and $a^*_i=\frac{1}{\sqrt{N}}\sum_k e^{ik\cdot
i}b^*_k$. One may check that as a direct consequence of the
commutation relations $[a_i^\#,a_j^\#]=0$ and
$[a_i,a^*_j]=\delta_{ij}$, $[b_k^\#,b_{k'}^\#]=0$ and $[b_k,
b^*_{k'}]=\delta_{kk'}$.

The total Hamiltonian for $a|k|\ll 1$ becomes
\begin{equation*}
H \approx -nNJS^2 - s_0^3 j_0 M + JS \sum_k a^2k^2b^*_kb_k
-\sqrt{\frac{S}{2N}}\sum_i\sum_k \{ j_i (s_0^+ e^{ik\cdot
i}b_k^*+s_0^- e^{-i k\cdot i}b_k)\}\; ,
\end{equation*}
and in the continuum limit $\sum_k\rightarrow
\frac{V}{(2\pi)^3}\int d^3k, \sum_i\rightarrow \frac{1}{V}\int
dx$, where $V$ is the volume of the system,
\begin{align}
H &\approx -nNJS^2 -s_0^3 j_0 M + \frac{JSV}{(2\pi)^3}\int d^3k k^2 b_k^* b_k -\frac{1}{(2\pi)^3}\sqrt{\frac{S}{2N}}\int d^3k (\hat{j}(k)s_0^+ b^*_k + \overline{\hat{j}}(k)s_0^- b_k) \\
&= H_{pp} + H_c +I \; ,
\end{align}
where
\begin{align*}
\hat{j}(k) &=\int dx j(x) e^{i k\cdot x}\; , \\
H_{pp} &= -nNJS^2 -s_0^3 j_0 M \; on \; {\mathbf C}^2, \\
H_c &= \frac{JSV}{(2\pi)^3}\int d^3k k^2 b_k^* b_k \; on \; \F_+ ; \\
I &=  -\frac{1}{(2\pi)^3}\sqrt{\frac{S}{2N}}\int d^3k
(\hat{j}(k)s_0^+ b^*_k + \overline{\hat{j}}(k)s_0^- b_k)=: \int d^3k
(G(k)b^*(k)+ \overline{G}(k)b(k))\; on \; {\mathbf C}^2\otimes \F_+ .
\end{align*}
Using the Pauli matrices $s_0^i, i=1,2,3$, it follows that
$$H_{pp}=diag(-nNJS^2-j_0M, -nNJS^2 +j_0M)$$

See remark after Proposition A2.2 for an explicit expression of
the level-shift operator for this model.

\section{Model $B$: a {\it quantum dot} coupled to a reservoir of non-relativistic fermions}

Consider a quantum mechanical system, say a quantum dot, composed of a small system
$\S$ with a finite dimensional Hilbert space coupled to an
infinitely extended reservoir $\R$ of free non-relativistic
fermions.\footnote{Mesoscopically, electrons in a normal metal are satisfactorily described by the Landau-Fermi liquid theory. This has been argued for heuristically and proven  rigorously using renormalization group analysis (see for example [CFS, FMRT, FLKT]).}

We assume that the small system can trap finitely many fermions. Its pure states are given by vectors in ${\mathbf C}^d$. Interpret $(1,0,\cdots,0)$ as the ground state (no fermions trapped by the quantum dot), $(0,1,\cdots,0)$ as the first excited state (one fermion trapped by the quantum dot), and so on, and assume that the Hamiltonian is given by
\begin{equation}
H^\S=diag (E_0,E_1,\cdots,E_{d-1})\; .
\end{equation}

Introduce the raising and lowering operators (which are $d\times d$ matrices and which correspond to adding or removing an electron in the quantum dot)
\begin{equation}
c_+=
\left(
\begin{matrix} 
0 & 0 &\cdots & 0\\
1 & 0 &\cdots & 0\\
0 & 1 & 0 & \cdots\\
\vdots & & & \\
0 & \cdots & 1 & 0\\
\end{matrix}
\right) \; ,
\end{equation}
and
\begin{equation}
c_- = (c_+)^* \; .
\end{equation}
They raise and lower the energy level by one. For $A\in\B({\mathbf C}^d)$, the dynamics is given by 
\begin{equation}
\a_t^\S (A) := e^{itH^\S}A e^{-itH^\S} \; .
\end{equation}

Initially, before the systems are coupled together, the state of
the fermionic reservoir is the KMS equilibrium state at inverse
temperature $\beta \in [\beta_0,\beta_*]$, $0<\beta_0<\beta_*$,
and chemical potential $\nu\in {\mathbf R}$.

Let $\h:= L^2 ({\mathbf R}^3, d^3x)$ be the Hilbert space of a
single fermion with $h$ its energy operator on $\h$. The fermionic
creation and annihilation operators $b^*(f), b(f), f\in \h$, on
the antisymmetric Fock space $\F_-(\h)$ satisfy the CAR relations
\begin{align}
&\{b^\# (f), b^\# (g) \} = 0 \; , \\
&\{b(f), b^* (g)\} = (f,g){\mathbf 1} \; ,
\end{align}
for $f,g\in L^2({\mathbf R}^3)$. Unlike in the bosonic case, it
follows from the CAR relations that $b^\#$ are bounded, since $\|
b^\# (f)\|=\|f\|$ for $f\in\h$.

The kinematical algebra of the Fermi gas $\O^f$ is the
$C^*$-algebra generated by the operators $\{ b^\# (f), f\in\h\}$
and the identity ${\mathbf 1}$. The field operators are defined by
$\varphi (f)=\frac{1}{\sqrt{2}}(b(f)+b^*(f))$.

The dynamics of the reservoir is specified by the Hamiltonian
$H^f=d\Gamma(h)$, the second quantization of the energy operator
$h$, such that
\begin{equation}
\a^f_t (b^\# (f))=e^{it H^f} b^\# (f) e^{-itH^f} = b^\#
(e^{ith}f)\; ,
\end{equation}
and $h=u=k^2$ the non-relativistic dispersion relation.\footnote{For non-zero chemical potential, an equivalent free dynamics of the reservoirs is generated by $d\Gamma (u-\nu) ,$ see for example [BR].}

For each inverse temperature $\beta>0$ and chemical potential $\nu\in {\mathbf R}$, there exists a unique
KMS state $\omega^f_{\beta,\nu}$ on $\O^f$, which is a quasi-free,
gauge-invariant state uniquely determined by the two-point
function
\begin{equation}
\omega^f_{\beta,\nu}(b^*(f)b(f))=(f,
(e^{\beta(u-\nu)}+1)^{-1}f) \; .
\end{equation}

The $C^*$-algebra of the combined system is the tensor product
algebra $\O=\O^\S\otimes \O^f$, and the free dynamics is generated
by the group of automorphisms $\a^0_t=\a^\S_t\otimes\a^f_t$, such
that, for $A \in \O$,
\begin{equation}
\a_t^0 (A) = e^{itH_0} A e^{-iH_0}\; ,
\end{equation}
and $H_0=H^\S\otimes{\mathbf 1}^f+{\mathbf 1}^\S\otimes H^f$.

Introduce the interaction between $\S$ and $\R$,
\begin{equation}
V=\sum_\a \{ c_- \otimes b^*(g_\a)+c_+ \otimes b(g_\a)\} \; ,
\end{equation}
where $g_\a\in \h$ are the form factors. Note that $V$ is a bounded
selfadjoint perturbation ($V\in\O$ and $V=V^*$), and the perturbation is invariant under gauge transformations of the first kind (ie, the total number of fermions is conserved, which is expected in the nonrelativistic regime).

The dynamics of the coupled system is generated by the Hamiltonian
\begin{equation}
H_g:= H_0+gV\; ,
\end{equation}
such that
\begin{equation}
\a^g_t (A)=e^{itH_g}A e^{-itH_g}\; ,
\end{equation}
for $A\in\O$.

As in the bosonic case, we will work directly in a concrete GNS
representation. Let $\hat{\Omega}_f$ be the Fock vacuum on $\F_-
(\h)$, $N$ the number operator, $\theta=\Gamma(-1)=(-1)^N$, and
$\rho_{\beta,\nu}(u):= (1+e^{\beta (u-\nu)})^{-1}$. Moreover, let
$\H^f=\F_-(\h)\otimes\F_-(\h)$ and
$\Omega_f=\hat{\Omega}_f\otimes\hat{\Omega}_f$.

The Araki-Wyss representation $\pi_\beta$ of $\O^f$ on $\H^f$ is
defined by [ArWy]

\begin{align}
\pi_\beta (b(f)) &:= b(\sqrt{1-\rho_{\beta,\nu}}f))\otimes {\mathbf 1}^f + \theta\otimes b^*({\sqrt{\rho_{\beta,\nu}}\; \overline{f}}) \; , \\
\pi_\beta^\# (b(f)) &:=
b^*(\sqrt{\rho_{\beta,\nu}}f)\theta\otimes\theta + {\mathbf
1}^f\otimes\theta b (\sqrt{1-\rho_{\beta,\nu}}\; \overline{f}) \;
.
\end{align}

The Liouvillean of the uncoupled system is
\begin{equation}
\L_0=\L^\S+\L^f \; ,
\end{equation}
where $\L^\S=H^\S\otimes {\mathbf 1}^\S-{\mathbf 1}^\S\otimes
H^\S$ and $\L^f=H^f\otimes {\mathbf 1}^f-{\mathbf 1}^f\otimes
H^f$. The {\it standard} Liouvillean of the coupled system is
\begin{equation}
\L_g=\L_0+g\pi_\beta (V)-g\pi_\beta^\# (V)\; .
\end{equation}

Associate to every function $f(u,\sigma)\in \h = L^2 ({\mathbf
R}^+, S^2)$ (in polar coordinates), two functions $f_\beta
(u,\sigma), f_\beta^\# (u,\sigma)\in L^2({\mathbf R},S^2)$, the
glued Hilbert space (with chemical potential $\nu\in {\mathbf R}$), such that

\begin{align}
f_\beta (u,\sigma) &:=
\begin{cases}
\sqrt{1-\rho_{\beta,\nu}(u)} \frac{u^{1/4}}{\sqrt{2}} f(u,\sigma)\; , u\ge 0 \\
\sqrt{\rho_{\beta,\nu}(-u)} \frac{|u|^{1/4}}{\sqrt{2}}  \overline{f}(-u,\sigma)\; , u< 0
\end{cases}\\
&=
\begin{cases}
\frac{e^{\beta (u-\nu)/4}}{2\cosh^{1/2}(\beta (u-\nu)/2)} u^{1/4} f(u,\sigma)\; , u\ge 0 \\
\frac{e^{-\beta (|u|-\nu)/4}}{2\cosh^{1/2}(\beta (|u|-\nu)/2)}
|u|^{1/4} \overline{f}(-u,\sigma)\; , u< 0
\end{cases} \; ,
\end{align}
and
\begin{align}
f_\beta^{\#} (u,\sigma) &:=
\begin{cases}
i\sqrt{\rho_{\beta,\nu}(u)} \frac{u^{1/4}}{\sqrt{2}} f(u,\sigma)\; , u\ge 0 \\
i\sqrt{1-\rho_{\beta,\nu}(-u)} \frac{|u|^{1/4}}{\sqrt{2}}  \overline{f}(-u,\sigma)\; , u< 0
\end{cases} \\
&=
\begin{cases}
\frac{i e^{-\beta (u-\nu)/4}}{2\cosh^{1/2}(\beta (u-\nu)/2)} u^{1/4} f(u,\sigma)\; , u\ge 0 \\
\frac{ie^{\beta (|u|-\nu)/4}}{2\cosh^{1/2}(\beta (|u|-\nu)/2)}
|u|^{1/4} \overline{f}(-u,\sigma)  , u< 0 
\end{cases} \\
&= i e^{-\beta \frac{|u|}{u}(|u|-\nu )/2}f_\beta (u,\sigma)\\
&=i\overline{f}_\beta(-u,\sigma) \; .
\end{align}

Map $\F_- (\h )\otimes\F_- (\h )\rightarrow \F_- (L^2 ({\mathbf R},
S^2))$ using the isomorphism between $L^2({\mathbf R}^+,
S^2)\oplus L^2 ({\mathbf R}^+, S^2)$ and $L^2({\mathbf R}, S^2)$ (see Appendix 2 for a discussion of the glued Hilbert space $L^2({\mathbf R}, S^2)$).

The perturbed Liouvillean acting on
$\H=\H^\S\otimes\H^\S\otimes\F_-(L^2 ({\mathbf R},S^2))$ is given
by
\begin{equation}
\L=\L^\S+\L^f+gI\; ,
\end{equation}
where $\L^\S$ is as before, $\L^f=d\Gamma (u)$, the second
quantization of the operator of multiplication by $u\in {\mathbf
R}$,
\begin{eqnarray}
I&=&\sum_\a \{ c_- \otimes {\mathbf 1}^\S \otimes b^*(g_{\alpha,\beta})+
c_+ \otimes {\mathbf 1}^\S \otimes b(g_{\alpha,\beta}) -  \nonumber \\
&-& {\mathbf 1}^\S \otimes C^\S c_- C^\S
\otimes (i\theta b^*(g_{\a,\beta}^\#))-{\mathbf 1}^\S \otimes C^\S c_+ C^\S
\otimes (i\theta b(g_{\a,\beta}^\#)) \} \; ,\label{intB}
\end{eqnarray}
and $C^\S$ is the operator corresponding to complex conjugation on
$\H^\S$.

Note that the perturbed Liouvillean is selfadjoint since the
perturbation is bounded and selfadjoint. Both the existence of the
perturbed dynamics and the perturbed KMS state follow from the
latter fact (see Appendix 1, chapter 3). We will make the following
assumptions on the form factors.

\begin{itemize}

\item[(B.1)] {\it Smoothness of the form factors.}
In order to apply the Virial theorem (see Proposition 5.10, section 5.2) in the proof of RTE for this model, impose that
\begin{equation}
\lim_{u\rightarrow 0}\partial_u^j g_{\a,\beta} (u,\sigma)= 0\; ,
j=0,1,2,3.
\end{equation}

This condition is satisfied if $g_\alpha (u,\sigma)=u^p
\tilde{g}_\a (u,\sigma)$, for $\tilde{g}\in C^3$ on $(0,\infty)$
and $p>11/4$.

\item[(B.2)] {\it Fermi Golden Rule.}
This condition ensures that the small system is coupled to the
reservoirs,
\begin{equation}
\min_{E_m\ne E_n}\int_{S^2} | \sum_\a (\varphi_m, G_\a \phi_n)
g_\alpha (|E_n-E_m|, \sigma)|^2 >0 \; .
\end{equation}
This condition is translated to the positivity of the so called
level-shift operator (Appendix 2).

\end{itemize}

%%%%%%%%%%%%%%%%%%%%%%%%%%%%%%%%%%%%%%%%%%%%%%%%%%%%%%%%%%%%%%%%%%%%%%%%%%%%%%

\section{Model $C_n$: a 2 level system coupled to $n$ fermionic reservoirs}

For the sake of concreteness, and without loss of generality,
assume that the small system $\S$ is a 2 level system (as described in section 4.3), coupled to
$n$ reservoirs of free fermions,$\R_1,\cdots,\R_n$, in equilibrium
at inverse temperatures
$\beta_1,\cdots,\beta_n\in[\beta_0,\beta_*]$, for fixed $\beta_0 ,\beta_*$, such that $0<\beta_0<\beta_*<\infty$, and chemical potentials $\nu_1,\cdots,\nu_n.$ For the sake of simplicity of exposion, we set all the chemical potentials of the reservoirs equal to $\nu\in {\mathbf R}.$\footnote{We exclude the case when $\nu=2$ so that the first nontrivial terms in perturbation theory for the coupled system are second order in the coupling parameter.} We will remark on how our results change if the chemical potentials are different in due course.

The kinematical algebra of the small system $\S$ is
$\O^\S=\M({\mathbf C}^2)$, the matrix algebra over ${\mathbf
C}^2$, and its Hamiltonian is $H^\S=\sigma_3$, where
$\sigma_i,i=1,2,3$ are the Pauli matrices.

Each dispersive and infinitely extended reservoir $\R_i,
i=1,\cdots , n$ is formed of free fermions, which are not
necessarily non-relativistic. We make the assumption that the
Hilbert space of a single fermion is $\h=L^2 ({\mathbf R}^+;\B)$,
where $\B$ is some auxilliary Hilbert space, and that the single
particle Hamiltonian $h$ is the operator of multiplication by
$u\in {\mathbf R}^+$. (In the previous section, $\B=L^2(S^2)$ and
$u=k^2$.)

The kinematical algebra of the coupled system is
$\O=\O^\S\otimes\O^{\R_1}\otimes\cdots\otimes \O^{\R_n}$. 
The interaction between the small system and the fermionic reservoirs
is given by a generally time-dependent perturbation 
\begin{equation}
V(t)=\sum_i \{ 
\sigma_-\otimes b^* (f_i(t))+ \sigma_+\otimes b(f_i(t)) \}, 
\end{equation}
where $t\in {\mathbf R},$
$\sigma_\pm = \sigma_1\pm i \sigma_2,$ the raising and lowering spin operators,
and $f_i(t)\in \h , i=1,\cdots , n$ are the form factors.\footnote{This form of interaction is invariant under gauge transformations of the first kind.}

The dynamics of the coupled system is generated by the Hamiltonian
\begin{equation}
H_g (t):= H_0+gV (t)\; ,
\end{equation}
where $H_0 = H^\S + H^{\R_1} + \cdots + H^{\R_n}$, such that
\begin{equation}
\a^g_t (A)=U(-t) A U(t)\; ,
\end{equation}
for $A\in\O$ and the propagator $U(t)$ satisfies the initial value
problem
\begin{align*}
\partial_t U(t) &= -i H_g(t) U(t) \; , \\
U(0) &= 1 .
\end{align*}
(Note that since the perturbation is bounded in the fermionic
case, the perturbed time evolution can be expanded in a Dyson
series (see [RS2]).)

The GNS representation of the system is similar to the one given
in the previous section (Model B). (In particular, each reservoir
is represented using the Araki-Wyss representation, and then
mapping the latter to the glued Hilbert space representation; see Appendix 2.)

For every function $f\in L^2 ({\mathbf R}^+;\B)$, associate the
functions $f_\beta,f_\beta^\# \in L^2({\mathbf R};\B)$ given by
\begin{equation}
f_\beta (u,\sigma,t) := \begin{cases}
\sqrt{1-\rho_{\beta,\nu}(u)} f(u,\sigma,t) \; , & u\ge 0 \\
\sqrt{\rho_{\beta,\nu}(-u)} \; \overline{f}(-u,\sigma,t) \; , & u<0 \; 
\end{cases} \; ,
\end{equation}
and
\begin{eqnarray}
f_\beta^{\#} (u,\sigma,t) &:=&
\begin{cases}
i\sqrt{\rho_{\beta,\nu} (u)} f(u,\sigma,t),  u\ge 0 \\
i\sqrt{1-\rho_{\beta,\nu} (-u)} \; \overline{f}(-u,\sigma,t), u<0 \end{cases} \\
&=& i\overline{f}_\beta(-u,\sigma,t) .
\end{eqnarray}

The interacting {\it standard} Liouvillean acting on the Hilbert
space $\H:=\H^\S\otimes\H^\S\otimes\F^{(1)}_- (L^2({\mathbf
R};\B))\otimes\cdots\otimes \F^{(n)}_- (L^2 ({\mathbf R}; \B))$ is

\begin{equation}
\L_g(t)= \L^\S + \L^f + gI(t)\; ,
\end{equation}
where $\L_0=\L^\S+\L^f$, $\L^\S=H^\S\otimes {\mathbf 1}^\S -
{\mathbf 1}^\S\otimes H^\S, \L^f=\sum_i d\Gamma (u_i)$, and
\begin{eqnarray*}
I(t)&=&\sum_{i=1}^n \{ \sigma_-\otimes {\mathbf 1}^\S\otimes b^*(f_{i,
\beta_i}(t))+\sigma_+\otimes {\mathbf 1}^\S\otimes b(f_{i,
\beta_i}(t)) \\
&-& i {\mathbf 1}^\S\otimes \sigma_-\otimes (-1)^{N_i} b^*(f_{i, \beta_i}^\#(t))-i {\mathbf 1}^\S\otimes \sigma_+\otimes (-1)^{N_i} b(f_{i, \beta_i}^\#(t))\} \; . 
\end{eqnarray*}

%%%%%%%%%%%%%%%%%%%%%%%%%%%%%%%%%%%%%%%%%%%%%%%%%%%%%%%%%%%%%%%%%%%
%Denote by $\N$ the set of states normal to an initial state
%$\omega=\omega^\S\otimes\omega^f$, where $\omega^\S\in
%E(\O^\S)$, the dual of $\O^\S$, and $\omega^f=\omega_{\beta_1}^{\R_1}\otimes\cdots\omega_{\beta_n}^{\R_n}$.
%%%%%%%%%%%%%%%%%%%%%%%%%%%%%%%%%%%%%%%%%%%%%%%%%%%%%%%%%%%%%%%%%%%

Let
\begin{align}
\label{I}
I(\delta)&:= \{ z\in {\mathbf C} : | \Im z | < \delta \} \; ; \\
I^-(\delta) &:= \{ z\in {\mathbf C} : -\delta < \Im z  < 0 \} \; .
\end{align}
Moreover, for every function $f\in L({\mathbf R}^+;\B)$ associate
the function $\tilde{f}\in \B$ such that
\begin{equation}
\tilde{f}(u,\sigma) := \begin{cases} f(u,\sigma) , u\ge 0\\

\overline{f}(|u|,\sigma) , u<0 \; ,
\end{cases}
\end{equation}
and denote by $H^2(\delta,\B)$ the Hardy class of all analytic
functions $g: I(\delta)\rightarrow \B$, such that
\begin{equation}
\label{Hardy}
\| g \|_{H^2 (\delta,\B)} :=
\sup_{|\theta |<\delta}\int_{\mathbf{R}}\| g(u+i\theta)\|^2_{\B} du <
\infty \; .
\end{equation}

In order to apply the method of complex translations, we make the following assumptions on the interaction.

\begin{itemize}

\item[($C_n.1$)] {\it Regularity of the form factors.}

Assume that $\exists \delta >0$, independent of $t$ and $i\in \{
1,\cdots ,n \}$, such that
\begin{equation}
\tilde{f}_{i}(t) \in H^2(\delta,\B) ,
\end{equation}
the Hardy class of analytic functions.

\item[($C_n.2$)] {\it Fermi Golden Rule.}

Assume that
\begin{equation}
\sum_i \| \tilde{f}_i (2,t) \|_\B >0 \; ,
\end{equation}
for almost all $t\in {\mathbf R}$, which is another way of saying
that the small system is coupled to at least one reservoir.

Note that for this model, $\sigma (\L^\S)=\{ E_j \}_{j=0}^3$,
where $E_0=E_1=0,E_2=-2$ and $E_3=2$.\footnote{A {\it concrete} example where the above assumptions are satisfied
is when $\h=L^2({\mathbf R}^3, d^3k), h=k^2$. In polar
coordinates, $\B=L^2(S^2,d\sigma)$. If the form factor
$f(k)=|k|^{1/2}e^{-|k|^4}$, then both assumptions ($C_n.1$) and
($C_n.2$) are satisfied.}

We make the following additional assumption which we will need in
discussing C-Liouvilleans in chapter 6.

%%%%%%%%%*******************

\item[($C_n.3$)] {\it Stronger Regularity of the form factors.}

Assume that $\exists \delta >0$, independent of $t$ and $i\in \{
1,\cdots ,n \}$, such that
\begin{equation}
e^{-\beta_i u_i/2}\tilde{f}_{i}(t) \in H^2(\delta,\B) \; ,
\end{equation}
the Hardy class of analytic functions. This assumption is stronger
than ($C_n.1$), and we shall need it in studying the spectrum of
the so called C-Liouvillean using complex deformation techniques. It implies that the mapping
\begin{equation}
{\mathbf R}\ni r\rightarrow \Delta^{ir} V(t)\Delta^{-ir}\in \M \; ,
\end{equation}
(where $\Delta = \Delta^\S\otimes \Delta^{\R_1}\otimes\cdots\otimes\Delta^{\R_n}$ is the modular operator of the coupled system,) 
has an analytic continuation to the strip $I(1/2)=\{ z\in {\mathbf
C}: |\Im z|<1/2 \}$, which is bounded and continuous on its
closure, $\forall t\in {\mathbf R}$.

%%%%%%%%%%%%%%%%%%%%%%%%%%%%%%%%%%%%%%%%%%%%%%%%%%%%%%%%%%%%%%%
%Note that we have the following estimates
%$$\| f_{i,\beta_i} \|_{H^2(\delta,\B)} \le C_i(\delta)\|\tilde{f_i}\|_{H^2(\de%lta,\B)} \; ,$$
%and
%$$\| f^{\#}_{i,\beta_i} \|_{H^2(\delta,\B)} \le C_i(\delta) \|e^{-\beta_i sgn %\; \Re (u_i)(|u_i|-\nu_i)/2}\tilde{f_i}\|_{H^2(\delta,\B)} .$$
%%%%%%%%%%%%%%%%%%%%%%%%%%%%%%%%%%%%%%%%%%%%%%%%%%%%%%%%%%%%%%%%

\item[($C_n.4$)] The perturbation is constant for $t<0$, $V(t)\equiv V(0)$,
and then {\it slowly} changes over a time interval $\tau$ such
that $V^\tau (t)=V(s)$, where $s=t/\tau\in [0,1]$ is the rescaled
time. We also assume that $V(s)$ is twice differentiable in $s\in
[0,1]$ as a bounded operator, such that
\begin{equation}
{\mathbf R}\ni r\rightarrow \Delta^{ir}
\partial_s^j V(s)\Delta^{-ir}\in \M\; , j=0,1,2
\end{equation}
has an analytic continuation to the strip $\{ z\in {\mathbf C}:
|\Im z|<1/2 \}$, which is bounded and continuous on its closure.
This follows if we assume that there exists $\delta >0$, independent of
$s$ and $i\in \{ 1,\cdots ,n \}$, such that
\begin{equation}
e^{-\beta_i u_i/2}\partial_s^j \tilde{f}_{i}(s) \in H^2(\delta,\B)
\; ,
\end{equation}
the Hardy class of analytic functions, for $j=0,1,2$. This
assumption is needed to prove an adiabatic theorem for states
close to NESS (chapter 8).

\item[($C_n.5$)] The perturbation is constant for $t<0$, $V(t)=V(0)$,
and then {\it slowly} changes over a time interval $\tau$ such
that $V^\tau (t)=V(s)$, where $s=t/\tau\in [0,1]$ is the rescaled
time. We also assume that $V(s)$ is twice differentiable in $s\in
[0,1]$ as a bounded operator. Moreover, we assume that there
exists $\delta>0$, independent of $s$, $i\in \{ 1,\cdots ,n \}$,
such that
\begin{equation}
\partial_s^j \tilde{f}_{i}(s) \in H^2(\delta,\B) \; ,
\end{equation}
the Hardy class of analytic functions, for $j=0,1,2$. This
assumption is needed in studying an explicit example of the
isothermal theorem (chapter 8).

\item[($C_n.6$)] The perturbation is periodic with
period $\tau_*<\infty$: $V(t)=V(t+\tau_*)$. This assumption is needed to
investigate cyclic thermodynamic processes (chapter 9).

\end{itemize}

%%%%%%%%%%%%%%%%%%%%%%%%%%%%%%%%%%%%%%%%%%%%%%%%%%%%%%%%%%%%%%%%

\section{Appendix 2}

\subsection{Selfadjointness of $\L_g$ and some relative bounds for
Model A}

Consider the positive operator $ \Lambda={\rm d}\Gamma(|u|) $ with
domain $\dom(\Lambda)=\{\psi\in{\cal H}: \|\Lambda\psi\|<\infty\}$
and the number operator
\begin{equation}
N={\rm d}\Gamma(1) \; , \label{nop}
\end{equation}
with domain $\dom(N)=\{\psi\in{\cal H}: \|N\psi\|<\infty\}$. \\
Without loss of generality, assume $u(k)=k^2.$

\vspace{0.5cm}

\noindent {\it Proposition A2.1 (Some Relative Bounds).}

Let $L^2=L^2({\mathbb R}\times S^2)$, and $0<\beta_0<\infty$ be a
fixed number. Then the following hold.

\begin{itemize}
\item[(i)] If $f\in L^2$, then $||a(f)N^{-1/2}||\leq ||f||_{L^2}$.

\item[(ii)] If $|u|^{-1/2}f\in L^2$, then
 $||a(f)\Lambda^{-1/2}||\leq ||\,|u|^{-1/2}f||_{L^2}$.

\item[(iii)] For $\psi\in\dom(N^{1/2})$ and $\psi\in\dom(\Lambda^{1/2})$
respectively, we have the following bounds, uniformly in
$\beta\geq \beta_0$:
\begin{eqnarray*}
||I\psi||^2&\leq& C\sum_\alpha||G_\alpha||\left( \|N^{1/2}\psi\|^2+\|\psi\|^2\right),\\
||I\psi||^2&\leq& C\sum_\alpha||G_\alpha||\left(
\|\Lambda^{1/2}\psi\|^2+\|\psi\|^2\right),
\end{eqnarray*}
where $C\leq C'(1+\beta_0^{-1})$, and $C'$ is a constant
independent of $\beta, \beta_0$.

\item[(iv)] For $\psi\in{\cal D}(N^{1/2})$, any $c>0$,
and uniformly in $\beta\geq \beta_0$, one has
 \vspace*{-.1cm}
\begin{equation*}
\left|\scalprod{\psi}{g I\psi}\right|\leq c
||N^{1/2}\psi||^2+\frac{16 g^2}{c}\sum_\alpha||G_\alpha||^2||\psi||^2
\int_{{\mathbb R}^3}(1+\beta_0^{-1}u^{-1})|g_\a|^2d^3k.
\end{equation*}
 \vspace*{-.75cm}

\item[(v)] For $\psi\in{\cal D}(\Lambda^{1/2})$, any $c>0$, and
uniformly in $\beta\geq \beta_0$, one has
 \vspace*{-.1cm}
\begin{equation*}
\left|\scalprod{\psi}{g I\psi}\right|\leq
c||\Lambda^{1/2}\psi||^2+\frac{32 g^2}{c}\sum_\alpha||G_\alpha||^2||\psi||^2
\int_{{\mathbb
R}^3}(1+\beta_0^{-1}u^{-1})\frac{|g_\a|^2}{u}d^3k.
\end{equation*}
\end{itemize}

{\it Proof.} The proof is standard (see for example [BFS], [JP1,2]). As
an illustration, we present the proof of
(iii). We know that $$ \|I\psi\|^2\leq \sum_\a 4\|G_\a\|^2\left(
\|a^*(g_\a)\psi\|^2+\|a(g_\a)\psi\|^2\right), $$ and using the CCR
$[a^*(f),a(g)]=\scalprod{f}{g}$, one gets
$$
\|a^*(g_\a)\psi\|^2=\scalprod{\psi}{a(g_\a)a^*(g_\a)\psi}=\|a(g_\a)\psi\|^2+\|g_\a\|^2_{L^2}\|\psi\|^2,
$$
and hence $ \|I\psi\|^2\leq \sum_\a 8\|G_\a\|^2\left(
\|a(g_\a)\psi\|^2 +\|g_\a\|^2_{L^2}\|\psi\|^2\right)$. (i) and
(ii) give
\begin{eqnarray*}
\|I\psi\|^2&\leq& 16 \sum_\a \|G_\a\|^2\|g_\a\|^2_{L^2}\left( \|N^{1/2}\psi\|^2+\|\psi\|^2\right),\\
\|I\psi\|^2&\leq& 16 \sum_\alpha\|G_\alpha\|^2\left\|
|u|^{-1/2}g_\alpha\right\|^2_{L^2}\left(
\|\Lambda^{1/2}\psi\|^2+\|\psi\|^2\right).
\end{eqnarray*}
We show that $\|g_\a\|_{L^2}\leq C$ and
$\|\,|u|^{-1/2}g_\a\|_{L^2}\leq C$, uniformly in $\beta\geq
\beta_0$. Note that $ \|g_\a\|_{L^2}^2=\int_{{\mathbf
R}^3}(1+2\rho)|g_\a(u,\sigma)|^2du d\sigma, $ where we
represented $g_\a$ in the integral in spherical coordinates, and
$u(k)=k^2$. Since we have $1+2\rho=1+2(e^{\beta
u}-1)^{-1}\leq 1+2\beta^{-1}u^{-1}\leq
1+2\beta_0^{-1}u^{-1}$, uniformly in $\beta\geq \beta_0$, one
has the following uniform bound in $\beta\geq \beta_0$:
\begin{equation}
\|g_\a\|_{L^2}^2\leq 2\int_{{\mathbf
R}^3}(1+\beta_0^{-1}u^{-1})|g_\a(k)|^2d^3k =C<\infty.
\label{*}
\end{equation}
Similarly, $ \|\,|u|^{-1/2}g_\a\|_{L^2}^2\leq2\int_{{\mathbf
R}^3}(1+\beta_0^{-1}u^{-1})u^{-1}
|g_\a(u,\sigma)|^2 d^3k=C<\infty, $ uniformly in $\beta\geq
\beta_0$. It is clear from the last two estimates that $C$
satisfies the bound indicated in the proposition. $\Box$

These relative bounds and Nelson's commutator theorem  (see [RS2])
yield essential selfadjointness of the standard Liouvillian
$\L_g$. (Essential selfadjointness of $\L_g$ also follows from the
GJN Theorem, see Appendix 3, chapter 5.)

\subsection{{\it Glued} Hilbert Space representation}

Consider a reservoir of non-relativistic bosons (Model A1 discussed in section 4.1). We want to show that 

\begin{equation*}
\F_+(L^2({\mathbf R}^3,d^3k))\otimes \F_+(L^2({\mathbf R}^3,d^3k)) \cong \F_+(L^2 ({\mathbf R}\times S^2, du d\sigma)) \; ,
\end{equation*}
where $S^2$ is the unit sphere in three dimensions, $d\sigma$ is the element of the solid angle, and $u=k^2$. 

For bosonic creation/annihilation operators on $\F_+(L^2({\mathbf R}^3, d^3k)),$ $$a^\# (f):= \int d^3k f(k)a^\# (k) \; , f \in L^2({\mathbf R}^3, d^3k),$$ define the creation/annihilation operators on $\F_+(L^2({\mathbf R}^3,d^3k))\otimes \F_+(L^2({\mathbf R}^3,d^3k))$ as
\begin{align*}
a_l^\# (f) := a^\# (f)\otimes \unit \; ; \\
a_r^\# (f) := \unit \otimes a^\# (\overline{f}) \; ,
\end{align*}
where $\overline{\cdot}$ corresponds to complex conjugation. 
An isomorphism between $\F_+(L^2({\mathbf R}^3,d^3k))\otimes \F_+(L^2({\mathbf R}^3,d^3k))$ and $\F_+ (L^2({\mathbf R}^3,d^3k)\oplus L^2({\mathbf R}^3,d^3k))$ follows by the identification 
\begin{equation*}
a_l^\# (f_1)\cdots a_l^\# (f_m)a_r^\# (g_1)\cdots a_r^\# (g_n) = a_l^\# ((f_1,0))\cdots a_l^\# ((f_m,0))a_r^\# ((0,g_1))\cdots a_r^\# ((0,g_n)) \; ,
\end{equation*}
where the RHS acts on $\F_+ (L^2({\mathbf R}^3,d^3k)\oplus L^2({\mathbf R}^3,d^3k))$. Now we claim that $\F_+ (L^2({\mathbf R}^3,d^3k)\oplus L^2({\mathbf R}^3,d^3k))$ is isomorphic to $\F_+(L^2 ({\mathbf R}\times S^2, du d\sigma))$. For $\phi, \psi \in {\mathbf R}$, consider the mapping 
\begin{equation*}
j_{\phi,\psi } : L^2({\mathbf R}^3,d^3k)\oplus L^2({\mathbf R}^3,d^3k) \ni (f,g)\rightarrow h \in L^2 ({\mathbf R}\times S^2, du d\sigma) \; ,
\end{equation*}
such that 
\begin{equation*}
h(u,\sigma) := 
\begin{cases}
\frac{e^{i\phi}}{\sqrt{2}} u^{1/4} f(u,\sigma) \; , u\ge 0 \\
\frac{e^{i\psi}}{\sqrt{2}} |u|^{1/4} g(|u|,\sigma) \; , u < 0 
\end{cases} \; .
\end{equation*}
This mapping is an isometry, since 
\begin{align*}
\| h \|^2_{L^2 ({\mathbf R}\times S^2, du d\sigma)} &= \| (f,g) \|^2_{ L^2 \oplus L^2 } \\
&= \int_{{\mathbf R}^+\times S^2} du d\sigma \frac{u^{1/2}}{2} |f(u,\sigma)|^2 + \int_{{\mathbf R}^+\times S^2} du d\sigma \frac{u^{1/2}}{2} |g(u,\sigma)|^2 \\
&= \int_{{\mathbf R}^+\times S^2} dk d\sigma k^2 |f(k,\sigma)|^2 + \int_{{\mathbf R}^+\times S^2} dk d\sigma k^2 |g(k,\sigma)|^2 \\
&= \| f \|^2_{L^2({\mathbf R}^3, d^3k)} + \| g \|^2_{L^2({\mathbf R}^3, d^3k)} \; ,
\end{align*}
where we have used the fact that $u=k^2$ for the non-relativistic reservoir.

Moreover, the mapping $j_{\phi,\psi }$ is an isomorphism, since, for given $h\in L^2 ({\mathbf R}\times S^2, du d\sigma)$, there exists a mapping $j_{\phi, \psi}^{-1}: h\rightarrow (f,g)\in L^2({\mathbf R}^3,d^3k)\oplus L^2({\mathbf R}^3,d^3k)$, such that 
\begin{align*}
f(u,\sigma)&:= \frac{\sqrt{2}e^{-i\phi}}{u^{1/4}}h(u,\sigma) , u>0 \; , \\
g(u,\sigma)&:= \frac{\sqrt{2}e^{-i\psi}}{|u|^{1/4}}h(|u|,\sigma) , u<0 \; .
\end{align*}

Using the Araki-Woods representation, section 4.1, and the mapping $\fer{gluedmap}$, one may write the interaction term on $L^2 ({\mathbf R}\times S^2, du d\sigma)$ as given in $\fer{interaction}$. 

Similarly, one may construct of the {\it glued} Hilbert space representation for fermionic reservoirs, as in Models B and $C_n$, sections 4.3 and 4.4, except that one needs to use the Araki-Wyss representation for fermionic reservoirs, section 4.3 (instead of the Araki-Woods representation). 

\subsection{Feshbach map}

Consider a closed operator $A$ acting on a Hilbert space $\H$, and a bounded (and not necessarily orthogonal) projection $P$ on $\H$ such that $Ran (P)\subset \D(A)$. Let $\bP:=1-P$, and define
\begin{align*}
A_P &:= PAP \; ,\\
A_\bP &:= \bP A \bP \; .
\end{align*}
We will view $A_\bP$ as an operator on $\bP\H$. We will make the following assumptions.
\begin{itemize}

\item[(F1)] $z\in \rho(A_\bP)$, where $\rho(A_\bP)$ is the resolvent set of $A_\bP$ (ie, $(A_\bP-z\unit)^{-1}$ exists and is bounded).

\item[(F2)] $\| \bP (A_\bP-z)^{-1}\bP A P \| <\infty$ and $\| PA\bP (A_\bP -z)^{-1}\bP\| <\infty$.

\end{itemize}

Define the Feshbach map 
\begin{equation}
\F_{P,z}(A):= (P(A-z)P-PA\bP ( A_\bP - z)^{-1} \bP AP )|_{Ran (P)}\; ,
\end{equation}
provided assumptions (F1) and (F2) are satisfied.

Moreover, define 
\begin{equation}
S_{P,z}:= P-\bP (A_\bP - z)^{-1}\bP A P \; .
\end{equation}
Since $P=(1+\bP (A_\bP - z)^{-1}\bP A P)S_{P,z}$, it follows that 
$$Ker (S_{P,z})=Ker (P).$$

The following theorem establishes a very useful property of the Feshbach map, which is its isospectrality (see, for example, [BFS,BFSS]). We will use this property in discussing the level-shift operator, which is relevant in a rigorous treatment of Fermi Golden Rule (for a review, see for example [DF1,2]), and in proving the property of RTE in chapter 5. 

\vspace{0.5cm}
\noindent {\it Theorem A2.2 (Isospectrality of the Feshbach map)}

Suppose (F1) and (F2) hold. Then
\begin{equation}
\label{iso1}
z\in \sigma_\# (A) \Leftrightarrow 0\in \sigma_\# (\F_{P,z}(A)) \; ,
\end{equation}
where $\sigma_\#=\sigma_c$ or $\sigma_{p}$ (continuous or pure point spectrum). Moreover, the eigenfunctions of $(A-z)$ and $\F_{P,z}(A)$ are related by 
\begin{equation}
\label{iso2}
Ker ((A-z)S_{P,z})=Ker (\F_{P,z}(A)) \; ,
\end{equation}
and 
\begin{equation}
\label{iso3}
P \; Ker (A-z)= Ker (\F_{P,z}(A)) \; .
\end{equation}

These imply that
$$dim \; Ker (A-z) = dim \; Ker (\F_{P,z}(A)) \; .$$ 

{\it Proof.}

Using the second resolvent identity and the fact that 
$$A= A_P + A_\bP + PA\bP + \bP A P \; ,$$
we have the following identities
\begin{equation}
\label{iso5}
(A-z)S_{P,z}=\F_{P,z}(A) \; ,
\end{equation}
and
\begin{equation}
\label{iso6}
P(A-z)^{-1}P=(\F_{P,z}(A))^{-1} 
\end{equation}
on $Ran (P)$, and 
\begin{align}
(A-z)^{-1}=&(\F_{P,z}(A))^{-1}P - (\F_{P,z}(A))^{-1}P A\bP (A_\bP -z)^{-1}\bP \nonumber \\
&-\bP (A_\bP-z)^{-1}\bP A P (\F_{P,z}(A))^{-1} + \bP (A_\bP-z)^{-1}\bP \nonumber \\
&+\bP (A_\bP -z)^{-1}\bP A P (\F_{P,z}(A))^{-1}PA\bP (A_\bP -z)^{-1}\bP \; \label{iso7}
\end{align}

Now, \fer{iso5} imply \fer{iso2}. Moreover, we claim that \fer{iso5} and \fer{iso6} imply \fer{iso1}. If $z\in \rho (A)\cap \rho (A_\bP)$, then the LHS of \fer{iso6} defines the inverse of $\F_{P,z}(A)$, and thus $0\in \rho(\F_{P,z}(A))$. Next suppose that $z\in \rho(A_\bP )$ and $0\in \rho(\F_{P,z}(A))$. The RHS of \fer{iso7} defines the inverse of $(A-z)$. Therefore, \fer{iso5} and \fer{iso6} imply \fer{iso1}.

It remains to show \fer{iso3}. Let $z\in \sigma_p (A)$, and $\psi\in Ker (A-z)$. Projecting $(A-z)\psi=0$ on $Ran (P)$ and $Ran (\bP)$ gives
\begin{equation}
\label{iso8}
(A_P -z)P\psi + PA\bP \psi = 0 \; ,
\end{equation}
and
\begin{equation}
\label{iso9}
(A_\bP-z)\bP\psi + \bP A P \psi = 0 \; .
\end{equation}

Since $z\in \rho(A_\bP)$, it follows from \fer{iso9} that 
\begin{equation}
\label{iso10}
\bP\psi = -(A_\bP - z)^{-1}\bP A P \psi \; .
\end{equation}
Substituting \fer{iso10} in \fer{iso8} gives $$\F_{P,z}(A)P\psi=0 \; .$$ Therefore, $P \; Ker (A-z)\subset Ker (\F_{P,z}(A))$. Conversely, if $\phi=P\psi\in Ker (\F_{P,z}(A))$, then $S_{P,z}\phi \in Ker (A-z)$ by \fer{iso8} and \fer{iso9}. Therefore, $$P\phi = PS_{P,z}\phi \in P \; Ker (A-z).$$
$\Box$

A further property of the Feshbach map is $$\F_{P_1}\circ \F_{P_2}=\F_{P_1P_2}, $$ if $[P_1,P_2]=0.$

\subsection{The Level-Shift Operator (LSO) and Fermi Golden Rule: a formal motivation}

In this subsection, we {\it formally} discuss the perturbation of the point spectrum of a selfadjoint operator $A$ (acting on a Hilbert space $\H$) using the Feshbach map, and we show how it relates to the so called level-shift operator (LSO) to second order in perturbation theory. Specific examples will be dealt with {\it rigorously} in chapters 5 and 6.

Consider a finitely degenerate eigenvalue $\lambda\in \sigma_p (A)$, with corresponding bounded projection $P$. Add a bounded perturbation, $gW$, to $A,$ 
$$A^g=A+gW,$$ 
and assume that $PWP=0$ (so that the first nontrivial perturbation of $\lambda$ is second order in $g$). Since the Feshbach map is isospectral (Theorem A2.2), we can use it to calculate the perturbation of $\lambda$. To second order in $g$,  this is related to the level-shift operator.

Consider the Feshbach map
\begin{align*}
\F_{P,\lambda}(A^g) &= P(A^g-\lambda)P -PA^g\bP ( A^g_\bP -\lambda)^{-1} \bP A^g P \\
&= -g^2 PW\bP (A^g_\bP - \lambda)^{-1}\bP W P \\
&= -g^2 \tilde{\Gamma} (\lambda) + O(g^3) \; ,
\end{align*}
where $\tilde{\Gamma} (\lambda) := PW\bP (A_\bP - \lambda)^{-1}\bP W P$ is the {\it level-shift} operator (LSO). Since $PW\bP=PW$ and $\bP WP=WP,$ it follows that 
$$\tilde{\Gamma}(\lambda) := PW(A_\bP - \lambda)^{-1}W P.$$

To second order in $g$, the real part of the shift of $\lambda$ is
\begin{align*}
-g^2 \Re \tilde{\Gamma} (\lambda) &= -\frac{1}{2} g^2 \lim_{\epsilon\rightarrow 0^+} PW\{ (A_\bP - \lambda + i\epsilon)^{-1} + (A_\bP -\lambda -i\epsilon)^{-1}\}W P \\
&= -g^2 PW {\cal PV} (A_\bP - \lambda)^{-1} WP \; ,
\end{align*}
where ${\cal PV}$ stands for the Cauchy principal value. Moreover, the imaginary part of the perturbation of $\lambda$ to second order in $g$ is 
\begin{align*}
-g^2\Im \tilde{\Gamma} (\lambda) &= -\frac{1}{2i} g^2 \lim_{\epsilon\rightarrow 0^+} PW\{ (A_\bP - \lambda + i\epsilon)^{-1} - (A_\bP -\lambda -i\epsilon)^{-1}\}W P \\
&= -g^2\pi PW \delta (A_\bP -\lambda) WP \; ,
\end{align*}
where we have used the fact that $\lim_{\epsilon\rightarrow 0}\frac{\epsilon}{x^2+\epsilon^2}=\delta (x)$. This last term is related to the {\it Fermi Golden Rule} for quantum resonances. 

To make these arguments rigorous, one uses spectral methods such as complex dilatation and RG analysis (see for example [BFS]) or complex translations (see for example [JP1,2] and chapter 6).

\subsection{Explicit calculation of LSO}

In this subsection we review some of the consequences of the {\it
Fermi Golden Rule} condition in proving RTE. For further details,
see, for example [BFSS,BFS], and previous subsection. In Models A1,A2 and B, the
Liouvillean of the small system $\L^\S=
H^\S\otimes\mathbf{1}^\S-\mathbf{1}^\S\otimes H^\S$ has discrete
spectrum $ \sigma(\L^\S)=\{e=E_i-E_j=:E_{ij}: E_i,E_j\in\sigma(H^\S)\}$.
For each eigenvalue $e\in\sigma(\L^\S)$, the (imaginary part of the) {\it
level-shift operator} acting on $\ran
P(\L^\S=e)\subset\H^\S\otimes\H^\S$ , is
\begin{equation}
\Gamma(e)=\int_{{\mathbf R}\times S^2} m^*(u,\sigma)P(\L^\S\neq
e)\delta(\L^\S-e+u) m(u,\sigma), \label{Gamma}
\end{equation}
where
\begin{equation}
m(u,\sigma)=\sum_{\alpha} \{ G_\alpha\otimes{\mathbf
1}^\S\otimes\; \tau_\beta (g_\alpha(u,\sigma))-{\mathbf
1}^\S\otimes C^\S G_\alpha C^\S \otimes \; \tau_\beta (e^{-\beta
u/2}g_\alpha(u,\sigma)) \} \; .
\label{m}
\end{equation}
Without loss of generality, set $\alpha=1$, and let
$G\otimes\unit^\S =G_l,\unit^\S\otimes C^\S G C^\S=G_r, \tau_\beta
(g)=g_1, \tau_\beta (e^{-\beta u/2}g)=g_2$. (The result can 
be easily generalized to arbitrary $\alpha\in\mathbf{N}$.) Note that
$\Gamma(e)$ is a non-negative selfadjoint operator. The Fermi
Golden Rule condition (assumptions A1.2 and B.2) is used to show
the instability of embedded eigenvalues away from zero:
\begin{equation}
\mbox{for $e\neq 0$,\ \ \ \ }
\gamma_e:=\inf\sigma\left(\Gamma(e)\upharpoonright \ran
P(\L^\S=e)\right)>0, \label{fgrc}
\end{equation}
while $\Gamma(0)$ has a simple eigenvalue at zero, with
corresponding eigenvector the Gibbs state of the small system,
$\Omega^\S_\beta$. Physically, this means that the zero eigenvalue
of $\L_0$ survives the perturbation, but its degeneracy is lifted:
\begin{equation}
\gamma_0:=\inf\sigma\left(\Gamma(0)\upharpoonright \ran
P(\L^\S=0)P^\perp_{\Omega^\S_\beta}\right)>0. \label{fgrc0}
\end{equation}
Here, $P_{\Omega^\S_\beta}$ is the projection onto ${\mathbf
C}\Omega^\S_\beta$, and
$P^\perp_{\Omega_\beta^\S}=\mathbf{1}-P_{\Omega_\beta^\S}$. In the
following, we consider Models A1,2, but the result holds when the
large system consists of free fermions (ie, for Model B) ;(see
chapter 5).

\vspace{0.5cm}

\noindent {\it Proposition A2.3}

Assume (A1.2), and let
$\Gamma_\S(e):=P(\L^\S=e)\Gamma(e)P(\L^\S=e)$. Then the following
hold.
\begin{itemize}
\item[(i)] Let $e\neq 0$. Then there is a non-negative number
$\delta_0$ (independent of $\beta,g$) such that
\begin{equation*}
\Gamma_\S(e)\geq \delta_0 \inf_{\{E_{ij}\neq
0\}}\left(|E_{ij}|\int_{S^2}dS(\omega,\sigma) \left|
g(|E_{ij}|,\sigma)\right|^2\right)  P(\L^\S=e) >0.
\end{equation*}

\item[(ii)]  $\Gamma_\S(0)$  has a simple eigenvalue at
zero, with the Gibbs state $\Omega^\S_\beta$ as
eigenvector:
\begin{equation}
\Omega^\S_{\beta}=Z^\S(\beta)^{-1/2}\sum_{i}e^{-\beta
E_i/2}\varphi_i\otimes\varphi_i, \label{particlegibbs}
\end{equation}
where $Z^\S(\beta)=Tr_{\H^\S}(e^{-\beta H^\S})$, and the spectrum
of $\Gamma_\S(0)$ has a gap at\ zero: $(0,2g_0
Z^\S)\cap\sigma(\Gamma_\S(0))=\emptyset$.
\end{itemize}

{\it Proof.}

For $e\neq 0$, let
\begin{eqnarray*}
\nr^{(i)}&:=&\{j| E_i-E_j=e\},\\
\nl^{(j)}&:=&\{i| E_i-E_j=e\},\\
\nr &:=&\cup_{i}\ \nr^{(i)}=\{ j| E_i-E_j=e\mbox{\ for some $i$}\},\\
\nl &:=&\cup_{j}\ \nl^{(j)}=\{ i| E_i-E_j=e\mbox{\ for some
$j$}\}.
\end{eqnarray*}
Moreover, let $P_i$ denote the rank-one projector onto ${\mathbf
C}\varphi_i$, where $\{\varphi_i\}_{i=0}^{d-1}$, the eigenvectors
of $H^\S$. For ${\cal N}\subset{\mathbf N}$, put
\begin{equation*}
P_{{\cal N}}:=\sum_{j\in{\cal N}} P_j,\mbox{\ \ \ and\ \ \
}P_{\cal N}:=0 \mbox{\ if
  ${\cal N}$ is empty.}
\end{equation*}
Let $E_{mn}:=E_m-E_n$, and for $e\in\sigma(\L^\S)\backslash
\{0\}$, $m\in{\cal N}_l$ and $n\in {\cal N}_r$, let
\begin{eqnarray}
\delta_m&:=&\inf\sigma\Big(P_{\nr^{(m)}} GP_{{\cal N}_r^c} G
P_{\nr^{(m)}}\upharpoonright P_{\nr^{(m)}}   \Big)\geq 0,
\label{105}\\
\delta'_n&:=&\inf\sigma\Big(P_{\nl^{(n)}} GP_{{\cal N}_l^c}G
P_{\nl^{(n)}} \upharpoonright  P_{\nl^{(n)}} \Big)\geq 0,
\label{106}
\end{eqnarray}
 where $\ ^{c}$ denotes the complement. If
$e=0$, then $\nr^c=\nl^c$ are empty, and $\delta_m,\delta_n'=0$.
Let
$\delta_0:=\inf_{m\in\nl}\{\delta_m\}+\inf_{n\in\nr}\{\delta_n'\}$.
From $P(\L^\S=e)=\sum_{\{i,j:E_{ij}=e\}}P_i\otimes P_j$, one has
\begin{equation}
\Gamma_\S(e)=\sum_{m,n}\left(1-\delta_{E_{mn},e}\right)\sum_{\{i,j:
E_{ij}=e\}}\sum_{\{k,l: E_{kl}=e\}}\int \delta(E_{mn}-e+u)P_{ij} \
m^* \ P_{mn}\  m\ P_{kl}. \label{110}
\end{equation}
We want to get a lower bound on $(m,n)\in\N\times\N$ by summing
only over a convenient subset of $\N\times\N$. Using (\ref{m}),
\begin{eqnarray*}
\lefteqn{
P_{ij}m^* P_{mn}m P_{kl}}\\
&=& P_{ij}\left(G_l\overline{g}_1 - G_r\overline{g}_2\right) P_{mn}\left( G_l g_1-G_r g_2\right) P_{kl}\\
&=& P_i GP_mGP_k\otimes P_n \delta_{jn}\delta_{nl} |g_1|^2 -P_i GP_m\otimes P_n C^\S G C^\S P_l \delta_{jn}\delta_{mk}\overline{g}_1 g_2\\
&& -P_m GP_k\otimes P_j C^\S G C^\S P_n \delta_{im}\delta_{nl}
\overline{g}_2 g_1 +P_m\otimes P_j C ^\S G C^\S P_n C^\S G C^\S
P_l \delta_{im}\delta_{mk} |g_2|^2.
\end{eqnarray*}
Summing over $i,j$ and $k,l$ gives
\begin{eqnarray}
\lefteqn{\sum_{\{i,j: E_{ij}=e\}}\sum_{\{ k,l: E_{kl}=e\}} P_{ij} m^* P_{mn} m P_{kl}}\nonumber\\
&&=\left(\overline{g}_1 P_{{\cal N}_l^{(n)}} GP_m\otimes P_n
-\overline{g}_2 P_m\otimes P_{{\cal N}_r^{(m)}} C^\S G C^\S
P_n\right)\cdot\mbox{\ adjoint}.\nonumber
\end{eqnarray}
For $(m,n)\in{\cal N}_l\times{\cal N}_r^c$, one has $P_{{\cal
N}_l^{(n)}}=0$ and $P_{{\cal N}_r^{(m)}}\neq 0$. For $(m,n)\in
{\cal N}_l^c\times{\cal N}_r$, $P_{{\cal N}_l^{(n)}}\neq 0$ and
$P_{{\cal N}_r^{(m)}}=0$. Sum only over the disjoint union
\begin{equation*}
(m,n)\in {\cal N}_l\times{\cal N}_r^c\ \dot{\cup}\ {\cal
N}_l^c\times{\cal N}_r.
\end{equation*}
After some calculation gives
\begin{eqnarray*}
\Gamma_\S(e)&\geq&\inf_{E_{ij}\neq 0}\left(\int_{S^2}dS \left|g_2(E_{ij},\alpha)\right|^2\right)\sum_{m\in{\cal N}_l}P_m\otimes C^\S P_{{\cal N}_r^{(m)}} G \ P_{{\cal N}_r^c} \ G P_{{\cal N}_r^{(m)}}C^\S\\
&&+\inf_{E_{ij}\neq 0}\left(\int_{S^2}dS
\left|g_1(E_{ij},\alpha)\right|^2\right)\sum_{n\in{\cal N}_r}
P_{{\cal N}_l^{(n)}} G \ P_{{\cal N}_l^c} \ G P_{{\cal
N}_l^{(n)}}\otimes P_n.
\end{eqnarray*}
Moreover,
\begin{equation*}
\int_{S^2}dS |g_{1,2}(E_{ij},\alpha)|^2 \geq |E_{ij}| \int_{S^2}dS
|g(|E_{ij}|,\alpha)|^2,
\end{equation*}
uniformly in $\beta\geq 1$, and together with \fer{105},
\fer{106}, it gives
\begin{equation*}
\Gamma_\S(e)\geq\inf_{E_{ij}\neq 0}\left( |E_{ij}|\int_{S^2}dS
|g(E_{ij},\alpha)|^2\right)\left(\inf_{m\in{\cal N}_l}\{\delta_m\}
+\inf_{n\in{\cal N}_r}\{\delta'_n\}\right) P(\L_\S=e),
\end{equation*}
since $\sum_{m\in{\cal N}_l} P_m\otimes P_{{\cal
N}_r^{(m)}}=\sum_{n\in{\cal N}_r} P_{{\cal N}_l^{(n)}}\otimes
P_n=P(\L^\S)$. This proves (i).

Consider now the case $e=0$. An element of $\ran P(\L^\S)$ is of
the form $\phi=\sum_ic_i\varphi_i\otimes\varphi_i$, with
$\sum_i|c_i|^2=1$, so
\begin{equation*}
\scalprod{\phi}{\Gamma(0)\phi}=\sum_{m,n}\left(
1-\delta_{E_{mn},0}\right)\sum_{i,j}\overline{c}_i c_j
\int\delta(E_{mn}+u)\scalprod{\varphi_i\otimes\varphi_i}{m^*
P_{mn} m\varphi_j\otimes\varphi_j}.
\end{equation*}

The fact that $\scalprod{\varphi_m}{C^\S G C^\S
\varphi_n}=\overline{\scalprod{\varphi_m}{G\varphi_n}}$, implies
\begin{equation}
\scalprod{\phi}{\Gamma(0)\phi}=\sum_{m,n}\left(1-\delta_{E_{mn},0}\right)\int
\delta(E_{mn}+u)\left|\scalprod{\varphi_n}{G\varphi_m}\right|^2
|c_ng_1-c_mg_2|^2, \label{114}
\end{equation}
and hence
\begin{eqnarray*}
\lefteqn{\int \delta(E_{mn}+u) |c_ng_1-c_mg_2|^2}\\
&=&\int_{{\mathbf
R}^3}\left\{\delta(E_{mn}+\omega)\left|\sqrt{1+\rho}c_n
g-\sqrt{\rho}c_m g\right|^2\right. +
\delta(E_{mn}-\omega)\left.\left|\sqrt{\rho}c_ng
-\sqrt{1+\rho}c_mg\right|^2\right\}.
\end{eqnarray*}
Together with \fer{114}, this implies
\begin{eqnarray}
\scalprod{\phi}{\Gamma(0)\phi}
&=&2\sum_{\{m,n: E_{mn}<0\}}\left|\scalprod{\varphi_n}{G\varphi_m}\right|^2\frac{e^{\beta E_n}}{e^{-\beta E_{mn}}-1} \nonumber\\
&&\ \ \ \times \left| e^{-\beta E_m/2}c_n-e^{-\beta
E_n/2}c_m\right|^2\int \delta(E_{mn}+\omega)|g|^2, \label{115}
\end{eqnarray}
where we used $\delta(E_{mn}+\omega)\rho=\delta(E_{mn}+\omega)
(e^{-\beta
  E_{mn}}-1)^{-1}$. Each term in the sum is
zero if one chooses $c_n=Z_\S^{-1/2}e^{-\beta E_n/2}$.
Furthermore,
$\scalprod{\Omega_\beta^\S}{\Gamma(0)\Omega_\beta^\S}=0$. Since
$\Gamma(0)\geq0$, this implies that $\Omega_\beta^\S$ is a zero
eigenvector of $\Gamma(0)$.

We still need to estimate the spectral gap at zero. Equation
\fer{115} imply
\begin{eqnarray*}
\scalprod{\phi}{\Gamma(0)\phi}&\geq& 2g_0\sum_{\{m,n: E_{mn}<0\}}|e^{-\beta E_m/2}c_n-e^{-\beta E_n/2}c_m|^2\\
&&=g_0\sum_{m,n}|e^{-\beta E_m/2}c_n-e^{-\beta E_n/2}c_m|^2\\
&&=g_0\sum_{m,n}\left( e^{-\beta E_m}|c_n|^2 +e^{-\beta E_n}|c_m|^2-e^{-\beta(E_m+E_n)/2}(\overline{c}_nc_m+c_n\overline{c}_m)\right)\\
&&=g_0\Big(Z_\S(\beta)+Z_\S(\beta)-2\Big|\sum_m e^{-\beta E_m/2}c_m\Big|^2\Big)\\
&&=2g_0Z_\S(\beta)\Big(
1-\Big|\scalprod{\Omega_\beta^\S}{\phi}\Big|^2\Big),
\end{eqnarray*}
where we used $\sum_n|c_n|^2=1$. Therefore, we obtain on $\ran
P_{\Omega_\beta^\S}^\perp$:\  $\Gamma(0)\geq 2g_0Z_\S(\beta)$.
$\Box$

\vspace{0.5cm}

{\it Remark.} As an explicit illustration of the results of the
previous Proposition, consider Model A2 of a spin impurity
interacting with magnons in a magnet. Recall that
\begin{align}
H &\approx -nNJS^2 -s_0^3 j_0 M + \frac{JSV}{(2\pi)^3}\int d^3k k^2 b_k^* b_k -\frac{1}{(2\pi)^3}\sqrt{\frac{S}{2N}}\int d^3k (\hat{j}(k)s_0^+ b^*_k + \overline{\hat{j}}s_0^- b_k) \\
&= H_{pp} + H_c +I \; ,
\end{align}
where
\begin{align*}
\hat{j}(k) &=\int dx j(x) e^{i k\cdot x}\; , \\
H_{pp} &= -nNJS^2 -s_0^3 j_0 M \; , \\
H_c &= \frac{JSV}{(2\pi)^3}\int d^3k k^2 b_k^* b_k \; ; \\
I &=  -\frac{1}{(2\pi)^3}\sqrt{\frac{S}{2N}}\int d^3k
(\hat{j}(k)s_0^+ b^*_k + \overline{\hat{j}}s_0^- b_k)=: \int d^3k
(G(k)b^*(k)+ \overline{G}(k)b(k))\; .
\end{align*}
Using the Pauli matrices $s_0^i, i=1,2,3$, it follows that
$$H_{pp}=diag(-nNJS^2-j_0M, -nNJS^2 +j_0M),$$ and hence
$\sigma(\L^\S)=\{-2j_0M, 0 ,2j_0 M\}$, with double degeneracy at
0. Direct computation (although lengthy) gives the following
result.

\begin{itemize}
\item[(1)] For $e=E_{01}=-2j_0M$,
\begin{equation}
\langle \varphi_0\otimes\varphi_1 ,\Gamma(-2j_0M)
\varphi_0\otimes\varphi_1 \rangle =
\frac{4}{(2\pi)^5}\frac{2S}{N}|\hat{j}(-2j_0M)|^2\frac{\sqrt{2j_0M}}{e^{-2\beta
j_0 M }-1} >0 \; .
\end{equation}

\item[(2)] For For $e=E_{10}=2j_0M$,
\begin{equation}
\langle \varphi_1\otimes\varphi_0 , \Gamma(2j_0M)
\varphi_1\otimes\varphi_0 \rangle =
\frac{4}{(2\pi)^5}\frac{2S}{N}|\hat{j}(2j_0M)|^2\frac{\sqrt{2j_0M}}{e^{2\beta
j_0 M }-1} >0 \; .
\end{equation}

\item[(3)] For $e=E_{11}=E_{00}=0$,
\begin{equation}
\Gamma(0) = \left(
\begin{matrix}
a & a^{\beta j_0M} \\
ae^{\beta j_0M} & ae^{2\beta j_0M}
\end{matrix}
\right)\; ,
\end{equation}
where
$$a=\frac{4\pi |G_{0,1}(2j_0M)|^2 \sqrt{2j_0M}}{e^{2\beta
j_0M}-1}.$$ The eigenvalues of $\Gamma(0)$ are
\begin{equation}
\lambda_0=0 \; , \lambda_1 = \frac{4\pi |G_{0,1}(2j_0M)|^2
\sqrt{2j_0M}(1+e^{2\beta j_0M})}{e^{2\beta j_0M}-1} > 0\; .
\end{equation}

\end{itemize}
 
%%%%%%%%%%%%%%%%%%%%%%%%%%%%%%%%%%%%%%%%%%%%%%%%%%%%%%%%%%%%%%%%%%%%%%%%%%
%\include{PCandRTE}

\chapter{RTE for a small quantum system coupled to non-relativistic reservoirs: PC method}

%%%%%%%%%%%%%%%%%%%%%%%%%%%%%%%%%%%%%%%%%%%%%%%%%%%%%%%%%%%%%%%%%%%%%%%%%%%%%%%
%%%% SECTION: INTRODUCTION
%%%%%%%%%%%%%%%%%%%%%%%%%%%%%%%%%%%%%%%%%%%%%%%%%%%%%%%%%%%%%%%%%%%%%%%%%%%%%%%

In this chapter, we investigate the property of return to
equilibrium (RTE) for a class of quantum systems composed of a
small system with a finite dimensional Hilbert space, weakly
coupled to an infinitely extended and dispersive heat bath by
studying the spectrum of the corresponding standard Liouvillean.
As discussed in chapter 2, this property is part of the zeroth
law of thermodynamics.

In [JP1,2], RTE is proven for a class of spin-boson system (by
studying the spectrum of the Liouvillian) using complex
deformation techniques (see also [DJ]). The proof is not uniform in
temperature. A stronger result of RTE, which is uniform in
temperature, has been shown when a toy atom is coupled to the
radiation field using the Feshbach map, complex dilatation, and an
operator theoretic renormalization group method in [BFS]. This has
been revisited again in [FM1] using Mourre's positive commutator
method [M1] and a suitable Virial theorem with an explicit zero
temperature limit. Similar methods have been used to investigate
thermal ionization in [FM2,FMS] and to prove the stability of
Bose-Einstein condensates in [M2]. We note that the approach of
proving RTE using Liouvillians is based on the insights of [HHW].

In this chapter, we extend the analysis of [FM1] to proving RTE
for a class of systems composed of a small system coupled to
(free) non-relativistic bosonic and fermionic reservoirs under
suitable assumptions on the perturbation (particularly, the form
factors and {\it Fermi golden rule}), as discussed in chapter 4,
Models A1,A2 and B. A physical example of the first model is a
spin impurity coupled to (free) magnons in a magnet, while an
example of the second is a {\it quantum dot} coupled to electrons
in a metal. The three essential elements that enter in our
analysis are a concrete representation of the free bosonic/fermionic reservoirs
(Araki-Woods and Araki-Wyss respectively), the positive commutator
method with a suitable Virial theorem, and an estimate on the norm of the difference between the
KMS-equilibrium states of the non-interacting and the interacting
systems (Appendix 2).

%%%remark: check recent work by Derezinski and Jacksic on RTE for spin-boson systems and perturbation of $W^*$-dynamical systems

%%%%%%%%%%%%%%%%%%%%%%%%%%%%%%%%%%%%%%%%%%%%%%%%%%%%%%%%%%%%%%%%%%%%%%%%%%%%%%%
%%%% SECTION: RTE FOR NON-RELATIVISTIC BOSONIC RESERVOIRS
%%%%%%%%%%%%%%%%%%%%%%%%%%%%%%%%%%%%%%%%%%%%%%%%%%%%%%%%%%%%%%%%%%%%%%%%%%%%%%%

\section{RTE for Models A1,2}

The main result of this section is Theorem 5.2, which claims the
property of return to equilibrium for quantum mechanical system
formed of a small system with a finite dimensional Hilbert space
coupled to a reservoir of free non-relativistic bosons. We refer
to Model A1 in chapter 4 for relevant details and assumptions.

\vspace{0.5cm}

\noindent {\it Proposition 5.1}

Assume (A1.2) (specified in section 4.1). There is an $\epsilon_0>0$, independent of
$\beta\geq\beta_0$ (for any $\beta_0$ fixed), such that if
$0<\epsilon<\epsilon_0$ then
\begin{equation}
\Pi I\frac{\epsilon}{\L_0^2+\epsilon^2} I\Pi \geq \Gamma_0\Pi
-C\epsilon^{1/4}, \label{126}
\end{equation}
where $C$ is a constant independent of the inverse temperature
$\beta$, $\Pi=P_0\otimes P_{\Omega_f}$, the projection onto the
kernel of $\L_0$, and $\Gamma_0$ is a bounded operator on
$\H=\H^{\S}\otimes\H^{\S}\otimes{\cal F}_+$, acting trivially on
the last factor, ${\cal F}_+$, and leaving $\ker \L_{\S}$
invariant. Furthermore, $\Gamma_0$ restricted to $\ker \L_{\S}$
has zero as a simple eigenvalue, with Gibbs state
$\Omega_\beta^{\S}$ as eigenvector, and is strictly positive on
the complement of $\cx\Omega_\beta^{\S}$. There is a constant
$\gamma_0>0$, independent of $0<\beta<\infty$, such that
\begin{equation}
\Gamma_0\upharpoonright_{\ran \Pbar_{\Omega_\beta^{\S}}}\geq
\gamma_0, \label{127}
\end{equation}
where, $\Pbar_{\Omega_\beta^{\S}}=\bbbone -P_{\Omega_\beta^{\S}}$
and $P_{\Omega_\beta^{\S}}$ is the projection onto $\cx
\Omega_\beta^{\S}$.

For a proof of this result (when the sum reduces to a single term and in the limit of $\epsilon\rightarrow 0$)
see Appendix 2, Proposition A2.2, which can be easily generalized.
One can show that,
\begin{eqnarray*}
\lefteqn{
\Gamma_0\upharpoonright_{\ran \Pbar_{\Omega_\beta^{\S}}}}\\
&&\geq \min_{E_m\neq E_n} \frac{(E_m-E_m)^{\frac{1}{2}}\ \tr
e^{-\beta H_{\S}} }{|e^{-\beta E_m}-e^{-\beta E_n}|}
\int_{S^2}d\sigma \left| \sum_\alpha
  \scalprod{\varphi_m}{G_\alpha \varphi_n}
  g_\alpha\left(|E_m-E_n|,\sigma\right)\right|^2,
\end{eqnarray*}
and $\gamma_0$ in \fer{127} is obtained by minimizing the RHS over
$0<\beta<\infty$.

%%%%%
\vspace{0.5cm}

\noindent {\it Theorem 5.2(RTE1)}

Suppose (A1.1) and (A1.2) (see section 4.1). Then there is a constant $g_1>0$,
independent of $\beta\geq \beta_0$, for any $\beta_0>0$ fixed,
such that, for
\begin{equation}
0<|g| <g_1\left\{
\begin{array}{cl}
\big(1+\log(1+\beta)\big)^{-9/2} & \mbox{if $p=1/4$} \\
1 &\mbox{if $p>1/4$} \; ,
\end{array}
\right. \label{f4}
\end{equation}
the kernel of $\L_g$ is spanned by the interacting KMS vector
$\Obl.$ (The system possesses the property of return to equilibrium in the ergodic sense.)

{\it Remark.} The analysis involves estimating $\|g I
(N+1)^{-1/2}\|$, which depends on $\beta$ since the interaction $I$ does. As we shall see, one has to be careful in taking the limit $\beta\rightarrow\infty$ in the infra-red singular regime $p=1/4$. Moreover, in order to estimate the norm of the
difference between the interacting and the non-interacting KMS
states, we need an upper bound on the expectation value of the
number operator $N$ in the interacting KMS state $\Obl$. In a
suitable infra-red regime $p\ge 1/4$, we expect the
KMS-equilibrium states of the non-interacting and interacting
systems to be close to each other for small enough $g$, which will turn out to be independent of $\beta$ for $p>1/4$, but which will go to zero in the limit $\beta\rightarrow\infty$ for $p=1/4$.

\vspace{0.5cm}

\noindent {\it Proposition 5.3}

Suppose (A1.1) holds, and let $P_\Obl$ and $P_\Obz$ denote the
projections onto the spans of the interacting and non-interacting
KMS states, $\Obl$ and $\Obz$, respectively. Then, for any
$\epsilon>0$ there is a $g_1(\epsilon)>0$, which does not depend
on $\beta>0$, such that for
\begin{equation}
|g|<g_1(\epsilon) \left\{
\begin{array}{cl}
\big(1+\log(1+\beta)\big)^{-1} &\mbox{if $p=1/4$}\\
1 &\mbox{if $p>1/4$} \; ,
\end{array}
\right.
\label{f1}
\end{equation}
the following estimate holds,
\begin{equation}
\left\| P_\Obl -P_\Obz\right\|<\epsilon. \label{1}
\end{equation}

%%%%%%%%%%%%%%%%%%%%%%%%%%%%%%%%%%%%%%%%%%%%%%%%%%%%%%%%%%%%%%%%%
%{\it Remark.\ } The constant $g_1(\epsilon)$ depends on the
%spectral gap $E_1-E_0>0$ of $H^\S$, and, if the norms
%$\|G_\alpha\|$ are assumed to satisfy a $d$-independent upper
%bound, then $g_1(\epsilon)$ can be chosen independently of the
%dimension $d$ of $\H^\S$.
%%%%%%%%%%%%%%%%%%%%%%%%%%%%%%%%%%%%%%%%%%%%%%%%%%%%%%%%%%%%%%%%%

We discuss first the proof of Theorem 5.2. Define the conjugate
operator to be $A_f=d\Gamma(i\partial_u)$, the second quantization
of the {\it generator of energy translation}, $i\partial_u$,
on ${\cal F}_+$ and let
\begin{equation}
A_0=i\theta g\left( \Pi I\repsilonbar^2-\repsilonbar^2
I\Pi\right), \label{201}
\end{equation}
where $\Pi=P_0\otimes P_{\Omega_f}$, $\repsilonbar=\Pibar\repsilon$,
$\Pibar=\bbbone-\Pi$, $\repsilon =(\L_0^2+\epsilon^2)^{-1/2}$, and
$\theta,\epsilon>0$ are parameters to be chosen later.

Note that $A_0$ is a bounded operator, and the commutator
$[\L_0,A_0]$ extends to a bounded operator with
\begin{equation}
\left\|[\L_g,A_0]\right\|\leq
C\left(\frac{\theta|g|}{\epsilon}+\frac{
\theta g^2}{\epsilon^2}\right). \label{202}
\end{equation}

Let $N=d\Gamma(\bbbone)$ be the number operator, and define on its
domain ${\mathcal D}(N)$ the operator
\begin{equation}
B=N+g I_1+i[\L_g,A_0], \label{203}
\end{equation}
where
\begin{equation}
I_1= \sum_\alpha \big( G_\alpha\otimes\bbbone^\S\otimes
\varphi(\partial_u\tau_\beta(g_\alpha)) -\bbbone^\S\otimes C^\S
G_\alpha C^\S\otimes\varphi (\partial_ue^{-\beta
u/2}\tau_\beta(g_\alpha))\big). \label{204}
\end{equation}
($B$ corresponds to the quadratic form $i[\L_g,A_f+A_0]$ in the
sense of Kato and $[\L^f,A_f]=N$; see [FM2].) Moreover, for any $\nu>1$, let
\begin{equation*}
{\frak B}_\nu=\{\psi\in \dom(N^{1/2}) : \|\psi\|=1,
\|(N+1)^{1/2}\psi\|\leq\nu\}.
\end{equation*}

%%%%%%%%%%%%%%%%%%%%%%%%%%%%%%%%%%%%%%%%%%%%%%%%%%%%%%%%%%%%%%%%%%%%%%%%%%%
\vspace{0.5cm}

\noindent {\it Theorem 5.4 (Positive commutator estimate)}

Suppose (A1.1) and (A1.2) (see section 4.1). Then there is a choice of the
parameters $\epsilon$ and $\theta$, and a constant
$g_1(\eta)=g_1>0$, independent of $\nu$ and $\beta\geq \beta_0$,
such that, for fixed $\eta$ and $\beta$, and
\begin{equation}
0<|g|<g_1 \left\{
\begin{array}{cl}
\min\left(\frac{1}{1+\log(1+\beta)}, \frac{\nu^{1/\eta-9/2}}{(1+\log(1+\beta))^\eta}\right) &\mbox{if $p=1/4$}\\
1& \mbox{if $p>1/4$} ,
\end{array}
\right. \label{f3}
\end{equation}
the following estimate holds,
\begin{equation} \Pbar_\Obl B\Pbar_\Obl \geq
|g|^{2-\eta}\nu^{3-9\eta/2} \gamma_0\Pbar_\Obl, \label{pc}
\end{equation}
in the sense of quadratic forms on $Ran E_\Delta(\L_g)\cap{\frak
B}_\nu$, where $\Delta$ is any interval around the origin such
that $\Delta \cap\sigma(\L^\S)=\{0\}$, $E_\Delta(\L_g)$ is the
spectral projection, and where $\gamma_0$ is given in \fer{127}.

We will examine $B$ as a quadratic form on ${\frak
B}_{\nu_0}\subset\dom(N^{1/2})$.  For $p>1/4$, $\nu_0$ is independent of $\beta\geq\beta_0$, while for
$p=1/4$, $\nu_0$ diverges logarithmically for large $\beta$.

\vspace{0.5cm}

\noindent {\it Theorem 5.5 (Regularity of eigenvectors and the
Virial Theorem)}

Assume (A1.1) (in section 4.1), and let $\psi_g$ be an eigenvector of $\L_g$. Then
there is a constant $C(p,\beta)<\infty$, not depending on $g$,
such that
\begin{equation}
\|N^{1/2}\psi_g\|\leq C(p,\beta) |g|\, \|\psi_g\|, \label{205}
\end{equation}
and such that for all $\beta\geq \beta_0$ (for any $\beta_0>0$
fixed),
\begin{equation}
\label{cpb1}
C(p,\beta)\leq c_1(p) 
\begin{cases}
1+\log(1+\beta) ,  p=1/4  \\
1 , p>1/4
\end{cases}
 , 
\end{equation}
where $c_1$ is independent of $\beta\geq \beta_0$. Furthermore,
\begin{equation}
\av{B}_{\psi_g}:=\scalprod{\psi_g}{B\psi_g}=0. \label{206}
\end{equation}

Note that the constant $C(p,\beta)$ may be expressed as
\begin{equation*}
\| I_1(N+1)^{-1/2}\|\leq 2\sum_\alpha \|G_\alpha\|\
\|\partial_u\tau_\beta(g_\alpha)\|_{L^2}=:C(p,\beta).
\end{equation*}

We defer the technical proofs to Appendix 3. However, it is
instructive at this point to understand them at a formal level.
Formally expanding the commutator gives,
\begin{equation}
\scalprod{\psi_g}{[\L_g,A_f+A_0] \psi_g} =2 i \Im
\scalprod{L_g\psi_g}{(A_f+A_0)\psi_g}=0. \label{206.1}
\end{equation}
Hence
\begin{eqnarray}
0\geq \av{N}_{\psi_g} -\left| \av{g I_1}_{\psi_g}\right|&\geq& \av{N}_{\psi_g} -C(p,\beta)|g|\, \|\psi_g\|\, \|N^{1/2}\psi_g\| \nonumber\\
&\geq& \frac{1}{2}
\av{N}_{\psi_g}-\frac{1}{2}C(p,\beta)^2g^2\|\psi_g\|^2,
\label{206.2}
\end{eqnarray}
which yields the bound \fer{205}.

We need to control multiple commutators of $\L_g$ with $A_f+A_0$
to make the above statements rigorous. Particularly, the second
and third commutator of $I$ with the dilatation generator $A_f$
need to be well-defined and relatively $N^{1/2}$-bounded. This is
satisfied if
\begin{eqnarray}
\partial_u^j\tau_\beta(g_\alpha) \mbox{\ \ is continuous in $u\in\r$ for $j=0,1,2$,
  and }\label{124}\\
\partial_u^j\tau_\beta(g_\alpha)\in L^2(\r\times
  S^2) \mbox{\ \ for $j=0,1,2,3$}.\ \ \ \ \ \ \ \label{125}
\end{eqnarray}
One can verify that \fer{124}, \fer{125} follow from (A1.1), section 4.1. Let
$p$ and $\phi_0$ be as in assumption (A1.1); then, for $p=3/4, 5/4$,
$p>11/4$, we choose $\phi=2\phi_0$, while for $p=1/4$, we take
$\phi=\pi+2\phi_0$.

The proof Theorem 5.2 (RTE1) follows directly from Theorems 5.4
and 5.5 by {\it reductio ad absurdum}, since, if for $g$
satisfying \fer{f3}, with $\nu=\nu_0$, there were an eigenvector
$\psi_g\in Ran \overline{P}_{\Omega_{\beta ,g}}$, orthogonal to $\Obl$, then
\begin{equation}
0=\av{B}_{\psi_g}\geq |g|^{2-\eta}\nu_0^{3-9\eta/2}\gamma_0 \; ,
\label{207}
\end{equation}
which is a contradiction, since the RHS is strictly positive. For
$p=1/4$ condition \fer{f3}, with $\nu=\nu_0=C[1+\log(1+\beta)]$,
gives \fer{f4}, independently of $\eta$.

Regarding Proposition 5.3, the high-temperature result for bounded
perturbation is relatively simple. For $\epsilon>0$, $\exists
\eta(\epsilon)>0$ such that, if
\begin{equation}
\beta|g|<\eta(\epsilon), \label{47'}
\end{equation}
then inequality \fer{f1} in Proposition 5.3 holds. A proof of this
fact can be given by using the explicit expression for the
perturbed KMS state (chapter 3), and using the Dyson series
expansion to estimate $\|\Obl-\Obz\|$. Condition \fer{47'} comes
from the fact that the term of order $g^n$ in the Dyson series is
given by an integral over an $n$-simplex of size $\beta$.
This result is extended to lower temperatures and unbounded
perturbation by using the decay in (imaginary) time of the field
propagators, chess-board estimates and the H\"older and
Peierls-Bogoliubov inequalities (see [Fr\"{o},DJP] and Appendices 1 and 3). \\

As for Theorem 5.4 (PC Theorem), we want to show that $\dim Ker
\L_g=1$. We know that $\dim Ker \L_g\ge 1$ since
$\Omega_{\beta,g}\in \ker \L_g$. To prove equality, we want to
show that
\begin{equation}
B+\delta P_\Obl\geq \gamma, \label{61}
\end{equation}
for some $\delta\geq \gamma>0$. For $\Delta\subset \r$ an interval
around the origin not containing any non-zero eigenvalue of the
$\L^\S$, we first prove \fer{61} in the sense of quadratic forms
on the spectral subspace of $\L_0$ associated with the interval
$\Delta$. Using this, we then show that
\begin{equation}
\Pbar_\Obl B\Pbar_\Obl \geq \frac{1}{2}\gamma\Pbar_\Obl,
\label{61.1}
\end{equation}
in the sense of quadratic forms on $Ran E_{\Delta'}(\L_g)\cap
{\frak B}_\nu$, where $E_{\Delta'}(\L_g)$ is the spectral
projection of $\L_g$ associated to an interval $\Delta'$,
an arbitrary interval properly contained in $\Delta$.\\

%%%%%%%%%%%%%%%%%%%%%%%%%%%%%%%%%%%%%%%%%%%%%%%%%%%%%%%%%%%%%%%%

\section{RTE for Model B}

We discuss RTE for Model B in chapter 4, section 3, composed of a small
quantum system coupled to reservoir of non-relativistic fermions
by using the PC method discussed in the previous section. The
result is {\it not} uniform in temperature since there is a
divergence in the derivative of the Fermi-Dirac distribution
function at the chemical potential when the temperature is zero.
For a result which is uniform in temperature, one may want to
extend the RG analysis developed in [BFS] to this case. Since {\it
all} of the proofs are almost identical to the previous section
(and those in Appendix 3), we only sketch the main steps.

\vspace{0.5cm}

\noindent {\it Proposition 5.6}

Assume (B.2) (in section 4.3). There is an $\epsilon_0>0$, independent of
$\beta\geq\beta_0$ (for any $\beta_0$ fixed), such that if
$0<\epsilon<\epsilon_0$ then
\begin{equation}
\Pi I\frac{\epsilon}{\L_0^2+\epsilon^2} I\Pi \geq \Gamma_0\Pi
-C\epsilon^{1/4}, \label{126'}
\end{equation}
where $C$ is a constant independent of the inverse temperature
$\beta$, $\Pi=P_0\otimes P_{\Omega_f}$, the projection onto the
kernel of $\L_0$, and $\Gamma_0$ is a bounded operator on
$\H=\H^{\S}\otimes\H^{\S}\otimes{\cal F}_-$, acting trivially on
the last factor, ${\cal F}_-$, and leaving $Ker \L^{\S}$
invariant. Furthermore, $\Gamma_0$ restricted to $Ker \L^{\S}$ has
zero as a simple eigenvalue, with Gibbs state $\Omega_\beta^{\S}$
as eigenvector, and is strictly positive on the complement of
$\cx\Omega_\beta^{\S}$. There is a constant $\gamma_0>0$,
independent of $0<\beta<\infty$, such that
\begin{equation}
\Gamma_0\upharpoonright_{Ran \Pbar_{\Omega_\beta^{\S}}}\geq
\gamma_0, \label{127'}
\end{equation}
where, $\Pbar_{\Omega_\beta^{\S}}=\bbbone -P_{\Omega_\beta^{\S}}$
and $P_{\Omega_\beta^{\S}}$ is the projection onto $\cx
\Omega_\beta^{\S}$.

The proof is similar to that of Proposition 5.1. Explicitly,
\begin{eqnarray*}
\lefteqn{
\Gamma_0\upharpoonright_{\ran \Pbar_{\Omega_\beta^{\S}}}}\\
&&\geq \min_{E_m\neq E_n} \frac{e^{\beta E_n}}{e^{-\beta(E_m-E_n-\nu)}}
\int du d\sigma \delta(E_m-E_n-u) \left| \sum_\alpha
  \scalprod{\varphi_m}{G_\alpha \varphi_n}
  g_\alpha\left(|u|,\sigma\right)\right|^2,
\end{eqnarray*}
and $\gamma_0$ in \fer{127'} is obtained by minimizing the RHS.

The main result of this section is the following theorem, which
says that Model B (with the corresponding assumptions B.1 and B.2)
possesses the property of return to equilibrium.

\vspace{0.5cm}

\noindent {\it Theorem 5.7(RTE2)}

Assume conditions (B.1) and (B.2) hold (see section 4.3). Then there is a constant
$g_1>0$, independent of $\beta_0\le\beta\le \beta_1$, for any
$\beta_1>\beta_0>0$ fixed such that, for
\begin{equation}
0<|g| <g_1 \; , \label{f4'}
\end{equation}
the kernel of $\L_g$ is spanned by the interacting KMS vector
$\Obl$, and the system has the property of return to equilibrium.

The proof relies on the following three propositions.

\vspace{0.5cm}

\noindent {\it Proposition 5.8}

Suppose (B.1) holds, and let $P_\Obl$ and $P_\Obz$ denote the
projections onto the spans of the interacting and non-interacting
KMS states, $\Obl$ and $\Obz$, respectively. Then, for any
$\epsilon>0$ there is a $g_1(\epsilon)>0$, which does not depend
on $\beta$, such that, for
\begin{equation}
|g|<g_1(\epsilon) \; , \label{f1'}
\end{equation}
the following estimate holds
\begin{equation}
\left\| P_\Obl -P_\Obz\right\|<\epsilon. \label{1}
\end{equation}

Since we are only interested in strictly positive temperatures,
the proof of this statement follows directly from a Dyson series
expansion (high-temperature result); (see remark after Theorem
5.5).

Recall $N=d\Gamma(\bbbone)$ is the number operator, and define on
its domain ${\mathcal D}(N)$ the operator
\begin{equation}
B=N+g I_1+i[\L_g,A_0], \label{203'}
\end{equation}
where
\begin{equation}
I_1= \sum_\alpha \big( G_\alpha \otimes {\mathbf 1}^\S \otimes
\varphi(\partial_u g_{\beta,\alpha}) -i {\mathbf 1}^\S\otimes C^\S
G_\alpha C^\S\otimes (-1)^N \varphi (\partial_u
 g_{\beta,\alpha}^{\#}) \big). \label{204'}
\end{equation}

\vspace{0.5cm}

\noindent {\it Proposition 5.9 (Positive commutator estimate)}

Suppose (B.1) and (B.2) (in section 4.3). Then there is a choice of the parameters
$\epsilon$ and $\theta$, and a constant $g_1(\eta)=g_1>0$,
independent of $\nu$ and $\beta\in [\beta_0, \beta_1]$, such that,
for fixed $\eta$ and $\beta$, and
\begin{equation}
0<|g|<g_1 \label{f3'}
\end{equation}
the following estimate holds,
\begin{equation} \Pbar_\Obl B\Pbar_\Obl \geq
|g|^{2-\eta}\nu^{3-9\eta/2} \gamma_0\Pbar_\Obl, \label{pc}
\end{equation}
in the sense of quadratic forms on $Ran E_\Delta(\L_g)\cap{\frak
B}_\nu$, where $\Delta$ is any interval around the origin such
that $\Delta \cap\sigma(\L^\S)=\{0\}$, $E_\Delta(\L_g)$ is the
spectral projection, and where $\gamma_0$ is given in \fer{127'}.

\vspace{0.5cm}

\noindent {\it Proposition 5.10 (Regularity of eigenvectors and the Virial
Theorem)}

Assume (B.1) (in section 4.3), and let $\psi_g$ be an eigenvector of $\L_g$. Then
\begin{equation}
\av{B}_{\psi_g}:=\scalprod{\psi_g}{B\psi_g}=0, \label{206'}
\end{equation}
where $B$ has been defined in \fer{203'}

The proof of Theorem 5.7 follows from Propositions 5.9 and 5.10 by
contradiction (similar to the proof of Theorem 5.3). Moreover, the
proof of Propositions 5.9 and 5.10 is very similar to the proof of
Theorems 5.4 and 5.5 (see Appendix 3).

%%%%%%%%%%%%%%%%%%%%%%%%%%%%%%%%%%%%%%%%%%%%%%%%%%%%%%%%%%%%%%%%%%%
\section{Appendix 3}

\subsection{Proof of Theorem 5.4} \vspace{0.5cm}

\noindent {\it Positive commutator estimate localized with respect
to $\L_0$.}

Consider the decomposition
\begin{equation}
\H_\Delta^0:=Ran E_\Delta^0=Ran E_\Delta^0\Pi\oplus Ran
E_\Delta^0\Pibar, \label{62}
\end{equation}
where
\begin{equation}
\Pi= P_0\otimes P_{\Omega_f} \label{125'}.
\end{equation}
and where $E_\Delta^0$ is the spectral projection of $\L_0$
associated with the interval $\Delta$. First, we to prove a 
lower bound on $E_\Delta^0\Pibar(B+\delta P_\Obl)\Pibar
E_\Delta^0$.

(Remark regarding the notation: $C$ will denote a constant
independent of $g, \theta,\epsilon, \beta\geq\beta_0>0$, and
$C(p,\beta)$ will denote a constant independent of
$g,\theta,\epsilon$, satisfying the bound given in \fer{cpb1}.) We
know that
\begin{equation}
\|\edeltanot\Pibar [\L_g,A_0]\Pibar\edeltanot\| \leq C\frac{\theta
g ^2}{\epsilon^2} \; , \label{64}
\end{equation}

where $\Pibar=\Pbar_0\otimes P_{\Omega_f} +\Pbar_{\Omega_f}$ (and
notice that $R E_\Delta^0\Pibar\subset \ran \Pbar_{\Omega_f}$). Hence,
\begin{eqnarray}
E_\Delta^0\Pibar(B+\delta P_\Obl)\Pibar E_\Delta^0&=&
E_\Delta^0\Pibar N^{1/2}\left( \bbbone +N^{-1/2}g
I_1N^{-1/2}\right) N^{1/2}\Pibar
E_\Delta^0\nonumber\\
&& + \edeltanot\Pibar \left( i[\L_g,A_0] +\delta P_\Obl\right)
\Pibar\edeltanot\nonumber\\
&\geq& \frac{1}{2}\edeltanot \Pibar +\edeltanot \Pibar
i[\L_g,A_0]\Pibar\edeltanot\nonumber\\
&\geq& \frac{1}{2}\left(1-C\frac{\theta g^2}{\epsilon^2}
\right)\edeltanot\Pibar, \label{63}
\end{eqnarray}
if
\begin{equation}
\|\Pbar_{\Omega_f} N^{-1/2}g I_1 N^{-1/2}\Pbar_{\Omega_f}\|\leq
C(p,\beta) |g| <1/2, \label{parcond1}
\end{equation}

Choosing the parameters such that
\begin{equation}
C\frac{\theta g^2}{\epsilon^2}<1/2 , \label{parcond2}
\end{equation}
it follows that
\begin{equation}
\edeltanot\Pibar(B+\delta
P_\Obl)\Pibar\edeltanot\geq\frac{1}{4}\edeltanot\Pibar. \label{65}
\end{equation}

%%%%%%%%%%%%%MOTIVATION FOR USING THE FESHBACH MAP

In order to proceed further, we apply the Feshbach map (as was done, for example, in [BFS, BFSS]). The so called isospectrality of the Feshbach map is a very useful property which is central for our application.\footnote{see Appendix 2, chapter 4 for a discussion and proof of the isospectrality of the Feshbach map.}

Consider the Feshbach map (with the spectral parameter
$m<1/8$), applied to
\begin{equation}
\edeltanot(B+\delta P_\Obl)\edeltanot \label{65.1}
\end{equation}
 as an operator on the Hilbert space $\H_\Delta^0$:
\begin{eqnarray}
\lefteqn{ F_{\Pi,m}(\edeltanot(B+\delta P_\Obl)\edeltanot)=
\edeltanot\Pi\Big( B+\delta P_\Obl}\nonumber\\
&& -(B+\delta P_\Obl) \edeltanot\Pibar
  \left(\overline{B+\delta P_\Obl}-m\right)^{-1} \Pibar \edeltanot(B+\delta
  P_\Obl)\Big) \Pi\edeltanot,\ \ \ \ \
\label{66}
\end{eqnarray}
(the barred operator is understood to be restricted to the
subspace $Ran \edeltanot\Pibar\subset \H_\Delta^0$). Since $\Pi
I_1\Pi=0$, it follows that
\begin{equation}
\Pi B \Pi = 2\theta g^2\Pi I\repsilonbar^2 I\Pi\geq 0. \label{66'}
\end{equation}

First, we want to show that the second term in \fer{66} is smaller
than $\Pi B\Pi$. Using the fact that $\| \left(\overline{B+\delta
P_\Obl}-m\right)^{-1} \|<8$ for $m<\frac{1}{8}$ and \fer{65},
$\|\L_0\repsilonbar\|\leq 1$, $\|\repsilonbar^2\|\leq
\epsilon^{-2}$ and $\Pibar i[\L_g,A_0]\Pi=\theta\lambda\Pibar
\L_g\repsilonbar^2I\Pi$, it follows that
\begin{eqnarray}
\lefteqn{ 8\| \edeltanot\Pibar (g
I_1 +i[\L_g,A_0]+\delta P_\Obl)\Pi\psi\|^2}\nonumber\\
&\leq& 16\theta^2 g^2 \|\repsilonbar I\Pi\psi\|^2 \nonumber\\
&&+ C\left(\delta^2\|\Pibar P_\Obl \Pi \|^2+
  C(p,\beta) g^2+\frac{\theta^2g^4}{\epsilon^4}\right)\|\psi\|^2.
\label{67}
\end{eqnarray}
It follows that
\begin{eqnarray}
\lefteqn{
\av{F_{\Pi,m}(\edeltanot(B+\delta P_\Obl)\edeltanot)}_\psi}\nonumber\\
&\geq& 2\theta g^2(1-8\theta)\av{\Pi I\repsilonbar^2I
  \Pi}_\psi +\delta \|P_\Obl \Pi\psi\|^2\nonumber \\
&& -C\frac{\theta
g^2}{\epsilon}\left(\frac{\epsilon}{\theta}C(p,\beta)+\frac{\theta
g^2}{\epsilon^3} +\frac{\epsilon}{\theta g^2}\delta^2\|\Pibar
  P_\Obl \Pi\|^2\right)\|\psi\|^2 \; ,
\label{68}
\end{eqnarray}
for any $\psi\in\H_\Delta^0$. Moreover,
\begin{equation}
\Pi I\repsilonbar^2 I\Pi\geq \frac{1}{\epsilon}\left(\Gamma_0
  -C\epsilon^{1/4}\right),
\label{69}
\end{equation}
if $\epsilon<\epsilon_0$. Choose $\epsilon$ and $\theta$ such that
\begin{equation}
\theta<1/16,\ \epsilon<\epsilon_0. \label{parcond3}
\end{equation}
For $\psi\in Ran\Pi$,
\begin{eqnarray}
\lefteqn{ \theta g^2\av{I\repsilonbar^2
  I+\frac{\delta}{\theta g^2} P_\Obl}_{\psi} \geq
  \frac{\theta g^2}{
  \epsilon}\av{\gamma_0\Pbar_{\Omega_\beta^\S}
  +\frac{\epsilon\delta}{\theta g^2} P_\Obl
  -C\epsilon^{1/4}}_\psi}\nonumber\\
&=&\frac{\theta g^2}{\epsilon}\gamma_0 \left[
  \left(1-C\epsilon^{1/4}/\gamma_0\right) \|\psi\|^2
  +\av{\frac{\epsilon\delta}{\theta g^2\gamma_0}P_\Obl
    -P_\Obz}_\psi\right],
\label{71}
\end{eqnarray}
since $P_{\Omega_\beta^\S}\psi=P_\Obz\psi$ for $\psi\in\ran \Pi$.
We choose
\begin{equation}
\delta\geq \frac{\theta g^2}{\epsilon}\gamma_0\geq \frac{\theta
g^2}{4\epsilon} \gamma_0=: \gamma, \label{parcond3'}
\end{equation}
and
\begin{equation}
C\frac{\epsilon^{1/4}}{\gamma_0}<1/4. \label{parcond4}
\end{equation}

Now, from Proposition 5.3 , $\| P_\Obl-P_\Obz \| < \frac{1}{4}$ if
\begin{equation} \mbox{$g$ satisfies condition \fer{f1}
(with $\epsilon=1/4$).} \label{cond}
\end{equation}
One may use this estimate to find a lower bound for the R.H.S. of
\fer{71}
\begin{equation}
\frac{\theta g^2}{\epsilon}\gamma_0\Big( 3/4- \|
P_\Obl-P_\Obz\|\Big) \|\psi\|^2\geq \frac{\theta
g^2}{2\epsilon}\gamma_0 \ \|\psi\|^2. \label{72}
\end{equation}
Furthermore,
\begin{equation*}
\|\Pibar P_\Obl\Pi\|^2 = \|\Pibar( P_\Obl-P_\Obz)\Pi\|^2\leq
\|P_\Obl-P_\Obz\|^2 \; ,
\end{equation*}
and with suitable bounds on the parameters  $\epsilon, g, \theta$,
one may choose them such that
\begin{equation}
C\left( \frac{\epsilon}{\theta}C(p,\beta)+ \frac{\theta
g^2}{\epsilon^3} +\frac{\epsilon\delta^2}{\theta
g^2}\right)<\gamma_0/4. \label{parcond5}
\end{equation}
Combining all of these estimates, we obtain
\begin{equation}
\av{F_{\Pi,m}(\edeltanot(B+\delta P_\Obl)\edeltanot)}_\psi\geq
\frac{\theta g^2}{4\epsilon}\gamma_0\ \|\psi\|^2 . \label{73}
\end{equation}

%%%% WE NEED TO PROVE THE ISOSPECTRALITY OF THE FESHBACH MAP

From the isospectrality of the Feshbach map (see [BFSS,BFS]), it
follows
\begin{equation}
\edeltanot (B+\delta P_\Obl)\edeltanot \geq\min\left(\frac{1}{8},
  \frac{\theta g^2}{4\epsilon}\gamma_0\right)\edeltanot =
  \frac{\theta g^2}{4\epsilon}\gamma_0\edeltanot.
\label{74}
\end{equation}
We will use the above estimate in the following.\\

\noindent {\it Positive commutator estimate localized with respect
to $\L_g$}

Denote by $0\leq \chi_\Delta\leq 1$ a smooth function with support
inside the interval $\Delta$ such that $\chi_\Delta(0)=1$,  and
let $\chi_\Delta^0=\chi_\Delta(\L_0)$ and
$\chi_\Delta=\chi_\Delta(\L_g)$ the operators obtained from the
spectral theorem. We will show that any vector $\psi\in Ran
\Pbar_\Obl\cap {\frak B}_\nu$, such that $\chi_\Delta\psi=\psi$,
satisfies
\begin{equation}
\av{B +\delta P_\Obl}_{\psi}= \av{B}_{\psi}\geq \frac{\theta g
^2}{8\epsilon}\gamma_0, \label{80}
\end{equation}
for appropriate bounds on the parameters $g, \epsilon$ and
$\theta$. One estimate from functional calculus that we use is
\begin{eqnarray}
\|(1-\chi_\Delta^0)\psi\|
=\|(\chi_\Delta-\chi_\Delta^0)\psi\|&\leq&
C|g| \ \|I(N+1)^{-1/2}\|\ \|(N+1)^{1/2}\psi\|\nonumber\\
&\leq& C \nu |g|, \label{81}
\end{eqnarray}
Decompose $\av{B}_{\psi}$ into three terms, which we will estimate
individually:
\begin{eqnarray}
\av{B}_{\psi}&=& \av{\chideltanot (B+\delta
  P_\Obl)\chideltanot}_{\psi} \label{82}\\
&& + \av{(1-\chideltanot)(B+\delta
  P_\Obl)(1-\chideltanot)}_{\psi} \label{83}\\
&&+ 2\, {\rm Re}\av{(1-\chideltanot)(B+\delta
P_\Obl)\chideltanot}_{\psi}. \label{84}
\end{eqnarray}

Using \fer{74} and the fact that
$\edeltanot\chideltanot=\chideltanot$, the first term is bounded
from below
\begin{equation}
\av{\chideltanot (B+\delta P_\Obl)\chideltanot}_{\psi} \geq
\frac{\theta g^2}{4\epsilon}\gamma_0 \| \chideltanot\psi\|^2 \geq
\frac{\theta g^2}{4\epsilon}\gamma_0(1-C\nu|g|) \|\psi\|^2.
\label{85}
\end{equation}
Since $N+\delta P_\Obl \ge0$ is non-negative, the second term is
also bounded from below by
\begin{eqnarray}
\fer{83} &\geq& - \left| \av{(1-\chideltanot)(g I_1 +i[\L_g,A_0])
    (1-\chideltanot)}_{\psi} \right| \nonumber\\
&\geq& -|g|\ \|(1-\chideltanot)\psi\|\ \|I_1(N+1)^{-1/2}\|\
\|(N+1)^{1/2}\psi\| \nonumber\\
&& -\|[\L_g,A_0]\|\ \|(1-\chideltanot)\psi\|^2\nonumber\\
&\geq&- C\nu^2 \frac{\theta g^2}{\epsilon} \left(
\frac{\epsilon}{\theta}C(p,\beta) +
  |g|+\frac{g^2}{\epsilon}\right).
\label{86}
\end{eqnarray}

As for the third term, since $N$ commutes (strongly) with
$\chideltanot$ and $\psi\in Ran\Pbar_{\Obl}$, it follows that
\begin{eqnarray}
\lefteqn{ {\Re}\av{(1-\chideltanot)(B+\delta
  P_\Obl)\chideltanot}_{\psi}}\nonumber\\
&\geq& \delta \av{(1-\chideltanot)P_\Obl
  (\chideltanot-1)}_{\psi}
+{\Re} \av{(1-\chideltanot)(g I_1
  +i[\L_g,A_0])\chideltanot}_{\psi}  \nonumber\\
&\geq& -\delta\|(1-\chideltanot)\psi\|^2 -C(p,\beta)\nu g^2
\|\psi\|^2 -\left|
\av{(1-\chideltanot)[\L_g,A_0]\chideltanot}_{\psi}\right|.
\label{87}
\end{eqnarray}
Moreover, $(1-\chideltanot)\Pi=0$ and $\|(1-\chideltanot)
\L_0^{-1}\|\leq C$, and hence the last term can be estimated by
\begin{eqnarray}
\lefteqn{ \left|
  \av{(1-\chideltanot)[\L_g,A_0]\chideltanot}_{\psi}\right|}\nonumber\\
&=&\theta|g|\, \left| \av{(1-\chideltanot)(g I\Pi I\repsilonbar^2
 -\L_g\repsilonbar^2 I\Pi +g \repsilonbar^2
 I\Pi I)\chideltanot}_{\psi}  \right| \nonumber\\
&&\leq C\nu\theta|g|\left( \frac{g^2}{\epsilon^2} +
  |g|\right)\|\psi\|^2 =C\nu\frac{\theta g^2}{\epsilon} \left(
  \frac{|g|}{\epsilon} +\epsilon\right) \|\psi\|^2.
\label{88}
\end{eqnarray}
Substituting back,
\begin{eqnarray}
\lefteqn{
\av{B}_{\psi}\geq}\label{89}\\
&& \frac{\theta g^2}{4\epsilon}\gamma_0 \left(
  (1-C\nu|g|) -\frac{C\nu}{\gamma_0}\left(\nu\frac{\epsilon}{\theta}C(p,\beta) +\nu |g| +\nu\frac{g^2}{\epsilon}
  +\frac{|g|}{\epsilon} +\epsilon\right)\right)\, \|\psi\|^2.
\nonumber
\end{eqnarray}
To arrive at inequality (5.55), choose the parameters such that
\begin{equation}
C\nu|g|<1/4 \mbox{\ \ \ and\ \ \ }
\frac{C\nu}{\gamma_0}\left(\nu\frac{\epsilon}{\theta}C(p,\beta)
+\nu |g| +\nu\frac{g^2}{\epsilon}
  +\frac{|g|}{\epsilon} +\epsilon\right) <1/4.
\label{parcond6}
\end{equation}

Indeed,
\begin{equation}
\mbox{\fer{parcond1}, \fer{parcond2}, \fer{parcond3},
\fer{parcond3'}, \fer{parcond4}, \fer{cond}, \fer{parcond5},
\fer{parcond6}} \label{conditions}
\end{equation}
 can be simultaneously satisfied: If we set
\begin{eqnarray}
g&=&\nu^{-9/2}g', \label{89.9}\\
\epsilon&=&\nu^{-3}|g'|^e, \mbox{\ \ \ \ some $0<e<1$,}\label{90}\\
\theta&=& |g'|^t, \mbox{\ \ \ \ some $0<t<e<1$ such that $t>3e-2$,}\label{91}\\
\delta&=&\frac{\theta g^2}{\epsilon}\gamma_0, \label{92}
\end{eqnarray}
there is a $g_1>0$, depending on $e,t$, but not on $\nu$,
$\beta\geq\beta_0$, such that if
\begin{equation}
0<|g|<g_1\min\left( C(p,\beta)^{-1},
\nu^{1/\eta-9/2}C(p,\beta)^{-1/\eta}\right), \label{f2}
\end{equation}
where $\eta=e-t>0$, then conditions \fer{conditions} are
satisfied.

\subsection{Proof of Proposition 5.3}

First, we use the fact that the trace-norm majorizes the
operator-norm,
\begin{eqnarray}
\|P_\Obl-P_\Obz\|^2&\leq& \|P_\Obl-P_\Obz\|^2_2=2\left(
1-\scalprod{\Obl}{P_\Obz
    \Obl}\right)\nonumber\\
&\leq& 2\scalprod{\Obl}{ \Pbar_{\Omega_\beta^\S}
  \Obl} +2\scalprod{\Obl}{ \Pbar_{\Omega_f} \Obl},
\label{47}
\end{eqnarray}
since $\bbbone- P_\Obz\leq \Pbar_{\Omega_\beta^\S}
+\Pbar_{\Omega_f}$. (Recall $\Omega_\beta^\S$ is the vector
corresponding to the Gibbs state of $\S$ at inverse temperature
$\beta$ , and $\Omega_f$ is the vacuum vector in $\cal F_+$.) We
know that
\begin{equation}
\scalprod{\Obl}{ \Pbar_\Omega \Obl}\leq \|N^{1/2}\Obl\|^2\leq
c(p,\beta)^2 |g|^2, \label{48}
\end{equation}
where $c(p,\beta)$ satisfies \fer{cpb1}. There is a
$\beta_1(\epsilon)\geq \beta_0$ such that if
$\beta>\beta_1(\epsilon)$ then
\begin{equation}
\|P_{\Omega_\beta^\S}-P_{\varphi_0\otimes\varphi_0}\| <
\epsilon/2, \label{49}
\end{equation}
where $\varphi_0$ is the groundstate eigenvector of $H^\S$ and
$P_{\varphi_0\otimes\varphi_0}\in{\cal B}(\H^\S\otimes \H^\S)$ is
the projection onto the span of $\varphi_0\otimes\varphi_0$. Hence
\begin{equation}
\|P_\Obl-P_\Obz\|^2 \leq
2\scalprod{\Obl}{\Pbar_{\!\!{\varphi_0\otimes
\varphi_0}}\Obl}+\epsilon +2c(p,\beta)^2 |g|^2, \label{50'}
\end{equation}
for $\beta>\beta_1(\epsilon)$. Let
\begin{equation}
Q=\Pbar_{\varphi_0}\in{\cal B}(\H^\S) \label{49'}
\end{equation}
be the projection onto the orthogonal complement of the
groundstate subspace of $H^\S$ so that
\begin{equation}
\Pbar_{\varphi_0\otimes\varphi_0}\leq Q\otimes\bbbone^\S +
\bbbone^\S\otimes Q. \label{51}
\end{equation}
Since $\scalprod{\Obl}{Q\otimes\bbbone^\S\
  \Obl}=\scalprod{\Obl}{\bbbone^\S\otimes Q\ \Obl}=\obl(Q)$,
\begin{equation}
\|P_\Obl-P_\Obz\|^2 \leq 4\obl(Q)+\epsilon +2c(p,\beta)^2 |g|^2,
\label{50}
\end{equation}
for $\beta>\beta_1(\epsilon)$.

We claim that for any $\epsilon>0$ there exist
$\beta_2(\epsilon)>0$ and $g_1(\epsilon)>0$ such that if $\beta
> \beta_2(\epsilon)$ and $|g|<g_1(\epsilon)$ then
\begin{equation}
\obl(Q)<\epsilon. \label{2}
\end{equation}
The proof is given in a while. For now, we use (\ref{2}) to prove
Proposition 5.3. Set
\begin{eqnarray}
\beta_3(\epsilon)&:=&\max(\beta_1(\epsilon),
\beta_2(\epsilon)),\nonumber\\
g'_0(\epsilon)&:=&\min\left(g_1(\epsilon),
c(p,\beta)^{-1}\sqrt{\epsilon/2},\eta(\epsilon)/\beta_3(\epsilon)
\right), \label{52}
\end{eqnarray}
When $p>1/4$, $c(p,\beta)$ has an upper bound which is uniform in
$\beta\geq \beta_0$, and we take $g_0(\epsilon)$ to be the RHS of
\fer{52} with $c(p,\beta)$ replaced by this upper bound. For
$p=1/4$ we can find a $g_0(\epsilon)$, independent of $\beta>0$,
satisfying
$(1+\log(1+\beta))^{-1}g_0(\epsilon)\leq g_0'(\epsilon)$. \\

Estimate \fer{50} and \fer{2} give
\begin{equation}
\|P_\Obl-P_\Obz\|^2 \leq 6\epsilon, \label{53}
\end{equation}
for $\beta>\beta_3(\epsilon)$. If $\beta\leq\beta_3(\epsilon)$
then $\beta|\lambda|< \eta(\epsilon)$, and \fer{1} follows from
the high-temperature result mentioned above. This completes
the proof of the theorem, given our claim (\ref{2}).\\

We now turn to proving claim \fer{2}. Without loss of generality,
we work with a finite volume approximation
\begin{equation}
\oblL(\cdot)=\frac{\tr \left(e^{-\beta\HlL}\,\cdot\,\right)}{\tr
e^{-\beta \HlL}}\; , \label{3}
\end{equation}
of the KMS state $\obl$, where $\Lambda= [-L/2,L/2]^3\subset\r^3$.
The trace is taken over the Hilbert space $\H^\S\otimes {\cal
F_+}(L^2(\Lambda,d^3x))$. For $n=(n_1,n_2,n_3)\in{\mathbb
  Z}^3$, let
\begin{equation}
e^\Lambda_n(x)=L^{-3/2}e^{2\pi i n x/L}, \
E_n^\Lambda=(\frac{2\pi}{L}|n|)^2=(\frac{2\pi}{L})^2(n_1^2+n_2^2+n_3^2)
\label{6}
\end{equation}
denote the eigenvectors and eigenvalues of the operator $-\Delta$
on $L^2(\Lambda,d^3x)$ with periodic boundary conditions at
$\partial \Lambda$. We identify the basis $\{e^\Lambda_n\}$ of
$L^2(\Lambda^3,d^3x)$ with the canonical basis of $l^2({\mathbb
Z}^3)$, and define the finite-volume Hamiltonian by
\begin{eqnarray}
\HlL&=& H^\S+\HfL+\lambda v^\Lambda,\label{4}\\
v^\Lambda&=&\sum_\alpha
G_\alpha\otimes\varphi(g_\alpha^\Lambda),\label{5}
\end{eqnarray}
where $g_\alpha^\Lambda\in l^2({\mathbb Z}^3)$ is given by
\begin{equation}
g_\alpha^\Lambda(n)=\left(\frac{2\pi}{L}\right)^{3/2} \left\{
\begin{array}{ll}
g_\alpha\left(\frac{2\pi n}{L}\right), & n\neq 0,\\
1,& n=0.
\end{array}
\right. \label{6.1}
\end{equation}
and the operator
\begin{equation}
\HfL=\d\Gamma(\hfL), \label{7}
\end{equation}
acting on ${\cal F}_+(l^2({\mathbb Z}^3))$, is the second
quantization of the one-particle Hamiltonian
\begin{equation}
\hfL e_n^\Lambda=\left\{
\begin{array}{cl}
E_n^\Lambda e_n^\Lambda, & \mbox{if $n\neq(0,0,0)$},\\
e_n^\Lambda, & \mbox{if $n=(0,0,0)$}.
\end{array}
\right. \label{8}
\end{equation}
On the complement of the zero-mode subspace $\hfL$ equals
$-\Delta$ with periodic boundary conditions. The thermodynamic
limit isn't affected by changing the action of $\hfL$ on finitely
many modes assuming $e^{-\beta \HfL}$ is trace-class. Similarly,
we may modify the definition of $g_\alpha^\Lambda$ on finitely
many modes without altering the thermodynamic limit. The existence
of the thermodynamic limit,
\begin{equation}
\lim_{L\rightarrow\infty} \oblL(A)=\obl(A), \label{9}
\end{equation}
can be proven by expanding $e^{-\beta \HlL}$ into a Dyson
(perturbation) series and using
\begin{equation}
\obzL(A)=\frac{\tr\left( e^{-\beta\HzL}A \right)}{\tr e^{-\beta
    \HzL}}
\label{9.1}
\end{equation}
has the expected thermodynamic limit for quasi-local observables $A$. \\

We want to show that $\oblL(Q)<\epsilon$, for $Q$ given in
\fer{49'}, provided $\beta$ and $g$ satisfy the conditions given
in our claim (\ref{2}), {\it uniformly} in the size of $\Lambda$.
We use the H\"older and Peierls-Bogoliubov inequalities. The
H\"older inequality (for traces) reads
\begin{equation}
\|A_1\ldots A_n\|_1\leq\prod_{j=1}^n \|A_j\|_{p_j}, \label{holder}
\end{equation}
where $1\leq p_j\leq\infty$, $\sum_j p_j^{-1}=1$, and the norms
are
\begin{equation}
\|A\|_p=\left(\tr |A|^p\right)^{1/p}, \mbox{for $p<\infty$, and }
\|A\|_\infty=\|A\| \mbox{\ (operator norm)}. \label{10}
\end{equation}
The Peierls-Bogoliubov inequality says that
\begin{equation}
\frac{\tr\left(e^{A+B}\right)}{\tr e^B}\geq \exp\left[ \tr
\left(Ae^B\right)/\tr e^B\right], \label{peierls}
\end{equation}
which implies that
\begin{equation}
\frac{\tr e^{-\beta\HzL}}{\tr e^{-\beta\HlL}}\leq e^{\beta|g
\obzL(v^\Lambda)|}=1, \label{applipeierls}
\end{equation}
since, by \fer{5}, $\obzL(v^\Lambda)=0$.\\

Using the H\"older inequality,
\begin{eqnarray}
\lefteqn{
\oblL(Q)=\frac{\tr\left(e^{-(\beta-2\tau)\HlL}e^{-\tau\HlL}Qe^{-\tau
      \HlL}\right) }{\tr e^{-\beta\HlL}}}\nonumber\\
&\leq& \left[\frac{\tr \left\{\left(e^{-\tau\HlL}Qe^{-\tau\HlL
      }\right)^{\frac{\beta}{2\tau}}\right\}}{\tr e^{-\beta\HlL}}\right]
      ^{\frac{2\tau}{\beta}}
= \left[\frac{\tr \left\{\left(Qe^{-\frac{\beta}{2M}\HlL}Q
      \right)^{2M}\right\}}{\tr e^{-\beta\HlL}}\right]
      ^{\frac{1}{2M}}\!\!\!,\ \ \ \ \ \ \ \
\label{11}
\end{eqnarray}
where we are choosing $\tau$ s.t.
\begin{equation}
\frac{\beta}{2\tau}=2M,\ \ \mbox{for some $M\in{\mathbb N}$}.
\label{12}
\end{equation}
Setting
\begin{equation}
v^\Lambda(t)=e^{-t\HzL}v^\Lambda e^{t\HzL} \label{12'}
\end{equation}
and using the Dyson series expansion one gets
\begin{equation}
Q e^{-\frac{\beta}{2M}\HlL}Q = A+B, \label{13}
\end{equation}
where the selfadjoint operators $A$ and $B$ are given by
\begin{eqnarray}
A&=& Qe^{-\frac{\beta}{2M}\HzL}Q\label{14}\\
B&=& \sum_{n\geq 1} (-g)^n\int_{0\leq t_n\leq \ldots\leq t_1\leq
  \frac{\beta}{2M}} Qv^g(t_n)\cdots
v^\Lambda(t_1)e^{-\frac{\beta}{2M}\HzL} Q\  dt_1\cdots dt_n.\ \ \
\ \ \ \ \ \label{15}
\end{eqnarray}

Substituting \fer{13} into \fer{11}, and using the H\"older
inequality,
\begin{equation}
\oblL(Q)\leq \left[ \frac{\tr \left(|A|^{2M}\right)}{\tr
    e^{-\beta\HlL}}\right]^{\frac{1}{2M}} +
\left[ \frac{\tr \left(|B|^{2M}\right)}{\tr
    e^{-\beta\HlL}}\right]^{\frac{1}{2M}}.
\label{16}
\end{equation}

Consider first the first term of \fer{16}. Let $\Delta=E_1-E_0>0$,
the spectral gap of $H^\S$. Then
\begin{eqnarray}
\frac{\tr\left( |A|^{2M}\right)}{\tr e^{-\beta\HzL}}&=&
\frac{\tr_{\H^\S}\left( Qe^{-\beta H^\S}\right)}{\tr_{\H^\S}
e^{-\beta H^\S}} =\frac{\sum_{j=1}^{d-1}
e^{-\beta(E_j-E_0)}}{1+\sum_{j=1}^{d-1} e^{-\beta
    (E_j-E_0)}}\nonumber\\
&\leq& \sum_{j=1}^{d-1} e^{-\beta(E_j-E_0)}\leq
2\int_{E_1-E_0}^\infty e^{-\beta
x}dx=2\frac{e^{-\beta\Delta}}{\beta}. \label{17}
\end{eqnarray}

Taking into account \fer{applipeierls} and \fer{12}, we obtain,
for $\beta\geq 1$,
\begin{equation}
\left[ \frac{\tr \left(|A|^{2M}\right)}{\tr
    e^{-\beta\HlL}}\right]^{\frac{1}{2M}}\leq
    2e^{-2\tau\Delta}.
\label{18}
\end{equation}
To make the RHS small, take $\tau\gg\Delta^{-1}$

Consider now the second term on the RHS of \fer{16}. From
\fer{applipeierls},
\begin{equation}
\frac{\tr\left( |B|^{2M}\right)}{\tr e^{-\beta \HlL}}\leq
\obzL\left(
  e^{\beta\HoL} |B|^{2M}\right)=\obzL\left( e^{\beta\HoL} B^{2M}\right).
\label{19}
\end{equation}
Expanding,
\begin{equation}
e^{\beta\HoL} B^{2M} =\sum_{k_1,\ldots, k_{2M}\geq 1}
T(k_1,\ldots,k_{2M}), \label{20}
\end{equation}
where
\begin{eqnarray}
\lefteqn{ T(k_1,\ldots,k_{2M}) =(-\lambda)^{k_1+\cdots +k_{2M}}
\int_0^{\frac{\beta}{2M}} dt^{(1)}_1\cdots
\int_0^{t^{(1)}_{k_1-1}}
dt^{(1)}_{k_1}} \nonumber\\
&& \times
\int_{\frac{\beta}{2M}}^{2\frac{\beta}{2M}}dt^{(2)}_1\cdots
\int_{\frac{\beta}{2M}}^{t^{(2)}_{k_2-1}} dt_{k_2}^{(2)} \ \cdots\
\int_{(2M-1)\frac{\beta}{2M}}^\beta dt_1^{(2M)}\cdots
\int_{(2M-1)\frac{\beta}{2M}}^{t_{k_{2M}-1}^{(2M)}}
dt_{k_{2M}}^{(2M)}
\nonumber\\
&& \times e^{\beta\HoL} Q v^\Lambda(t_{k_1}^{(1)})\cdots
v^\Lambda(t_1^{(1)})Q \
Qv^\Lambda(t^{(2)}_{k_2})\cdots v^\Lambda(t_1^{(2)}) Q\times \cdots \nonumber\\
&& \cdots \times Qv^\Lambda(t^{(2M)}_{k_{2M}})\cdots
v^\Lambda(t_1^{(2M)}) Q e^{-\beta\HoL}. \label{21}
\end{eqnarray}
The time variables in the integrand are ordered,
\begin{equation}
0\leq t_{k_1}^{(1)}\leq\cdots\leq t_1^{(1)}\leq
t_{k_2}^{(2)}\leq\cdots \leq t_1^{(2M)}\leq \beta. \label{22'}
\end{equation}
We want to estimate an upper bound on
$|\obzL(T(k_1,\ldots,k_{2M}))|$, sharp enough so that
\begin{equation}
\sum_{k_1,\ldots,k_{2M}\geq 1}
\left|\obzL(T(k_1,\ldots,k_{2M}))\right| \label{22}
\end{equation}
converges, and to estimate the value of the series. The
expectation value of the integrand in the state
$\obzL=\omega_\beta^\S\otimes \omega_\beta^{f,\Lambda}$ (see
\fer{9.1}) splits into a sum over products
\begin{eqnarray}
\lefteqn{ \sum_{\alpha_1^{(1)},\ldots,\alpha_{k_1}^{(1)}}\cdots
\sum_{\alpha_1^{(2M)},\ldots,\alpha_{k_{2M}}^{(2M)}}
\omega_\beta^\S\left(
  QG_{\alpha_{k_1}^{(1)}}(t^{(1)}_{k_1}) \cdots
G_{\alpha_1^{(2M)}}(t^{(2M)}_1)Q\right)}\nonumber\\
&&\ \ \ \ \ \ \ \ \ \times \omega_\beta^{f,\Lambda}\left(
  \varphi_{\alpha_{k_1}^{(1)}}^\Lambda(t_{k_1}^{(1)})\cdots
\varphi_{\alpha_1^{(2M)}}^\Lambda(t_1^{(2M)})\right),\ \ \ \ \ \ \
\ \ \label{23} ,
\end{eqnarray}
where
\begin{eqnarray}
G_\alpha(t)&=&e^{-tH^\S}G_\alpha e^{tH^\S}\label{24}\\
\varphi_\alpha^\Lambda(t)&=&e^{-t\HfL}\varphi(g^\Lambda_\alpha)e^{t\HfL}
= a^*\left(e^{-t\hfL}g_\alpha^\Lambda\right) +a\left(e^{t\hfL}
  g_\alpha^\Lambda\right).
\label{25}
\end{eqnarray}
Using the H\"older inequality \fer{holder},
\begin{equation}
\left| \omega^\S_\beta\left(
  QG_{\alpha_{k_1}^{(1)}}(t^{(1)}_{k_1}) \cdots
G_{\alpha_1^{(2M)}}(t^{(2M)}_1)Q\right)\right|\leq
\prod_{j=1}^{2M}
  \|G_{\alpha_1^{(j)}}\| \cdots \|G_{\alpha_{k_j}^{(j)}}\|.
\label{26}
\end{equation}
Since $\omega^{f,\Lambda}_\beta$ is a quasi-free state we can
estimate the second factor in \fer{23} with the help of Wick's
theorem:
\begin{equation}
\omega_\beta^{f,\Lambda} \left( \varphi_{\alpha_1}^\Lambda (t_1)
\cdots \varphi_{\alpha_{2N}}^\Lambda (t_{2N})\right) =\sum_{\cal
P} \prod_{(l,r)\in{\cal P}} \omega_\beta^{f,\Lambda} \left(
\varphi_{\alpha_l}^\Lambda (t_l) \varphi_{\alpha_{r}}^\Lambda
(t_{r})\right), \label{26.1}
\end{equation}
where the sum extends over all {\it contraction schemes}, i.e.,
decompositions of $\{1,\ldots,2N\}$ into $N$ disjoint, ordered
pairs $(l,r)$, $l<r$. Applying  the latter equation to
\begin{equation}
\omega_\beta^{f,\Lambda} \left(
  \varphi_{\alpha_{k_1}^{(1)}}^\Lambda(t_{k_1}^{(1)})\cdots
\varphi_{\alpha_1^{(2M)}}^\Lambda(t_1^{(2M)})\right) \label{27}
\end{equation}
one finds that all resulting terms can be organized in {\it
graphs} $\cal G$, constructed in the following way. (This can be
done by partition the circle of circumference $\beta$ into $2M$
segments parametrized by the arc length
$\Delta_j=[(j-1)\frac{\beta}{2M}, j\frac{\beta}{2M}]$,
$j=1,\ldots,2M$. Putting $k_j$ ``dots'' into the interval
$\Delta_j$, each dot representing a time variable
$t^{(j)}_\cdot\in \Delta_j$ (increasing times are ordered
according to increasing arc length). Pick any arbitrary dot in any
interval and pair it with an arbitrary different dot in any
interval. Then pick any unpaired dot (i.e., one not yet paired up)
and pair it with any other unpaired dot. Continue this procedure
until all dots in all intervals are paired. The graph $\cal G$
associated to such a pairing consists of all pairs -- including
multiplicity -- of intervals $(\Delta,\Delta')$ with the property
that some dot in $\Delta$ is paired with some dot in $\Delta'$.
``Including multiplicity'' means that if, say, three dots of
$\Delta$ are paired with three dots in $\Delta'$, we understand
that $\cal G$ contains the pair $(\Delta, \Delta')$ three times.)
Denote the class of all pairings $\cal P$ leading to a given graph
$\cal G$ is by $C_{\cal G}$, and let
\begin{equation}
A_{\cal P}=\prod_{(l,r)\in{\cal P}} \omega_\beta^{f,\Lambda}
\left( \varphi_{\alpha_l}^\Lambda (t_l)
\varphi_{\alpha_{r}}^\Lambda (t_{r})\right) \label{26.2}
\end{equation}
denote the contribution to \fer{26.1} corresponding to the pairing
$\cal P$. The value $|{\cal G}|$ corresponding to a graph $\cal G$
is given by
\begin{equation}
|{\cal G}|=\left| \sum_{{\cal P}\in C_{\cal G}} A_{\cal P}\right|.
\label{26.3}
\end{equation}
It follows from \fer{26.1}, \fer{26.2} and \fer{26.3} that
\begin{equation}
\left|\omega^\Lambda_{\beta,f}\left(\varphi_{\alpha_{k_1}^{(1)}}^\Lambda(t_{k_1}^{(1)})\cdots
\varphi_{\alpha_1^{(2M)}}^\Lambda(t_1^{(2M)}) \right)\right| \leq
\sum_{\cal G}|{\cal G}|. \label{28}
\end{equation}
In order to give an upper bound on the RHS of \fer{28}, one needs
to estimate the imaginary-time propagators (two-point functions)
\begin{eqnarray}
\lefteqn{ \omega_\beta^{f,\Lambda}\left( e^{-t_l \HfL}
\varphi(g_{\alpha_l}^\Lambda) e^{t_l \HfL}
e^{-t_r\HfL} \varphi(g_{\alpha_r}^\Lambda)e^{t_r \HfL}\right)}\nonumber\\
&=&\scalprod{g_{\alpha_r}^\Lambda}{e^{-(\beta
+t_l-t_r)\hfL}\frac{e^{\beta \hfL}}{e^{\beta\hfL}-1}
g_{\alpha_l}^\Lambda}
+\scalprod{g_{\alpha_l}^\Lambda}{e^{-(t_r-t_l)\hfL}\frac{e^{\beta
\hfL}}{e^{\beta\hfL}-1} g_{\alpha_r}^\Lambda} \label{28.1}\ \ \ \
\
\end{eqnarray}
where the $g_{\alpha_{l,r}}^\Lambda\in l^2({\mathbb Z}^3)$ are
given in \fer{6.1}, and where $t_l\in\Delta_l$, $t_r\in \Delta_r$
s.t. $0\leq t_l\leq t_r\leq \beta$. Hence, the RHS of \fer{28.1}
is
\begin{eqnarray}
\lefteqn{ \left(\frac{2\pi}{L}\right)^3\sum_{n\neq (0,0,0)}
  \Big[ \overline{g_{\alpha_r}(2\pi n/L)} g_{\alpha_l}(2\pi n/L)
  e^{-(\beta+t_l-t_r)E_n^\Lambda}}\nonumber\\
&&+\overline{g_{\alpha_l}(2\pi n/L)} g_{\alpha_r}(2\pi n/L) e^{-(t_r-t_l)E_n^\Lambda} \Big] \times \frac{e^{\beta E_n^\Lambda}}{e^{\beta E_n^\Lambda}-1}  \nonumber\\
&+& \left(\frac{2\pi}{L}\right)^3\left[ e^{-(\beta+t_l-t_r)}
+e^{-(t_r-t_l)}\right]\frac{e^\beta}{e^\beta-1}. \ \ \ \ \ \ \ \ \
\ \ \ \label{32}
\end{eqnarray}
In the thermodynamic limit $L\rightarrow\infty$, the sum in
\fer{32} converges to
\begin{equation}
\int_{\r^3} d^3k \ \left[ \overline{g_{\alpha_r}(k)}
g_{\alpha_l}(k)
  e^{-(\beta+t_l-t_r)k^2} + \overline{g_{\alpha_l}(k)} g_{\alpha_r}(k) e^{-(t_r-t_l)k^2}\right]
  \frac{e^{\beta k^2}}{e^{\beta k^2}-1}.
\label{33}
\end{equation}

For arbitrary $\Delta_l, \Delta_r$ and $t_l\in\Delta_l$,
$t_r\in\Delta_r$,
\begin{equation}
|t_l-t_r|\geq d_-(\Delta_l,\Delta_r):=\frac{\beta}{2M}\left\{
\begin{array}{ll}
0, & \mbox{if $l=r$}\\
|l-r|-1, & \mbox{if $l\neq r$}
\end{array}
\right. \label{32'}
\end{equation}
and
\begin{equation}
\beta -|t_l-t_r|\geq d_+(\Delta_l,\Delta_r):=\beta
-\frac{\beta}{2M}\left(|l-r|+1\right). \label{33'}
\end{equation}
Define
\begin{equation}
d(\Delta, \Delta'):=\min(d_-(\Delta, \Delta'), d_+(\Delta,
\Delta')). \label{34}
\end{equation}
It follows from \fer{28.1} and \fer{33}, and for $L$ large enough,
\begin{eqnarray}
\lefteqn{ \left| \omega_\beta^{f,\Lambda}\left( e^{-t_l \HfL}
\varphi(g^\Lambda_{\alpha_l}) e^{t_l \HfL}
e^{-t_r\HfL} \varphi(g_{\alpha_r}^\Lambda)e^{t_r \HfL}\right)\right|}\nonumber\\
&&\leq 2\scalprod{g_{\alpha_l}^\Lambda}{\frac{e^{-d(\Delta_l,
    \Delta_r) k^2}}{1-e^{-\beta k^2}} g_{\alpha_l}^\Lambda}^{1/2}
\scalprod{g_{\alpha_r}^\Lambda}{\frac{e^{-d(\Delta_l,
    \Delta_r) k^2}}{1-e^{-\beta k^2}}g_{\alpha_r}^\Lambda}^{1/2}.
\label{35}
\end{eqnarray}
Given any two intervals $\Delta, \Delta'$, let
\begin{equation}
C(\Delta,\Delta'):=4\max_{\alpha}\scalprod{
  g_\alpha}{\frac{e^{-d(\Delta,  \Delta') k^2}}{1-e^{-\beta k^2}}
  g_\alpha}.
\label{36}
\end{equation}
For $L\ge C$, where $C$ is some constant, \fer{36} is a
volume-independent upper bound on the (finite-volume) two-point
functions arising from contractions in the graph expansion. We are
now ready to give an upper bound on the RHS of \fer{28}. Start the
procedure of pairing dots in the interval with the highest order
$k$. Let $\pi$ be a permutation of $2M$ objects, such that
\begin{equation}
k_{\pi(1)}\geq k_{\pi(2)}\geq\cdots\geq k_{\pi(2M)}. \label{29'}
\end{equation}
There are $k_{l_1^{(\pi(1))}}$ possibilities of pairing the dot
  $t^{(\pi(1))}_1$ with
  some dot in an interval $\Delta_{l_1^{(\pi(1))}}$. We associate to each such pairing the value
\begin{equation}
 k_{l_1^{(\pi(1))}} \ C(\Delta_{\pi(1)},\Delta_{l_1^{(\pi(1))}})\leq
\sqrt{k_{\pi(1)}} \ \sqrt{k_{l_1^{(\pi(1))}}}\
C(\Delta_{\pi(1)},\Delta_{l_1^{(\pi(1))}}), \label{30'}
\end{equation}
where we use \fer{29'}. Next, we pair the dot labelled by
$t_2^{(\pi(1))}$ with a dot in $\Delta_{l_2^{(\pi(1))}}$ and
associate to this pairing the value
\begin{equation}
\sqrt{k_{\pi(1)}} \ \sqrt{k_{l_2^{(\pi(1))}}}\
C(\Delta_{\pi(1)},\Delta_{l_2^{(\pi(1))}}). \label{37}
\end{equation}
We continue this procedure until all dots are paired. This gives
\begin{equation}
\sum_{\cal G}|{\cal G}|\leq\prod_{j=1}^{2M}
(k_j)^{k_j/2}\sum_{\cal G}
 \prod_{(\Delta,\Delta')\in{\cal G}}
 C(\Delta,\Delta').
\label{38}
\end{equation}
Using that
\begin{equation}
|g_\alpha(k) |\leq C |k|^{2p}, \label{39}
\end{equation}
for some constant $C$, and for all $\alpha$, provided $|k|$ is
small enough and $p>-1/4$, it follows that
\begin{equation}
C(\Delta,\Delta')\leq C\left\{
\begin{array}{ll}
d(\Delta,\Delta')^{-\frac{3}{2}-2p},& d(\Delta,\Delta')\neq 0\\
1/\beta +1,& d(\Delta,\Delta')=0
\end{array}
\right. \label{40}
\end{equation}
Moreover, using \fer{34} and \fer{40},
\begin{equation}
\sum_{\Delta'} C(\Delta,\Delta')\leq\Gamma:=
C\left(1+\frac{1}{\beta}+\frac{1}{p+1/4}\left(\frac{\beta}{2M}\right)^{-2p-1/2}
  \right)<\infty,
\label{41}
\end{equation}
provided $p>-1/4$. Hence,
\begin{equation}
\sum_{\cal G} \prod_{(\Delta,\Delta')\in{\cal G}}
C(\Delta,\Delta')\leq \Gamma^{k_1+\cdots +k_{2M}}. \label{42}
\end{equation}
Now carry out the integral over the simplex in \fer{21}, and using
\fer{23}, \fer{26}, \fer{28}, \fer{38}, \fer{42},
\begin{equation}
\left|\obzL(T(k_1,\ldots,k_{2M}))\right|\leq \left(C'|g|\Gamma
  \frac{\beta}{2M}\right)^{k_1+ \cdots + k_{2M}}\  \prod_{j=1}^{2M}
  \frac{(k_j)^{k_j/2}}{k_j!},
\label{43}
\end{equation}
where $C'=\sum_\alpha \|G_\alpha\|$, and where the factor
$(\frac{\beta}{2M})^{k_j}\frac{1}{k_j!}$ is the volume of the
simplex $\{t\leq t_{k_j}\leq \cdots\leq t_1\leq
t+\frac{\beta}{2M}\}$. Therefore, \fer{22} converges for all
values of $g$ and $\beta>0$, and
\begin{equation}
\left[ \obzL\left(e^{\beta\HoL}
B^{2M}\right)\right]^{\frac{1}{2M}} \leq
 C'|g|\Gamma\frac{\beta}{2M} \sum_{k\geq 0} \left(
 C'|g|\Gamma\frac{\beta}{2M}\right)^k \frac{(k+1)^{\frac{k+1}{2}}}{
 (k+1)!}.
\label{44}
\end{equation}
From \fer{16}, \fer{18}, \fer{19}, \fer{44}, \fer{12}, one
obtauns, for $L$ is large enough
\begin{equation}
\oblL(Q)\leq 2e^{-2\tau\Delta} + C'|g|\Gamma\tau \sum_{k\geq 0}
\left( C'|g|\Gamma\tau\right)^k
\frac{(k+1)^{\frac{k+1}{2}}}{(k+1)!}. \label{45}
\end{equation}

Choose $\beta_2(\epsilon)>1$ such that
$e^{-\beta_2(\epsilon)\Delta}<\epsilon/2$, and for $\beta\geq
\beta_2(\epsilon)$, choose $\tau=\beta_2(\epsilon)/2\leq\beta/2$.
From the definition of $\Gamma$, \fer{41}, and the relation
$\frac{\beta}{2M}=2\tau$, see \fer{12}, one obtains
$\Gamma\tau\leq C(\epsilon)$, uniformly in
$\beta\geq\beta_2(\epsilon)$. Therefore $\exists g_1(\epsilon)>0$
such that if $|g|<g_1(\epsilon)$ then the second term on the RHS
of \fer{45} $\le\epsilon/2$. This proves our claim \fer{2}.$\Box$

\subsection{The Virial Theorem}

In this section, we state two Virial theorems proven in [FM2,FMS].
Since the Theorems are applied {\it without} any alteration or
extension, we just sketch the main steps of their proof.

We start by discussing the {\it Glimm-Jaffe-Nelson} condition.
Consider a Hilbert space $\H$, a selfadjoint operator
$Y\geq\bbbone$ with core
 $\dom\subset\D$, and $X$ a symmetric operator on
$\dom$. The triple $(X,Y,\dom)$ satisfies the {\it GJN
(Glimm-Jaffe-Nelson) Condition}, or $(X,Y,\dom)$ is a {\it
GJN-triple}, if there is a constant $k<\infty$, such that, for all
$\psi\in\dom$:
\begin{eqnarray}
\|X\psi\|&\leq& k\|Y\psi\| \label{nc1}\\
\pm
i\left\{\scalprod{X\psi}{Y\psi}-\scalprod{Y\psi}{X\psi}\right\}&\leq&
k\scalprod{\psi}{Y\psi}. \label{nc2}
\end{eqnarray}
Since $Y\geq\bbbone$, inequality \fer{nc1} is equivalent to
\begin{equation*}
\| X\psi\|\leq k_1\|Y\psi\|+k_2\|\psi\|,
\end{equation*}
for some $k_1, k_2<\infty$.

\vspace{0.5cm}

\noindent {\it Theorem A3.1 (GJN commutator theorem)}

If $(X,Y,\dom)$ satisfies the GJN Condition, then $X$ determines a
selfadjoint operator (again  denoted by $X$), such that
$\dom(X)\supset\dom(Y)$. Moreover, $X$ is essentially selfadjoint
on any core for $Y$, and \fer{nc1} is valid for all
$\psi\in\dom(Y)$.\\

Suppose one is given a selfadjoint operator $\Lambda\geq\bbbone$
with core $\dom\subset\H$, and operators  $L, A, N, D, C_n$,
$n=0,1,2,3$, all symmetric on $\dom$, and satisfying
\begin{eqnarray}
\scalprod{\varphi}{D\psi}&=&i\left\{
  \scalprod{L\varphi}{N\psi}-\scalprod{N\varphi}{L\psi}\right\} \label{44}\\
C_0&=& L\nonumber\\
\scalprod{\varphi}{C_n\psi}&=&i\left\{\scalprod{C_{n-1}\varphi}{A\psi}-\scalprod{A\varphi}{C_{n-1}\psi}\right\},\
  \ n=1,2,3,
\label{45}
\end{eqnarray}
where $\varphi, \psi\in\dom$. Assume that
\begin{itemize}
\item[$\bullet$]  $(X,\Lambda,\dom)$ satisfies the GJN Condition, for
$X=L,N,D,C_n$. Consequently, all these operators determine
selfadjoint operators, which we denote by the same letters.
\item[$\bullet$] $A$ is selfadjoint, $\dom\subset\dom(A)$, and
$e^{itA}$ leaves $\dom(\Lambda)$ invariant.
\end{itemize}

\vspace{0.5cm}

{\it Remark.} From the invariance condition
$e^{itA}\dom(\Lambda)\subset \dom(\Lambda)$, it follows that for
some $0\leq k,k'<\infty$, and all $\psi\in\dom(\Lambda)$,
\begin{equation}
\|\Lambda e^{itA}\psi\|\leq ke^{k'|t|}\|\Lambda\psi\|.
\label{alambda}
\end{equation}
(A proof can be found in [ABG], Propositions 3.2.2 and 3.2.5.)

\vspace{0.5cm}

\noindent {\it Theorem A3.2 (First Virial Theorem)}

Assume that, in addition to \fer{44}, \fer{45}, we have, in the
sense of Kato on $\dom$,
\begin{eqnarray}
D&\leq& kN^{1/2},\label{41}\\
e^{itA}C_1 e^{-itA}&\leq& ke^{k'|t|}N^p, \mbox{\ \ \ some $0\leq
p<\infty$},
\label{40'} \\
e^{itA} C_3 e^{-itA}&\leq& ke^{k'|t|} N^{1/2}, \label{40}
\end{eqnarray}
for some $0\leq k,k'<\infty$, and all $t\in\r$. Let $\psi$ be an
eigenvector of $L$. Then there is a one-parameter family
$\{\psi_\alpha\}\subset\dom(L)\cap \dom(C_1)$, s.t.
$\psi_\alpha\rightarrow\psi$, $\alpha\rightarrow 0$, and
\begin{equation}
\lim_{\alpha\rightarrow
0}\scalprod{\psi_\alpha}{C_1\psi_\alpha}=0. \label{42'}
\end{equation}\\

\vspace{0.5cm}

{\it Remark.} Formally, $C_1$ is the commutator $i[L,A]=i(LA-AL)$,
and \fer{42'} as $\scalprod{\psi}{i[L,A]\psi}=0$, which is a
standard way of stating the virial theorem, see also [ABG], and
[GG].

The result of Theorem A3.2 is still valid if we add to the
operator $A$ a suitably small perturbation $A_0$:\\

\vspace{0.5cm}

\noindent {\it Theorem A3.3 (Second Virial Theorem)}

Suppose that we are in the situation of Theorem A3.2
and that $A_0$ is a bounded operator on $\H$ such that $\ran
A_0\subset\dom(L)\cap\ran P(N\leq n_0)$, for some $n_0<\infty$.
Then $i[L,A_0]=i(LA_0-A_0L)$ is well defined in the strong sense
on $\dom(L)$, and we have, for the same family of approximating
eigenvectors as in Theorem A3.2
\begin{equation}
\lim_{\alpha\rightarrow
0}\scalprod{\psi_{\alpha}}{(C_1+i[L,A_0])\psi_{\alpha}}=0.
\label{43}
\end{equation}

\vspace{0.5cm}

{\it Theorem A3.4 (Regularity of eigenfunctions)}

Suppose $C$ is a symmetric operator on a domain $\dom(C)$ such
that, in the sense of quadratic forms on $\dom(C)$, we have that
$C\geq {\cal P}-B$, where ${\cal P}\geq 0$ is a selfadjoint
operator, and $B$ is a bounded (everywhere defined) operator. Let
$\psi_\alpha$ be a family of vectors in $\dom(C)$, with
$\psi_\alpha\rightarrow \psi$, as $\alpha\rightarrow 0$, and s.t.
\begin{equation}
\lim_{\alpha\rightarrow 0}\scalprod{\psi_\alpha}{C\psi_\alpha}=0.
\label{wtl}
\end{equation}
Then $\scalprod{\psi}{B\psi}\geq 0$, $\psi\in\dom({\cal
P}^{1/2})$, and
\begin{equation}
\|{\cal P}^{1/2}\psi\|\leq \scalprod{\psi}{B\psi}^{1/2}.
\end{equation}

\subsubsection{Concrete setting of Virial Theorem}

The Hilbert space is the GNS representation
$\H=\H^\S\otimes\H^\S\otimes \F_+(L^2(\mathbf{R}\times S^2))$. Let
\begin{equation}
\dom=C_0^\infty\otimes C_0^\infty\otimes\dom_f, \label{a39}
\end{equation}
where
\begin{equation*}
\dom_f={\cal F}\left( C_0^\infty(\r\times S^2)\right)\cap {\cal
F}_0, \label{45'}
\end{equation*}
and ${\cal F}_0$ denotes the finite-particle subspace of Fock
space. The operator $\Lambda$ is given by
\begin{eqnarray}
\Lambda&=&\Lambda^\S\otimes\mathbf{1}^\S+\mathbf{1}^\S\otimes\Lambda^\S+\Lambda^f,\label{a40}\\
\Lambda^\S&=&H^\S P_+ (H^\S)+\mathbf{1}^\S,\label{a41}\\
\Lambda^f&=& \d\Gamma(u^2+1)+\mathbf{1}^f, \label{a42}
\end{eqnarray}
$P_+(H^\S)$ is the projection onto the spectral interval $\r_+$ of
$H^\S$. $\Lambda$ is essentially selfadjoint on $\dom$, and
$\Lambda\geq\mathbf{1}$.  The operator $L$ is the interacting
Liouvillian $\L_g$, and
\begin{equation}
N=\d\Gamma(1) \label{en}
\end{equation}
is the particle number operator in ${\cal F}_+\equiv {\cal
F}_+(L^2(\mathbf{R}\times S^2))$. $X=L,N$ are symmetric operators on
$\dom$, and the symmetric operator $D$ on $\dom$ is given by
\begin{eqnarray}
D&=&\frac{ig}{\sqrt{2}}\sum_\alpha\big\{
G_{\alpha}\otimes\mathbf{1}^\S
    \otimes\left(-a^*(\tau_\beta(g_\alpha))+  a(\tau_\beta(g_\alpha))\right)\nonumber\\
&&-\mathbf{1}^\S\otimes C^\S G_{\alpha} C^\S \otimes\left(
-a^*(e^{-\beta u/2}\tau_\beta(g_\alpha))+a(e^{-\beta
   u/2}\tau_\beta(g_\alpha))\right)\big\}.
\label{operatorD}
\end{eqnarray}
The operator $A$ is identified by $A^f=$ and $A_0$ is given by
$i\theta g (\Pi I \overline{R}^2_\epsilon -
\overline{R}^2_\epsilon I \Pi)$, as before.

%%%%%%%%%%%%%%%%%%%%%%%%%%%%%%%%%%%%%%%%%%%%%%%%%%%%%%%%%%%%%%%%%%%

\subsubsection{Some functional analysis}

In this section, we collect some results which are useful in the
previous analysis, particularly the first two theorems, whose
proof can be found in [Fr\"o].

\vspace{0.5cm}

\noindent {\it Theorem A3.5 (Invariance of domain, [Fr\"o]) }

Suppose $(X,Y,\dom)$ satisfies
    the GJN Condition, \fer{nc1}, \fer{nc2}. Then the unitary group,
    $e^{itX}$, generated by the selfadjoint operator
    $X$ leaves $\dom(Y)$ invariant, and
\begin{equation}
\|Ye^{itX}\psi\|\leq e^{k|t|}\|Y\psi\|, \label{80}
\end{equation}
for some $k\geq 0$, and all $\psi\in\dom(Y)$.

\pagebreak

\vspace{0.5cm}

\noindent {\it Theorem A3.6 (Commutator expansion, [Fr\"o]) }

Suppose $\dom$ is a core for the selfadjoint operator
$Y\geq\bbbone$. Let $X,Z, {\rm ad}_X^{(n)}(Z)$
  be symmetric operators on $\dom$, where
\begin{eqnarray*}
{\rm ad}_X^{(0)}(Z)&=&Z,\\
\scalprod{\psi}{{\rm ad}_X^{(n)}(Z)\psi}&=&i\left\{\scalprod{{\rm
      ad}_X^{(n-1)}(Z)\psi}{X\psi} -\scalprod{X\psi}{{\rm
      ad}_X^{(n-1)}(Z)\psi}\right\},
\end{eqnarray*}
for all $\psi\in\dom$, $n=1,\ldots,M$. We suppose that the triples
$({\rm ad}_X^{(n)}(Z),Y,\dom)$, $n=0,1,\ldots,M$, satisfy the GJN
Condition \fer{nc1}, \fer{nc2}, and that $X$ is selfadjoint, with
$\dom\subset\dom(X)$, $e^{itX}$ leaves $\dom(Y)$ invariant, and
\fer{80} holds. Then
\begin{eqnarray}
e^{itX}Ze^{-itX}&=& Z-\sum_{n=1}^{M-1}\frac{t^n}{n!}
{\rm ad}_X^{(n)}(Z)\nonumber\\
&&-\int_0^tdt_1\cdots\int_0^{t_{M-1}} dt_M e^{it_MX}{\rm
  ad}_X^{(M)}(Z)e^{-it_MX},
\label{cmexp}
\end{eqnarray}
as operators on $\dom(Y)$.

Two corollaries follows from \fer{80}.

\vspace{0.5cm}

\noindent {\it Corollary A3.7}

Suppose that the unitary group $e^{itX}$ leaves $\dom(Y)$
invariant, for some operator $Y$, and that estimate \fer{80}
holds. For a function $\chi$ on $\r$ with Fourier transform
$\what{\chi}\in L^1(\r)$, we define $\chi(X)=\int_\r
\what{\chi}(s) e^{isX}ds$. If $\what{\chi}$ has compact support,
then $\chi(X)$ leaves $\dom(Y)$ invariant, and, for
$\psi\in\dom(Y)$,
\begin{equation}
\|Y\chi(X)\psi\|\leq e^{k R}\|\what{\chi}\|_{L^1(\r)}\ \|Y\psi\|,
\label{easybutusefullabel}
\end{equation}
for any $R$ such that $\supp\what{\chi}\subset [-R,R]$.

\vspace{0.5cm}

\noindent {\it Corollary A3.8}

Suppose $(X,Y,\dom)$ satisfies the GJN Condition, and so do the
triples $({\rm ad}_X^{(n)}(Y),Y,\dom)$, for $n=1,\ldots,M$, and
for some $M\geq 1$. Moreover, assume that, in the sense of Kato on
$\dom(Y)$, $\pm{\rm ad}_X^{(M)}(Y)\leq k X$, for some $k\geq 0$.
For $\chi\in C_0^\infty(\r)$, a smooth function with compact
support, define $\chi(X)=\int\what{\chi}(s)e^{isX}$, where
$\what{\chi}$ is the Fourier transform of $\chi$. Then $\chi(X)$
leaves $\dom(Y)$ invariant.

\vspace{0.5cm}

\noindent {\it Lemma A3.9}

Let $\chi\in C_0^\infty(\r)$, $\chi=F^2\geq 0$. Suppose
$(X,Y,\dom)$ satisfies the GJN condition. Suppose $F(X)$ leaves
$\dom(Y)$ invariant. Let $Z$ be a symmetric operator on $\dom$
such that, for some $M\geq 1$, and $n=0,1,\ldots,M$, the triples
$({\rm ad}_X^{(n)}(Z),Y,\dom)$ satisfy the GJN condition.
Moreover, assume that the multiple commutators, for
$n=1,\ldots,M$, are relatively $X^{2p}$-bounded in the sense of
Kato on $\dom$, for some $p\geq 0$. In other words, there is some
$k<\infty$, s.t. $\forall \psi\in \dom$,
\begin{equation*}
\|{\rm ad}_X^{(n)}(Z)\psi\|\leq k\left(
\|\psi\|+\|X^{2p}\psi\|\right) ,\ \ \ \ n=1,\ldots,M.
\end{equation*}
Then the commutator $[\chi(X),Z]=\chi(X)Z-Z\chi(X)$ is well
defined on $\dom$ and extends to a bounded operator.

\vspace{0.5cm}

\noindent {\it Lemma A3.10}

Suppose $(X,Y,\dom)$ is a GJN triple. Then the resolvent
$(X-z)^{-1}$ leaves $\dom(Y)$ invariant, for all $z\in\{{\mathbf
C} : \ |\Im z|>k \}$, for some $k>0$.

%%%%%%%%%%%%%%%%%%%%%%%%%%%%%%%%%%%%%%%%%%%%%%%%%%%%%%%%%%%%%%%%%%%%%%%%%%%%55

\subsubsection{Proof of the Virial Theorems and regularity of
eigenfunctions}

Let $g_1 \in C_0^\infty( (-1,1) )$ be a real valued function, such
that $g_1(0)=1$, and $g=g_1^2\in C_0^\infty( (-1,1) )$, $g(0)=1$.
Choose a real valued function $f$ on $\r$ with the properties that
$f(0)=1$ and $\what{f}\in C_0^\infty(\r)$ (Fourier transform). Let
\begin{equation*}
f_1(x)=\int_{-\infty}^x f^2(y)dy,
\end{equation*}
so that $f_1'(x)=f^2(x)$. Note that $\what{f_1}$ has compact
support, and is smooth except at $s=0$, where it behaves like
$s^{-1}$; $\what{f^{(n)}_1}=(is)^n\what{f_1}\in C_0^\infty$, for
$n\geq 1$. Let $\alpha, \nu>0$ be two parameters and define
\begin{eqnarray*}
g_{1,\nu}&=&g_1(\nu N)=\int_\r \what{g_1}(s) e^{is\nu N} ds,\\
g_\nu&=&g_{1,\nu}^2\\
f_\alpha&=& f(\alpha A)=\int_\r \what{f}(s) e^{is\alpha A}ds,\\
\end{eqnarray*}
For $\eta>0$, define
\begin{equation*}
f^\eta_{1,\alpha}=\frac{1}{\alpha}\int_{\r\backslash (-\eta,\eta)}
ds \what{f_1}(s) e^{is\alpha A}= (f^\eta_{1,\alpha})^*.
\end{equation*}
$f^\eta_{1,\alpha}$ leaves $\dom(\Lambda)$ invariant, and
$\|f^\eta_{1,\alpha}\|\leq k/\alpha$, where $k$ is a constant

Suppose $\psi$ is an eigenfunction of $L$ with eigenvalue $e$.
$\psi=(L+i)^{-1}\varphi$, for some $\varphi\in\H$. Let
$\{\varphi_n\}\subset\dom(\Lambda)$ be a sequence of vectors
converging to $\varphi$. Then
\begin{equation}
\psi_n:=(L+i)^{-1}\varphi_n\rightarrow\psi,\ \ \
n\rightarrow\infty, \label{*}
\end{equation}
 and, $\psi_n\in\dom(\Lambda)$. Assume
that $k=1$. We know that $f_\alpha$ leaves $\dom(\Lambda)$
invariant, and $g_\nu$ leaves $\dom(\Lambda)$ invariant ($\Lambda$
commutes with $N$ in the strong sense on $\dom$). Hence, the
regularized eigenfunction
\begin{equation*}
\psi_{\alpha,\nu,n}=f_\alpha g_\nu\psi_n
\end{equation*}
satisfies $\psi_{\alpha,\nu,n}\in\dom(\Lambda)$,
$\psi_{\alpha,\nu,n}\rightarrow\psi$, as $\alpha, \nu\rightarrow
0$,
$n\rightarrow \infty$. \\

We claim
\begin{equation}
\left|\av{if^\eta_{1,\alpha}(L-e)}_{g_\nu\psi_n}\right|\leq
k\frac{1}{\alpha}\left(\sqrt{\nu} +o(n)\right), \label{a0}
\end{equation}
where $k$ is some constant independent of $\eta,\alpha,\nu, n$.
This follows from
\begin{equation}
\|(L-e)g_\nu\psi_n\|\leq k\left(\sqrt{\nu} +o(n)\right),
\label{a1}
\end{equation}
which is proven as follows.
\begin{eqnarray}
(L-e)g_\nu\psi_n&=& g_\nu (L-e)\psi_n\label{a2}\\
&&+g_{1,\nu}[L,g_{1,\nu}]\psi_n\label{a3}\\
&&+[L,g_{1,\nu}]g_{1,\nu}\psi_n,\label{a4}
\end{eqnarray}
and the RHS of \fer{a2} is $o(n)$. Both \fer{a3} and
\fer{a4} are bounded above by $k\sqrt{\nu}$, uniformly in $n$. The
commutator expansion of Theorem gives
\begin{equation}
g_{1,\nu}[L,g_{1,\nu}]=\nu\int_\r ds\ \what{g_1}(s) e^{is\nu
N}\int_0^s ds_1e^{-is_1\nu N}g_{1,\nu}D e^{is_1\nu N}, \label{a5'}
\end{equation}
as operators on  $\dom(\Lambda)$. We use that $g_{1,\nu}$ commutes
with $e^{is\nu N}$. From \fer{41}, $\phi\in\dom(\Lambda)$,
\begin{eqnarray*}
\lefteqn{\|g_{1,\nu}De^{is_1\nu N}\phi\|=\sup_{\varphi\in \dom,
\varphi\neq
  0}\frac{\left|\scalprod{\varphi}{g_{1,\nu}De^{is_1\nu
  N}\phi}\right|}{\|\varphi\|}}\\
&\leq& \sup_{\varphi\in \dom, \varphi\neq
  0}\frac{\|Dg_{1,\nu}\varphi\|\ \|\phi\|}{\|\varphi\|}
\leq k \sup_{\varphi\in \dom, \varphi\neq
  0}\frac{\|N^{1/2}g_{1,\nu}\varphi\|}{\|\varphi\|}\ \|\phi\|
\leq k\frac{1}{\sqrt{\nu}}\|\phi\|,
\end{eqnarray*}
and hence,
\begin{eqnarray}
\|g_{1,\nu}[L,g_{1,\nu}]\phi\|&\leq&\nu\int_\r ds
|\what{g_1}(s)|\int_0^s ds_1\,
\|g_{1,\nu}De^{is_1\nu N}\phi\|\nonumber\\
&\leq & k\sqrt{\nu}\int_\r ds\, |s\what{g_1}(s)| \ \|\phi\|.
\label{a5}
\end{eqnarray}

Similarly for \fer{a4}.This establishes \fer{a1}.\\

Since $f^\eta_{1,\alpha}$ leaves $\dom(\Lambda)$ invariant,
$[f^\eta_{1,\alpha},L]$ is defined (in the strong sense) on
$\dom(\Lambda)$, and the commutator theorem gives
\begin{eqnarray}
\lefteqn{[f^\eta_{1,\alpha},L]}\nonumber\\
&=&\int_{\r\backslash (-\eta,\eta)} ds \what{f_1}(s)e^{is\alpha
A}\left(sC_1
  +\alpha\frac{s^2}{2}C_2\right)\nonumber\\
&&+\alpha^2\int_{\r\backslash (-\eta,\eta)} ds\what{f_1}(s)
e^{is\alpha
  A}\int_0^sds_1\int_0^{s_1}ds_2\int_0^{s_2}ds_3 \ e^{-is_3\alpha
  A}C_3e^{is_3\alpha A}.\nonumber\\
\label{a6'}
\end{eqnarray}
For $n\geq 1$,
\begin{equation*}
f_1^{(n)}(\alpha A)=\int_\r ds (is)^n\what{f_1}(s) e^{is\alpha
  A}=\int_{\r\backslash (-\eta,\eta)} ds (is)^n\what{f_1}(s) e^{is\alpha
  A}-{\cal R}_{\eta, n},
\end{equation*}
where
\begin{equation*}
{\cal R}_{\eta,n}=-\int_{-\eta}^\eta ds\ (is)^n \what{f_1}(s)
e^{is\alpha A}
\end{equation*}
satisfies ${\cal R}_{\eta,n}=({\cal R}_{\eta,n})^*$, and $\|{\cal
  R}_{\eta,n}\|\leq k_n\eta$, with a constant $k_n$ that does not depend on
  $\alpha,\eta$. From \fer{a6'}
\begin{eqnarray}
[f^\eta_{1,\alpha},L]&=&-i\left(f_1'(\alpha A)+{\cal
  R}_{\eta,1}\right)C_1-\frac{\alpha}{2}\left( f''_1(\alpha A)+{\cal
  R}_{\eta,2}\right)C_2\nonumber\\
&&+\alpha^2\int_{\r\backslash (-\eta,\eta)} ds\what{f_1}(s)
e^{is\alpha
  A}\int_0^sds_1\int_0^{s_1}ds_2\int_0^{s_2}ds_3 \ e^{-is_3\alpha
  A}C_3e^{is_3\alpha A}.\nonumber\\
\label{a6}
\end{eqnarray}
Moreover,
\begin{eqnarray}
-if_\alpha^2C_1&=& -if_\alpha C_1f_\alpha\nonumber\\
&&-if_\alpha\int_\r ds\what{f}(s) e^{is\alpha A}\left(\alpha s
  C_2+\alpha^2\int_0^sds_1\int_0^{s_1}ds_2 e^{-is_2\alpha A} C_3 e^{is_2\alpha
    A}\right)\nonumber\\
&=&-if_\alpha C_1f_\alpha -\alpha f_\alpha f'_\alpha C_2\nonumber\\
&&-i\alpha^2 f_\alpha \int_\r ds\what{f}(s) e^{is\alpha A}
 \int_0^sds_1\int_0^{s_1}ds_2 e^{-is_2\alpha A} C_3 e^{is_2\alpha
    A},
\label{a8}
\end{eqnarray}
where $f_\alpha'=f'(\alpha A)$. Since
 $f_\alpha f'_\alpha=\frac{1}{2}(f^2)'(\alpha
A)=\frac{1}{2}f_1''(\alpha A)$, one has from \fer{a6}, \fer{a8}:
\begin{eqnarray}
[f^\eta_{1,\alpha},L]&=&-if_\alpha C_1f_\alpha -\alpha
f_1''(\alpha
A)C_2-i{\cal R}_{\eta,1}C_1-\frac{\alpha}{2}{\cal R}_{\eta,2} C_2 \nonumber\\
&&+\alpha^2\int_{\r\backslash (-\eta,\eta)} ds\what{f_1}(s)
e^{is\alpha
  A}\int_0^sds_1\int_0^{s_1}ds_2\int_0^{s_2}ds_3 \ e^{-is_3\alpha
  A}C_3e^{is_3\alpha A}\nonumber\\
&&-i\alpha^2f_\alpha \int_\r ds\what{f}(s) e^{is\alpha A}
 \int_0^sds_1\int_0^{s_1}ds_2 e^{-is_2\alpha A} C_3 e^{is_2\alpha
    A}.
\label{a9}
\end{eqnarray}
Using \fer{40}, one obtains
\begin{eqnarray}
\av{i[f^\eta_{1,\alpha},L]}_{g_\nu\psi_n}&=&\av{C_1}_{\psi_{\alpha,\nu,n}}
-\Re  i\alpha \av{ f''(\alpha A)C_2}_{g_\nu\psi_n} +\Re\av{{\cal
    R}_{\eta,1}C_1}_{g_\nu\psi_n}\nonumber\\
&&-\Re \ i\frac{\alpha}{2}\av{{\cal R}_{\eta,2}C_2}_{g_\nu\psi_n}+
\O{\frac{\alpha^2}{\sqrt{\nu}}}. \label{a10}
\end{eqnarray}

Note that
\begin{eqnarray*}
\lefteqn{-\Re\av{i\alpha f''(\alpha
 A)C_2}_{g_\nu\psi_n}=-\frac{\alpha}{2}\av{i[f''(\alpha
 A),C_2]}_{g_\nu\psi_n}}\\
&=&-\frac{\alpha^2}{2}\av{\int_\r ds\ \what{f''}(s)e^{is\alpha
A}\int_0^s ds_1 \
 e^{-is_1\alpha A} C_3 e^{is_1\alpha
 A}}_{g_\nu\psi_n}=\O{\frac{\alpha^2}{\sqrt{\nu}}},
\end{eqnarray*}
where \fer{40} has been used again. Similarly,
\begin{equation*}
\Re \ i\frac{\alpha}{2}\av{{\cal
    R}_{\eta,2}C_2}_{g_\nu\psi_n}=-i\frac{\alpha}{4}\av{[{\cal R}_{\eta,2},
    C_2]}_{g_\nu\psi_n}= \O{\frac{\alpha^2\eta}{\sqrt{\nu}}},
\end{equation*}
and
\begin{equation*}
\av{{\cal R}_{\eta,1}C_1}_{g_\nu\psi_n}=\O{\frac{\eta}{\nu^p}}.
\end{equation*}
\fer{a10} and \fer{a0} imply
\begin{equation}
\left|\av{C_1}_{\psi_{\alpha,\nu,n}}\right|\leq
k\left(\frac{\sqrt{\nu}+o(n)}{\alpha} +
\frac{\alpha^2}{\sqrt{\nu}}+\frac{\eta}{\nu^p}\right).
\label{dagger}
\end{equation}

By taking the limit $n\rightarrow\infty$ in \fer{dagger}
\begin{equation*}
\left|\av{C_1}_{\psi_{\alpha,\nu}}\right|\leq
k\left(\frac{\sqrt{\nu}}{\alpha}+
\frac{\alpha^2}{\sqrt{\nu}}+\frac{\eta}{\nu^p}\right).
\end{equation*}
For example, choose $\nu=\alpha^3$, $\eta=\alpha^{3p+\delta}$, for
any $\delta>0$, then
\begin{equation*}
\lim_{\alpha\rightarrow 0} \av{C_1}_{\psi_{\alpha,\alpha^3}} =0.
\end{equation*}
This proves Theorem A3.2. To show Theorem A3.3, it is sufficient
to
\begin{equation*}
\lim_{\alpha\rightarrow
0}\scalprod{\psi_\alpha}{i[L,A_0]\psi_\alpha}=0,
\end{equation*}
where $\psi_\alpha=\psi_{\alpha,\nu}|_{\nu=\alpha^3}$. Moreover,
\begin{eqnarray*}
\left|\scalprod{\psi_\alpha}{i[L,A_0]\psi_\alpha}\right|&\leq&
2\left|
  \scalprod{(L-e)\psi_\alpha}{A_0\psi_\alpha}\right|\\
& \leq& 2\|P(N\leq
  n_0)(L-e)\psi_\alpha\|\ \|A_0\psi_\alpha\|.
\end{eqnarray*}
We have
\begin{eqnarray}
P(N\leq n_0)(L-e)\psi_{\alpha,\nu}&=&\lim_{n\rightarrow\infty}
P(N\leq
n_0)[L,f_\alpha] g_\nu\psi_n\label{a11}\\
&&+\lim_{n\rightarrow\infty} P(N\leq
n_0)f_\alpha[L,g_\nu]\psi_n.\label{a12}
\end{eqnarray}
Note that $\|P(N\leq n_0)[L,f_\alpha] g_\nu\psi_n\|\leq
k_{n_0}\alpha$ and $\| P(N\leq n_0)f_\alpha[L,g_\nu]\psi_n \|\leq
k\sqrt{\nu}$. It follows that $ \|P(N\leq n_0)
(L-e)\psi_{\alpha}\|\leq k_{n_0}\alpha$.

We still need to prove the regularity of the eigenfunctions. Using
the inequality $C\geq\pp-B$, the continuity of $B$, and \fer{wtl}
one has, for any $\epsilon>0$, an $\alpha_0(\epsilon)$, such that
if $\alpha<\alpha_0(\epsilon)$ then
\begin{equation}
\scalprod{\psi_\alpha}{\pp\psi_\alpha}\leq\scalprod{\psi}{B\psi}+\epsilon.
\label{star}
\end{equation}
Denote by $\mu_\phi$ be the spectral measure of $\pp$
corresponding to some $\phi\in\H$. Then
\begin{equation*}
\scalprod{\psi_\alpha}{\pp\psi_\alpha}=\int_{\r_+}
pd\mu_{\psi_\alpha}(p)=\lim_{R\rightarrow\infty}
\int_{\r_+}p\chi(p\leq R)d\mu_{\psi_\alpha}(p),
\end{equation*}
where $\chi(p\leq R)$ is the indicator of $[0,R]$. It follows from
\fer{star} that
\begin{equation*}
  \lim_{R\rightarrow\infty}\scalprod{\psi_\alpha}{\chi(\pp\leq
  R)\pp\psi_\alpha}=\lim_{R\rightarrow\infty}\left\|\chi(\pp\leq R)\pp^{1/2}
  \psi_\alpha\right\|^2\leq\scalprod{\psi}{B\psi} +\epsilon \equiv k.
\end{equation*}
One has $ \left\|\chi(\pp\leq R)\pp^{1/2} \psi\right\|\leq
R^{1/2}\|\psi-\psi_\alpha\| +\sqrt{k}$, and taking
$\alpha\rightarrow 0$ yields $\|\chi(\pp\leq R)\pp^{1/2}\psi\|\leq
\sqrt{k}$, uniformly in $R$, so $ \lim_{R\rightarrow\infty}
\int_0^R pd\mu_{\psi}(p) $ is finite (by the monotone convergence
theorem). Since $ \dom(\pp^{1/2})=\left\{\psi\left| \int_0^\infty
    pd\mu_\psi(p)<\infty\right.\right\}$, one has $\psi\in\dom(\pp^{1/2})$, and
$\|\pp^{1/2}\psi\|\leq\scalprod{\psi}{B\psi}$. $\Box$

%%%%%%%%%%%%%%%%%%%%%%%%%%%%%%%%%%%%%%%%%%%%%%%%%%%%%%%%%%%%%%%%%%%%%%%%%%%%%%%%%%

\subsection{Some operator calculus}

For the sake of completeness, we very briefly review operator calculus for
functions of selfadjoint operators used in the previous sections. For a more detailed review, see for example [HS].

Let $f\in C_0^k({\mathbf R})$, $k\geq 2$, and define the compactly
supported complex measure
$$d\tilde{f}(z)=-\frac{1}{2\pi}\left(\partial_x+i\partial_y\right)\tilde{f}(z)
dxdy,$$ where $z=x+iy$ and $\tilde{f}$ is an almost analytic
complex extension of $f$ in the sense that
$$\left(\partial_x+i\partial_y\right)\tilde{f}(z)=0,\ \ \
z\in{\mathbb R}. $$

Then, for a selfadjoint operator $A$,
\begin{equation*}
f(A)=\int d\tilde{f}(z) (A-z)^{-1},
\end{equation*}
which is absolutely convergent. Given $f$, one can construct
explicitely an almost analytic extension $\tilde{f}$ supported in
a complex neighbourhood of the support of $f$, and for $p\leq
k-2$,
\begin{equation}
\int\left|d\tilde{f}(z)\right|\,|\Im z|^{-p-1}\leq C\sum_{j=0}^k
\|f^{(j)}\|_{j-p-1}, \label{norms}
\end{equation}
where $ \|f\|_n=\int dx \langle x\rangle^n |f(x)|, $ and $\langle
x\rangle=(1+x^2)^{1/2}$. Moreover, the derivatives of $f(A)$ are
given by
\begin{equation}
f^{(p)}(A)=p!\int d\tilde{f}(A) (A-z)^{-p-1}. \label{deriv}
\end{equation}
These results extend by a limiting argument to functions $f$ that
do not have compact support, as long as the norms in the RHS of
\fer{norms} are finite.

%%%%%%%%%%%%%%%%%%%%%%%%%%%%%%%%%%%%%%%%%%%%%%%%%%%%%%%%%%%%%%%%
%\include{paradigm}

\chapter{The $C_n$ paradigm}

The results of this chapter are an extension of the analysis in
[JP1,2,3] to fermionic reservoirs and to models with
{\it time-dependent} interaction (since we are interested in thermodynamic processes). The analysis presented here is important for later chapters, in particular, to prove an adiabatic theorem for NESS in chapter 8, section 2, and to apply the isothermal theorem to a concrete example in chapter 8, section 3. Moreover, the methods developed here are applied in chapter 9 to study the spectrum of the so called Floquet Liouvillean, which we use to prove convergence to time-periodic states in cyclic thermodynamic processes. 

We first study the spectrum of the {\it standard}
Liouvillean using complex deformations (translations) for Model $C_n$
introduced in chapter 4, section 4: a two level system coupled to $n$ free
fermionic reservoirs. For the case $n=1$, and for time-independent
interaction, we show that the system possesses the property of RTE
in the mixing sense. We then introduce the C-Liouvillean of the coupled system, which is related to non-equilibrium steady states, and we study its
spectrum using complex deformation techniques. Finally we establish the existence of the so called deformed time evolution, which we use in chapter 8 to prove adiabatic theorems in NEQSM.

\section{Complex translations and the spectrum of $\L_g$}

We recall the basic properties of complex translations, and how
they relate to the problem at hand. We are a bit pedantic
in the presentation, so that the material is
self-contained as much as possible. A reader who is familiar with the
method may opt to skip this section in a first reading.

We work in the Araki-Wyss representation of the fermionic reservoirs for Model $C_n$ (see section 4.4 for a discussion of this model). Assume ($C_n.1$). For $\theta\in {\mathbf R}$, let
$\u_i(\theta )$ be the unitary transformation generating
translations in energy for the $i^{th}$ reservoir, $i=1,\cdots
,n$,

\begin{equation}
(\u_i(\theta )f_i)(u_i)=f_i^{(\theta )}(u_i)=f_i(u_i+\theta) \; ,
\end{equation}
and let $U_i(\theta )=d\Gamma (\u_i(\theta ))$ the second
quantization of $\u_i(\theta )$,
\begin{align}
U_i(\theta ) \varphi (f_{i, \beta}) U_i(-\theta ) &=\varphi (f_{i, \beta}^{(\theta )}) \; , \\
U_i (\theta ) d\Gamma (u_i) U_i (\theta ) &= d\Gamma (u_i) + N_i
\theta\; ,
\end{align}
where $N_i$ is the number operator of the $i^{th}$-reservoir
$\R_i$. Let $N=\sum_i N_i$, the total number of particles, and
$U(\theta)={\mathbf 1}^\S\otimes {\mathbf 1}^\S \otimes U_1
(\theta) \otimes\cdots\otimes U_n (\theta)$. It follows that the
deformed {\it standard} Liouvillean is
\begin{equation}
\L_g (t, \theta):= U(\theta )\L_g (t) U(-\theta) = \L_0 +
N\theta + V_g^{tot}(t,\theta) \; , \label{standardL}
\end{equation}
where $\L_0=\L^\S+\sum_i \L^{\R_i}$, $\L^{\R_i}=d\Gamma (u_i),
i=1,\cdots ,n$, and
\begin{eqnarray*}
V_g^{tot}(t,\theta)&=&gV^{tot}(t,\theta)= 
g \sum_{i=1}^n \{ \sigma_-\otimes {\mathbf 1}^\S\otimes b^*(f_{i,
\beta_i}^{(\theta )}(t))+\sigma_+\otimes {\mathbf 1}^\S\otimes b(f_{i,
\beta_i}^{(\theta )}(t))  \\
&-& i {\mathbf 1}^\S\otimes \sigma_-\otimes (-1)^{N_i} b^*(f_{i, \beta_i}^{\# (\theta )}(t))-i {\mathbf 1}^\S\otimes \sigma_+\otimes (-1)^{N_i} b(f_{i, \beta_i}^{\# (\theta)}(t))\} \; . 
\end{eqnarray*}

Explicitly, $U_i (\theta )= e^{i A_{\R_i}\theta }, i=1,\cdots,n$,
where $A_{\R_i}=d\Gamma (i\partial_{u_i})$, the second
quantization of the energy translations in the $i^{th}$ reservoir.

The follwoing lemmas are needed for applying complex deformation
techniques.

\vspace{0.5cm}

\noindent {\it Lemma 6.1}

The following holds for $0<\delta' <\delta$.

\begin{itemize}

\item[(i)] If $f\in H^2 (\delta, \B)$, the Hardy class of analytic functions as defined
in section 4.4, then $f'\in H^2(\delta',\B)$, where the prime stands for differentiation with respect to $u$. Furthermore, the following inequality holds
\begin{equation}
\| f' \|_{H^2 (\delta' ,\B)} \le \frac{1}{\delta-\delta'} \| f
\|_{H^2 (\delta,\B)}\; .
\end{equation}

\item[(ii)] If $f\in H^2 (\delta, \B)$. then the map $I(\delta )\ni \theta \rightarrow f^{(\theta )}\in L^2({\mathbf R};\B)$
is analytic in $\theta$, where the strip $I(\delta )$ has been
defined in chapter 4. Moreover, $\frac{d f^{(\theta )}}{d
\theta}=f'^{(\theta)}$.

\item[(iii)] If $\theta_1 ,\theta_2\in I(\delta')$, then, for any $f\in H^2(\delta, \B)$, one has the following estimate
\begin{equation}
\| f^{(\theta_1)}-f^{(\theta_2)} \|_{L^2 ({\mathbf R}; \B)}\le
\frac{|\theta_1 - \theta_2 |}{\delta-\delta'} \| f \|_{H^2
(\delta, \B)}\; .
\end{equation}

\end{itemize}

{\it Proof.} Since $f\in H^2 (\delta, \B)$, this implies that $f:
I(\delta)\rightarrow\B$ is analytic. We will prove the boundedness
of $f'$ by looking at its Fourier transform. Denote by $\hat{f}\in
L^2 ({\mathbf R};\B)$ the Fourier transform of $f$, then
\begin{equation*}
\| f \|_{H^2 (\delta, \B)}=
\sup_{|\eta|<\delta}\int_{-\infty}^{\infty} \| f(u+i\eta )
\|_{\B}^2 du =\sup_{|\theta|<\delta}\| e^{\theta r}\hat{f}(r)\|.
\end{equation*}
Therefore,
\begin{align*}
\| f' \|_{H^2 (\delta' , \B)} &= \sup_{|\theta | < \delta'} \| r e^{r\theta} \hat{f} \| \\
& \le \sup_{r\in {\mathbf R}} | re^{-(\delta-\delta')|r|}| \sup_{|\theta|<\delta'} \| e^{\theta r + (\delta-\delta')|r|\hat{f}} \| \\
& \le \frac{1}{e (\delta-\delta')} \sup_{|\theta |< \delta} \| e^{|\theta r|}\hat{f} \| \\
&\le \frac{\sqrt{2}}{e}\frac{1}{\delta-\delta'} \| f \|_{H^2
(\delta ,\B)},
\end{align*}
which implies claim (i).

Assume that $| \Im \theta |< \delta' < \delta$, then for small
enough $\epsilon\in {\mathbf C}$,
\begin{align*}
\| f^{(\theta + \epsilon)}-f^{(\theta )}-\epsilon f'^{(\theta)} \|
&= \| e^{i\theta r} (e^{i\theta \epsilon}-1-i\epsilon r)\hat{f}\| \\
&\le \sup_{r\in {\mathbf R}} | e^{-\Im \theta r -\delta' |r|}(e^{i\epsilon r}-1-i\epsilon r)| \| e^{\delta' |r|}\hat{f}\| \\
&\le O(\epsilon) \| f \|_{H^2 (\delta, \B)},
\end{align*}

as $\epsilon\rightarrow 0$. This implies (ii).

It follows from (ii) that
\begin{align*}
\| f^{(\theta_1)}-f^{(\theta_2)} \| &= \| (\theta_1 - \theta_2) \int_0^1 (f')^{(\theta_2 + t (\theta_1 - \theta_2))}dt \| \\
&\le |\theta_1 - \theta_2 | \sup_{0\le t \le 1} \|(f')^{(\theta_2 + t (\theta_1 - \theta_2))} \| \\
&\le | \theta_1 - \theta_2 | \| f' \|_{H^2 (\delta', \B)} \\
&\le \frac{\theta_1 -\theta_2}{\delta - \delta'} \| f \|_{H^2
(\delta', \B)},
\end{align*}

which is (iii). 
$\Box$

For $\theta \in I(\delta)$, it follows from ($C_n.1$) and
(\ref{standardL}) that the perturbed standard Liouvillean is
well-defined on the domain
$\D:=\D(N)\cap\D(\L^{\R_1})\cap\cdots\cap \D(\L^{\R_n})$. As an
operator on $\D$, $V_g^{tot} (\theta)$ satisfies $V_g^{tot 
}(\theta)^*\supset V_{\overline{g}}^{tot}(\overline{\theta})$, and
$V_g^{tot}(\theta )$ is closable for each $(g,\theta )\in {\mathbf
C}\times I(\delta)$. We will denote the closure of operators by
the same symbol.

\vspace{0.5cm}

\noindent {\it Lemma 6.2}

For $\theta\in {\mathbf C}$, the following hold:

\begin{itemize}
\item[(i)] For any $\psi\in \D$, one has
\begin{equation}
\| \L_0 (\theta ) \psi \|^2 = \| \L_0 (\Re\theta) \psi \|^2 + |
\Im \theta |^2 \| N \psi \|^2 \; .
\end{equation}

\item[(ii)] If $\Im \theta \ne 0$, then $\L_0 (\theta )$ is a normal operator satisfying
\begin{equation}
\L_0(\theta)^*=\L_0 (\overline{\theta}) \; ,
\end{equation}
and $\D(\L_0(\theta))=\D$.

\item[(iii)] The spectrum of $\L_0 (\theta )$ is
\begin{equation}
\sigma (\L_0 (\theta )) = \{ n\theta + s : n\in {\mathbf N} \;
{\mathrm and} \;  s\in {\mathbf R} \} \cup \sigma (\L^\S) \; .
\end{equation}
\end{itemize}

{\it Proof.} The first claim follows directly by looking at the
sector $N=n$, since $\L_0 (\theta)$ restricted to this sector is
reduced to
\begin{equation*}
\L_0^{n}(\theta )=\L^\S + s_1 + \cdots + s_n + n\theta\; .
\end{equation*}

Since $\D = \{ \psi = \{ \psi^{(n)} \} : \psi^{(n)} \in
\D(\L_0^{(n)}(\theta)) \; {\mathrm and} \; \sum_n \|
\L_0^{(n)}(\theta) \psi^{(n)}\| < \infty \}$, it follows that
$\L_0 (\theta )$ is a closed normal operator on $\D$. Claims (ii)
and (iii) follow from the corresponding statements on
$\L_0^{(n)}(\theta )$. $\Box$

Note that $V_g^{tot}(\theta )$ is bounded as a 
consequence of assumption ($C_n.1$).

\vspace{0.5cm}

\noindent {\it Lemma 6.3}

 Assume that $(g, \theta)\in {\mathbf
C}\times I^- (\delta )$, then

\begin{itemize}

\item[(i)] $\D(\L_g (t, \theta ))=\D$ and $\L_g (t, \theta)^*=\L_{\overline{g}}(t, \overline{\theta})$.
\item[(ii)] The spectrum of $\L_g (t, \theta )$ satisfies
\begin{equation}
\sigma (\L_g (t, \theta)) \subset \{ z\in {\mathbf C} : \Im z \le
 C(g, \theta ) \} \; ,
\end{equation}
where
\begin{equation}
C(g,\theta ):= \sup_{t\in {\mathbf R}} \{ \frac{|\Re g|}{\delta -
\Im g} | \Im \theta |^{1/2} + | \Im g | |\Im \theta |^{-1/2} \}
\sum_i \| f_{i,\beta_i} (t) \|_{H^2 (\delta, \B)}.
\end{equation}
Furthermore, if $\Im z > C(g, \theta )$, then
\begin{equation}
\| (\L_g (t, \theta) - z)^{-1} \| \le \frac{1}{\Im z -
C(g,\theta)} \; .
\end{equation}

\item[(iii)] The map $(g, \theta) \rightarrow \L_g(t, \theta)$ from ${\mathbf C}\times I^- (\delta)$
to the set of closed operators on $\H$ is an analytic family in
each variable separately; (see [Ka1], chapter V, section 3.2).

\end{itemize}

{\it Proof.} The first claim (i) follows from the fact that
$V_g^{tot}(t, \theta)$ is infinitesimally small with respect to
$\L_0 (\theta)$ for $\theta\in I(\delta)$. To establish the second
assertion, let $\hat{C}(t, g, \theta):= \sup \Im (\eta (\L_g (t,
\theta))$, where $\eta(\L_g (t, \theta))$ is the numerical range
of $\L_g(t, \theta)$. We know that $\sigma (\L_g (t,
\theta))\subset \eta (\L_g (t, \theta))$, and that $\| (\L_g (t,
\theta)-z)^{-1} \| \le \frac{1}{dist (z, \eta (\L_g (t,
\theta)))}$ for $z\in {\mathbf C}\backslash\eta (\L_g (t,\theta))$
(see [Hu]). One may readily check that $\hat{C}(t,g,\theta)\le
C(g,\theta)$ from Lemma 6.1 and (\ref{standardL}). This suffices
to prove the second claim. Since $\| \L_g (t,
\theta+\epsilon)-\L_g(t, \theta) -\epsilon \frac{\partial \L_g (t,
\theta)}{\partial\theta}\| = O(\sum_i
\|f_{i,\beta_i}^{(\theta+\epsilon)}-f_{i,\beta_i}^{(\theta)}-\epsilon
f'^{(\theta)}_{i,\beta_i}\| )$, it follows that the map $(g,
\theta)\rightarrow \L_g(t, \theta)$ is analytic in $\theta$.
Analyticity in $g$ is obvious from definition (\ref{standardL}).

$\Box$

\noindent Denote by 
$$\Xi (\eta):= \{ z\in {\mathbf C} : \Im z > \eta \} ,$$
which is an open half-plane.

\vspace{0.5cm}

\noindent {\it Proposition 6.4}

Suppose ($C_n.1$) and ($C_n.2$). Then there is a constant $g_1>0$,
independent of $\beta_i\in [\beta_0, \beta_*], i=1,\cdots ,n,
0<\beta_0<\beta_*<\infty$ fixed, such that the following hold.

\begin{itemize}

\item[(i)] If $|g| < g_1 |\Im \theta |$, then the spectrum of the operator $\L_g(t, \theta)$
in the half-plane $\Xi (\Im \theta + \frac{|g|}{g_1})$ is purely
discrete and independent of $\theta$.

\item[(ii)] If $|g|< \frac{1}{4} g_1 | \Im \theta |$, then the spectral projection
$P_g (t, \theta)$ associated to the spectrum of $\L_g (t, \theta)$
in the half-plane
 $\Xi (\Im \theta + \frac{|g|}{g_1})$ is analytic in $g$ and satisfies the estimate
\begin{equation}
\| P_g (t, \theta) - P_0  \| < \frac{3|g|}{g_1 | \Im \theta |}\; .
\end{equation}

\end{itemize}

{\it Proof.} The resolvent formula
\begin{equation}
\label{resolvent} (\L_g (t, \theta)-z)^{-1} = (\L_0 (\theta
)-z)^{-1}(1+ V_g^{tot}(t,\theta )(\L_0 (\theta )-z)^{-1})^{-1} \; ,
\end{equation}
holds for small $g$, as long as $z$ belongs to the cone $\{ z\in
{\mathbf C}: 0< c_1 < |z| < c_2 \Im z \}$, since
$V_g^{tot}(t,\theta)$ is infinitesimally small with respect to
$\L_0 (\theta )$ for $\theta\in I(\delta)$. The strategy of the
proof is as follows: We extend the domain of validity of
(\ref{resolvent}) by refining the estimate on $V_g^{tot}(t,
\theta)(\L_0(\theta)-z)^{-1}$, and then use analytic perturbation
theory as developed in [HP].

Note that
\begin{equation}
\| V_g^{tot} (t, \theta) (\L_0 (\theta)-z)^{-1} \| \le g \sum_i \|
f_{i,\beta_i}(t)\|_{H^2 (\delta, \B)} \| (\L_0
(\theta)-z)^{-1} \|.
\end{equation}

Since $N$ and $\L_0 (\theta)$ are commuting operators, one may
apply the spectral theorem to evaluate the norm of $A(z):=
V^{tot}(t, \theta)(\L_0 (\theta)-z)^{-1}$. On the sector $N=0$,
\begin{equation}
\| A^{(0)}(z) \| = \| (\L^\S - z)^{-1} \| = \frac{1}{dist (z,
\sigma (\L^\S))}\; .
\end{equation}
Without loss of generality, assume that $\Im \theta =: -\mu <0$,
ie, $\theta\in I^-(\delta)$. On the sector $N=n>0$,
\begin{align}
\| A^{(n)}(z)\} &= \| \frac{\sqrt{n+1}}{(\L^\S+s_1+\cdots +s_n +n\Re \theta -\Re z)-i (n\mu +\Im z)}\| \\
&= \frac{\sqrt{n+1}}{| n\mu + \Im z|}\; .
\end{align}

Since $\| A(z) \| = \sup_{n\ge 0} \| A^{(n)}(z) \|$,
\begin{equation}
\label{Az} \| A(z) \| \le
\begin{cases}
\frac{\sqrt{2}}{dist(z,\sigma (\L_0 (\theta)))}  ,  -\mu<\Im z < 3\mu, \\
\frac{1}{2\sqrt{\mu (\Im z -\mu)}} , 3\mu\le\Im z
\end{cases}.
\end{equation}

Let
\begin{equation}
\label{g1} g_1:= \sup_{t}\frac{1}{4\sum_i \|
f_{i,\beta_i}(t)\|_{H^2(\delta,\B)}}\; ,
\end{equation}
and $G(\theta, \epsilon):= \{ z\in {\mathbf C}: \Im z > \Im \theta
; dist(z,\sigma (\L_0 (\theta))>\epsilon)\}$. It follows from
(\ref{Az}) and (\ref{g1}) that, for $\epsilon < \mu$,
\begin{equation}
\sup_{z\in G(\theta,\epsilon)}\| V_g^{tot}(t, \theta)(\L_0
(\theta)-z)^{-1}\| \le \frac{|g|}{g_1\epsilon}.
\end{equation}
If $|g| < g_1\epsilon$, the resolvent formula (\ref{resolvent})
holds on $G(\theta, \epsilon)$, and for $N\ge 0$,
\begin{equation*}
\sup_{z \in G( \theta , \epsilon )} \| (z-\L_g (t,
\theta))^{-1}-\sum_{j=0}^{N-1}(z-\L_0 (\theta))^{-1}(V_g^{tot}(t,
\theta)(z-\L_0(\theta))^{-1})^j \|
\le\frac{1}{\epsilon}\frac{(\frac{|g|}{g_1\epsilon})^N}{1-\frac{|g|}{g_1\epsilon}}.
\end{equation*}
For small enough $|g|$, it follows that
\begin{equation}
\Xi (\Im \theta) \backslash \sigma (\L^\S) = \bigcup_{\epsilon>0}
G(\theta, \epsilon)\subset \rho (\L_g(t, \theta)) \; ,
\end{equation}
where $\rho (\L_g(t, \theta))$ is the resolvent set of $\L_g (t,
\theta)$. Therefore, the discrete spectrum of $\L_g(t,\theta)$ is
stable, and one may apply analytic perturbation theory. We still
need to prove the independence of (i) from $\theta\in I^-
(\delta)$. Fix $(g_0, \theta_0)\in {\mathbf C}\times I^- (\delta)$
such that $|g_0| < g_1 | \Im \theta |$. The discrete eigenvalues
of $\L_{g_0}(t, \theta)$ are analytic functions with at most
algebraic singularities in the neighbourhood of $\theta_0$, since
$\L_{g_0}(t, \theta)$ is analytic in $\theta$. Moreover, since
$\L_{g_0}(t, \theta_0)$ and $\L_{g_0}(t, \theta)$ are unitarily
equivalent if $(\theta - \theta_0)\in {\mathbf R}$, it follows
that the discrete spectrum of $\L_{g_0}(t, \theta)$ is independent
of $\theta$.

To prove (ii), assume that $2\epsilon < | \Im \theta |$ and
$|g|<g_1 \epsilon$, and let the contour $\gamma=\gamma_+ -
\gamma_-$, where $\gamma_{\pm}:= \{ z\in {\mathbf C}: \Im z = \pm
\Im \theta /2 \}$. At the formal level (for now!), let
\begin{equation}
P_g (t, \theta):= \oint_{\gamma} \frac{dz}{2\pi i}(z-\L_g (t,
\theta))^{-1}\; ,
\end{equation}
 which is the spectral projection onto the discrete spectrum of $\L_g (t, \theta)$.

Iterating the resolvent identity (\ref{resolvent}), one obtains
\begin{equation}
P_g(t, \theta) = P_0 +\Pi_g^{(1)}(t, \theta) + \Pi_g^{(2)}(t,
\theta)\; ,
\end{equation}
where
\begin{align}
P_0 &= P_0 (\theta) \; , \\
\Pi_g^{(1)}(t, \theta) &= \oint_{\gamma}\frac{dz}{2\pi i} (\L_0 (\theta)-z)^{-1}V_g^{tot}(t, \theta) (\L_0 (\theta )-z)^{-1} \; ,\\
\Pi_g^{(2)}(t, \theta) &= -\oint_{\gamma}\frac{dz}{2\pi i} (\L_0
(\theta)-z)^{-1}V_g^{tot}(t, \theta) (\L_g (t, \theta
)-z)^{-1}V_g^{tot}(t, \theta)(\L_0(\theta)-z)^{-1} \; .
\end{align}
Applying the spectral theorem (with respect to $\L_0 (\theta)$,
which is normal),
\begin{equation}
\Pi_g^{(1)}(t, \theta ) = -i \int_{-\infty}^\infty e^{-\mu
|s|}\Theta_s ds\; ,
\end{equation}
where
\begin{equation}
\Theta_s :=
\begin{cases}
P_0 e^{i\L_0 s} V_g^{tot} e^{-i\L_0 s} , s<0 \\
e^{i\L_0 s} V_g^{tot} e^{-i\L_0 s} P_0 , s>0 .
\end{cases}
\end{equation}

Since $\| \Theta_s \| \le \frac{|g|}{2\sqrt{2} g_1}$, we have the
following estimate
\begin{equation}
\| \Pi_g^{(1)}(t, \theta) \| \le \frac{|g|}{g_1 \mu} \; ,
\label{pi1}
\end{equation}
uniformly in $t\in {\mathbf R}$.

We still need to estimate the norm of $\Pi^{(2)}_g$. Consider two
vectors $\varphi, \psi\in \H$. By the Cauchy-Schwarz inequality,
\begin{equation}
| \langle \varphi , \Pi_g^{(2)} \psi\rangle |\le \sup_{z\in
\gamma} \| V_g^{tot}(t, \theta) (\L_g (t,
\theta)-z)^{-1}V_g^{tot}(t, \theta)\| \nu(\varphi)\nu(\psi)\; ,
\end{equation}
where $\nu(\cdot):= \{ \int_\gamma \frac{d|z|}{2\pi}\|
(\L_0(\theta)-z)^{-1}\cdot \|^2 \}^{1/2}$. By the spectral
theorem,
\begin{equation*}
\nu (\varphi) \le \sqrt{2/\mu}\|\varphi\|\;,
\end{equation*}
and
\begin{equation*}
\sup_{z\in \gamma}\| V_g^{tot}(t,\theta)(\L_0 (\theta)-z)^{-1}
V_g^{tot}(t,\theta)\| \le \frac{2g^2}{g_1^2 \mu} \; .
\end{equation*}
Again using the resolvent of the identity and the previous
estimates,
\begin{equation*}
\sup_{z\in \gamma}\| V_g^{tot}(t,\theta) (\L_g (t,
\theta)-z)^{-1}V_g^{tot}(t,\theta)\| \le \frac{2g^2}{g_1^2 \mu}
(1-\frac{|g|}{g_1\epsilon})^{-1}.
\end{equation*}
Optimizing the latter equation with respect to $\epsilon$ gives
$\epsilon=\mu/2$, and hence
\begin{equation}
\| \Pi_g^{(2)}(t, \theta) \| \le
\frac{4g^2}{g_1^2\mu^2}(1-\frac{2|g|}{g_1\mu})^{-1}\; ,
\label{pi2}
\end{equation}
uniformly in $t\in {\mathbf R}$.

Estimates (\ref{pi1}) and (\ref{pi2}) imply that
\begin{equation}
\label{projdif} \| P_g(t, \theta) - P_0 \| \le
x(\frac{1+2x}{1-2x}) \; ,
\end{equation}
with $x=\frac{|g|}{g_1\mu}<1/2$, independent of $t\in\mathbf{R}$.
$\Box$

Proposition 6.4 allows one to apply reduction theory to the
discrete spectrum of resonances as developed in [HP,JP1,2,3], and to
construct a quasi-Liouvillean acting on $\H^\S\otimes\H^\S$ by
transforming the reduced Liouvillean $\tilde{\L_g}:= P_g \L_g P_g$
from $Ran (P_g (\theta))$ to $\H^\S\otimes \H^\S$.

If $|g|<\frac{g_1\mu}{4}$, estimate (\ref{projdif}) implies that
$\| P_g(t, \theta) -P_0\| <1$. It follows that the maps
\begin{align*}
P_0 : Ran (P_g(t, \theta)) &\rightarrow \H^\S\otimes\H^\S \; ; \\
P_g(t, \theta) : \H^\S\otimes\H^\S &\rightarrow Ran (P_g(t,
\theta)) \;
\end{align*}
are isomorphisms for each fixed $t$.

Let $T_g (t):= P_0 P_g (t, \theta)P_0$, then
\begin{equation}
S_g(t, \theta):= T_g(t)^{-1/2} P_0 P_g (t, \theta): Ran(P_g(t,
\theta))\rightarrow \H^\S\otimes\H^\S
\end{equation}
has an inverse
\begin{equation}
S_g^{-1}(t,\theta):= P_g(t, \theta) P_0 T_g^{-1/2}(t):
\H^\S\otimes\H^\S \rightarrow Ran(P_g(t,\theta)) .
\end{equation}
Let
\begin{equation}
M_g(t):= P_0 P_g (t, \theta)\L_g(t, \theta)P_g(t, \theta)P_0\;,
\end{equation}
and define the {\it quasi-Liouvillean}
\begin{equation}
\label{qL} \Sigma_g(t) := S_g (t, \theta)P_g(t, \theta)\L_g(t,
\theta)P_g(t, \theta)S_g^{-1}(t, \theta)=
T_g^{-1/2}(t)M_g(t)T^{1/2}_g(t)\; ,
\end{equation}
which is nothing but the mapping of the reduced Liouvillean
$\tilde{\L_g}$ from $Ran(P_g)$ to $\H^\S\otimes\H^\S$.[HP,JP1,2,3]

Since $U(\theta)P_0=P_0U(\theta)=P_0 \; , \forall \theta\in {\mathbf
C}$, the operators $T_g(t)$ and $M_g(t)$ are independent of
$\theta$ for $|g|<\frac{g_1 \mu}{4}$.

\vspace{1cm}

\pagebreak

\noindent {\it Proposition 6.5}

Suppose assumptions ($C_n.1$) and ($C_n.2$) hold, and that
$|g|<\frac{g_1\mu}{4}$. Then, for each fixed time $t$, the
quasi-Liouvillean $\S_g(t)$ defined in (\ref{qL}) depends
analytically on $g$, and has a Taylor expansion of the form
\begin{equation}
\label{qLTaylor} \S_g(t)=\L^\S + \sum_{j=1}^{\infty} g^{2j}
\S^{(2j)} \; .
\end{equation}
The first non-trivial coefficient in (\ref{qLTaylor}) is
\begin{equation}
\S^{(2)}(t)= \frac{1}{2}\oint_\gamma \frac{dz}{2\pi i}(K(t,z)(z-\L^\S)^{-1}+ (z-\L^\S)^{-1}K(t,z))\; ,
\label{qLTaylor2}
\end{equation}
where
$K(t,z)=P_0 V_g^{tot}(t,\theta )(z-\L_0(\theta ))^{-1}V_g^{tot}(t,\theta )P_0.$

{\it Proof.} Fix the time $t\in {\mathbf R}$. Analyticity of $T_g$
directly follows from the previous proposition and the definition
of $T_g$. Since $\| T_g - 1 \| < 1$ for $|g|<\frac{g_1\mu}{4}$,
$T_g^{-1/2}$ is also analytic in $g$. Inserting the Neumann series
for the resolvent of $\L_g(t, \theta)$, gives
\begin{equation}
T_g = 1 + \sum_{j=1}^{\infty} g^j T^{(j)} \; ,
\end{equation}
with
\begin{equation}
T^{(j)} (t)=\oint_\gamma \frac{dz}{2\pi i} (z-\L^\S)^{-1} P_0
V^{tot}(t, \theta) ((z-\L_0(\theta))^{-1}V^{tot}(t,
\theta))^{j-1}P_0 (z-\L^\S)^{-1}\; .
\end{equation}

Similarly,
\begin{equation}
M_g(t)=\L^\S+\sum_{j=1}^{\infty}g^j M^{(j)}\; ,
\end{equation}
with
\begin{equation}
M^{(j)}(t)=\oint_\gamma \frac{dz}{2\pi i}z (z-\L^\S)^{-1} P_0
V^{tot}(t, \theta) ((z-\L_0(\theta))^{-1}V^{tot}(t,
\theta))^{j-1}P_0 (z-\L^\S)^{-1}\; .
\end{equation}

The odd terms in the above two expansions are zero due to the fact
that $P_0$ projects onto the $N=0$ sector. The first non-trivial
coefficient in the Taylor series of $\S_g$ is
\begin{align}
\S^{(2)}(t) &= M^{(2)}(t)-\frac{1}{2} (T^{(2)}(t)\L^\S + \L^\S T^{(2)}(t)) \\
&= \frac{1}{2} \oint_\gamma \frac{dz}{2\pi
i}(K(t,z)(z-\L^\S)^{-1}+(z-\L^\S)^{-1}K(t,z))\; , \label{2orderqL}
\end{align}
with $K(t, z)=P_0 V_g^{tot}(t,
\theta)(z-\L_0(\theta))^{-1}V_g^{tot}(t, \theta)P_0.$ $\Box$

Up until now we have assumed that $\Im \theta \ne 0$. The following
lemma asserts that under certain assumptions, one can take the
limit $\Im \theta \uparrow 0$ in the resolvent.

\vspace{0.5cm}

\noindent {\it Lemma 6.6}

Suppose that assumptions ($C_n.1$) and ($C_n.2$) hold. Then, for
$g\in {\mathbf R}$ and $\Im z$ large enough,
\begin{equation}
s-\lim_{\Im \theta\uparrow 0}(\L_g(t, \theta)-z)^{-1}= (\L_g(t,
\Re \theta)-z)^{-1}\; , \label{removeCD}
\end{equation}
for each fixed time $t\in {\mathbf R}$.

{\it Proof.} Without loss of generality, assume that $\Re \theta =
0$. Since $(\L_g(t, \theta)-z)^{-1}$ is uniformly bounded as $\Im
\theta\uparrow 0$ for $g\in {\mathbf R}$ and $\Im z$ large enough,
it is enough to prove (6.48) on a dense subspace. Let
$\B_v:= \{\psi\in\H : N\psi <v \}$, the subspace of finite
particle vectors and let $\D_v := \{ (\L_g(\theta)-z)\psi :
\psi\in \B_v \}$. For $\varphi\in D_v$,
\begin{align*}
&[(\L_g(t, \theta)-z)^{-1}-(\L_g(t)-z)^{-1}]\varphi \\
&= (\L_g(t)-z)^{-1}(\L_g(t)-\L_g (t,\theta))(\L_g(t,\theta)-z)^{-1}\varphi \\
&=(\L_g(t)-z)^{-1}(V_g^{tot}(t)-V_g^{tot}(t,\theta)-N\theta)(\L_g(t,\theta)-z)^{-1}\varphi
\;.
\end{align*}
Using this and the fact that $\D_v$ is dense in
$\D=\D(N)\cap\D(\L_0)$, it follows that
\begin{align*}
&\| [(\L_g(t, \theta)-z)^{-1}-(\L_g(t)-z)^{-1}](N+1)^{-1}\| \\
&\le \frac{|\theta |}{\Im z} (1+ \frac{2\sqrt{2}|g|}{\delta-\mu}
\sum_{i=1}^n \| f_{i,\beta_i}(t)\|_{H^2(\delta,\B )}) \|(N+1)(\L_g
(t, \theta)-z)^{-1}(N+1)^{-1}\| \; .
\end{align*}
Moreover, since $(N+1)(\L_g (t, \theta)-z)^{-1} (N+1)^{-1}$ is
uniformly bounded as $\Im \theta \uparrow 0$, it follows that
\begin{equation}
\lim_{\Im \theta\uparrow 0} \| [(\L_g(t,
\theta)-z)^{-1}-(\L_g(t)-z)^{-1}](N+1)^{-1}\| = 0 \; .
\end{equation}
$\Box$

\vspace{0.5cm}

Recall that for model $C_n$, we set all the chemical potentials equal to $\nu\in {\mathbf R}.$ The result can be generalized to arbitrary chemical potentials (see remark after Theorem 6.7).

\vspace{0.5cm}

\noindent {\it Theorem 6.7 (Spectrum of the {\it standard}
Liouvillean of $C_n$)}

Suppose assumptions ($C_n.1$) and ($C_n.2$) hold. Then there is a
constant $g_1$ independent of $\beta_i\in [\beta_0,\beta_*], 0<\beta_0<\beta_*<\infty,
i=1,\cdots, n$, such that the following holds for $|g|<g_1$ and
fixed $t$.
\begin{itemize}
\item[(i)] If there are at least two reservoirs to which the small system is
connected with different temperatures (ie, $\exists i,j\in
{1,\cdots, n} \; {\rm such \ \ that} \; \beta_i\ne\beta_j$) \footnote{(or different chemical potentials $\nu_i\ne \nu_j$; see remark after Theorem 6.7)}, then
the spectrum of $\L_g(t)$ is absolutely continuous for all $t\in
{\mathbf R}$. All the eigenvalues of $\L_g$ become resonances in
this case.

\item[(ii)] If there is only one reservoir at inverse temperature
$\beta$ and chemical potential $\nu$, then $dim \; Ker \L_g (t) = 1$. In particular, if
the coupling of the small system $\S$ to the reservoir $\R$ is
{\it time-independent}, ie, $V(t)=V$, then the system possesses
the property of return to equilibrium. For all states $\rho\in
\N$, the set of states normal to the initial state
$$\omega=\omega^\S \otimes \omega^\R_\beta , $$
we have
\begin{equation}
\lim_{t\rightarrow\infty} \rho (\a^t_g (A)) =
\omega_{\beta, \nu,g}(A)\; ,
\end{equation}
where $A\in \O=\O^\S\otimes\O^\R$ and $\omega_{\beta,\nu ,g}$ is the
($\a^t_g,\beta, \nu$)-KMS state of the coupled system. The limit is
exponentially fast, in the sense that $\exists \gamma (g)>0$, a
set of states $\N_0$ dense in $\N$, and a norm-dense subalgebra
$\O_0\subset\O$, such that for all $\rho\in\N_0$, $A\in\O_0$, and
$t>0$,
\begin{equation}
|\rho (\a^t_g(A)) - \omega_{\beta, \nu ,g}(A)| \le
C(A,\beta,g)e^{-\gamma (g)t} \; ,
\end{equation}
and $\omega_{\beta,g}$ is analytic in the coupling $g$.
\end{itemize}
 {\it Proof.}
The proof relies on the method of complex translations as developed in [JP1,2,3]. Fix $t\in {\mathbf R}$, and choose $k$ such that $0<k<\min
(\frac{\pi}{\beta_1}, \cdots, \frac{\pi}{\beta_n}, \delta)$, where
$\beta_i, i=1,\cdots,n,$ are the inverse temperatures of the
reservoirs and $\delta$ appears in assumption ($C_n.1$), chapter
4. Recall that
\begin{equation}
\L_g(t, \theta) := U(\theta)\L_g(t)U(-\theta) = \L_0 + N\theta +
V_g^{tot}(t, \theta) \; .
\end{equation}
The function $I^-(k)\otimes {\mathbf C}\ni (\theta , g)\rightarrow
\L_g(t, \theta)$ with values in the closed operators on $\H,$  is
an analytic family of type A in each variable separately (see for example
[Ka1]). The spectrum of $\L_0(\theta)$ consists of two simple
eigenvalues $E_2=-2$ and $E_3=2$, a doubly degenerate eigenvalue
at 0, $E_0=E_1=0$, and a sequence of lines $\{ i n \Im \theta +
{\mathbf R} : n\in {\mathbf N}^+ \}$.

Let $C(\delta,\beta_i):= \sup_{|\Im z|<\delta}|1+e^{-\beta_i
(z-\nu)}|^{-1/2}$, which is finite. Then
\begin{equation}
C:= \sup_{\theta\in I^-(k)} \| V^{(tot)}(t,\theta)\| \le 2\sqrt{2}
\sum_{i=1}^n C(k,\beta_i) \| f_{i,\beta_i}(t) \|_{H^2(k)} <\infty\;
,
\end{equation}
due to ($C_n.1$). Choose $g_1$ such that $g_1 C < (k-\mu)/4$,
where $\mu=|\Im \theta|$. Then, for $|g|<g_1$ and $-k<\Im \theta <
-(k+\mu)/2$, the essential spectrum $\sigma_{ess}(\L_g(t,\theta))$
is contained in the half-plane $\{ z\in {\mathbf C}: \Im z <-\mu
\}$. The location of the discrete spectrum of $\L_g(t,\theta )$
inside the half-plane $\Xi (-\mu)$ can be computed using regular
perturbation theory. Taking $g_1$ small enough, the discrete
spectrum of $\L_g(t, \theta)$ consists of four points (resonances)
$\{ E_k (g)\}_{k=0}^3$, such that $E_0(t,g), E_1(t,g)$ are
localized near 0, while $E_2(t,g),E_3(t,g)$ are localized near
$\mp 2$ respectively.

%%%%%%%%%%%%%%%%%%%%%%%%%%%%%%%%%%%%%%%%%%%%%%%%%%%%%%%%%%%%%%%%%%%%
%DETAILS OF THE CALCULATION
%%%%%%%%%%%%%%%%%%%%%%%%%%%%%%%%%%%%%%%%%%%%%%%%%%%%%%%%%%%%%%%%%%%%

From Proposition 6.5, we know that
\begin{align}
\S^{(2)} =& \frac{1}{2}\oint_\gamma \frac{dz}{2\pi i}  \{ P_0 V^{tot}(t, \theta )(z-\L_0 (\theta ))^{-1}V^{tot}(t, \theta )P_0 (z-\L^\S)^{-1} \nonumber \\
&+ (z-\L^\S)^{-1}P_0 V^{tot}(t, \theta )(z-\L_0 (\theta ))^{-1}V^{tot}(t, \theta )P_0 \} \; . \label{2orderPer}
\end{align}
We now apply regular perturbation theory to compute the shift of the eigenvalues of $\L_g(\theta )$ to second order in $g$. Let $P_k, k=0,\cdots , 3$ be the spectral projection corresponding to the eigenvalues $E_k$ of $\L^\S$ (recall that $E_0=E_1=0, E_2=-2; E_3=2$), and let 
\begin{equation}
\Gamma^{(2)}_k (t):= P_k \S^{(2)}(t)P_k \; .
\end{equation}

To compute the shift in $E_3=2$, look at $\Gamma^{(2)}_3 (t) = P_3 \S^{(2)}(t)P_3.$ Applying the Cauchy integration formula to \fer{2orderPer} gives
\begin{align*}
\Gamma^{(2)}_3 (t) = & \frac{1}{2}\{ P_3 P_0 \sigma_+ \otimes \unit^\S\otimes b(f_{i,\beta_i}^{(\theta )}(t))(2-\L_0(\theta))^{-1}\sigma_-\otimes\unit^\S \otimes b^*(f_{i,\beta_i}^{(\theta )}(t))P_0P_3 \\
&+  P_3 P_0 \unit^\S \otimes \sigma_+ \otimes b(f_{i,\beta_i}^{\#(\theta )}(t))(2-\L_0(\theta))^{-1}\unit^\S\otimes\sigma_-\otimes  b^*(f_{i,\beta_i}^{\# (\theta )}(t))P_0P_3 \} \; .
\end{align*}
Using the fact that 
\begin{eqnarray*}
\lim_{\epsilon\searrow 0}\Re \frac{1}{x-i\epsilon} &=& \PV \frac{1}{x}\; ,\\
\lim_{\epsilon\searrow 0}\Im \frac{1}{x-i\epsilon} &=& i\pi \delta (x) \; ,
\end{eqnarray*}
where $\PV$ stands for the Cauchy principal value,
it follows that 
\begin{eqnarray}
\Re \Gamma^{(2)}_3 &=& \sum_{i=1}^n \PV \int_{{\mathbf R}} du \frac{\| f_{i,\beta_i}(u,t)\|^2_\B}{2-u} \; , \\
\Im \Gamma^{(2)}_3 &=& -\pi \sum_{i=1}^n \int_{{\mathbf R}} du \| f_{i,\beta_i}(u,t)\|^2_\B \delta (u-2) = -\pi \sum_{i=1}^n \| f_{i,\beta_i}(2,t)\|^2_\B \; .
\end{eqnarray}

Similarly, 
\begin{eqnarray}
\Re \Gamma^{(2)}_2 &=& -\sum_{i=1}^n \PV \int_{{\mathbf R}} du \frac{\| f_{i,\beta_i}(u,t)\|^2_\B}{2-u} \; , \\
\Im \Gamma^{(2)}_2 &=& -\pi \sum_{i=1}^n \int du_{{\mathbf R}} \| f_{i,\beta_i}(u,t)\|^2_\B \delta (u-2) = -\pi \sum_{i=1}^n \| f_{i,\beta_i}(2,t)\|^2_\B \; .
\end{eqnarray}

We need to apply degenerate perturbation theory for the zero eigenvalue of $\L^\S.$ Using the definition of $f_{i,\beta_i}, f^\#_{i,\beta_i}$ and $\tilde{f}_i$ given in section 4.4,
\begin{eqnarray}
\Re \Gamma^{(2)}_{0,1} &=& 0 \; , \\
\Im \Gamma^{(2)}_{0,1} &=& -\pi \sum_{i=1}^n 
\frac{\|\tilde{f}_i(2,t)\|^2_\B}{2\cosh(\beta_i(2-\nu)/2)}
\left(
\begin{matrix}
e^{\beta_i (2-\nu)/2} & -1 \\
-1 & e^{-\beta_i (2-\nu)/2} 
\end{matrix}
\right) \; .
\end{eqnarray}
Therefore, to
second order in the coupling $g$,
\begin{equation}
E_{2,3}(t,g)=\pm (2 + g^2 \PV\int_{\mathbf R} du \frac{1}{2-u}
\sum_{i=1}^n \| f_{i,\beta_i}(u,t)\|^2_{\B}) -i\pi g^2
\sum_{i=1}^n \|f_{i,\beta_i}(2,t)\|^2_\B + O(g^4) \; .
\end{equation}
Note that $\Im E_{2,3}(t,g)<0$. Furthermore,
\begin{equation}
E_{0,1}(t,g)=g^2 a_{0,1}(t) + O(g^4)\; ,
\end{equation}
where $a_{0,1}(t)$ are the eigenvalues of the $2\times 2$ matrix
\begin{equation}
\label{secondorder0} 
 -i \pi \sum_{i=1}^n 
\frac{\|\tilde{f}_i(2,t)\|^2_\B}{cosh(\beta_i(2-\nu)/2)}
\left(
\begin{matrix}
e^{\beta_i (2-\nu)/2} & -1 \\
-1 & e^{-\beta_i (2-\nu)/2} 
\end{matrix}
\right) \; .
\end{equation}

\begin{itemize}
\item[(i)] If $\exists \beta_i\ne\beta_j$, for $i,j\in
\{1,\cdots,n\}$, it follows from (\ref{secondorder0}) that $\Im
E_{0,1}(t,g)<0$, and together with Lemma 6.6, it follows that all
the eigenvalues of the $\L_g (t)$ are pushed to the lower
half-plane, ie, they all become resonances, and the spectrum of
$\L_g(t)$ is purely absolutely continuous.

\item[(ii)] Suppose there is only one reservoir at inverse
temperature $\beta$(or equivalently several reservoirs with the
same temperature), then $\Omega_V(t)\in Ker \L_g(t)$ (see chapter
3). It follows from (\ref{secondorder0}) that $\Im E_1(t)=-\pi g^2
\| \tilde{f} (2,t) \|^2_\B +O(g^4) <0$, while
$E_0(t,g)=0$ with $\psi_\beta = \left( \begin{matrix} e^{-\beta(2-\nu)/4} \\
e^{\beta(2-\nu)/4}
\end{matrix}\right)$ the corresponding eigenvector of
$\S_g^{(2)}$ (which is consistent with $Ker \L_g \ge 1$). Together
with Lemma 6.6, this implies that 0 is a simple eigenvalue of
$\L_g(t)$, and that the spectrum of the Liouvillean away from zero
is absolutely continuous, $\sigma (\L_g (t))\backslash
\{0\}=\sigma_{ac}(\L_g (t))$.

In particular, if the perturbation is time-independent
($V(t)\equiv V$), the system possesses the property of return to
equilibrium in the mixing sense; (see chapter 3), with an explicit
rate of convergence to the perturbed equilibrium state
$\omega_{\beta ,\nu ,g}$ as $t\rightarrow \infty$.

Let
$${\mathcal E}\subset \F_- (L^2 ({\mathbf R};\B))$$
be the set of entire functions for the group $U(\theta )$, ie, for
$\psi\in \F_-(L^2({\mathbf R};\B))$,  $U(\theta )\psi$ has an
entire extension. Denote by $\C := \H^\S\otimes \H^\S \otimes \E$,
which is dense in $\H$. For $\Phi,\Psi\in \C, g\in {\mathbf R}$,
and $\Im z$ large enough, the function
\begin{equation}
\theta\rightarrow h(\theta) = \langle U(\overline{\theta})\Phi ,
(\L_g(\theta)-z)^{-1} U(\theta)\Psi\rangle
\end{equation}
is analytic on $I^-(k)$, and continuous on $I^-(k)\cup {\mathbf
R}$ due to Lemma 6.6. Hence, the map
\begin{equation}
z\rightarrow \langle U(\overline{\theta})\Phi ,
(\L_g(\theta)-z)^{-1}U(\theta)\Psi\rangle \; ,
\end{equation}
provides the meromorphic extension to the half-plane $\Xi (\Im \theta +
\frac{|g|}{g_1} )$. To see the exponential convergence to the
perturbed equilibrium state, define
\begin{equation}
f(t):= \langle \Phi, e^{-i\L_g t}\Psi \rangle \; .
\end{equation}
For $\Im z>0$, the transform (Fourier-Laplace)
\begin{equation}
\hat{f}(z):= \int_0^\infty f(t)e^{izt}dt = -i \langle \Phi,
(\L_g-z)^{-1}\Psi \rangle
\end{equation}
is well-defined, with inverse
\begin{equation}
\label{transform} f(t)=\frac{1}{2\pi} \int_{-\infty}^\infty
\hat{f}(u+i\eta)e^{-i(u+i\eta)t}du \; ,
\end{equation}
where $t,\eta>0$. Rewriting (\ref{transform}),
\begin{equation}
\label{ft} f(t)=\frac{1}{2\pi} \oint_\gamma
\hat{f}(z)e^{-izt}+\frac{1}{2\pi}\int_{-\infty}^\infty
\hat{f}(u-i(\mu-\epsilon'))e^{-i(u-i(\mu-\epsilon'))t}du \; ,
\end{equation}
where the contour $\gamma$ is as before, $\mu=|\Im \theta |$ and
$\mu>\epsilon'>\epsilon$. The first term in (\ref{ft}) is
\begin{equation}
\frac{1}{2\pi} \oint_\gamma \hat{f}(z)e^{-izt}dz = \langle
U(\overline{\theta})\Phi , S_g^{-1}(\theta) e^{-i\S_g
t}S_g(\theta) U(\theta) \Psi \rangle\; ,
\end{equation}
which converges to the ($\a^t_g,\beta, \nu$)-KMS state with a life-time
$\tau_\R=O(g^{-2})$. The second term in (\ref{ft}) is of order
$O(e^{-(\mu-\epsilon'')})$, for $\mu>\epsilon''>\epsilon$; (see,
for example, Theorem 19.2 in [Rud]).
\end{itemize}
$\Box$
\vspace{0.5cm}

\noindent {\it Remark.}
One may show using a similar computation that if all the temperatures of the reservoirs are equal, but the chemical potential of at least two reservoirs are different, the spectrum of the Liouvillean is absolutely continuous.

\vspace{0.5cm}

The fact that the kernel of the standard Liouvillean is empty if
at least two reservoirs have {\it different} temperatures (or chemical potentials) is consistent with expectation that the property of return to
equilibrium does not hold if one starts with asymmetric boundary
conditions. This {\it motivates} the introduction of the so-called
C-Liouvillean, which has a non-trivial kernel even when one starts
with asymmetric initial conditions, for the study of
non-equilibrium steady states (NESS). The C-Liouvillean was
introduced for the first time in [JP3].

%%%%%%%%%%%%%%%%%%%%%%%%%%%%%%%%%%%%%%%%%%%%************

\section{NESS and the C-Liouvillean}

Consider the Model $C_n, n>1$, such that the unperturbed initial
state of the system is
\begin{equation}
\Omega = \Omega^\S\otimes \Omega_{\beta_1}\otimes \cdots
\Omega_{\beta_n} \; ,
\end{equation}
where, without loss of generality, $\Omega^\S$ is the vector in
$\H^\S\otimes\H^\S$ corresponding to the trace state, and
$\Omega_{\beta_i}$ the vacuum of $\F^{\R_i}_-(L^2({\mathbf
R};\B))$.\footnote{The results of this section hold for initial states that are normal to $\Omega.$} Construct the Banach space $C(\O,\Omega)$, which is the
vector space $\O\Omega=\{ A\Omega : A\in \O\}$ with norm $\|
A\Omega \|_\infty = \|A\|$. There is a Banach space isomorphism,
\begin{equation}
\label{BSiso} \O\ni A \rightarrow A\Omega \in C(\O,\Omega)\; .
\end{equation}
Under the isomorphism (\ref{BSiso}), the time evolution from time $t'$ to $t$ is mapped
to
\begin{equation}
\a^{t,t'}_g (A) \rightarrow \tilde{U}_g(t',t)A\Omega\; ,
\end{equation}
such that $\tilde{U}_g(t',t)\Omega = \Omega$. Let $L_g$ be the
generator of $\tilde{U}_g$, such that
\begin{align}
\partial_t \tilde{U}_g(t,t') &= -i L_g(t) \tilde{U}_g(t,t') \; \\
\tilde{U}_g(t,t) &= 1 \; .
\end{align}
Differentiating both sides of the equation
$$\a^{t,t'}_g (A)\Omega = \tilde{U}_g(t',t) A\Omega$$
with respect to $t$ and setting $t=t'$ gives
\begin{align*}
[(\L_0+gV(t))A-A(\L_0+gV(t))]\Omega &= [(\L_0+gV(t))A-(V(t)A^*)^*]\Omega \\
&= (\L_0+gV(t)-gJ\Delta^{1/2}V(t)\Delta^{-1/2}J)A\Omega \\
&\equiv L_g(t)A\Omega \; ,
\end{align*}
where $J=J^\S\otimes J^{\R_1}\otimes \cdots \otimes J^{\R_n}$ and
$\Delta={\mathbf 1}^\S\otimes
\Delta^{\R_1}\otimes\cdots\otimes\Delta^{\R_n}$ are the modular
conjugation and the modular operator of the Model $C_n$.

We define the C-Liouvillean to be
\begin{equation}
\label{CL} L_g(t):= \L_0 + gV(t) -
gJ\Delta^{1/2}V(t)\Delta^{-1/2}J \; .
\end{equation}

Assumption ($C_n.3$) on the perturbation is sufficient to show
that $L_g(t)$ satisfy the conditions of the Yosida-Hille-Phillips
Theorem, and Theorem X.70 in [RS2], and hence $\tilde{U}_g^t\equiv \tilde{U}_g(t,0)$ can be extended to a strongly continuous group on the Banach space
$C(\O,\Omega)$.

We will show later in this chapter that when the perturbation is
time-independent, the state of the coupled system converges to a
non-equilibrium steady state. The NESS is related to the zero
energy resonance of the adjoint of the C-Liouvillean. Before doing
so, we study the spectrum of $L_g(t)$, which is generally
time-dependent. (As an application of the latter, we will prove a
novel adiabatic theorem for states close to NESS in chapter 8.)

The results of the previous section regarding complex deformations
are directly translated to this case.

%%%%%%%%%%%%%%%%%%%%%%%%%%%%%%%%%%%%%%%%%%%%%%%%%%********************

\section{Spectrum of the C-Liouvillean for $C_n$}

We study the spectrum of the (adjoint of the) C-Liouvillean for Model $C_n$ in the
time-dependent case using the method of complex deformations as developed earlier in this chapter; (see also [JP1,2,3]). The deformed
adjoint of the C-Liouvillean is
\begin{equation}
L^*_g (t, \theta):= U(\theta )L^*_g (t) U(-\theta) = \L_0
+ N\theta + g\tilde{V}^{tot}(t,\theta) \; , \label{CL}
\end{equation}
where $\L_0=\L^\S+\sum_i \L^{\R_i}$, $\L^{\R_i}=d\Gamma (u_i),
i=1,\cdots ,n$, and
\begin{eqnarray*}
\tilde{V}^{tot}(t,\theta) &=& \sum_i \{ \sigma_+\otimes {\mathbf
1}^\S \otimes b (f_{i,\beta}^{(\theta )}(t)) + \sigma_-\otimes {\mathbf
1}^\S \otimes b^* (f_{i,\beta}^{(\theta )}(t)) \\
&-& i {\mathbf
1}^\S \otimes \sigma_+ \otimes (-1)^{N_i} b(e^{-\beta_i (u_i-\nu)/2
}f_{i, \beta_i}^{\# (\theta )} (t)) - i {\mathbf
1}^\S \otimes \sigma_- \otimes (-1)^{N_i} b^*(e^{\beta_i (u_i-\nu)/2}f^{\#
(\theta )}_{i,\beta_i}) \} \; .
\end{eqnarray*}
Let
\begin{equation}
P_g' (t, \theta):= \oint_{\gamma} \frac{dz}{2\pi i}(z-L^*_g (t,
\theta))^{-1}\; ,
\end{equation}
where the contour $\gamma$ as in section 6.1, and
$T_g' (t):= P_0 P_g' (t, \theta)P_0$, and
\begin{equation}
\label{quasiCL}
S_g'(t,\theta ):= (T_g'(t))^{-1/2} P_0 P_g' (t, \theta):
Ran(P_g'(t, \theta))\rightarrow \H^\S\otimes\H^\S
\end{equation}
which we will show has an inverse
\begin{equation}
(S_g')^{-1}(t, \theta):= P_g'(t, \theta) P_0 (T_g')^{-1/2}(t):
\H^\S\otimes\H^\S\rightarrow Ran(P'_g(t,\theta)) \; .
\end{equation}
Moreover, let
\begin{equation}
M_g'(t):= P_0 P_g' (t, \theta)L_g(t, \theta)P_g'(t, \theta)P_0\;,
\end{equation}
and define the quasi-C-Liouvillean
\begin{equation}
\label{qL2} \Sigma_g'(t) := S_g' (t, \theta)P_g'(t, \theta)L^*_g(t,
\theta)P_g'(t, \theta)(S_g')^{-1}(t, \theta)=
(T_g')^{-1/2}(t)M_g'(t)(T_g')^{1/2}(t)\; ,
\end{equation}
which is nothing but the mapping of the reduced C-Liouvillean
$\tilde{L_g}=P_g'L_gP_g'$ from $Ran(P_g')$ to $\H^\S\otimes\H^\S$.

\vspace{0.5cm}

\noindent {\it Theorem 6.8}

Suppose ($C_n.2$) and ($C_n.3$) (see section 4.4). Then there is a constant $g_1>0$ such
that the following holds.
\begin{itemize}
\item[(i)] Assume that $(g, \theta)\in {\mathbf C}\times I^- (\delta )$, then $\D(L^*_g (t, \theta ))=\D$ and $(L^*_g (t,
\theta))^*=L_{\overline{g}}(t, \overline{\theta})$. Moreover, the spectrum of
$L^*_g (t, \theta )$ satisfies
\begin{equation}
\sigma (L^*_g (t, \theta)) \subset \{ z\in {\mathbf C} : \Im z \le
 C(g, \theta ) \} \; ,
\end{equation}
where
\begin{align*}
C(g,\theta )&:= \sup_{t\in {\mathbf R}} \{2 \frac{|\Re g|}{\delta -
\Im g} | \Im \theta |^{1/2} + | \Im g | |\Im \theta |^{-1/2} \}
\times\\
&\times \sum_i \{ \| f_{i,\beta_i} (t) \|_{H^2 (\delta, \B)} +
\| e^{-\beta_i (u_i-\nu)/2} f_{i,\beta_i}(t) \|_{H^2(\delta,\B)} \} \; .
\end{align*}
Furthermore, if $\Im z > C(g, \theta )$, then
\begin{equation}
\| (L^*_g (t, \theta) - z)^{-1} \| \le \frac{1}{\Im z -
C(g,\theta)} \; ,
\end{equation}
and the map $(g, \theta) \rightarrow  L^*_g(t, \theta)$ from ${\mathbf
C}\times I^- (\delta)$ to the set of closed operators on $\H$ is
an analytic family in each variable separately.

\item[(ii)] If $|g| < g_1 |\Im \theta |$, then the spectrum of the operator
$L^*_g(t, \theta)$ in the half-plane $\Xi (\Im \theta +
\frac{|g|}{g_1})$ is purely discrete and independent of $\theta$.
If $|g|< \frac{1}{4} g_1 | \Im \theta |$, then the spectral
projection $P_g' (t, \theta)$ associated to the spectrum of $L^*_g
(t, \theta)$ in the half-plane
 $\Xi (\Im \theta + \frac{|g|}{g_1})$ is analytic in $g$ and satisfies the estimate
\begin{equation}
\| P_g' (t, \theta) - P_0 \| < \frac{3|g|}{g_1 | \Im \theta |}\; .
\end{equation}

\item[(iii)]If $|g|<\frac{g_1 |\Im \theta |}{4}$, then, for each fixed time $t$, the
quasi-C-Liouvillean $\S_g'(t)$ defined in (\ref{qL2}) depends
analytically on $g$, and has a Taylor expansion of the form
\begin{equation}
\label{qLTaylor2} \S_g'(t)=\L^\S + \sum_{j=1}^{\infty} g^{2j}
(\S')^{(2j)} \; .
\end{equation}
The first non-trivial coefficient in (\ref{qLTaylor2}) is
\begin{align*}
(\S')^{(2)}(t) &= \frac{1}{2} \oint_{\gamma} \frac{dz}{2\pi i} \{
P_0 \tilde{V}^{tot}(\theta,t)(z-\L_0(\theta))^{-1}
\tilde{V}^{tot}(\theta , t) P_0 (z-\L^{\S})^{-1} + \\
&+ (z-\L^{\S})^{-1}P_0 \tilde{V}^{tot}(\theta ,t)
(z-\L_0(\theta))^{-1}\tilde{V}^{tot}(\theta, t)P_0 \} \; .
\end{align*}

\item[(iv)] For
$g\in {\mathbf R}$ and $\Im z$ large enough,
\begin{equation}
s-\lim_{\Im \theta\uparrow 0}(L^*_g(t, \theta)-z)^{-1}= (L^*_g(t,
\Re \theta)-z)^{-1}\; , \label{removeCD}
\end{equation}
for each fixed time $t\in {\mathbf R}$.

\end{itemize}

The proof of the claims of Theorem 6.8 are very similar to those in section 6.1: the proof of claim (i) is similar to that of Lemma 6.3, of claim
(ii) to Proposition 6.4, of claim (iii) to Proposition 6.5 and of claim (iv)
to Lemma 6.6. In order to avoid redundancy, we will not repeat the proofs.

We are interested in the spectrum of the adjoint of the
C-Liouvillean in order to prove convergence to NESS and an
adiabatic theorem for states close to NESS. Let
$k=min(\frac{\pi}{\beta_1}, \cdots ,\frac{\pi}{\beta_n},\delta)$,
where $\delta$ appears in assumption ($C_n.3$), and choose
$\theta\in I^-(k)$.

\vspace{0.5cm}

\noindent {\it Proposition 6.9 (Spectrum of $L_g^*(\theta)$)}

Assume ($C_n.2$) and ($C_n.3$), choose $\theta\in I^-(k)$ and fix
$t\in{\mathbf R}$. Then there exists a constant $g_1>0$
independent of $t$ and $\theta$, such that, for $0<|g|<g_1$, the
essential spectrum of $L_g^*(\theta, t)$,
$\sigma_{ess}(L_g^*(\theta,t))\in {\mathbf C}\backslash
\Xi(-\mu)$, where $\mu=|\Im \theta |$, and the discrete spectrum
$\sigma_{disc}(L_g^*(\theta))\in \Xi(-\mu)$, with all eigenvalues
in the lower half-plane {\it except} for {\it one} eigenvalue at
zero.

{\it Proof.} Since we assumed ($C_n.2$) and ($C_n.3$), the results of Theorem 6.8 hold. Similar to the
proof of Theorem 6.7, the essential spectrum $\sigma(L^*_g(\theta,
t))\in {\mathbf C}\backslash \Xi(-\mu)$ for $|g|<g_1$. The
perturbation of the discrete spectrum can be studied by mapping
the reduced operator $P_g'(\theta,t)L_g^*(\theta,t)P_g'(\theta,t)$
to the quasi-C-Liouvillean $\S_g'(t)$, defined in (6.76),
which acts on $\H^\S\otimes\H^\S$, as in Theorem 6.7, section 6.7.

%%%%%%%%%%%%%%%%%%%%%%%%%%%%%%%%%%%%%%%%%%%%%%%%%%%%%%%%%%%%%%%%%%%%%%
%DETAILS OF THE CALCULATION
%%%%%%%%%%%%%%%%%%%%%%%%%%%%%%%%%%%%%%%%%%%%%%%%%%%%%%%%%%%%%%%%%%%%%%

The remaining discussion is very similar to the proof of Theorem 6.7. Let 
\begin{equation}
\label{gamma2}
(\Gamma')^{(2)}_k (t):= P_k (\S')^{(2)}(t)P_k \; .
\end{equation}

Applying the Cauchy integration formula to \fer{gamma2} gives
\begin{align*}
&(\Gamma')^{(2)}_3 (t) = \frac{1}{2}\{ P_3 P_0 \sigma_+ \otimes \unit^\S\otimes b(f_{i,\beta_i}^{(\theta )}(t))(2-\L_0(\theta))^{-1}\sigma_-\otimes\unit^\S\otimes b^*(f_{i,\beta_i}^{(\theta )}(t))P_0P_3 \\
&+  P_3 P_0 \unit^\S \otimes \sigma_+ \otimes b(e^{-\beta_i (u_i-\nu)/2}f_{i,\beta_i}^{\#(\theta )}(t))(2-\L_0(\theta))^{-1}\unit^\S\otimes\sigma_- \otimes b^*(e^{\beta_i (u_i-\nu)/2}f_{i,\beta_i}^{\# (\theta )}(t))P_0P_3 \} \; .
\end{align*}
It follows that 
\begin{eqnarray}
\Re (\Gamma')^{(2)}_3 &=& \sum_{i=1}^n \PV \int_{{\mathbf R}} du \frac{\| f_{i,\beta_i}(u,t)\|^2_\B}{2-u} \; , \\
\Im (\Gamma')^{(2)}_3 &=& -\pi \sum_{i=1}^n \int_{{\mathbf R}} du \| f_{i,\beta_i}(u,t)\|^2_\B \delta (u-2) = -\pi \sum_{i=1}^n \| f_{i,\beta_i}(2,t)\|^2_\B \; .
\end{eqnarray}

Similarly, 
\begin{eqnarray}
\Re (\Gamma')^{(2)}_2 &=& -\sum_{i=1}^n \PV \int_{{\mathbf R}} du \frac{\| f_{i,\beta_i}(u,t)\|^2_\B}{2-u} \; , \\
\Im (\Gamma')^{(2)}_2 &=& -\pi \sum_{i=1}^n \int du_{{\mathbf R}} \| f_{i,\beta_i}(u,t)\|^2_\B \delta (u-2) = -\pi \sum_{i=1}^n \| f_{i,\beta_i}(2,t)\|^2_\B \; .
\end{eqnarray}

Now apply degenerate perturbation theory for the zero eigenvalue. Using the definition of $f_{i,\beta_i}$ and $f^\#_{i,\beta_i}$ given in section 4.4,
\begin{eqnarray}
\Re (\Gamma')^{(2)}_{0,1} &=& 0 \; , \\
\Im (\Gamma')^{(2)}_{0,1} &=& -\pi \sum_{i=1}^n 
\frac{\|\tilde{f}_i(2,t)\|^2_\B}{cosh(\beta_i(2-\nu)/2)}
\left(
\begin{matrix}
e^{\beta_i (2-\nu)/2} & -e^{\beta_i(2-\nu)/2} \\
-e^{-\beta_i(2-\nu)/2} & e^{-\beta_i (2-\nu)/2} 
\end{matrix}
\right) \; .
\end{eqnarray}
Therefore, to
second order in the coupling $g$,
\begin{equation}
E'_{2,3}(t,g)=\pm (2 + g^2 \PV\int_{\mathbf R} du \frac{1}{2-u}
\sum_{i=1}^n \| f_{i,\beta_i}(u,t)\|^2_{\B}) -i\pi g^2
\sum_{i=1}^n \|f_{i,\beta_i}(2,t)\|^2_\B + O(g^4) \; ,
\end{equation}
while
\begin{equation}
E_{1,2}'(g,t)= g^2 a_{1,2}(t) + O(g^4) \; ,
\end{equation}
where $a_{1,2}(t)$ are the eigenvalues of the matrix
\begin{equation}
-i\pi \sum_{i=1}^n 
\frac{\|\tilde{f}_i(2,t)\|^2_\B}{2\cosh(\beta_i(2-\nu)/2)}
\left(
\begin{matrix}
e^{\beta_i (2-\nu)/2} & -e^{\beta_i(2-\nu)/2} \\
-e^{-\beta_i(2-\nu)/2} & e^{-\beta_i (2-\nu)/2} 
\end{matrix}
\right) \; .
\end{equation}

Since $\Omega$ is an eigenvector corresponding to the isolated
zero eigenvalue of $L_g(\theta,t)$ (by construction,
$L_g(t, \theta)\Omega=0$), then zero is also an
isolated eigenvalue of $L_g^*(\theta,t)$. (One way of seeing this
is to take the adjoint of the spectral projection of $L_g(\theta,t)$
corresponding to $\Omega$, which is defined using the resolvent and functional calculus.) In fact, $\psi=\left(\begin{matrix} 1\\ 1 \end{matrix}\right)$ is the eigenvector corresponding to the
zero eigenvalue of $(\S_g')^{2}(t)$. Hence,
\begin{align}
E_0'(g,t)&=0 \; ,\\
E_1'(g,t)&=-i\pi g^2 \sum_{i=1}^n \| \tilde{f}_i (2,t) \|^2_\B
+O(g^4)\; .
\end{align}
Moreover, we know from Theorem 6.8, (iv), that $s-\lim_{\Im \theta
\uparrow 0} (z-L^*_g(\theta,t))^{-1} = (z-L^*_g(\Re \theta,t))^{-1}$
for $\Im z$ big enough, and hence the claim of this proposition.
$\Box$

The following corollary says that when the interaction is {\it
time-independent}, $V(t)=V$, the coupled system converges to a
non-equilibrium steady-state exponentially fast. {\it This result has been proven in [JP3], but we mention it for the sake of completeness.} The NESS will
correspond to the zero energy resonance of $L_g^*$. This provides
some of the background for stating and proving a novel adiabatic
theorem for states close to NESS in chapter 8.

Let $D$ be a positive bounded operator on $\H$ such that $Ran (D)$
is dense in $\H$ and $D\Omega=\Omega$.

\vspace{0.5cm}

\noindent {\it Corollary 6.10 (NESS)}

Suppose assumptions ($C_n.2$) and ($C_n.3$), and that the perturbation
$V(t)\equiv V$ is {\it time-independent}. Then there exists
$g_1>0$ such that, for $|g|<g_1$ and $a\in \D(D^{-1})$, the
following limit exists,
\begin{equation}
\lim_{t\rightarrow\infty} \langle \Omega ,\a^t_g (a) \Omega
\rangle = \langle \Omega_g, D^{-1} a \Omega \rangle \; ,
\end{equation}
where $\Omega_g$ corresponds to the zero-energy resonance of
$L_g^*$, and $\a^g_t$ is the perturbed dynamics. This limit is
exponentially fast, with relaxation time $\tau_R=O(g^{-2})$.

{\it Proof.} Choose $k=\min(\frac{\pi}{\beta_1}, \cdots ,
\frac{\pi}{\beta_n},\delta)$, where $\delta$ appears in assumption
($C_n.3$), and let $\theta\in I^-(k)$. We already know the
spectrum of $L_g^*(\theta)$ from Proposition 6.9. Let
\begin{equation}
D:= {\mathbf 1}^\S\otimes {\mathbf 1}^\S \otimes e^{-k
\tilde{A}_{\R_1}}\otimes\cdots\otimes e^{-k\tilde{A}_{\R_n}} \; ,
\end{equation}
where $\tilde{A}_{\R_j}=d\Gamma(\sqrt{p_j^2+1})$, and
$p_j=i\partial_{u_j}$ is the generator of energy translations for
the $\R_j$ reservoir, $j=1,\cdots ,n$. Moreover, let $\h^{test}=
D(e^{k\sqrt{p^2+1}})$, and $\O^{test}=\F_-(\h^{test})$, which is
dense in $\F_-(L^2({\mathbf R};\B))$. For $a\in
\O^\S\otimes\O^{test}_1\otimes\cdots\O^{test}_n$,
\begin{align}
\lim_{t\rightarrow\infty} \langle \Omega , \a^t_g (a) \Omega
\rangle &= \lim_{t\rightarrow\infty}\langle \Omega, e^{itL_g}a
e^{-itL_g}\Omega \rangle \\
&= \lim_{t\rightarrow\infty} \langle e^{-itL_g^*} \Omega , a
\Omega
\rangle \\
&= \lim_{t\rightarrow\infty} \langle e^{-itL_g^*(\theta)} \Omega ,
a(\overline{\theta}) \Omega
\rangle \\
&= \lim_{t\rightarrow\infty} \frac{1}{2\pi i} \langle
\int_{-\infty}^\infty du (u+i\eta -
L_g^*(\theta))^{-1}e^{-i(u+i\eta)t} \Omega ,a (\overline{\theta})
\Omega\rangle \; ,
\end{align}
for $\eta>0$. One may decompose the last integral into two parts
(as in the proof of Theorem 6.7,(ii)). The first part is
\begin{equation}
\lim_{t\rightarrow\infty} \frac{1}{2\pi i} \langle
\oint_{\gamma}dz (z -L_g^*(\theta ) )^{-1} e^{-izt}\Omega,
a(\overline{\theta}) \Omega \rangle = \langle \Omega_g , D^{-1} a
\Omega \rangle \; ,
\end{equation}
where the zero energy resonance is
\begin{equation}
\Omega_g:=DU(-\theta)P_g'(\theta)U(\theta)D\Omega=
DU(-\theta)P_g'(\theta)\Omega \; .
\end{equation}
The second term converges to zero exponentially fast as
$t\rightarrow\infty$, since
\begin{equation}
\frac{1}{2\pi i} \langle \int_{-\infty}^\infty
(u-i(\mu-\epsilon)-L_g^*(\theta))^{-1}
e^{-i(u-i(\mu-\epsilon))t}\Omega, a(\overline{\theta}) \Omega
\rangle = O(e^{-(\mu-\epsilon')t}) \; ,
\end{equation}
where $0<\epsilon'<\epsilon<|\Im\theta|=\mu$.

$\Box$

\section{Existence of the {\it deformed} time evolution}

In the following we let
\begin{equation}
A(t):= \L_0 + g\mathbf{V}^{tot}(t) \; \,
\end{equation}
where $\mathbf{V}^{tot}(t)$ equals $V(t)-JV(t)J$ or
$V(t)-J\Delta^{-1/2}V(t)\Delta^{1/2}J$, ie, $A(t)=\L_g(t)$ or
$L^*_g(t)$. We also make the following assumption.

\begin{itemize}

\item[($C_n.7$)] Assume ($C_n.1$) if $A(t)=\L_g$, and
assume ($C_n.3$) if $A(t)=L^*_g$. (see section 4.4)

\end{itemize}

Let $\mathbf{U}(t)$ be the propagator generated by $A(t)$, and
which satisfies the initial value problem
\begin{equation}
\partial_t {\mathbf U}(t)=-iA(t){\mathbf U}(t) \; ,
{\mathbf U}(t=0)=1 \; ,
\end{equation}
ie, ${\mathbf U}(t)=U_g(t)$ if $A(t)=\L_g(t)$, and ${\mathbf
U}(t)=\tilde{U}_g(t)$ if $A(t)=L^*_g(t)$. (Remark regarding the
notation: In the previous section, $\tilde{U}_g$ was generated by
$L_g$, but it will denote the propagator generated by $L_g^*$ here.)

Choose $\theta\in I^-(\delta)$, where $\delta$ appears in
assumption ($C_n.7$), and let $A(t,\theta):=
U(\theta)A(t)U(-\theta)$. The {\it deformed} time evolution is
given by the propagator ${\mathbf U}(t,t',\theta)$ which satisfies
the initial value problem
\begin{equation}
\label{dte}
\partial_t {\mathbf U}(t,t',\theta) = -i A(t,\theta)
{\mathbf U}(t,t',\theta)\; , {\mathbf U}(t,t,\theta)=1 \; .
\end{equation}

The following two Lemmas guarantee the existence of ${\mathbf
U}(t,t',\theta)$. Let $\E_i\subset\F^{\R_i}_-(L^2(\mathbf{R};\B))$
the set of entire states for $U_i(\theta)$,
$\C=\H^\S\otimes\H^\S\otimes (D(\L^{\R_1})\cap\E_1)\otimes \cdots
\otimes (D(\L^{\R_n})\cap\E_n)$, and
\begin{equation}
\label{supInt} C:= \sup_{t\in\mathbf{R}}\sup_{\theta\in
I^-(\delta)} \| \mathbf{V}^{tot}(t,\theta) \| \; .
\end{equation}

\vspace{0.5cm}

\noindent {\it Lemma 6.11}

Assume ($C_n.7$), choose  $\theta\in I^-(\delta) \cup \mathbf{R}$
and $|g|< g_1 $, and fix $t \in \mathbf{R}$. Then

\begin{itemize}
\item[(i)] $A(t,\theta )$ with domain $\D$ generates a contraction semi-group
$e^{-i\sigma A(t, \theta)},\sigma\ge 0$ on $\H$, such that

\begin{equation}
\| (iA(t,\theta)-z)^{-1} \| \le | \Re (z+C) |^{-1} \; ,
\end{equation}
for $\Re z<0.$

\item[(ii)] The $\B(\H)$-valued function $e^{-i\sigma
A(t,\theta)}$ is analytic in $\theta\in I^-(\delta)$. For
$\theta'\in\mathbf{R}$ and $\theta\in I^-(\delta)\cup\mathbf{R}$,
\begin{equation}
\label{twicedeformation} U(\theta') e^{-i\sigma
A(t,\theta)}U(-\theta')= e^{-i\sigma A(t,\theta + \theta')} \; .
\end{equation}
\end{itemize}

{\it Proof.} Claim (i) follows from Phillip's Theorem for the
perturbation of semigroups (see [Ka1] chapter IX). Analyticity of
$e^{-i\sigma A(t,\theta)}$ and (\ref{twicedeformation}) follow
from assumption ($C_n.7$), the resolvent equation
\begin{equation*}
(A(t,\theta)-z)^{-1} = (\L_0(\theta)-z)^{-1}
(1+\mathbf{V}^{tot}(t,\theta)(\L_0(\theta)-z)^{-1})^{-1} \; ,
\end{equation*}
\begin{equation*}
U(\theta')A(t,\theta)U(-\theta')=A(t,\theta + \theta') \; ,
\end{equation*}
and the fact that
\begin{equation*}
e^{-i\sigma A(t,\theta)}=\frac{1}{2\pi i}\int_{\Gamma} e^{-\sigma
z} (iA(t,\theta)-z)^{-1}dz \; ,
\end{equation*}
where $\Gamma$ is a contour encircling the spectrum of $A(t)$.
$\Box$

\vspace{0.5cm}

\noindent {\it Lemma 6.12}

Assume ($C_n.7$) and let $\theta\in I^-(\delta)\cup \mathbf{R},
|g|<g_1$. Then (\ref{dte}) generates a unique propagator
$\mathbf{U}(t,t',\theta)$ such that the following hold.

\begin{itemize}

\item[(i)] ${\mathbf U}(t,t',\theta){\mathbf U}(t',t'',\theta)={\mathbf U}(t,t'',\theta)$
for $t\ge t'\ge t''$.

\item[(ii)] ${\mathbf U}(t,t',\theta)\D\subset\D$, and for
$\psi\in\D$, ${\mathbf U}(t,t',\theta)\psi$ is differentiable in
$t$ and $t'$ such that
\begin{align}
\partial_t {\mathbf U}(t,t',\theta) \psi&= -i
A(t,\theta){\mathbf U}(t,t',\theta)\psi\; , \\
\partial_{t'}{\mathbf U}(t,t',\theta)\psi &=
i{\mathbf U}(t,t',\theta)A(t',\theta) \psi\; .
\end{align}

\item[(iii)] For $\theta'\in\mathbf{R}$,
\begin{equation}
U(\theta'){\mathbf U}(t,t',\theta) U(-\theta') = {\mathbf
U}(t,t',\theta + \theta') \; .
\end{equation}
Moreover, ${\mathbf U}(t,t',\theta)$ is analytic in $\theta\in
I^-(\delta)$.

\end{itemize}

{\it Proof.} Claims (i) and (ii) are consequences of Kato's
Theorem [Ka2], to which we refer the reader. Without loss of
generality, rescale time such that $t=\tau s, s\in [0,1]$, and let
$A^{n}(s\tau,\theta)=A(\tau\frac{k}{n},\theta)$ for $n\in
\mathbf{N}^+$ and $s\in[\frac{k}{n},\frac{k+1}{n}],
k=0,\cdots,n-1$. Moreover, define ${\mathbf U}^{n}(\tau s,\tau
s',\theta)=e^{-i\tau(s-s')A^n(\tau\frac{k}{n},\theta)}$ if
$\frac{k}{n}\le s'\le s \le \frac{k+1}{n}$, and ${\mathbf
U}^n(\tau s,\tau s',\theta)={\mathbf U}^n(\tau s,\tau
s'',\theta){\mathbf U}^n(\tau s'',\tau s',\theta)$ if $0\le s'\le
s'' \le s \le 1$. It follows from Lemma 6.11 that, for
$\theta'\in\mathbf{R}$,
\begin{equation*}
U(\theta'){\mathbf U}^n (\tau s, \tau s' ,\theta) U(-\theta) =
{\mathbf U}^n(\tau s, \tau s', \theta +\theta') \; ,
\end{equation*}
and that ${\mathbf U}^n(\tau s, \tau s', \theta)$ is analytic in $\theta\in
I^-(\delta)$, where $\delta$ appears in ($C_n.7$). Claim (iii) follows by taking the $n\rightarrow\infty$ limit. 
$\Box$

%%%%%%%%%%%%%%%%%%%%%%%%%%%%%%%%%%%%%%%%%%%%%%%%%%%%%%%%%%%%%%%
%\include{isothermal}
\chapter{Isothermal theorem and (reversible) isothermal processes}

%%%%%%%%%%%%%%%%%%%%%%%%%%%%%%%%%%%%%%%%%%%%%%%%%%%%%%%%%%%%%%%%%%%%%%%%%%%%%%
%INTRODUCTION
%%%%%%%%%%%%%%%%%%%%%%%%%%%%%%%%%%%%%%%%%%%%%%%%%%%%%%%%%%%%%%%%%%%%%%%%%%%%%%

In this chapter, we investigate isothermal processes of a finitely
extended, driven quantum system $\S$ in contact with an infinite
heat bath ${\mathcal R}$ at inverse temperature $\beta$ from the
point of view of quantum statistical mechanics. A theorem
characterizing {\it reversible} isothermal processes as {\it
quasi-static} processes (``{\it isothermal theorem}'') is
described, which is an adiabatic theorem for states close to
thermal equilibrium at constant temperature. We also discuss
corollaries of this theorem and their physical significance
pertaining to the changes of entropy and free energy in reversible
isothermal processes and on the $0^{th}$ law of thermodynamics.

As discussed in chapter 2, the dynamics of a system composed of a
small system $\S$ with a finite dimensional Hilbert space coupled
to an infinitely extended dispersive reservoir is generated by a
(generally time-dependent) thermal Hamiltonian, or {\it standard }
Liouvillean,
\begin{equation}
\label{standL} \L_g(t):= \L_0(t)+g(t)I \; ,
\end{equation}
where
\begin{equation}
\label{freeL} \L_0(t)= ({\mathbf 1}\otimes {\mathbf 1})\otimes
\L_\beta^{\mathcal R} + (H_0^{\Sigma}(t)\otimes {\mathbf
1})\otimes {\mathbf 1} - ({\mathbf 1}\otimes
H_0^{\Sigma}(t))\otimes {\mathbf 1}
\end{equation}
is the Liouvillean of the uncoupled system, $\L_\beta^{\mathcal
R}$ is the Liouvillean of the heat bath, $H_0^{\Sigma}(t)$ is the
 Hamiltonian of the small system, and
where $g(t)I$ is a spatially localized term describing the
interactions between ${\mathcal R}$ and $\Sigma$, with a
time-dependent coupling constant $g(t)$.

We will only consider heat baths with a unique equilibrium
$\Omega_\beta^{\mathcal R}$ state at each temperature (ie, with no
phase coexistence). We have seen in chapters 3 that if
 $\L_0(t)\equiv \L_0$ and $g(t)\equiv g$ are independent of $t$, for $t\ge t_*$,
``return to equilibrium'' holds true if we can prove that $\L_g$
has a simple eigenvalue at $0$ and that the spectrum, $\sigma
(\L_g)$, of $\L_g$ is purely continuous away from $0$. The
eigenvector, $\Omega_\beta\equiv\Omega_\beta^{\S\vee\R}$, of
$\L_g$ corresponding to the eigenvalue $0$ is the thermal
equilibrium state of the coupled system $\S\vee\R$ at temperature
$(k_B \beta)^{-1}$. Under suitable hypotheses on ${\mathcal R}$
and $\Sigma$, we have shown the property of RTE in the mixing case
is described by an {\it exponential law} involving a finite {\it
relaxation time}, $\tau_{\mathcal R}$; (see chapters 5 and 6 where
the property of RTE has been verified for physical models).

If the interacting Liouvillean $\L_g(t)$ of $\S\vee\R$ depends on
time $t$, but with the property that, for all times $t$, $\L_g(t)$
has a simple eigenvalue at $0$ corresponding to an eigenvector
$\Omega_\beta(t)$, then $\Omega_\beta(t)$ can be viewed as an {\it
instantaneous equilibrium} (or {\it reference}) {\it state}, and
$\tau_{\mathcal R}(t)$ is called instantaneous relaxation time of
$\S\vee\R$. Let $\tau$ be the time scale over which $\L_g(t)$
changes appreciably. Assuming that, at some time $t_0$, the state
, $\Psi(t_0)$, of $\S\vee\R$ is given by $\Omega_\beta(t_0)$, it
is natural to compare the state $\Psi(t)$ of $\S\vee\R$ at a {\it
later} time $t$ with the instantaneous equilibrium state
$\Omega_\beta(t)$ and to estimate the norm of the difference
$\Psi(t)-\Omega_\beta(t)$. One would expect that if $\tau \gg
\sup_{t} \tau_{\mathcal R}(t)$, then
\begin{equation*}
\Psi(t)\simeq \Omega_\beta (t) \; .
\end{equation*}

In this chapter, we prove an adiabatic theorem for states close to
thermal equilibrium, which we call ``{\it isothermal theorem}'',
saying that
\begin{equation}
\Psi(t)\stackrel{\tau\rightarrow\infty}\rightarrow \Omega_\beta(t)
\; ,
\end{equation}
for all times $t\ge t_0$. The physical consequences of this
theorem have been sketched in chapter 2; e.g., to show that {\it
quasi-static} ($\tau\rightarrow\infty$) isothermal processes are
{\it reversible} and that, in the quasi-static limit, a variant of
the $0^{th}$ {\it law} holds. We propose general definitions of
heat flux and of entropy for trajectories of states of $\S\vee\R$
sampled in arbitrary isothermal processes and use the isothermal
theorem to relate these definitions to more common ones from
equilibrium statistical mechanics.

%%%%%%%%%%%%%%%%%%%%%%%%%%%%%%%%%%%%%%%%%%%%%%%%%%%%%%%%%%%%%%%%%%%%%%%%%%%%%
%ADIABATIC THEOREM
%%%%%%%%%%%%%%%%%%%%%%%%%%%%%%%%%%%%%%%%%%%%%%%%%%%%%%%%%%%%%%%%%%%%%%%%%%%%%

\section{An adiabatic theorem}

In this section we carefully state and prove an adiabatic
theorem, which is a slight improvement of results in [AE,Te]
concerning adiabatic theorems for Hamiltonians without spectral
gaps. Our simplest result follows from those in [AE,Te] merely by
eliminating the superfluous hypothesis of semiboundedness of the
generator of time evolution.

Let ${\mathcal H}$ be a separable Hilbert space, and let $\{ L(s)
\}_{s\in I}$, with $I\subset {\mathbf R}$ a compact interval, be a
family of selfadjoint operators on ${\mathcal H}$ with the
following properties:

\begin{itemize}
\item[(H7.1)] The operators $L(s),s\in I,$ are selfadjoint on a
common domain, ${\mathcal D}$, of definition dense in ${\mathcal
H}$.

\item[(H7.2)] The resolvent $R(i,s):=(L(s)-i)^{-1}$ is bounded and
differentiable, and $L(s)\dot{R}(i,s)$ is bounded uniformly in
$s\in I$, where $\dot{( \; )}$ denotes the derivative with respect
to $s$.

\end{itemize}

\vspace{0.5cm}

\noindent {\it Lemma 7.1 (Existence of time evolution).} If
assumptions (H7.1) and (H7.2) hold then there exist unitary
operators $\{ U(s,s')| s,s' \in I \}$ with the properties:

For all $s,s',s''$ in $I$,
\begin{equation*}
U(s,s)={\mathbf 1} \; , U(s,s')U(s',s'')=U(s,s'') \; ,
\end{equation*}
$U(s,s')$ is strongly continuous in $s$ and $s'$, and
\begin{equation}
i\frac{\partial}{\partial s}U(s,s')\Psi=L(s)U(s,s')\Psi \; ,
\end{equation}
for arbitrary $\Psi\in {\mathcal D}$, $s,s'$ in $I$; ($U$ is
called a ``{\it propagator}'').

%%%%
{\it Proof.} Note that $A(s):=iL(s)+1$ is a generator of a
contraction semigroup, which follows from the fact that $A(s)$
satisfies the conditions of the Yosida-Hille theorem:
$(-\infty,0]\in \rho (A(s))$ and $|| (\lambda+A(s))^{-1} || \le
\frac{1}{\lambda}$ for all $\lambda > 0$ and every $s\in I$, where
$\rho (A(s))$ is the resolvent of $A(s)$ (see, for example, Theorem
X.47 in [RS2]). Furthermore, $0\in \rho (A(s))$, for $s,s' \in I$,
$A(s')A(s)^{-1}$ is bounded by the closed graph theorem and
(H7.1), and for small $|s-s'|$, $\| (s'-s)A(s')^{-1}A(s)-{\mathbf
1}\|=\| \dot{A}^{-1}(s)A(s)\|+o(|s-s'|)$ is bounded due to (H7.2).
Hence, by Theorem X.70 in [RS2] (or Theorem 2 in Chp XIV [Yo],
section 4), $\tilde{U}^\tau (s,s')$, which satisfies the initial
value problem
\begin{equation*}
\partial_s \tilde{U}(s,s')=-\tau A(s) \tilde{U} (s,s') \; ;  \; \tilde{U} (s,s)=1 \; ,
\end{equation*}
exists uniformly in $s,s'\in I$ and $s' < s$. The existence of the
unitary evolution $U (s,s')$ follows by noting that $U(s,s')
=\tilde{U}(s,s') e^{(s-s')\tau}$.
$\Box$

In order to prove an adiabatic theorem, one must require some
additional assumptions on the operators $L(s)$.

\begin{itemize}

\item[(H7.3)] We assume that $L(s)$ has an eigenvalue $\lambda (s)$,
that $\{ P(s) \}$ is a family of finite rank projections such that
$L(s)P(s)=\lambda (s)P(s)$, $P(s)$ is twice continuously
differentiable in $s$ with bounded first and second derivatives,
for all $s\in I$, and that $P(s)$ is the spectral projection of
$L(s)$ corresponding to the eigenvalue $\lambda (s)$ for almost
all $s\in I$.

\end{itemize}

We consider a quantum system whose time evolution is generated by
a family of operators
\begin{equation}
L_\tau (t):= L(\frac{t}{\tau}) ,\; \frac{t}{\tau}=: s\in I \; ,
\end{equation}
where $\{ L(s) \}_{s\in I}$ satisfies assumptions (H7.1)-(H7.3).
The propagator of the system is denoted by $U_\tau(t,t')$. We
define
\begin{equation}
U^{(\tau )}(s,s'):= U_\tau (\tau s, \tau s')
\end{equation}
and note that $U^{(\tau )}(s,s')$ solves the equation
\begin{equation}
\label{realtimeevol} i\frac{\partial}{\partial
s}U^{(\tau)}(s,s')\Psi=\tau L(s) U^{(\tau)}(s,s')\Psi\; , \Psi\in
{\mathcal D}\;,
\end{equation}
and $U^{(\tau)}(s,s)=1$.

Next, we define
\begin{equation}
L_a(s):= L(s) +\frac{i}{\tau }[\dot{P}(s),P(s)]
\end{equation}
and the corresponding propagator, $U_a^{(\tau )}(s,s')$, which
solves the equation
\begin{equation}
\label{adiabtimeevol} i\frac{\partial }{\partial s} U_a^{(\tau
)}(s,s')\Psi = \tau L_a(s)U_a^{(\tau )}(s,s')\Psi , \; \Psi\in
{\mathcal D}\; ,
\end{equation}
and $U_a^{(\tau)}(s,s)=1$.
The propagator $U_a^{(\tau )}$ describes what one calls the {\it
adiabatic time evolution}. (Note that the operators $L_a(s), s\in
I$, satisfy (H7.1) and (H7.2), since, by (H7.3), $\frac{i}{\tau
}[\dot{P}(s),P(s)]$ are bounded, selfadjoint operators with
bounded derivative in $s$.)

\vspace{0.5cm}

\noindent {\it Theorem 7.2 (Adiabatic Theorem).}

If assumptions (H7.1)-(H7.3) hold then
\begin{align*}
&\; \; \; \;(i) \; U_a^{(\tau )}(s',s)P(s)U_a^{(\tau )}(s,s')=P(s') \; (intertwining \; property)\; , \\
&{\mathrm for \; arbitrary \; } s,s'\; {\mathrm in \;} I , {\mathrm  \;and} \\
&\; \; \; \;(ii) \lim_{\tau\rightarrow\infty} \sup_{s,s'\in I}
||U^{(\tau )}(s,s')-U_a^{(\tau )}(s,s')||=0 \; .
\end{align*}

{\it Proof.} The proof essentially follows that of [Te].
\begin{itemize}
\item[(i)] Equality trivially holds when $s=s'$. Moreover, the derivative of the LHS with respect to $s$ is zero
\begin{align*}
&\partial_s  [U_a^{(\tau )}(s',s)P(s)U_a^{(\tau )}(s,s')] \\
&=  \dot{U}_a^{(\tau )}(s',s)P(s)U_a^{(\tau )}(s,s')+ U_a^{(\tau )}(s',s)\dot{P}(s)U_a^{(\tau )}(s,s')+ U_a^{(\tau )}(s',s)P(s)\dot{U}_a^{(\tau )}(s,s') \\
&= - U_a^{(\tau )}(s',s)[\dot{P}(s)P(s)+P(s)\dot{P}(s)]U_a^{(\tau )}(s,s')+ U_a^{(\tau )}(s',s)\dot{P}(s)U_a^{(\tau )}(s,s') \\
&= 0 \; ,
\end{align*}
using equations (\ref{realtimeevol}) and (\ref{adiabtimeevol}),
assumption (H7.3) and the fact that
$P(s)\dot{P}(s)+\dot{P}(s)P=\dot{P}(s)$ (the latter follows from
differentiating $P^2(s)=P(s)$).

\item[(ii)]
We will use Cook's argument and an extension of Kato's commutator
equation for cases when there is no spectral gap.

Consider $\psi\in {\mathcal D}$. Using the {\it Duhamel} formula,
\begin{align*}
(U^{(\tau)} (s,s') - U^{(\tau)}_a (s,s'))\psi 
&= -\int_{s'}^s du \partial_u (U^{(\tau)} (s,u)U^{(\tau)}_a (u,s')) \psi \\
&= - \int_{s'}^s du U^{(\tau)} (s,u)
[\dot{P}(u),P(u)] U^{(\tau)}_a (u,s') \psi \; .
\end{align*}

Since ${\mathcal D}$ is dense in ${\mathcal H}$, it follows that
\begin{equation*}
|| U^{(\tau)} (s,s')-U^{(\tau)}_a (s,s') || = || \int_{s'}^s du
U^{(\tau)} (s,u) [\dot{P}(u),P(u)] U^{(\tau)}_a (u,s')|| \; .
\end{equation*}

Now, using the commutator equation
\begin{equation*}
\label{commutatoreq} [\dot{P}(u),P(u)] = [L(u),X_\epsilon (u)] +
i\epsilon X_\epsilon (u) \; ,
\end{equation*}
where $X_\epsilon (u)= R(\lambda (u) - i \epsilon,
u)\dot{P}(u)P(u)+P(u)\dot{P}(u)R(\lambda (u) + i \epsilon, u)$,
one can write the integrand as a total derivative plus a
remainder. One may check that the commutator equation is satisfied
by direct substitution.

We claim that there is a constant $C < \infty$ such that, for
small enough $\epsilon$,
\begin{align*}
&(a) || X_\epsilon (u) || < \frac{C}{\epsilon} \; , \\
&(b) || \dot{X}_\epsilon (u) || < \frac{C}{\epsilon^2} \; ,\\
&(c) \lim_{\epsilon\rightarrow 0}\epsilon^{1/2} || X_\epsilon (u) || = 0
\; .
\end{align*}

Inequality $(a)$ is a direct consequence of assumptions (H7.1) and
(H7.3),
\begin{align*}
|| R(\lambda (u) - i \epsilon, u) \dot{P}(u) P(u) || &\le || R(\lambda (u) - i\epsilon,u)|| ||\dot{P}(u)P(u)||  \\
                                                 & < \frac{C}{\epsilon} \; .
\end{align*}

To prove inequality $(b)$, note that, using the resolution of the
identity,

\begin{align*}
& \dot{R}(\lambda (u) - i\epsilon,u ) = - R(\lambda (u) -i\epsilon,u )(\dot{L}(u)-\dot{\lambda}(u))R(\lambda (u)-i\epsilon,u) \\
&= -R(i,u)(L(s)-i-\lambda (u)+\lambda (u) -i\epsilon +i \epsilon) \times \\
& \ \ \ \times R(\lambda (u)-i\epsilon,u)(\dot{L}(u)-\dot{\lambda}(u))R(\lambda (u)-i\epsilon,u) \\
&= -R(i,u) (\dot{L}(u)-\dot{\lambda}(u) ) R(\lambda (u)-i\epsilon,u) -\\
& \ \ \  -(\lambda
(u)-i\epsilon-i)R(i,u)R(\lambda(u)-i\epsilon,u)(\dot{L}(u)-\dot{\lambda}(u))R(\lambda(u)-i\epsilon,u)\;
,
\end{align*}
and that $\lambda (u)=\frac{Tr(L(u)P(u))}{Tr(P(u))}$ is
continuously differentiable since $P(u)$ and $L(u)$ are
differentiable.

Inequality $(b)$ now follows from assumptions (H7.1)-(H7.3) for
small enough $\epsilon$.

In order to prove $(c)$, consider $\psi\in {\mathcal D}$, and let
$\varphi (u):=\dot{P}(u)P(u)\psi$. Since $P(u)\dot{P}(u)P(u)=0$,
it follows that
\begin{equation*}
P(u)\varphi(u)=P(u)(1-P(u))\dot{P}(u)P(u)\psi = 0 \; .
\end{equation*}

Now, using the spectral theorem,
\begin{align*}
\lim_{\epsilon^{1/2}\rightarrow 0} ||\epsilon^{1/2} X_\epsilon (u) \psi ||^2
&= \lim_{\epsilon\rightarrow 0} 2\epsilon \langle R(\lambda(u)-i\epsilon,u)\varphi (u), R(\lambda (u)-i\epsilon,u)\varphi(u) \rangle \\
&= \lim_{\epsilon\rightarrow 0} 2 \epsilon \langle \varphi (u), R(\lambda (u)+i\epsilon,u)R(\lambda(u)-i\epsilon,u)\varphi(u) \rangle \\
&= \lim_{\epsilon\rightarrow 0} 2 \int_{\sigma (L(u))} d\mu_{\varphi (u)}(\eta) \frac{\epsilon}{(\eta-\lambda(u))^2+\epsilon^2} \\
&= 2\mu_{\varphi(u)}(\lambda(u))=0 \; ,
\end{align*}
since $\varphi(u)\in Ker\{ P(s) \}$. 

%try to find an alternative proof not based on the spectral theorem, if possible

Furthermore, using (\ref{realtimeevol}) and (\ref{adiabtimeevol}),
\begin{equation}
\label{normcom} \| \int_{s'}^s du U^{(\tau)}(s,u) [L(u),
X_{\epsilon}(u) ]U_a^{(\tau)} (u,s') \| \le \frac{1}{\tau}
\sup_{s}\{ (2\| \dot{P}(s)P(s) \| + 1) \|X_{\epsilon}(s)\| + \|
\dot{X_{\epsilon}}(s)\| \} \; .
\end{equation}

From (H7.3), equation (\ref{commutatoreq}), and estimates
$(a)-(c)$,(\ref{normcom}), it follows that

\begin{equation}
\| U^\tau (s,s')-U^\tau_a(s,s') \| < \frac{A}{\tau \epsilon} +
\frac{B}{\tau \epsilon^2} + \epsilon^{1/2} C(\epsilon) \; ,
\end{equation}
where $A,B<\infty$ are constants independent of $\epsilon$, and
$\lim_{\epsilon\rightarrow 0} C(\epsilon)=0$. The second
claim in the Theorem follows if we choose $\epsilon$ such that
$\epsilon\rightarrow 0$ and $\tau \epsilon^2\rightarrow \infty$ as
$\tau\rightarrow \infty$. $\Box$

\end{itemize}

\vspace{0.5cm}

{\it Remarks.}

(1) We note that $U^{(\tau )}(s',s)=U^{(\tau )}(s,s')^*$.

(2) With more precise knowledge about the nature of the spectrum
of $\{ L(s) \}$, one can obtain information about the speed of
convergence in $(ii)$, as $\tau \rightarrow \infty$; see chapter
8.

(3) {\it Adiabatic Theorem for Resonances}. This result resembles
the adiabatic theorem described above, but eigenstates of $L(s)$
are replaced by resonance states, and one must require the
adiabatic time scale $\tau$ to be small as compared to the life
time, $\tau_{res}(s)$, of a resonance of $L(s)$, uniformly in
$s\in I$. (For shape resonances, the techniques in [FP] are
useful; see [A-SF1].)

%%%%%%%%%%%%%%%%%%%%%%%%%%%%%%%%%%%%%%%%%%%%%%%%%%%%%%%%%%%%%%%%%%%%%%%%%
%ISOTHERMAL THEOREM
%%%%%%%%%%%%%%%%%%%%%%%%%%%%%%%%%%%%%%%%%%%%%%%%%%%%%%%%%%%%%%%%%%%%%%%%%

\section{The isothermal theorem}

In this section, we turn to the study of isothermal processes of
``small'' driven quantum systems, $\Sigma$, in diathermal contact
with a heat bath, ${\mathcal R}$, at a fixed temperature
$(k\beta)^{-1}$.

Let $\L_g^\tau(t):=\L_g(\frac{t}{\tau})$ denote the Liouvillian of
the coupled system $\S\vee\R$, where $\{ \L_g(s) \}_{s\in I}$ is
as in eqs.(\ref{standL}) and (\ref{freeL}) of section 7.1 and satisfies
assumptions (H7.1) and (H7.2) of section 7.2. The interval
\begin{equation*}
I_{\tau }=\{ t \; |\;  \frac{t}{\tau }\in I\subset {\mathbf R} \}
\end{equation*}
is the time interval during which an isothermal process of
$\S\vee\R$ is studied.

We assume that $\Sigma$ is driven ``slowly'', i.e., that $\tau$ is
large as compared to the relaxation time $\tau_{\mathcal
R}=\max_{s\in I}\tau_{\mathcal R}(s)$ of $\S\vee\R$.

Assumption (H7.3) of section 7.2 is supplemented with the following
more specific assumption.

\begin{itemize}

\item[(H7.4)] For all $s\in I\equiv [s_0,s_1]$, the operator
$\L_g(s)$ has a {\it single, simple eigenvalue} $\lambda(s)=0$,
the spectrum, $\sigma(\L_g(s))\backslash \{ 0 \}$, of $\L_g(s)$
being {\it purely continuous} away from 0. It is also assumed
that, for $s\le s_0$, $\L_g(s)\equiv \L_g$ is independent of $s$
and has spectral properties sufficient to prove return to
equilibrium, as discussed in chapters 5 and 6.

\end{itemize}

Let $\Omega_\beta (s)\in {\mathcal H}$ denote the eigenvector of
$\L_g(s)$ corresponding to the eigenvalue 0, for $s\le s_1$. Then
$\Omega_\beta(\frac{t}{\tau })$ is the {\it instantaneous
equilibrium state} of $\S\vee\R$ at time $t$. Let
\begin{equation}
\label{Proj} P_g(s)= | \Omega_\beta (s)\rangle\langle
\Omega_\beta(s) |
\end{equation}
denote the orthogonal projection onto $\Omega_\beta (s)$; $P_g(s)$
is assumed to satisfy (H7.3).

%%%%%%%%%%%%%%%%%%%%%%%%%%%%%%%%%%%%%%%%%%%%%%%%%%%%%%%%%%
%Define the generator of the adiabatic time evolution to be
%\begin{equation}
%\label{adiabaticL}
%\L_a(s):=\L_g (s) +
%\frac{i}{\tau}[\dot{P}_g(s),P_g(s)] \; ,
%\end{equation}
%and the corresponding propagator $U_a^{(\tau)}$ satisfying the
%initial value problem
%\begin{align}
%i\partial_s U_a^{(\tau)}(s) &= \tau \L_a(s)U_a^{(\tau)}(s) \; , \\
%U_a^{(\tau)}(0)&=1 \; .
%\end{align}
%%%%%%%%%%%%%%%%%%%%%%%%%%%%%%%%%%%%%%%%%%%%%%%%%%%%%%%%%%%

Let $\Psi(t)$ be the ``true'' state of $\S\vee\R$ at time $t$; in
particular
\begin{equation*}
\label{truestate} \Psi(t)=U_\tau (t,t')\Psi(t')\; ,
\end{equation*}
where $U_\tau(t,t')$ is the propagator corresponding to $\{ L_\tau
(t) \}$; see eqs.(7.5)-(7.7), section 7.2. By the property of return
to equilibrium and assumption (H7.4),
\begin{equation}
\Psi(t)=\Omega_\beta\; , t\le \tau s_0 \; ,
\end{equation}
for an arbitrary initial condition $\Psi (-\infty )\in {\mathcal
H}$ at $t=-\infty$.

We set
\begin{equation}
\Psi^{(\tau)}(s)=\Psi(\tau s) \; ,
\end{equation}
and note that,
\begin{equation}
\Psi^{(\tau )}(s)=U^{(\tau)}(s,s_0)\Omega_\beta \; ,
\end{equation}
for $s\in I$.

\vspace{0.5cm}

\noindent {\it Theorem 7.3 (Isothermal Theorem).} Suppose that
$\L_g(s)$ and $P_g(s)$ satisfy assumptions (H7.1)-(H7.4). Then
\begin{equation*}
\lim_{\tau\rightarrow\infty}\sup_{s\in I} || \Psi^{(\tau )}(s)
-\Omega_\beta(s) || = 0\; .
\end{equation*}

{\it Proof.} This follows readily from equations (7)-(9), (12) and
(15), and the Adiabatic Theorem (Theorem 7.2) of Sect.7.2. $\Box$

{\it Remarks}

(1) We define the expectation values (states)
\begin{equation}
\omega_t^\beta (a):= \langle \Omega_\beta(\frac{t}{\tau }), a
\Omega_\beta(\frac{t}{\tau }) \rangle
\end{equation}
and
\begin{equation}
\rho_t(a):= \langle \Psi(t), a \Psi(t)\rangle \; ,
\end{equation}
where $a$ is an arbitrary bounded operator on ${\mathcal
H}={\mathcal H}^{\S\vee\R}$. Then the isothermal theorem says that
\begin{equation}
\rho_t(a)=\omega_t^\beta (a) + \epsilon_t^{(\tau )}(a)\; ,
\end{equation}
where
\begin{equation}
\lim_{\tau\rightarrow\infty}\frac{| \epsilon_t^{(\tau )} (a) | }{
|| a ||} = 0 \; ,
\end{equation}
for all times $t\in I_\tau$.

(2) In chapter 8, we show that if the complex spectral deformation
techniques, as developed in chapter 6, are applicable to the
analysis of the coupled system $\S\vee\R$ then
\begin{equation}
|\epsilon_t^{(\tau )}(a)|\le {\mathcal O}(\tau^{-1})
||a|| \; ;
\end{equation}.

(3) All the assumptions (H7.1)-(H7.4), are admissible for the
classes of systems for which RTE has been established, such as a
{\it quantum dot} interacting with electrons in a metal or a spin
impurity interacting with magnons in a magnet; see chapters 5 and
6. For example, suppose that, for a system which is at equilibrium
at $s=0$, the perturbation Hamiltonian $V(s)$ is such that
\begin{itemize}
\item[(a)] $\sup_{s\in I}||V(s)-V(0)|| < \infty$ ,
\item[(b)] $\dot{V}(s)$ and $\ddot{V}(s)$ are bounded ,
\item[(c)] $V(s)$ is such that the system possesses the property of RTE at each fixed $s\in I,$
\end{itemize}
then all the assumptions are satisfied.
\begin{itemize}
\item[(i)] Assumption (H7.1) trivially follows from (a), since the perturbation is
bounded.

\item[(ii)] Assumption (H7.2) follows from (a) and (b) and the
resolvent equation
\begin{equation*}
(\L_g-i)^{-1}= (\L_0 -i)^{-1}(1+g I(\L_0-i)^{-1})^{-1} \; .
\end{equation*}

\item[(iii)] Assumptions (H7.3) and (H7.4) follow from (b), (c), (\ref{Proj}) and
the Dyson expansion of the instantaneous equilibrium state when
the perturbation has finite norm (see chapter 3).

\end{itemize}

%%%%%%%%%%%%%%%%%%%%%%%%%%%%%%%%%%%%%%%%%%%%%%%%%%%%%%%%%%%%%%%%%%%%%%%%%%%%%
%ISOTHERMAL PROCESSES
%%%%%%%%%%%%%%%%%%%%%%%%%%%%%%%%%%%%%%%%%%%%%%%%%%%%%%%%%%%%%%%%%%%%%%%%%%%%%

\section{(Reversible) isothermal processes}

In this section, we study general isothermal processes and use the
so called isothermal theorem to characterize {\it reversible}
isothermal processes.

It will be convenient to view the heat bath ${\mathcal R}$ as the
{\it thermodynamic limit} of an increasing family of quantum
systems confined to compact subsets of physical space, as
discussed in [Ru1,BR] and in chapter 3. The pure states of a
quantum mechanical system confined to a bounded region of physical
space are unit rays in a separable Hilbert space, while its mixed
states are described by density matrices, which are positive
trace-class operators with unit trace. Before passing to the
thermodynamic limit of the heat bath, the dynamics of the coupled
system, $\S\vee\R$, is generated by a family of time-dependent
Hamiltonians
\begin{equation}
H(t)\equiv H^{\S\vee\R}(t):=H^\Sigma(t)+ H^{\mathcal R}\; ,
\end{equation}
where
\begin{equation}
H^\Sigma (t)=H_0^\Sigma (t) + g(t)W\; ,
\end{equation}
$H_0^\Sigma (t)$ is as in section 7.1, and $W$ is the interaction
Hamiltonian (as opposed to the interaction Liouvillian, $I=ad_W$,
introduced in section 7.1).

Let ${\mathsf P}(t)$ denote the density matrix describing the
state of the coupled system, $\S\vee\R$, at time $t$, ({\it
before} the thermodynamic limit for ${\mathcal R}$ is taken). Then
${\mathsf P}(t)$ satisfies the Liouville equation
\begin{equation}
\dot{{\mathsf P}}(t)=-i[H(t),{\mathsf P}(t)] \; .
\end{equation}

The instantaneous equilibrium-, or reference state of the coupled
system is given, in the {\it canonical} ensemble, by the density
matrix
\begin{equation}
{\mathsf P}^\beta (t)=Z^\beta(t)^{-1}e^{-\beta H(t)} \; ,
\end{equation}
where
\begin{equation}
Z^\beta (t)= \tr (e^{-\beta H(t)})
\end{equation}
is the partition function, and $\tr$ denotes the trace. We assume
that the thermodynamic limits
\begin{eqnarray}
\rho_t(\cdot)=TD\lim_{\mathcal R}\tr ({\mathsf P}(t)(\cdot)) \\
\omega_t^\beta (\cdot) = TD\lim_{\mathcal R}\tr ({\mathsf P}^\beta
(t)(\cdot))
\end{eqnarray}
exist on a suitable kinematical algebra of operators describing
$\S\vee\R$.

The equilibrium state and partition function of a finitely
extended heat bath are given by
\begin{eqnarray}
{\mathsf P}_{\mathcal R}^\beta &=& (Z_{\mathcal R}^\beta)^{-1}e^{-\beta H^{\mathcal R}}\; ,\\
Z_{\mathcal R}^\beta &=& \tr (e^{-\beta H^{\mathcal R}}) \; ,
\end{eqnarray}
respectively.

Next, we introduce {\it thermodynamic potentials} for the small
system $\Sigma$: The {\it internal energy} of $\Sigma$ in the
``true'' state, $\rho_t$, of $\S\vee\R$ at time $t$ is defined by
\begin{equation}
U^\Sigma (t):= \rho_t (H^\Sigma (t))
\end{equation}
and the {\it entropy} of $\Sigma$ in the state $\rho_t$ at time
$t$ by
\begin{equation}
S^\Sigma (t):= -k \; TD\lim_{\mathcal R} \tr ({\mathsf P}(t)[ln
{\mathsf P}(t)-ln{\mathsf P}_{\mathcal R}^\beta]) \; .
\end{equation}
Note that we here define $S^\Sigma (t)$ as a {\it relative}
entropy (with the aim of subtracting the divergent contribution of
the heat bath to the {\it total} entropy). It follows from a general inequality
for traces that\footnote{\noindent{\it Lemma.} 
Consider $f$ a real convex function on ${\mathbf R}$, and $A$ and $B$ selfadjoint operators on a Hilbert space $\H$ with discrete spectrum such that $f(A)$ and $f(B)$ are trace-class. Then
\begin{equation*}
\tr (f(B)-f(A))\ge \tr (f'(A)(B-A)) \; .
\end{equation*}
If $f$ is strictly convex, equality holds only when $A=B$.

{\it Proof.}
Let $\{ \psi_j \}_{j=0}^\infty$ be an eigenbasis of $B$, such that $B\psi_j=b_j \psi_j$. Moreover, for $\psi\in\H$, let $c_j:= \langle \psi_j , \psi \rangle.$ Then
\begin{eqnarray}
\langle \psi, f(B) \psi \rangle &=& \sum_{j} |c_j |^2 f(b_j)\nonumber \\
&\ge& f(\sum_{j}|c_j|^2 b_j)=f(\langle \psi, B \psi \rangle) \label{con1}\; ,
\end{eqnarray}
where we have used the convexity of $f$ in the last line. 
Convexity of $f$ also gives
\begin{equation}
\label{con2}
f(\langle \psi, B\psi\rangle)\ge f(\langle \psi, A\psi\rangle)+f'(\langle\psi, A \psi\rangle)\langle \psi, (B-A)\psi\rangle \; .
\end{equation}
When $\psi$ is an eigenvector of $A$, the RHS of the above inequality becomes
\begin{equation}
\langle \psi, [f(A)+f'(A)(B-A)]\psi\rangle \; .
\label{con3}
\end{equation}
Summing over an eigenbasis of $A$ gives 
\begin{equation*}
\tr (f(B)-f(A))\ge\tr (f'(A)(B-A)) \; .
\end{equation*}

%%%%%%%%%%%%%%%%%%%%%%%%%%%%%%%%%%%%*************************
In particular, for $f(x)=x\log (x)$ and $A\ge 0,B>0,$ we have
$$\tr (A\log A -A\log B)\ge \tr (A-B).$$}

\begin{equation}
S^\Sigma (t)\le 0 \; .
\end{equation}

The {\it free energy} of $\Sigma$ in an {\it instantaneous
equilibrium state}, $\omega^\beta_t$, of $\S\vee\R$ is defined by
\begin{equation}
F^\Sigma (t):= -kT \; TD\lim_{\mathcal R}ln (\frac{Z^\beta
(t)}{Z^\beta_{\mathcal R}}) \; .
\end{equation}

Next, we define quantities associated not with states but with the
{\it thermodynamic process} carried out by $\S\vee\R$: the {\it
heat flux} into $\Sigma$ and the {\it work rate}, or {\it power},
of $\Sigma$. Let $\delta$ denote the so called inexact
differential. Then
\begin{equation}
\frac{\delta Q^\Sigma}{dt}(t):= TD\lim_{\mathcal
R}-\frac{d}{dt}\tr ({\mathsf P}(t)H^{\mathcal R})\; ,
\end{equation}
and
\begin{equation}
\frac{\delta A^\Sigma}{dt}(t):= \rho_t(\dot{H}^\Sigma (t)) \; .
\end{equation}

We are now prepared to summarize our main results on isothermal
processes. The first two results are general and concern the first
law of thermodynamics and the relationship between the rate of
change of entropy and the heat flux into $\Sigma$. The remaining
three results are corollaries pertaining to free energy and
changes of entropy in reversible isothermal processes, i.e.,
processes in which states are sampled at equilibrium, and on the
zeroth law of thermodynamics.

(1) From definitions (7.28), (7.32) and (7.33) and the Liouville
equation (7.21) it follows that
\begin{equation}
\dot{U}^\Sigma (t)=\frac{\delta Q^\Sigma}{dt}(t)+\frac{\delta
A^\Sigma}{dt}(t)\; ,
\end{equation}
which is the {\it first law of thermodynamics}.

(2) Note that, by the unitarity of time evolution and the cyclic
invariance of the trace,
\begin{equation*}
\frac{d}{dt}\tr ({\mathsf P}(t)ln {\mathsf P}(t))=0 \; ,
\end{equation*}
and
\begin{equation*}
\frac{d}{dt} \tr ({\mathsf P}(t)ln Z^\beta_{\mathcal
R})=\frac{d}{dt}lnZ^\beta_{\mathcal R}=0\; .
\end{equation*}
Together with definitions (7.26), (7.29) and (7.32), this implies
that
\begin{equation}
\dot{S}^\Sigma (t)=\frac{1}{T} \frac{\delta Q^\Sigma}{dt} (t)\; ,
\end{equation}
for {\it arbitrary} isothermal processes at temperature
$T=(k\beta)^{-1}$.

(3) Next, we consider an isothermal process of $\S\vee\R$ during a
finite time interval $I_\tau = [\tau s_0, \tau s_1]$, with $s_0$
and $s_1$ fixed. The initial state $\rho_{\tau s_0}$ of $\S\vee\R$
is assumed to be an equilibrium state $\omega^\beta_{\tau s_0}$ of
the Liouvillian $L_\tau (\tau s_0)=L(s_0)$. We are interested in
the properties of such a process when $\tau$ becomes large, i.e.,
when the process is {\it quasi-static}.

\vspace{0.5cm}

{\it Result.} Quasi-static isothermal processes are {\it
reversible} (in the sense that all intermediate states $\rho_t$ of
$\S\vee\R$, $t\in I_\tau$, converge in norm to {\it instantaneous
equilibrium states} $\omega^\beta_t$, as $\tau\rightarrow\infty$).

This result is an immediate consequence of the isothermal theorem.
It means that, for all practical purposes, an isothermal process
with time scale $\tau$ is reversible if $\tau \gg \tau_{\mathcal
R} = \max_{s\in I}\tau_{\mathcal R}(s)$.

(4) For reversible isothermal processes, the usual {\it
equilibrium definitions} of internal energy and entropy of the
small system $\Sigma$ can be used:
\begin{equation}
U_{rev}^\Sigma (t) := \omega^\beta_t (H^\Sigma (t))\; ,
\end{equation}
\begin{equation}
S_{rev}^\Sigma (t):= -k \; TD\lim_{\mathcal R}\tr ({\mathsf
P}^\beta (t)[ln {\mathsf P}^\beta (t)-ln {\mathsf P}^{\mathcal
R}])=\frac{1}{T}(U_{rev}^\Sigma (t)-F^\Sigma (t))\; ,
\end{equation}
where the free energy $F^\Sigma (t)$ has been defined in (7.31),
and the second equation in (7.37) follows from (7.22), (7.26),
(7.31) and (7.36). Eqs.(7.37) and (7.31) then imply that
\begin{equation*}
\dot{S}^\Sigma_{rev}(t)=\frac{1}{T} (\frac{d}{dt}\omega^\beta_t
(H^\Sigma (t))-\omega^\beta_t (\dot{H}^\Sigma (t)))\; .
\end{equation*}
Recalling (7.34) and (7.35), and applying the isothermal theorem,
we find that
\begin{align}
\dot{S}^\Sigma (t) &\rightarrow \dot{S}^\Sigma_{rev}(t) \; , \\
\frac{\delta A^\Sigma}{dt}(t) &\rightarrow \dot{F}^\Sigma (t) \; ,
\end{align}
as $\tau\rightarrow\infty$.

(5) We conclude this chapter by considering a quasi-static
isothermal process of $\S\vee\R$ with $H^\Sigma
(s)\rightarrow H^\Sigma_0, \; g(s)\rightarrow 0$, as $s\nearrow
s_1$, i.e., the interactions between ${\mathcal R}$ and $\Sigma$
are switched off at the end of the process. Then the isothermal
theorem implies that
\begin{equation}
\lim_{\tau\rightarrow\infty}\lim_{s\nearrow s_1}\rho_{\tau s} =
\omega^\beta_\S\otimes \omega^\beta_\R \; ,
\end{equation}
where $\omega^\beta_{\mathcal R}(\cdot)=TD\lim_{{\mathcal
R}}(Z^\beta_{\mathcal R})^{-1}\tr (e^{-\beta H^{\mathcal R}}\cdot
)$, see (7.26), and
\begin{equation}
\omega^\beta_\Sigma(\cdot )=(Z^\beta_\Sigma)^{-1}\tr (e^{-\beta
H_0^\Sigma} \cdot )
\end{equation}
is the Gibbs state of the small system $\Sigma$ at the temperature
$(k \beta)^{-1}$ of the heat bath, {\it independently} of the
properties of the diathermal contact (i.e., of the interaction
Hamiltonian $W$), assuming that (H7.1)-(H7.4) hold for $s<s_1$.

This result is part of the zeroth law of thermodynamics, as discussed in chapter 2.
%%%%%%%%%%%%%%%%%%%%%%%%%%%%%%%%%%%%%%%%%%%%%%%%%%%%%%%%%%
%\include{at}

\chapter{Adiabatic theorems in nonequilibrium quantum statistical mechanics}

In this chapter, we prove a novel adiabatic theorem which is
general enough to handle generators of time evolution that are not
necessarily normal or bounded. We discuss two applications of this
theorem in non-equilibrium quantum statistical mechanics: an
adiabatic theorem for states close to non-equilibrium steady
states, and an isothermal theorem with an {\it explicit} rate of
convergence to the instantaneous equilibrium state in the {\it
quasi-static} limit.

\section{A general adiabatic theorem for (non-)normal and
(un)bounded generators of time evolution}

Consider a family of closed operators $\{ A (t) \}, t\in {\mathbf
R}$ acting on a Hilbert space $\H$. We make the following
assumptions on $A(t)$ in order to prove the existence of a time
evolution and to prove an adiabatic theorem. {\it All} of these
assumptions will be verified in the applications we consider later
in this chapter.

\begin{itemize}

\item[(H8.1)] $A(t)$ is a generator of a contraction semi-group for all $t\in {\mathbf R}$.

\item[(H8.2)] $A(t)$ have a common dense domain $\D\subset\H$ for all $t\in {\mathbf R}$.

\item[(H8.3)] For $z\in
\rho(A(t)),$ the resolvent set of $A(t)$, let $R(z,t):= (z-A(t))^{-1}$. Assume that $R(-1,t)$ is bounded and differentiable
as a bounded operator on $\H$, and that $A(t)\dot{R}(-1,t)$ is bounded,
where the $(\dot{} )$ stands for differentiation with respect to
$t$. Moreover, assume that for every $\epsilon > 0,$ $-\epsilon\in \rho(A(t)).$

\end{itemize}

Let $U(t)$ be the propagator that satisfies
\begin{equation}
\partial_t U(t) \psi= -A(t)U(t) \psi\; ,U(t=0) =1 \; , \; \psi\in\D \; ,
\label{timeevolution}
\end{equation}
for $t\ge 0.$

\vspace{0.5cm}

\noindent {\it Lemma 8.1}

Suppose that assumptions (H8.1)-(H8.3) hold. Then the
propagator $U(t)$ satisfying (\ref{timeevolution}) exists and is
unique, and $\| U(t)\psi \| \le \|\psi\|.$

{\it Proof.} For $t,t' \in {\mathbf R}$, $A(t')A(t)^{-1}$ is
bounded by the closed graph theorem and (H8.2) (see [RS2]).
Moreover, for small $|t-t'|$, $|| (t'-t)A(t')^{-1}A(t)-{\mathbf
1}||=|| \dot{A}^{-1}(t)A(t)||+o(|t-t'|)$, which is bounded due to
(H8.3).By Theorem X.70 in [RS2] (or Theorem 2 in Chp XIV [Yo],
section 4), this implies, together with (H8.1), that $U(t)$ exists
and is unique. Furthermore, choose $\epsilon > 0$ and let $\tilde{U}(t)$ be the propagator generated by $A(t)+\epsilon.$ Then from (H8.3), $0\in\rho(A(t)+\epsilon)$, and hence by Theorem X.70 in [RS2], $\tilde{U}(t)$ is a contraction semigroup for $t>0.$ In particular, $\|\tilde{U}(t)\psi \|\le 1$ (for $\|\psi\|=1$). We also have $\| U(t)\| = e^{\epsilon t}\|\tilde{U}(t)\|.$ Taking the limit $\epsilon\rightarrow 0$ gives $\|U(t)\psi\|\le 1.$  
$\Box$

Assume that $A(t)\equiv A(0)$ for $t\le 0$, and that it is
perturbed {\it slowly} over a time $\tau$ such that
$A^{(\tau)}(t)\equiv A(s)$, where $s:=\frac{t}{\tau}\in [0,1] $ is
the reduced time. The following two assumptions are needed to
prove an adiabatic theorem.

\begin{itemize}

\item[(H8.4)] The eigenvalue $\ls\in \sigma(A(s))$ is isolated and
simple, such that
$$dist(\ls, \sigma(A(s))\backslash \{ \ls\})>\delta,$$
where $\delta>0$ is a constant independent of $s\in [0,1]$, and
$\ls$ is continuously differentiable in $s\in [0,1]$.

\item[(H8.5)] The projection onto $\ls$,
\begin{equation}
P_\ls:= \frac{1}{2\pi i}\oint_{\gamma_\ls} R(z,s) dz \; ,
\end{equation}
where $\gamma_\ls$ is a contour enclosing $\ls$ only, is twice
differentiable as a bounded operator.

%%%%%%%%%%%%%%%%%%%%%%%%%%%%%%%%%%%%%%%%%%%%%%%%%%%%%%%%%%%%%%%%%%%%%%%
%\item[(H8.6)] For $z\in \gamma_\ls$, the resolvent $R(z,s)$ is bounded
%as a differentiable operator.
%%%%%%%%%%%%%%%%%%%%%%%%%%%%%%%%%%%%%%%%%%%%%%%%%%%%%%%%%%%%%%%%%%%%%%%

\end{itemize}

Note that, since $\ls$ is simple, the resolvent of $A(s)$ in a neighborhood $\N$ of $\ls$
contained in a ball ${\mathbf B}(\ls,r)$ centered at $\ls$ with
radius $r<\delta$ is
\begin{equation}
\label{res} R(z,s)=\frac{P_\ls}{z-\ls} + R_{analytic} (z,s) \; ,
\end{equation}
where $R_{analytic}(z,s)$ is analytic in $\N$. We list some useful
properties of the resolvent and the spectral projection $P_\ls$.

\begin{itemize}
\item[(i)] It follows by direct application of the contour
integration formula that
\begin{equation}
\label{sqrproj} (P_\ls)^2=P_\ls \; ,
\end{equation}
and hence
\begin{equation}
\label{dotP} 
P_\ls \dot{P}_\ls P_\ls=0 \; .
\end{equation}

\item[(ii)]
\begin{equation}
\label{projection} A(s)P_\ls = P_\ls A(s)=\ls P_\ls \;.
\end{equation}
{\it Proof.}
\begin{align*}
A(s)P_\ls &= \frac{1}{2\pi i}\oint_{\gamma_\ls}
(A(s)-z+z)(z-A(s))^{-1} dz \\
&= \frac{1}{2\pi i} \{ -\oint_{\gamma_\ls}dz
+\oint_{\gamma_\ls} (\frac{z P_\ls}{z-\ls} + zR_{analytic})dz \} \\
&= \ls P_\ls\; ,
\end{align*}
and similarly, $P_\ls A(s)=\ls P_\ls$.

\item[(iii)] It follows from (\ref{res}) and (H8.4) that, for $\eta\in {\mathbf C}$ and $\frac{\delta}{2}\le |\eta| < \delta,$
there exists a constant $C<\infty$, independent of $\eta$,
such that
\begin{equation}
\| R(\ls+\eta ,s) \| < C \; , \label{ResIneq}
\end{equation}
uniformly in $s\in [0,1]$. Moreover, since $(\ls+\eta ) \in \rho(A(s)),$ it follows by the spectral mapping theorem (see for example [Yo], Chp. VIII, section 7) and (H8.3) that $R(\ls+\eta ,s)$ is twice differentiable as a bounded operator.\footnote{This follows from the fact that, for $z,\omega\in \rho(A),$ $$(z-A)^{-1}=(1+(z-\omega)(\omega-A)^{-1})^{-1}(\omega-A)^{-1}.$$}

\end{itemize}

We now discuss our general adiabatic theorem. Let $U_\tau(s,s')$ be the propagator satisfying
\begin{equation}
\label{rtevol2}
\partial_s U_\tau (s,s') = -\tau A(s) U_\tau (s,s')\; , U_\tau(s,s)=1 \; ,
\end{equation}
for $s\ge s'$.
Moreover, define the generator of the {\it adiabatic time
evolution},
\begin{equation}
A_a(s):= A(s)-\frac{1}{\tau}[\dot{P}_\ls,P_\ls] \; ,
\end{equation}
with the corresponding propagator $U_a(s,s')$ which satisfies
\begin{equation}
\label{ate2}
\partial_s U_a(s,s') = - \tau A_a(s) U_a (s,s') \; ; U_a(s,s)=1 \; ,
\end{equation}
for $s\ge s'$.

By Lemma 8.1 and (H8.1)-(H8.3) and (H8.5), both propagators
$U_\tau(s,s')$ and $U_a(s,s')$ exist and are unique, and $\|U_\tau (s,s')\| ,\|U_a(s,s')\| < C$ for $s\ge s'$, where $C$ is a finite constant independent of $s,s'\in [0,1]$.

We are in a position to state our adiabatic theorem.

\vspace{0.5cm}

\noindent {\it Theorem 8.2 (A general adiabatic theorem)}

Assume (H8.1)-(H8.5). Then the following holds.

\begin{itemize}
\item[(i)]
\begin{equation}
\label{intertwining}P_\ls U_a (s,0) =  U_a (s,0)P_\lambda (0) \;
,
\end{equation}
for $s\ge 0$ (the intertwining property).
\item[(ii)] $\sup_{s\in [0,1]} \| U_\tau (s,0)-U_a(s,0) \| =
O(\tau^{-1})$ as $\tau\rightarrow\infty$.
\end{itemize}

{\it Proof.}
\begin{itemize}
\item[(i)] Equality holds trivially for $s=0$, since $U_a(s,s)=1$. Let 
\begin{equation}
h(s,s'):= U_a(s,s')P_\lsp U_a(s',0) \; ,
\end{equation}
for $0\le s' \le s$,
and 
\begin{equation}
g(s):= U_a(s,0)P_\lambda (0) \; ,
\end{equation}
for $0\le s$.

%%%%%%%%%%%%%%%%%%%%%%%%%%%%%%%%%%%%%%%%%%%%%%%%%%%%%%%%%%%%%%%%%%%%%

Using (\ref{projection}), (\ref{ate2}) , the definition of $A_a(s)$ and
the fact that $\dot{P}_\ls P_\ls + P_\ls \dot{P}_\ls =
\dot{P}_\ls$, it follows that
\begin{align*}
\partial_{s'}h(s,s') &= \partial_{s'}(U_a(s,s')P_\lsp U_a(s',0)) \\
&= \tau U_a(s,s') \{ A_a(s')P_\lsp
-P_\lsp A_a(s') \} U_a(s',0) + U_a (s,s') \dot{P}_\lsp U_a (s',0) \\
&= U_a(s,s') \{ -\dot{P}_\lsp P_\lsp - P_\lsp \dot{P}_\lsp + \dot{P}_\lsp
\} U_a(s',0) \\
&= 0 \; .
\end{align*}
Note also that $\partial_{s'}g(s)=0$ and $h(s,s'=0)=g(s)$. Furthermore, 
\begin{align}
\partial_s h(s,s')&=-\tau A_a h(s,s') \; , \\
\partial_s g(s)&=-\tau A_a g(s) \; ,\\
h(s=0,s'=0)&=g(s=0) \; .
\end{align}
Together with assumptions (H8.1)-(H8.3) and (H8.5), it follows that 
\begin{equation*}
h(s,s')=g(s)\; .
\end{equation*}
In particular, when $s'=s$,
\begin{equation}
P_\ls U_a(s,0) = U_a (s,0) P_\lambda (0)\; .
\end{equation}

\item[(ii)] Consider $\psi\in \D$, where the dense domain $\D$
appears in assumption (H8.2). We are interested in estimating the
norm of the difference $(U_\tau (s)-U_a (s))\psi$ as
$\tau\rightarrow \infty$. Using (\ref{rtevol2}), (\ref{ate2}) and the Duhamel formula,
\begin{align}
\label{diff} (U_\tau (s,0)-U_a (s,0))\psi  &=  - \int_0^s ds'
\partial_{s'} (U_\tau(s,s') U_a(s',0))\psi \\
&= \int_0^s  ds' (U_\tau(s,s')[\dot{P}_\lsp,P_\lsp]U_a(s',0))\psi \; .
\end{align}

Let\begin{equation}
X(s):= \frac{1}{2\pi i} \oint_{\gamma_\ls}dz R(z,s)\dot{P}_\ls R(z,s) \; ,
\end{equation}
where $\gamma_\ls$ is a contour of radius $\delta /2$ centered at $\ls,$ and where $\delta$ appears in (H8.4).
Then
\begin{eqnarray}
[X(s), A(s)] &=& \frac{1}{2\pi i}\oint_{\gamma_\ls}dz [z-A(s),X(s)] \\
&=& \dot{P}_\ls P_\ls - P_\ls \dot{P}_\ls = [\dot{P}_\ls , P_\ls] \; .\label{commutator3}
\end{eqnarray}

Assumptions (H8.3),(H8.4) and the spectral mapping theorem imply that, for $z\in \gamma_\ls\subset\rho(A(s)), R(z,s)$ is differentiable as a bounded operator.
Together with (H8.5), this implies that,
\begin{eqnarray}
\|X(s)\| &<& C_1 \; , \label{est1}\\ 
\|\dot{X}(s)\| &<& C_2 \; , \label{est2}
\end{eqnarray} 
where $C_1$ and $C_2$ are finite constants independent of $s\in [0,1]$.
Moreover,
\begin{align*}
&U_\tau (s,s')[X(s'),A(s')]U_a(s',0) =\frac{1}{\tau}\{ -\partial_{s'}U_\tau(s,s')X(s')U_a(s',0) \\
&+U_\tau (s,s')(X(s')[\dot{P}_\lambda (s'),P_\lambda (s')])U_a(s',0)+U_\tau(s,s')\dot{X}(s')U_a(s',0)\} \; .
\end{align*}
Together with \fer{commutator3}, one may write the integrand in \fer{diff} as a total derivative plus a remainder term. Using the fact that $\D$ is dense in $\H$ and (H8.5),
\begin{equation}
\label{diff2} \| U_\tau (s,0) - U_a(s,0) \| \le 
\frac{1}{\tau }[C_1\|X(s)\| + C_2\| \dot{X}(s) \| ] \; ,
\end{equation}
where $C_i,i=1,2$ are finite constants independent of $s\in [0,1].$ 

Together with \fer{est1} and \fer{est2}, this implies
\begin{equation}
\|U_\tau (s,0) - U_a (s,0)\| \le \frac{C}{\tau} \; ,
\end{equation}
as $\tau\rightarrow\infty,$ where $C<\infty$ is independent of $s\in [0,1].$ 
$\Box$

\vspace{1cm}

{\it Remarks.}

\item[(1)] One may improve the results of Theorem 8.2 (also to cover the case of eigenvalue crossing) by making
further smoothness assumptions on $A(s)$ and applying methods
developed, for example in [J,Ne].

\item[(2)] An application of Theorem 8.2 other than in NEQSM is an adiabatic theorem for
quantum resonances.[A-SF1]

\end{itemize}

\section{Application 1: an adiabatic theorem for non-equilibrium steady states (NESS)}

In this section, we consider again the $C_n$ paradigm of a two level system coupled to $n$ fermionic reservoirs (see chapter 4, section 4). It is important to
note that the result can be generalized to bosonic reservoirs, but
the analysis will be technically more cumbersome since the
interaction in the bosonic case is unbounded.

Together with assumptions ($C_n.2$), we assume ($C_n.4$), ie, that
$V^{\tau}(t)=V(s)$, where $s\in[0,1]$ is the rescaled time with
sufficient smoothness properties to apply Theorem 8.2. From
Proposition 6.9, we know the spectrum of the deformed adjoint of
the C-Liouvillean $L^*_g(t,\theta)=U(\theta)L^*_g(t)U(-\theta)$
for $\theta\in I^-(k)=\{ z\in\mathbf{C} : -k<\Im z <0\}$, where
$k=\min (\frac{\pi}{\beta_1},\cdots,\frac{\pi}{\beta_n},\delta)$,
and $\delta$ appears in assumption ($C_n.5$). Let $\gamma_0$ be a
contour only enclosing the zero eigenvalue of $L^*_g(s,\theta)$,
for all $s\in [0,1]$, and
\begin{equation}
\label{dproj1} P'_g(\theta,s):= \oint_{\gamma_0}\frac{dz}{2\pi i}
(z-L^*_g(s,\theta))^{-1},
\end{equation}
the spectral projection onto the state corresponding to the zero
eigenvalue of $L^*_g(s,\theta)$, and let $D$ be the positive operator
as defined in Corollary 6.10, section 6.3 (ie, $Ran D$ dense in $\H$ and
$D\Omega=\Omega$)
\begin{equation}
D:= {\mathbf 1}^\S\otimes {\mathbf 1}^\S \otimes e^{-k
\tilde{A}_{\R_1}}\otimes\cdots\otimes e^{-k\tilde{A}_{\R_n}} \; ,
\end{equation}
where $\tilde{A}_{\R_j}=d\Gamma(\sqrt{p_j^2+1})$, and
$p_j=i\partial_{u_j}$ is the generator of energy translations for
the $\R_j$ reservoir, $j=1,\cdots ,n$.

Let $\h^{test}=
D(e^{k\sqrt{p^2+1}})$, and $\O^{test}= \F_-(\h^{test} )$, which
is dense in $\F_-(L^2({\mathbf R};\B))$, and define 
\begin{equation}
{\mathcal C}:= \O^\S\otimes\O^{\R_1 ,test}\otimes \cdots \otimes \O^{\R_n, test} \; ,
\end{equation}
which is dense in $\O$.
We will make the following additional assumption.

\begin{itemize}
\item[(H8.6)] The perturbation of the Hamiltonian $V(s)\in {\mathcal C}$ for $s\in [0,1]$. 
\end{itemize}

In order to characterize the quasi-static evolution of nonequilibrium steady states, we introduce the new notion of {\it instantaneous} NESS.
Define the {\it instantaneous} NESS vector to be
\begin{equation}
\Omega_g(s):= DU(-\theta)P'_g(s,\theta)U(\theta)D \Omega \; .
\end{equation}
It is important to note that introducing the operator $D$ is needed so that one can remove the
complex deformation. 

%%%%%%%%%%%%%%%%%%%%%%%%%%%%%%%%%%%%%%%%%%%%%%%%%%%%%%%%%%%%%%%%%
%Note that
%\begin{equation}
%\label{deformedprojection}
%P'_g(s,\theta)=
%|\Omega_g(s,\theta)\rangle\langle \Omega|\; .
%\end{equation}
%%%%%%%%%%%%%%%%%%%%%%%%%%%%%%%%%%%%%%%%%%%%%%%%%%%%%%%%%%%%%%%%%

We have the following Theorem, which effectively says that if a
system, which is initially in a NESS, is perturbed slowly over a
time scale $\tau\gg \tau_R$, where $\tau_R$ is some generic time
scale ($\tau_R=\max_{s\in [0,1]}\tau_{R(s)}$, where $\tau_{R(s)}$
is the relaxation time to a NESS, see Corollary 6.10), then the
real state of the system is infinitesimally close to the {\it
instantaneous} NESS, and the difference of the two states is
bounded form above by a term of order $O(\tau^{-1})$.

\vspace{0.5cm}

\noindent {\it Theorem 8.3 (Adiabatic Theorem for NESS)}

Suppose ($C_n.2$), ($C_n.4$), as specified in section 4.4, and (H8.6). Then there exists $g_1>0$,
independent of $s\in [0,1]$, such that, for $a\in {\mathcal C}, s\in [0,1]$, 
and $|g|<g_1$, the following estimate
holds
\begin{equation}
\sup_{s\in [0,1]} | \langle \Omega_g(0), D^{-1} \a_g^{\tau s}(a)
\Omega\rangle - \langle \Omega_g(s), D^{-1} a \Omega\rangle | =
O(\tau^{-1}) \; ,
\end{equation}
as $\tau\rightarrow\infty$.

{\it Proof.} The proof is reduced to mainly showing that the
assumptions of Theorem 8.2 are satisfied. Choose $\theta\in
I^-(k)$. It follows from assumption ($C_n.4$) and Lemmas 6.11 and
6.12 in chapter 6, section 3, that the deformed C-Liouvillean $L^*_g(s,\theta)$
with common dense domain $\D=\D(\L_0)\cap \D(D^{-1})\subset
C(\O,\Omega)$ generates the propagator
$\tilde{U}^{(\tau)}_g(s,s',\theta), s'\le s$ which satisfied the initial value
problem
\begin{equation}
\partial_s \tilde{U}^{(\tau)}_g(s,s',\theta) = -i \tau L^*_g(s,\theta)
\tilde{U}^{(\tau )}_g(s,s',\theta) \; ,\; {\rm for} \; s'\le s ; \tilde{U}^{(\tau)}(s,s,\theta)= 1 .
\end{equation}
This implies that (H8.1) and (H8.2) are satisfied. Furthermore,
(H8.3) follows from the second resolvent identity
\begin{equation}
\label{dr} (L^*_g (s, \theta)-z)^{-1} = (\L_0 (\theta
)-z)^{-1}(1+g \tilde{V}^{tot}(s,\theta)(\L_0 (\theta
)-z)^{-1})^{-1} \; ,
\end{equation}
where $\tilde{V}^{tot}$ has been defined in section 6.3, and assumption ($C_n.4$). The results of Proposition 6.9 (section 6.3) follow from ($C_n.2$) and ($C_n.4$), and hence we know that
zero is an isolated simple eigenvalue of $L^*_g(s,\theta )$ such
that $dist(0, \sigma(L^*_g(s,\theta))\backslash \{ 0 \})>\delta $,
where $\delta >0$ is a constant independent of $s\in [0,1]$. This
implies that assumption (H8.4) holds. Again using the resolvent
equation (\ref{dr}) and assumption ($C_n.4$), $P'_g(s,\theta)$
defined in (\ref{dproj1}) is twice differentiable as a bounded
operator for all $s\in [0,1]$, which imply (H8.5). Since
(H8.1)-(H8.5) are satisfied, the result of Theorem 8.2 holds.
\begin{equation}
P'_g(s,\theta) \tilde{U}^{(\tau)}_a
(s,0,\theta) = \tilde{U}^{(\tau)}_a(s,0,\theta)P'_g (0,\theta) \; ,
\end{equation}
and
\begin{equation}
\sup_{s\in [0,1]}\| \tilde{U}^{(\tau
)}_g(s,0,\theta)-\tilde{U}^{(\tau )}_a(s,0,\theta) \| = O(\tau^{-1})
\; ,
\end{equation}
as $\tau\rightarrow\infty$, where $\tilde{U}^{(\tau
)}_a(s,s',\theta)$ is the propagator of the {\it deformed} adiabatic
evolution satisfying
\begin{equation}
\label{dae}
\partial_s \tilde{U}^{(\tau )}_a(s,\theta)=
-i\tau L^*_a(s,\theta)\tilde{U}^{(\tau )}_a(s,s',\theta)\; {\rm for} \; s'\le s \; ;
\tilde{U}^{(\tau )}_a(s,s,\theta)=1 \; ,
\end{equation}
and
\begin{equation}
L^*_a(s,\theta) = L^*_g(s,\theta)
+\frac{i}{\tau}[\dot{P}'_g(s,\theta),P'_g(s,\theta)] \; .
\end{equation}
(Here, the $\dot{()}$ stands for differentiation with respect to
$s$.)

For $h$ the single particle Hamiltonian of the free fermions (see section 4.4),$e^{ih t}$ leaves $D(e^{k\sqrt{p^2+1}})$ invariant. Therefore, for $a\in {\mathcal C}$, $\a_0^t (a)\in {\mathcal C}$, where $\a_0^t$ corresponds to the free time evolution. Moreover, together with assumption (H8.6) and the boundedness of $V$, this implies (using a Dyson series expansion) that $\a_g^{\tau s}(a)\in {\mathcal C}=\D(D^{-1})$. 

Now, applying the time evolution on $C(\O,\Omega)$, and remembering
that $D\Omega=\Omega$, $U(\theta)\Omega=\Omega$, the fact that
$U(\theta)$ and $D$ commute,  and the definition of the
instantaneous NESS,
\begin{equation}
\langle \Omega_g (0) ,D^{-1}\a_g^{(\tau s)}(a)\Omega\rangle 
=\langle \tilde{U}_g^{(\tau )}(s,0,\theta) P_g'(0,\theta) \Omega, a(\overline{\theta})\Omega \rangle \; .
\end{equation}
Using the results of Theorem 8.2, it follows that 
\begin{align*}
&\langle \tilde{U}_g^{(\tau )}(s,0,\theta) P_g'(0,\theta) \Omega, a(\overline{\theta})\Omega \rangle \\
&= \langle \tilde{U}_a^{(\tau )}(s,0,\theta) P_g'(0,\theta) \Omega, a(\overline{\theta})\Omega \rangle + O(\tau^{-1}) \\
&= \langle P_g'(s,\theta) \tilde{U}_a^{(\tau )}(s,0,\theta) \Omega, a(\overline{\theta})\Omega \rangle + O(\tau^{-1}) \\
&= \langle P_g'(s,\theta) \tilde{U}_g^{(\tau )}(s,0,\theta) \Omega, a(\overline{\theta})\Omega \rangle + O(\tau^{-1}) \; .
\end{align*}
The fact that $(\tilde{U}_g^{(\tau )}(s,0,\theta))^*\Omega=\Omega$ implies that
\begin{align*}
D P_g'(s,\theta) \tilde{U}_g^{(\tau )}(s,0,\theta) &= | \Omega_g(s,\theta) \rangle\langle \Omega | \tilde{U}^{(\tau )}(s,0,\theta) \\
&= | \Omega_g(s,\theta) \rangle\langle (\tilde{U}^{(\tau )}(s,0,\theta))^*\Omega | \\
&= | \Omega_g(s,\theta) \rangle\langle \Omega | = D P_g'(s,\theta) \; .
\end{align*}
It follows that 
\begin{equation*}
\langle \Omega_g (0) ,D^{-1}\a_g^{(\tau s)}(a)\Omega\rangle 
= \langle \Omega_g(s) , D^{-1} a \Omega \rangle +O(\tau^{-1})\;,
\end{equation*}

%%%%%%%%%%%%%%%%%%%%%%%%%%%%%%%%%%%%%%%%%%%%%%%%%%%%%%%%%%%%%%%%%%%%%%%%%%%
%USE VECTOR NOTATION
%&=\tr (|\Omega\rangle\langle\Omega_g(0)| D^{-1}\a_g^{\tau s}(a)) \\
%&=\tr (DU(-\theta)(P_g'(0,\theta))^*U(\theta)D D^{-1}\a_g^{\tau s}(a)) \\
%&=\tr (U(-\theta)(P_g'(0,\theta))^*U(\theta)\a_g^{\tau s}(a)) \\
%&=\tr (U(-\theta)(\tilde{U}_g^{(\tau )}(s,\theta)P_g'(0,\theta)\tilde{U}_g^{(\tau )}(-s,\theta))^*U(\theta)a) \\
%&=\tr (U(-\theta)(\tilde{U}_a^{(\tau )}(s,\theta)P_g'(0,\theta)\tilde{U}_a^{(\tau )}(-s,\theta))^*U(\theta)a) + O(\tau^{-1/2})\\
%&=\tr (U(-\theta)(P_g'(s,\theta))^*U(\theta)a) + O(\tau^{-1/2})\\
%&=\tr (DU(-\theta)(P_g'(s,\theta))^*U(\theta)D D^{-1}a) +O(\tau^{-1/2})\\
%&=\tr (|\Omega\rangle\langle\Omega_g(s)| D^{-1}a) +O(\tau^{-1/2})\\
%%%%%%%%%%%%%%%%%%%%%%%%%%%%%%%%%%%%%%%%%%%%%%%%%%%%%%%%%%%%%%%%%%%%%%%%%%%%
for large $\tau$.
$\Box$

\vspace{0.5cm}

{\it Remark.}

A weaker adiabatic theorem for states close to NESS using
scattering theory can be proven using the approach 
developed in [Ru2,3] and extending the adiabatic theorem in [NT].
However, this theorem relies on the rather strong assumption of
asymptotic abelianness. For further details about adiabatic
theorems in NEQSM, see [A-S].

\section{Application 2: a concrete example of the isothermal theorem}

As a second application of Theorem 8.2, we consider in this
section a concrete example of an isothermal process for model $C_n$ (see section 4.4), and calculate an explicit rate of convergence ($O(\tau^{-1})$)
between the instantaneous equilibrium state and real state of the
system in the quasi-static limit $\tau\rightarrow\infty$; (see
chapter 7 for our notation and the relevant definitions).

\vspace{0.5cm}

\noindent {\it Theorem 8.4 (Isothermal Theorem revisited)} Suppose
($C_n.2$), ($C_n.5$), as specified in section 4.4, and (H8.6). Then there exists a constant $g_1>0$ such
that, for $a\in \D(D^{-1})$ and $|g|<g_1$, the following
estimate holds
\begin{equation}
| \rho_{\tau s}(a) - \omega_{\tau s}^\beta (a) | = O(\tau^{-1})
\; ,
\end{equation}
as $\tau\rightarrow\infty$, where $\rho_{\tau s}$ is the real
state of the system at time $t=\tau s$, and $\omega_{\tau
s}^\beta$ is the instantaneous equilibrium state defined in chapter
7.

{\it Proof.} Choose $\theta\in I^-(k)$. It follows from assumption
($C_n.5$) and Lemmas 6.11 and 6.12 in section 6.3 that the deformed Liouvillean
$\L_g(s,\theta)$ with common dense domain $\D=\D(\L_0)\cap
\D(D^{-1})\subset C(\O,\Omega)$ generates the propagator
$U^{(\tau)}_g(s,s',\theta)$ which satisfied the initial value problem
\begin{equation}
\partial_s U^{(\tau)}_g(s,s',\theta) = -i \tau \L_g(s,\theta)
U^{(\tau )}_g(s,s',\theta) \; ; U^{(\tau)}(s,s,\theta) = 1 ,
\end{equation}
which implies (H8.1) and (H8.2). Moreover (H8.3) follows from the second resolvent identity
\begin{equation}
\label{drL} (\L_g (s, \theta)-z)^{-1} = (\L_0 (\theta
)-z)^{-1}(1+g V^{tot}(s,\theta)(\L_0 (\theta )-z)^{-1})^{-1} \; ,
\end{equation}
and assumption ($C_n.5$). The results of Theorem 6.7 follow from
($C_n.2$) and ($C_n.5$), and hence we know that zero is an
isolated simple eigenvalue of $\L_g(s,\theta)$ such that $dist(0,
\sigma(\L_g(s,\theta))\backslash \{ 0 \})>\delta'$, where
$\delta'>0$ is a constant independent of $s\in [0,1]$. This
implies assumption (H8.4). Using the resolvent equation
(\ref{drL}) and assumption ($C_n.5$),
$P_g(s,\theta):=\frac{1}{2\pi
i}\oint_{\gamma_0}dz(z-\L_g(s,\theta))^{-1}$ is twice
differentiable as a bounded operator for all $s\in [0,1]$, which
implies (H8.5). Since (H8.1)-(H8.5) are satisfied, the result of
Theorem 8.2 holds.
\begin{equation}
P_g(s,\theta) U^{(\tau)}_a (s,0,\theta) = U^{(\tau)}_a(s,0,\theta)P_g
(0,\theta) \; ;
\end{equation}
\begin{equation}
\sup_{s\in [0,1]}\| U^{(\tau)}_g(s,0,\theta)-U^{(\tau)}_a(s,0,\theta)
\| = O(\tau^{-1}) \; ,
\end{equation}
as $\tau\rightarrow\infty$, where $U^{(\tau)}_a(s,s',\theta)$ is the
propagator which satisfies
\begin{equation}
\label{dae}
\partial_s U^{(\tau)}_a(s,s',\theta)=
-i\tau\L_a(s,\theta) U^{(\tau)}_a(s,s',\theta) \; {\rm for} s'\le s \; ;
U^{(\tau)}_a(s,s,\theta)=1 \; ,
\end{equation}
and
\begin{equation}
\L_a(s,\theta) = \L_g(s,\theta)
+\frac{i}{\tau}[\dot{P}_g(s,\theta),P_g(s,\theta)] \; .
\end{equation}
Assumption (H8.6) implies that, for $a\in {\mathcal C}$, $\a_g^{\tau s}(a)\in {\mathcal C}$. 

Using the fact that $U(\pm\theta)|\Omega \rangle= | \Omega \rangle$ and the results of Theorem 8.2, it follows that
\begin{align*}
U(\theta)U_g^{(\tau)}(s,0)|\Omega\rangle &= U_g^{(\tau)}(s,0,\theta)P_g(0,\theta)|\Omega\rangle \\
&= U_a^{(\tau)}(s,0,\theta)P_g(0,\theta)|\Omega\rangle + O(\tau^{-1}) \\
&= P_g(s,\theta)U_a(s,0,\theta)|\Omega\rangle + O(\tau^{-1}) \\
&= P_g(s,\theta)U_g(s,0,\theta)|\Omega\rangle + O(\tau^{-1}) \; .
\end{align*}

We know that $P_g(s,\theta)U_g(s,0,\theta)|\Omega\rangle = c|\Omega(s,\theta)\rangle$, where $\Omega(s,\theta)$ is the vector corrsponding to the instantaneous equilibrium state and $c\in {\mathbf C}$. Since 
\begin{align*}
\langle P_g(s,\overline{\theta})U_g^{(\tau)}(s,0,\overline{\theta}) \Omega, P_g(s,\theta)U_g^{(\tau)}(s,0,\theta) \Omega\rangle &= \langle U_g^{(\tau)}(s,0)\Omega , U_g^{(\tau)}(s,0)\Omega\rangle + O(\tau^{-1}) \\
&= 1 + O(\tau^{-1}) \; ,
\end{align*}
and 
\begin{equation*}
\langle \Omega(s,\overline{\theta}), \Omega(s, \theta) \rangle =1 \; ,
\end{equation*}
it follows that $c$ is a phase up to an error of order $O(\tau^{-1})$. Therefore, 
\begin{equation}
\rho_{\tau s}(a) = \langle\Omega, U^{(\tau)}_g(0,s)aU^{(\tau)}(s,0) \Omega\rangle = \langle \Omega (s), a \Omega(s) \rangle + O(\tau^{-1}) \; .
\end{equation}

Note that unlike in the proof of Theorem 8.3,
zero is an {\it eigenvalue} of the undeformed Liouvillean
$\L_g(s)$, which is selfadjoint, and hence $P_g(s)$ is
well-defined as a spectral projection by the spectral theorem.
$\Box$

\chapter{Cyclic thermodynamic processes and Floquet
theory}

In this chapter, we study cyclic thermodynamic processes of a
small quantum system coupled to $n$ fermionic reservoirs by
extending Floquet theory for quantum mechanical systems that are
driven by periodic forcing. We introduce a new operator, which we
call the Floquet Liouvillean, and we show that under certain
assumptions, the time-periodic state to which the coupled system
converges is related to the zero-energy resonance of the adjoint
of the Floquet Liouvillean. We study the spectrum of the Floquet
operator using complex deformation techniques, as developed in
chapter 6; (see also [JP1,2,3]). Although technically more
complicated, since the perturbation will be unbounded , the
analysis is in principle applicable to the case when the
reservoirs are bosonic. On the mathematical side, it is
interesting to extend the strong spectral methods developed in
[BFS] based on operator theoretic renormalization group or the
positive commutator method developed in [M1, M2, FM1](see also chapter 5) to study the spectrum of the Floquet operator.

One result that follows from our analysis is that one can compute
entropy production per cycle and the degree of efficiency $\eta$
to arbitrary orders in the weak coupling (see remark after Theorem
9.3).

We again consider the same paradigm: Model $C_n$ of a two level
system coupled to $n$-fermionic reservoirs that are not necessarily
at the same temperature. Together with assumptions
($C_n.2$) and ($C_n.3$), chapter 4, section 4, we assume ($C_n.6$), which
pertains to the periodicity of the perturbation,
$V(t+\tau_*)=V(t), \tau_* <\infty.$

Assume further that the initial state of the system is
\begin{equation*}
\Omega=\Omega^\S\otimes\Omega^{\R_1}\otimes\cdots\otimes\Omega^{\R_n}
\; ,
\end{equation*}
where $\Omega^{\R_i},i=1,\cdots,n$ are the KMS-states of the
uncoupled reservoirs, and, without loss of generality, $\Omega^\S$
is the vector in $\H^\S\otimes\H^\S$ corresponding to the trace state on $\S$.
The C-Liouvillean of the coupled system $L_g(t)$, which has been
introduced in chapter 6, generates the dynamics on the Banach
space $C(\O,\Omega)$. It is obviously time-periodic with period
$\tau_*$ if assumption ($C_n.6$) holds. By construction,
\begin{equation*}
L_g(t)\Omega = 0 \; .
\end{equation*}
Now denote by $\tilde{U}_g(t,t')$ the propagator generated by
$L_g^*(t)$ (note the change in notation), which satisfies the
initial value problem
\begin{equation*}
\partial_t \tilde{U}_g(t,t')=-iL_g^*(t)\tilde{U}_g(t,t') \; ; \tilde{U}_g(t,t)=1 \; .
\end{equation*}
The existence of $\tilde{U}_g$ follows from assumption ($C_n.3$), section 4.4, and the Yosida-Hille-Phillips Theorem, as
discussed in chapter 6, section 4.

%%%%%%%%%%%%%%%%%%%%%%%%%%%%%%%%%%%%%%%%%%%%%%***********

\section{The {\it Floquet Liouvillean}}

Introduce the {\it Floquet Liouvillean}
\begin{equation}
\label{FloquetL} K_g^* := -i\partial_t + L_g(t)^* \; ,
\end{equation}
acting on $\tilde{\H}=L^2([0,\tau_*])\otimes \H$, where the GNS Hilbert space $\H$ for Model $C_n$ has been defined in section 4.4, with periodic boundary conditions in $t$. Note that
under the previous assumptions, $K_g^*$ is a closable operator (since the perturbation is bounded), and
we denote its closure by the same symbol. 

Let $\omega:= \frac{2\pi}{\tau^*}.$ By Fourier transformation, $\tilde{\H}$ is isomorphic to $\bigcup_{n\in {\mathbf Z}} \langle e^{in\omega t}\rangle \otimes\H= \bigcup_{n\in {\mathbf Z}} \h^{(n)}\otimes\H .$

According to Floquet theory of quantum mechanical systems driven
by periodic perturbation [Ho,Ya1,Ya2], the semi-group generated by
$K_g^*$ is given by
\begin{equation}
\label{FSM} (e^{-i\sigma K_g^*} f)(t) =
\tilde{U}_g(t,t-\sigma)(f)(t-\sigma) \;,
\end{equation}
where $f \in \tilde{\H}$ and $\sigma\in
{\mathbf R}$. Relation (\ref{FSM}) can be seen by differentiating
both sides with respect to $\sigma$ and setting $\sigma=0$ (see
[Ho]). (Alternatively, use the Trotter product formula.[RS1]) Note
that if
\begin{equation}
K_g^* \phi(t)=\lambda \phi (t) \; ,
\end{equation}
for $\phi (t) \in \tilde{\H}$ and
$\lambda\in {\mathbf C}$, then $\phi(t)$ satisfies
\begin{equation}
\tilde{U}_g(t,0)\phi(0)=e^{-i\lambda t}\phi(t) \; .
\end{equation}
Conversely, if
\begin{equation}
\tilde{U}_g(\tau_*,0)\phi (0)=e^{-i\lambda \tau_*}\phi_0 \; ,
\end{equation}
then
\begin{equation}
\phi(t)=e^{i\lambda t} \tilde{U}_g(t,0)\phi_0
\end{equation}
is an eigenfunction of $K_g^*$ with eigenvalue $\lambda$.

Before proving the convergence to a time-periodic state of the
coupled system, we first study the spectrum of the adjoint of the
Floquet Liouvillean.

%%%%%%%%%%%%%%%%%%%%%%%%%%%%%%%%%%%%%%%%%%%%%%%%%%%%%%%%%%%%%%%%%%
%SECTION: SPECTRUM OF THE FLOQUET LIOUVILLEAN
%%%%%%%%%%%%%%%%%%%%%%%%%%%%%%%%%%%%%%%%%%%%%%%%%%%%%%%%%%%%%%%%%%

\section{Spectrum of $K_g^*$}

We study the spectrum of $K_g^*$ using complex deformation techniques developed in chapter 6. This is why we only sketch the main steps of the proofs.

Let $k=min (\frac{\pi}{\beta_1}, \cdots,
\frac{\pi}{\beta_n},\delta)$, where $\delta$ appears in assumption
($C_n.3$). For $\theta\in I^- (k)$, let
\begin{align}
K^*_g(\theta) &:= U(\theta)K^*_g U(-\theta) \; \\
&= -i\partial_t + L_g^*(\theta,t) \; ,
\end{align}
where $L^*_g(\theta,t)$ is given in section 6.3 by
\begin{equation}
L^*_g (t, \theta):= U(\theta )L^*_g (t) U(-\theta) = \L_0
+ N\theta + g\tilde{V}^{tot}(t,\theta) \; , \label{CL}
\end{equation}
$\L_0=\L^\S+\sum_i \L^{\R_i}$, $\L^{\R_i}=d\Gamma (u_i),
i=1,\cdots ,n$, and
\begin{align*}
\tilde{V}^{tot}(t,\theta)&= \sum_i \{ \sigma_+\otimes {\mathbf
1}^\S \otimes b (f_{i,\beta_i}^{(\theta )}(t)) + \sigma_-\otimes\unit^\S \otimes b^* (f_{i,\beta_i}^{(\theta )}(t)) \\ 
&- i\unit^\S \otimes \sigma_+ \otimes (-1)^{N_i}(b(e^{-\beta_i
(u_i-\nu)/2 }f_{i, \beta_i}^{\# (\theta )}(t)) -i\unit^\S\otimes \sigma_-\otimes (-1)^{N_i} b^*(e^{\beta_i
(u_i-\nu)/2}f_{i,\beta_i}^{\# (\theta )}(t)) \} \; .
\end{align*}

For $n\in {\mathbf Z}$, let
\begin{equation}
\tilde{P}_{g,\bn} (\theta):= \oint_{\gamma_n} \frac{dz}{2\pi i}(z-K^*_g
(\theta))^{-1} \; , 
\end{equation}
such that $\gamma_n$ is a contour that encloses only the eigenvalues $E_j^{(n)}(g)=n\omega+E_j'(g), j=0,\cdots,3,n\in {\mathbf Z},$ where $E_j'(g), j=0,\cdots,3$ are the eigenvalues of $L_g^*(\theta)$ (see section 6.3). Moreover, let
$\tilde{T}_{g,\bn}:= \tilde{P}_{0,\bn} \tilde{P}_{g,\bn} (\theta)\tilde{P}_{0,\bn}$, then
we will show that the isomorphism
\begin{equation}
\label{quasiFL} \tilde{S}_{g,\bn}(\theta ):= \tilde{T}_{g,\bn}^{-1/2} \tilde{P}_{0,\bn}
\tilde{P}_{g,\bn} (\theta): Ran(\tilde{P}_{g,\bn}(\theta))\rightarrow
\h^{(n)}\otimes\H^\S\otimes\H^\S
\end{equation}
has an inverse
\begin{equation}
\tilde{S}_{g,\bn}^{-1}(\theta):= \tilde{P}_{g,\bn}(\theta) \tilde{P}_{0,\bn}
\tilde{T}_{g,\bn}^{-1/2}(t):
\h^{(n)}\otimes\H^\S\otimes\H^\S\rightarrow
Ran(\tilde{P}_{g,\bn}(\theta)) .
\end{equation}
Let
\begin{equation}
\tilde{M}_{g,\bn}(t):= \tilde{P}_{0,\bn} \tilde{P}_{g,\bn}
(\theta)K^*_g(\theta)\tilde{P}_{g,\bn}(\theta)\tilde{P}_{0,\bn}\;,
\end{equation}
and define the quasi-Floquet Liouvillean by
\begin{equation}
\label{qFL} \tilde{\S}_{g,\bn} := \tilde{S}_{g,\bn}
(\theta)\tilde{P}_{g,\bn}(\theta)K^*_g(\theta)\tilde{P}_{g,\bn}(\theta)\tilde{S}_{g,\bn}^{-1}(\theta)=
\tilde{T}_{g,\bn}^{-1/2}\tilde{M}_{g,\bn}\tilde{T}_{g,\bn}^{1/2}\; ,
\end{equation}
which is nothing but the mapping of the reduced Floquet
Liouvillean $\tilde{K_g}_\bn$ from $Ran(\tilde{P}_{g\bn})$ to
$\h^{(n)}\otimes\H^\S\otimes\H^\S .$

\vspace{0.5cm}

\noindent {\it Proposition 9.1}

Suppose ($C_n.2$),($C_n.3$) and ($C_n.6$) (see section 4.4). Then there is a
constant $g_1>0$ such that the following holds.

\begin{itemize}

\item[(i)] Assume that $(g, \theta)\in {\mathbf C}\times I^- (\delta )$, then $\D(K^*_g (\theta ))=\D$, $(K^*_g (\theta))^*=K_{\overline{g}}(\overline{\theta})$ and
the spectrum of $K_g (\theta )$ satisfies
\begin{equation}
\sigma (K^*_g (\theta)) \subset \{ z\in {\mathbf C} : \Im z \le
 C(g, \theta ) \} \; ,
\end{equation}
where
\begin{align*}
&C(g,\theta ):= \sup_{t\in {\mathbf R}} \{ 2\frac{|\Re g|}{\delta -
\Im g} | \Im \theta |^{1/2} + | \Im g | |\Im \theta |^{-1/2} \}
\times \\
& \times \sum_i \{ \| f_{i,\beta_i} (t) \|_{H^2 (\delta, \B)}+\|
e^{-\beta_i (u_i-\nu)/2}f_{i,\beta_i} (t) \|_{H^2 (\delta, \B)}\}.
\end{align*}
Furthermore, if $\Im z > C(g, \theta )$, then
\begin{equation}
\| (K^*_g (\theta) - z)^{-1} \| \le \frac{1}{\Im z - C(g,\theta)}
\; ,
\end{equation}
and the map $(g, \theta) \rightarrow  K^*_g(\theta)$ from
${\mathbf C}\times I^- (\delta)$ to the set of closed operators on
$\tilde{\H}$ is an analytic family (of
type A) in each variable separately.

\item[(ii)] If $|g| < g_1 |\Im \theta |$, then the spectrum of the operator
$K^*_g(\theta)$ in the half-plane $\Xi (\Im \theta +
\frac{|g|}{g_1})$ is purely discrete and independent of $\theta$.
If $|g|< \frac{1}{4} g_1 | \Im \theta |$, then the spectral
projections $\tilde{P}_{g,\bn} (\theta), n\in {\mathbf Z}$ associated to the spectrum of
$K^*_g (\theta)$ in the half-plane
 $\Xi (\Im \theta + \frac{|g|}{g_1}):= \{ z\in {\mathbf C} : \Im z > \Im \theta + \frac{|g|}{g_1} \}$ are analytic in $g$ and satisfy the estimate
\begin{equation}
\| \tilde{P}_{g,\bn}(\theta) - \tilde{P}_{0,\bn} \| < \frac{3|g|}{g_1 | \Im
\theta |}\; .
\end{equation}

\item[(iii)]If $|g|<\frac{g_1|\Im \theta |}{4}$, then the
quasi-Floquet Liouvillean $\tilde{\S}_{g,\bn}$ defined in (\ref{qFL})
depends analytically on $g$, and has a Taylor expansion
\begin{equation}
\label{FLTaylor2} \tilde{\S}_{g,\bn}=K^\S_\bn + \sum_{j=1}^{\infty} g^{2j}
\tilde{\S}_\bn^{(2j)} \;
\end{equation}
where
\begin{equation}
K^\S_\bn:= n\omega +\L^\S \;, n\in {\mathbf Z} .
\end{equation}

%%%%%%%%%%%%%%%%%%%%%%%%%%%%%%%%%
%modification of this expression
%%%%%%%%%%%%%%%%%%%%%%%%%%%%%%%%%

The first non-trivial coefficient in (\ref{FLTaylor2}) is

\begin{align*}
\tilde{\S}_\bn^{(2)}& = \frac{1}{2} \oint_\gamma \frac{dz}{2\pi
i}(\xi_\bn(z)(z-K_\bn^\S)^{-1}+(z-K_\bn^\S)^{-1}\xi_\bn(z))\; , \label{2orderqFL}
\end{align*}
where $\xi_\bn(z)=\tilde{P}_{0,\bn}
\tilde{V}^{tot}(\theta)(z-K_0(\theta))^{-1}\tilde{V}^{tot}(\theta)\tilde{P}_{0,\bn}.$

\item[(iv)] For
$g\in {\mathbf R}$ and $\Im z$ large enough,
\begin{equation}
s-\lim_{\Im \theta\uparrow 0}(K^*_g(\theta)-z)^{-1}= (K^*_g(\Re
\theta)-z)^{-1}\; . \label{removeCD2}
\end{equation}

\end{itemize}

{\it Proof.} To evade redundancy, we refer to chapter 6 for
details of the proof, since they are very similar to the ones
detailed there (with $K^*_g$ instead of $\L_g$). The proof of (i)
is similar to that of Lemma 6.3, and the proof of claim (ii) is
similar to that of Proposition 6.4, and (iii) to Proposition 6.5.
Since $\tilde{T}_{g,\bn}$ is analytic and $\| \tilde{T}_{g,\bn} - 1 \| < 1$
for $|g|<\frac{g_1\mu}{4}$, $\tilde{T}_{g,\bn}^{-1/2}$ is also analytic
in $g$. Inserting the Neumann series for the resolvent of
$K^*_g(\theta)$,
\begin{equation}
\tilde{T}_{g,\bn} = 1 + \sum_{j=1}^{\infty} g^j \tilde{T}_\bn^{(j)} \; ,
\end{equation}
with
\begin{equation}
\tilde{T}_\bn^{(j)} =\oint_\gamma \frac{dz}{2\pi i} (z-K_\bn^\S)^{-1}
\tilde{P}_{0,\bn} \tilde{V}^{tot}(\theta)
((z-K_0(\theta))^{-1}\tilde{V}^{tot}(\theta))^{j-1}\tilde{P}_{0,\bn}
(z-K_\bn^\S)^{-1}\; .
\end{equation}

Similarly,
\begin{equation}
\tilde{M}_{g,\bn}=K_\bn^\S+\sum_{j=1}^{\infty}g^j \tilde{M}_\bn^{(j)}\; ,
\end{equation}
with
\begin{equation}
\tilde{M}_\bn^{(j)}=\oint_\gamma \frac{dz}{2\pi i}z (z-K_\bn^\S)^{-1}
\tilde{P}_0 \tilde{V}^{tot}(\theta)
((z-K_0(\theta))^{-1}\tilde{V}^{tot}(\theta))^{j-1}\tilde{P}_{0,\bn}
(z-K_\bn^\S)^{-1}\; .
\end{equation}

The odd terms in the above two expansions are zero due to the fact
that $\tilde{P}_{0,\bn}$ projects onto the $N=0$ sector. The first
non-trivial coefficient in the Taylor series of $\S_{g,\bn}$ is
\begin{align}
\tilde{\S}_\bn^{(2)} &= \tilde{M}_\bn^{(2)}-\frac{1}{2} (\tilde{T}_\bn^{(2)}K_\bn^\S + K_\bn^\S \tilde{T}_\bn^{(2)}) \\
&= \frac{1}{2} \oint_\gamma \frac{dz}{2\pi
i}(\xi_\bn(z)(z-K_\bn^\S)^{-1}+(z-K_\bn^\S)^{-1}\xi_\bn(z))\; . \label{2orderqFL}
\end{align}

The proof of (iv) is similar to Lemma 6.6, section 6.1. $\Box$

Let $\tilde{P}_j^{(n)}$ be the spectral projections onto $E_j^{(n)}=n\omega+E_j, E_j\in\sigma (\L^\S),j=0,\cdots, 3, n\in
\mathbf{Z}.$

\vspace{1cm}

\pagebreak

\noindent {\it Proposition 9.2 (Spectrum of $K_g^*(\theta)$)}

Assume ($C_n.2$),($C_n.3$) and ($C_n.6$) (see section 4.4), and choose $\theta\in
I^-(k)$. Then there exists a constant $g_1>0$ independent of $t$
and $\theta$, such that, for $|g|<g_1$, the essential spectrum of
$K_g^*(\theta)$, $\sigma_{ess}(K_g^*(\theta ))\in {\mathbf
C}\backslash \Xi(\Im\theta)$, and the discrete spectrum
$\sigma_{disc}(K_g^*(\theta ))\in \Xi(\Im\theta)\cap \{ z\in
{\mathbf C} : \Im z\le 0\}$, with at least countably infinite
eigenvalues on the real line, $\{ \omega n: \; n\in
{\mathbf Z}\}$.

{\it Proof.} When the coupling $g=0$, $\sigma_{pp}(\L_0)=\{
-2,0,2\}$, with double degeneracy at 0, and
$\sigma_{c}(\L_0)=\mathbf{R}$. By Fourier transformation,
$L^2([0,\tau_*])\otimes \H$ is isomorphic to
$\bigcup_{n=-\infty}^\infty \langle e^{in\omega t} \rangle \otimes\H$, where $\omega=\frac{2\pi}{\tau_*}$. It follows that
\begin{equation}
\label{K0pps} \sigma_{pp}(K_0)=\{ E_j^{(n)}=E_j+n\omega:
E_j\in\sigma(\L^\S), j=0,\cdots,3; n\in\mathbf{Z}\} \; .
\end{equation}
From Proposition 9.1, (i), the map
\begin{equation*}
(I^-(k),{\mathbf C})\ni (\theta,g) \rightarrow K_g^*(\theta) \;
\end{equation*}
with values in the set of closed operators on
$L^2([0,\tau_*])\otimes \H$ is an analytic family in each variable separately. Let
\begin{equation}
C(\beta_i,k):= \sup_{|\Im z|<k} \frac{1}{\sqrt{|1+e^{-\beta_i
(z-\nu)}|}} , i=1,\cdots,n \; ,
\end{equation}
which is finite, and let
\begin{equation}
\label{tildeC}
\tilde{C}:= \sup_{t}\sup_{\theta\in I^-(k)}\|
\tilde{V}^{tot}(\theta,t)\| \;,
\end{equation}
which is also finite due to assumption ($C_n.4$) and the estimate
\begin{align*}
&\tilde{C} \le \sqrt{2}\sum_{i=1}^n C(\beta_i,k)\{\|
f_{i,\beta_i}\|_{H^2(k)} +\\
&+ \|e^{-\beta_i (u_i-\nu)/2}f_{i,\beta_i}\|_{H^2(k)}\} < \infty \; .
\end{align*}
The remaining part of the proof is similar to the proof of Theorem
6.7, section 6.1. The essential spectrum of $K^*_0(\theta)$ are lines
$\{ z\in {\mathbf C}: \Im z = in\Im \theta, n\in {\mathbf N}^+
\}$, while $\sigma_{disc}(K_0(\theta))=\sigma(K^\S)$.

Choose $g_1>0$ such that $g_1 \tilde{C} < (k-\mu)/4$, where
$\mu=|\Im \theta|$ and $\tilde{C}$ is defined in \fer{tildeC}.
Then, for $|g|<g_1$ and $-k<\Im \theta < -(k+\mu)/2$, the
essential spectrum $\sigma_{ess}(K_g(\theta))$ is contained in the
half-plane $\{ z\in {\mathbf C}: \Im z <-\mu \}$. The location of
the discrete spectrum of $K_g(\theta )$ inside the half-plane $\Xi
(-\mu)$ can be computed using regular perturbation theory (see
Proposition 9.1). 

%%%%%%%%%%%%%%%%%%%%%%%%%%%%%%%%%%%%%%%%%%%%%%%%%%%%%%%%%%%%%%%%%%%%%
%DETAILS OF THE COMPUTATION TO SECOND ORDER IN THE COUPLING
%%%%%%%%%%%%%%%%%%%%%%%%%%%%%%%%%%%%%%%%%%%%%%%%%%%%%%%%%%%%%%%%%%%%%

We know that
\begin{align*}
\tilde{\S}_\bn^{(2)} &= \frac{1}{2}\oint_\gamma \frac{dz}{2\pi i} \{ \tilde{P}_{0,\bn} \tilde{V}^{tot}(\theta )(z-K_0(\theta ))^{-1}\tilde{V}^{tot}(\theta )\tilde{P}_{0,\bn} (z-K_\bn^\S)^{-1} \\
&+(z-K_\bn^\S) \tilde{P}_{0,\bn} \tilde{V}^{tot}(\theta )(z-K_0(\theta ))^{-1}\tilde{V}^{tot}(\theta )\tilde{P}_{0,\bn} \} \; .
\end{align*}
Let 
\begin{equation}
\tilde{\Gamma}^{n(2)}_{j}:= \tilde{P}^{(n)}_j \tilde{\S}_\bn^{(2)} \tilde{P}^{(n)}_j \; ,
\end{equation}
where $\tilde{P}^{(n)}_j, j=0,\cdots,3, n\in {\mathbf Z}$ are the spectral projections onto $E_j^{(n)}=n\omega+E_j, $ the eigenvalues of $K^\S.$

Applying the Cauchy integration formula to the expression of $\tilde{\Gamma}^{n(2)}_j$ (as in the proof of Proposition 6.9, section 6.3) gives

\begin{equation}
E^{(n)}_{2,3}(g)=n\omega \pm (2 + g^2 \PV\int_{\mathbf R} du \int_{0}^{\tau_*} dt
\frac{1}{2-u} \sum_{i=1}^n \| \tilde{f}_{i}(u,t)\|^2_{\B}) -i\pi
g^2 \sum_{i=1}^n \int_0^{\tau_*} dt \|\tilde{f}_{i}(2,t)\|^2_\B + O(g^4) \; ,
\end{equation}
where $\PV$ stands for the Cauchy prinicipal value, and
$n\in\mathbf{Z}$. Note that $\Im E^{(n)}_{2,3}(g)<0, n\in {\mathbf
Z}$. Furthermore,
\begin{equation}
E^{(n)}_{0,1}(g)=n\omega+ g^2 a_{0,1} + O(g^4)\; ,
\end{equation}
where $a_{0,1}(t)$ are the eigenvalues of the $2\times 2$ matrix
\begin{equation}
-i \pi \sum_{i=1}^n \int_0^{\tau_*} dt \frac{\| \tilde{f}_i(2,t)\|^2_\B}{2\cosh \beta_i(2-\nu)/2}
\left(
\begin{matrix}
e^{\beta_i(2-\nu)/2}  & -e^{\beta_i(2-\nu)/2}  \\
-e^{-\beta_i(2-\nu)/2} & e^{-\beta_i(2-\nu)/2} 
\end{matrix}
\right) \; .
\end{equation}
By construction, $K_g e^{i n\omega t}\Omega=n\omega e^{in\omega
t}\Omega$ and $U(\theta)e^{in\omega t}\Omega= e^{in\omega t
}\Omega$, so $n\omega,n\in\mathbf{Z}$ are also isolated
eigenvalues of $K_g^*(\theta)$. In this case, this can be seen by defining the spectral projection to the real isolated eigenvalues of $K_g(\theta)$ using the resolvent, and taking the adjoint to define the corresponding spectral projections for the real isolated eigenvalues of $K_g^*(\theta)$.

In fact, $\psi=\left( \begin{matrix} 1
\\ 1
\end{matrix}\right)$ is the eigenvector corresponding to
zero eigenvalue of $\tilde{\S}_{g,\bn}^{2}$. Hence,
\begin{align}
E^{(n)}_0&=n\omega \; ,\\
E^{(n)}_1&=n\omega -i\pi g^2 \sum_{i=1}^n \int_0^{\tau_*} dt \| \tilde{f}_{i} (2,t)
\|^2_\B +O(g^4) \; .
\end{align}
Note that $\Im E^{(n)}_j <0, j=1,2,3$ and $n\in {\mathbf Z}$.
Moreover, we know from Proposition 9.1, (iv), that $s-\lim_{\Im
\theta \uparrow 0} (z-K^*_g(\theta))^{-1} = (z-K^*_g(\Re \theta
))^{-1}$ for $\Im z$ big enough, and hence the claim of this
theorem. $\Box$

\section{Convergence to time-periodic states}

The following theorem claims that under suitable sufficient
assumptions, the real state of the system converges to a
time-periodic state with the same period of the perturbation
$\tau_*$.

Let $\h^{test}=
D(e^{k\sqrt{p^2+1}})$, and $\O^{test}= \F_-(\h^{test} )$, which
is dense in $\F_-(L^2({\mathbf R};\B))$, and define 
\begin{equation}
{\mathcal C}:= \O^\S\otimes\O^{\R_1 ,test}\otimes \cdots \otimes \O^{\R_n, test} \; ,
\end{equation}
which is dense in $\O$.

We make the following additional assumption.
\begin{itemize}
\item[(H9.1)] The perturbation of the Hamiltonian $V(t)\in {\mathcal C}$ for $t\in {\mathbf R}$. 

\end{itemize}

\vspace{0.5cm}
\noindent {\it Theorem 9.3 (Convergence to
time-periodic states)}

Assume ($C_n.2$),($C_n.3$), ($C_n.6$) and (H9.1), and let $D$ be a
positive operator on $\H$ such
that $Ran D$ is dense in $\H$ and $D\Omega=\Omega$.
Assume further that $a\in {\mathcal C}$. Then there is a
constant $g_1>0$, such that, for $|g|<g_1$, the following holds
\begin{equation}
\lim_{n\rightarrow\infty} \langle \Omega, \a_g^{n\tau_*+t}(a)
\Omega \rangle = \langle \tilde{\Omega}_g(0) , D^{-1} \a_g^t (a)
\Omega \rangle \; ,
\end{equation}
where $\tilde{\Omega}_g$ corresponds to the zero-energy resonance of the
adjoint of the Floquet Liouvillean, $K_g^*$.

{\it Proof.} Choose $k$ as in Proposition 9.2, and let
\begin{equation}
D:=\mathbf{1}^\S\otimes
e^{-k\sqrt{A_{\R_1}^2+1}}\otimes\cdots\otimes
e^{-k\sqrt{A_{\R_n}^2+1}} \; ,
\end{equation}
where $A_{\R_i}=d\Gamma(i\partial_{u_i}), i=1,\cdots,n$ is the
second quantization of the generator of energy translations for
the $i^{th}$ reservoir. The operator $D$ is positive such that
$Ran D$ is dense in $\H$ and $D\Omega=\Omega$. It follows from assumption (H9.1) and the fact that $a\in {\mathcal C}$ that $\a_g^t (a)\in \D(D^{-1})$ (as discussed in the proof of Theorem 8.3, section 8.2).

The remainder of the proof relies on the result of Proposition 9.2 and equation
\fer{FSM}. It follows from \fer{FSM} and the time periodicity of $f\in\tilde{\H}$ that
\begin{equation}
\label{evolK} 
(e^{-iK_g^* n \tau_*} \unit\otimes\Omega)(0)=\tilde{U}_g(n\tau_*,0) (\unit\otimes\Omega)(0)=\tilde{U}_g(n\tau_*,0)\Omega \; .
\end{equation}

Let $\unit\otimes\Omega=: \overline{\Omega}\in \tilde{\H},$ and $\overline{D}:=\unit\otimes D.$

Without loss of generality, we assume $\frac{2\pi}{\tau_*}\ne 2$; however, if
$\frac{2\pi}{\tau_*}=2$, the state of the system typically
oscillates between two resonance states until it finally converges
to a time-periodic state; see remark 1. Using the dynamics on
$C(\O,\Omega)$, \fer{evolK}, and knowledge of
$\sigma(K_g^*(\theta))$, it follows that
\begin{align}
&\lim_{n\rightarrow\infty}\langle \Omega, \a_g^{n\tau_*+t}(a) \Omega \rangle \\
&= \lim_{n\rightarrow\infty}\langle \tilde{U}_g(n\tau_*, 0)\Omega, \a_g^t(a)\Omega\rangle \\
&= \lim_{n\rightarrow\infty} \langle (e^{-iK_g^* n\tau_*} \unitOmega)(0),\a_g^{t}(s)\Omega\rangle \\
&= \lim_{n\rightarrow\infty} \langle (\unitD U(-\theta)e^{-iK^*_g(\theta)n \tau_*}U(\theta) \unitD \unitOmega)(0), D^{-1} \a^t_g(a) \Omega \rangle
\\
&= \lim_{n\rightarrow\infty} \langle (\unitD U(-\theta)\int_{-\infty}^\infty du (u+i\eta-K_g^*(\theta
))^{-1}e^{-i(u+i\eta)n\tau_*} \nonumber \\
&\unitOmega)(0),D^{-1} \a_g^t(a)\Omega \rangle \\
&= \lim_{n\rightarrow\infty} \langle  (\unitD U(-\theta)
\oint_\gamma dz (z -K_g^*(\theta))^{-1}e^{-iz n\tau_* } \unitOmega)(0) ,D^{-1} \a_g^t(a) \Omega \rangle + \nonumber \\
&+\lim_{n\rightarrow\infty} \langle (\unitD U (-\theta )
\int_{-\infty}^\infty du (u+i(\mu -\epsilon )-K_g^*(\theta
))^{-1}e^{-i(u+i(\mu-\epsilon ))n\tau_*}\unitOmega)(0), \nonumber \\
& D^{-1} \alpha_g^t(a) \Omega \rangle , \label{2term}
\end{align}
where $\eta>0$ and $0<\epsilon<\mu$.

Using the results of Propositions 9.1 and 9.2, the first term in the last equation
converges to the time-periodic expression,
\begin{align*}
&\lim_{n\rightarrow\infty} \langle (\unitD U(-\theta)
\oint_\gamma dz (z-K_g^*(\theta))^{-1} e^{-iz n\tau_*}\unitOmega )(0),D^{-1}
\a_g^t(a)\Omega \rangle \\
&= \sum_{k\in{\mathbf Z}}\lim_{n\rightarrow\infty} \langle 
(\unitD U(-\theta)\tilde{S}_{g,(k)}^{-1}(\theta)e^{-i\tilde{\S}_{g,(k)}(\theta)n\tau_*}\tilde{S}_{g,(k)}(\theta )\unitOmega )(0), D^{-1}\a_g^t(a)\Omega \rangle\\
&=\sum_{k\in {\mathbf Z}} \langle (\unitD \tilde{P}_{g,(k)}(\theta) \unitOmega ) (0), \a_g^t(a) \Omega \rangle .
\end{align*}
Let
\begin{equation}
\tilde{\Omega}_{g,(k)}:= \unitD \tilde{P}_{g,(k)}(\theta)(e^{ik\omega t}\otimes\Omega) \; ,
\end{equation}
then 
\begin{equation}
\unitD\tilde{P}_{g,(k)} \unitD = \{ (e^{ik\omega t} \otimes\Omega), \cdot \} \tilde{\Omega}_{g,(k)} \; ,
\end{equation}
where $\{ \cdot, \cdot \}$ denotes the scalar product on $\tilde{H}.$
Therefore,
\begin{align}
\unitD \tilde{P}_{g,(k)}(\theta) \unitOmega 
&= \{ e^{ik\omega t} \otimes \Omega, \unit\otimes\Omega \}\tilde{\Omega}_{g,(k)} \\
&= \tilde{\Omega}_{g,(0)} \; ,
\end{align}
and hence
\begin{equation}
\sum_{k\in {\mathbf Z}} \langle (\unitD \tilde{P}_{g,(k)} \unitOmega)(0),D^{-1}\a_g^t (a)\Omega \rangle 
=\langle (\tilde{\Omega}_{g,(0)})(0) , D^{-1} \a_g^t (a) \Omega \rangle \; ,
\end{equation}
where $\tilde{\Omega}_{g,(0)}$ is the zero-energy resonance of the
Floquet Liouvillean.
The second term \fer{2term} converges exponentially fast to zero since
\begin{equation}
\langle (\unitD U(-\theta) \int_{-\infty}^\infty du
(u+i(\mu-\epsilon)-K_g^*(\theta))^{-1}e^{-i(u+i(\mu-\epsilon))n\tau_*}
\unitOmega)(0),D^{-1} \a_g^t(a)\Omega\rangle = O(e^{-(\mu-\epsilon')n\tau_*}) ,
\end{equation}
where $0<\epsilon'<\epsilon<\mu$. 
$\Box$

\vspace{0.5cm}

{\it Remarks.}
\begin{itemize}

\item[(1)]When $\frac{2\pi}{\tau_*}=2$, the system exhibits the
phenomena of resonance: The state of the system oscillates between
two resonances until it finally converges to the time periodic
state corresponding to $\tilde{\Omega}_g$.[A-SF2] (This can be
verified by a second order time-dependent perturbation theory
calculation; (see also [Ya2]).

\item[(2)]Note that
\begin{align*}
\langle (\tilde{\Omega}_{g,(0)})(0), D^{-1}\a^t_g(a)\Omega \rangle
= \langle (\tilde{\Omega}_{g,(k)})(0), D^{-1}\a^t_g(a)e^{ik\omega t}\Omega\rangle \; ,
\end{align*}
where $\tilde{\Omega}_{g,(k)}$ is the state corresponding to the
$k\omega$-energy resonance of the adjoint of the
Floquet Liouvillean.

\item[(3)] For small enough coupling $g$, it follows from
Proposition 9.1,(ii) that $\tilde{P}_g(\theta )$ is analytic in
$g$. This is of especially {\it practical} importance, since in
principle one can expand $\tilde{\Omega}_g$ in a Taylor series to
arbitrary order in $g$, and compute the entropy production per
cycle. In particular, when there are only two reservoirs, one may
explicitly compute the degree of efficiency $\eta$ and compare it
to $\eta^{rev}$ (see chapter 2). Further details will appear in
[A-SF2].

\end{itemize}

\pagebreak

%%%%%%%%%%%%%%%%%%%%%%%%%%%%%%%%%%%%%%%%%%%%%%%%%%%%%%%%%%%%%%%%%
\noindent {\LARGE Acknowledgments}\\[10pt]

\noindent First and foremost, I thank my supervisor, J\"urg Fr\"ohlich, 
for his constant support and useful criticism.

I also thank Gian Michele Graf for accepting to be my co-referee
and for the pleasant interaction and the useful discussions we had during my stay at ETHZ.

Many thanks to my loving family: my parents, Moukarram and Khaled,
for all that they have taught me, and my sisters, Fatima, Bushra
and Dina, for their sincere friendship.

\pagebreak

%%%%%%%%%%%%%%%%%%%%%%%%%%%%%%%%%%%%%%%%%%%%%%%%%%%%%%%%%%%%%%%%%


\addcontentsline{toc}{chapter}{Bibliography}

\begin{thebibliography}{ABCD}



\bibitem[A-S]{} Abou-Salem,W., in preparation

\bibitem[A-SF1]{} Abou-Salem,W., Fr\"ohlich, J., {\it Adiabatic theorems for quantum
resonances}, in preparation

\bibitem[A-SF2]{} Abou-Salem,W., Fr\"ohlich, J.,{\it Cyclic thermodynamic processes and Floquet
theory}, in preparation

\bibitem[A-SF3]{} Abou-Salem,W., Fr\"ohlich, J., {\it Adiabatic theorems and reversible isothermal
processes}, Lett. Math. Phys. {\bf 72}, 153-163 (2005)

\bibitem[A-SF4]{} Abou-Salem,W., Fr\"ohlich, J.,{\it Status of the fundamental laws of thermodynamics},
in preparation.


\bibitem[ABG]{} Amrein, W., Boutet de Monvel, A., Georgescu, V., {\it
    $C_0$-Groups, Commutator Methods and Spectral Theory of $N$-Body
    Hamiltonians}, Basel-Boston-Berlin, Birkh\"auser, 1996


\bibitem[Ar1]{} Araki, H.,{\it Relative Hamiltonian for faithful normal states
    of a von Neumann algebra}. Pub. R.I.M.S., Kyoto Univ. {\bf 9}, 165-209 (1973)

\bibitem[Ar2]{} Araki, H. , {\it Relative entropy of states of von
Neumann algebras}, Pub. R.I.M.S., Kyoto Univ. {\bf 11}, 809
(1976); {\it Relative entropy of states of von Neumann algebras
II},Pub. R.I.M.S., Kyoto Univ. {\bf 13}, 173 (1977)


\bibitem[AWo]{} Araki, H., Woods, E., {\it Representations of the canonical
    commutation relations describing a non-relativistic infinite free bose
    gas}. J. Math. Phys. {\bf 4}, 637-662 (1963)

\bibitem[AWy]{} Araki, H., Wyss, W., {\it Representations of canonical anticommutation
relations}, Helv. Phys. Acta {\bf 37}, 136 (1964)

\bibitem[AE]{} Avron,J.E., Elgart, A., {\em Adiabatic theorem
without a gap condition}, Commun. Math. Phys. {\bf 203},
445-463 (1999)


\bibitem[BFS]{} Bach, V., Fr\"ohlich, J., Sigal, I.M., {\it Return to
    Equilibrium}. J. Math. Phys. {\bf 41 no 6}, 3985-4061 (2000)


\bibitem[BFSS]{} Bach, V., Fr\"ohlich, J., Sigal, I.M., Soffer, A., {\it
    Positive Commutators and the spectrum of Pauli-Fierz hamiltonian of atoms
    and molecules}, Commun. Math. Phys. {\bf 207 no. 3}, 557-587 (1999)

\bibitem[Boy]{} Boyling, J.B., {\it An axiomatic approach to classical thermodynamics}, Proc. R. Soc. London A {\bf 329}, 35-70 (1972)

\bibitem[BR]{} Bratteli, O., Robinson, D., {\it Operator Algebras and
    Quantum Statistical Mechanics 1,2}, Texts and Monographs in Physics,
    Springer-Verlag Berlin, 1987

\bibitem[CFS]{} Chen, T., Fr\"ohlich, J. and Seifert, M., {\it Renormalization group methods: Landau-Fermi liquid and BCS superconductors}, in {\it Fluctuating geometries in statistical mechanics and field theory}, Les Houches session LXII (1994), David, F., Ginsparg, P., Zinn-Justin, J. (eds.), Amsterdam, North-Holland, 1996 

\bibitem[DF1]{} Derezinski, J. and Fr\"uboes, R., {\it Level-Shift operator and 2nd order perturbation theory}, to appear in Journ. Math. Phys.

\bibitem[DF2]{} Derezinski, J. and Fr\"uboes, R., {\it Fermi Golden Rule and Open Quantum Systems}, to appear in proceedings of Summer School {\it ``Open Quantum Systems''}, Grenoble 2003

\bibitem[DJ]{} Derezi\'nski, J., Jaksi\'c, V., {\it Spectral theory of
    Pauli-Fierz operators}, preprint (2000)

\bibitem[DJP]{}Derezi\'nski, J., Jaksi\'c, V.,Pillet, C.-A., {\it Perturbation theory of KMS
states}, preprint (2003)

\bibitem[Don]{} Donald, M.J., {\it Relative Hamiltonians which are not bounded from
above}, J. Fun. Anal. {\bf 91}, 143 (1990)

\bibitem[FLKT]{} Feldman, J., Lehmann, D., Kn\"orrer, H. and Trubowitz, E., {\it Fermi Liquids in two space dimensions}, in {\it Constructive Physics}, Rivasseau, V. (ed.), Lecture notes in Physics, vol {\bf 446}, Berlin, Springer-Verlag, 1995 

\bibitem[FM1]{} Fr\"ohlich, J. and Merkli, M.,
{\em Another return of ``return to equilibrium''}, Commun. Math.
Phys. {\bf 251}, 235-262 (2004)

\bibitem[FM2]{} Fr\"ohlich, J., Merkli, M., {\it Thermal Ionization.} Math. Phys. Analysis and Geometry {\bf 7}, 239-287 (2004)


\bibitem[FMRT]{} Feldman, J., Magnen, J., Rivasseau, V., and Trubowitz, E., {\it Fermionic many-body models} in {\it Mathematical quantum theory I}, Feldman, J. et al (eds), CRM Proceedings and Lecture Notes, 1994

\bibitem[FMS]{}Fr\"ohlich, J., Merkli, M., Sigal, I.M., {\it Ionization of atoms
in a thermal field.}, J. Stat. Phys. {\bf 116}, 311-359 (2004)

\bibitem[FMUe]{} Fr\"ohlich, J., Merkli, M. and Ueltschi, D., {\em Dissipative
transport: thermal contacts and tunnelling junctions}, Ann. Henri
Poincar\'e {\bf 4}, 897-945 (2003)

\bibitem[FMSU]{}Fr\"ohlich, J., Merkli, M.,  Schwarz, S., and Ueltschi, D.,
 {\it Statistical mechanics of thermodynamic processes},
in {\it A garden of quanta}, 345-363, World Sci. Publishing, River
Edge, New Jersey, 2003

\bibitem[FP]{}Fr\"ohlich, J. and Pfeifer, P.,
{\em Generalized time-energy uncertainty relations and bounds on
lifetimes of resonances}, Rev. Mod. Phys. {\bf 67}, 759-779 (1995)

\bibitem[Fr\"o]{}Fr\"ohlich, J., {\it An introduction to some topics in
constructive quantum field theory.} NATO advanced study institutes
seires: Series B, Physics; v.30. International Summer Institute on
Theoretical Physics, 8th, University of Bielefeld, 1976. Many
degrees of freedom in field theory



\bibitem[GG]{} Georgescu, V., G\'erard, C., {\it On the Virial Theorem in
    Quantum Mechanics}, Commun. Math. Phys. {\bf 208}, 275-281 (1999)

\bibitem[H]{} Haag, R., {\it Local Quantum Physics. Fields, Particles,
    Algebras}. Text and Monographs in Physics. Springer-Verlag Berlin (1992)

\bibitem[HHW]{}Haag, R., Hugenholtz, N. M., Winnink, M., {\it On the equilibrium
states in quantum statistical mechanics.}, Commun. Math. Phys.
{\bf 5}, 215--236 (1967)


\bibitem[He]{} Hepp, K., {\it Rigorous results on the $s-d$ model of the Kondo
effect.} Solid State Commun. {\bf 8} , 2087--2090 (1970)


\bibitem[Ho]{} Howland, J.S., {\it Stationary scattering theory for time dependent
Hamiltonians}, Math. Ann. {\bf 207}, 315-335 (1974)

\bibitem[HP]{}Hunziker, W., Pillet, C.-A., {\it Degenerate asymptotic perturbation
theory}, Commun. Math. Phys. {\bf 90}, 219 (1983)

\bibitem[HS]{} Hunziker, W., Sigal, I.M., {\it Time-dependent scattering
    theory of $N$-body quantum systems}, Rev. Math. Phys. {\bf 12} (2000)

\bibitem[Hu]{} Hunziker, W., {\it Notes on asymptotic perturbation theory for Schr\"odinger eigenvalue
problems}, Helv. Phys. Acta {\bf 61}, 257-304 (1988)

\bibitem[JOP]{} Jaksic, V., Ogata, Y., Pillet, C.-A., {\it The Green-Kubo formula and the Onsager reciprocity relations in quantum statistical mechanics}, preprint

\bibitem[JP1]{} Jaksi\'c, V., Pillet, C.A., {\it On a Model for Quantum
    Friction II. Fermi's Golden Rule and Dynamics at Positive
    Temperature}, Commun. Math. Phys. {\bf 176}, 619-644 (1996)

\bibitem[JP2]{} Jaksi\'c, V., Pillet, C.A., {\it On a Model for Quantum
    Friction III. Ergodic Properties of the Spin-Boson
    System}, Commun. Math. Phys.  {\bf 178}, 627-651 (1996)

\bibitem[JP3]{} Jaksi\'c, V. and Pillet, C.-A.:{\em Non-equilibrium
steady states of finite quantum systems coupled to thermal
reservoirs}, Commun. Math. Phys. {\bf 226}, 131-162 (2002)

\bibitem[J]{} Joye, A., {\it Proof of Landau-Zener formula},
Asymptotic Analysis {\bf 9}, 209-258 (1994)


\bibitem[Ka1]{} Kato, T., {\it Perturbation theory for linear
operators}, Berlin: Springer, 1980

\bibitem[Ka2]{} Kato, T., {\it Linear evolution equations of hyperbolic
type}, I.J. Fac. Sci. Univ. Tokyo Sect. IA {\bf 17}, 241-258
(1970)

\bibitem[LR]{} Lieb, E., Ruskai, M.-B, {\it Proof of the strong subadditivity of quantum-mechanical
entropy}, J. Math. Phys. {\bf 14}, 1938-1941 (1973)

\bibitem[LY]{} Lieb, E., Yngvason, J., {\it The mathematical structure of the second law of thermodynamics}, in {\it Current Developments in Mathematics, 2001}, International Press, Cambridge, 2002

\bibitem[M1]{} Merkli, M., {\it Positive Commutator Method in Non-Equilibrium Statistical Mechanics}, Commun. Math. Phys. {\bf 223}, 327-362 (2001)

\bibitem[M2]{} Merkli, M.,{\it Stability of equilibria with a condensate}, Commun. Math. Phys. {\bf 257}, 621-640 (2005)

\bibitem[NT]{}Narnhofer, H.,  Thirring, W. {\it Adiabatic theorem in quantum statistical mechanics}, Phys. Rev. A {\bf 26}, 3646-3652 (1982)


\bibitem[Ne]{} Nenciu, G., {\it Linear adiabatic theory and
applications}, preprint (1991)



\bibitem[OP]{} Ohya, M., Petz, D., {\it Quantum entropy and its
use}, Berlin: Springer-Verlag, 1993

\bibitem[RS1,2]{}
Reed, M., Simon, B., {\it Methods of Modern Mathematical Physics},
Vol. I (Functional Analysis), Vol. II (Fourier Analysis,
Self-Adjointness), Academic Press, New York 1975


\bibitem[Rob]{} D. Robinson, {\it Return to Equilibrium}, Commun. Math. Phys. {\bf
31}, 171--189 (1973)

\bibitem[Rud]{} Rudin, W., {\it Real and Complex Analysis}, 3rd
ed., Mc-Graw-Hill, New York, 1987


\bibitem[Ru1]{}Ruelle, D., {\it Statistical Mechanics. Rigorous results}, Reprint
of the 1989 edition. World Scientific Publishing Co., Inc., River
Edge, NJ; Imperial College Press, London, 1999


\bibitem[Ru2]{} Ruelle, D., {\it Entropy production in quantum spin systems},
Comm. Math. Phys. {\bf 224} , no. 1, 3-16 (2001)

\bibitem[Ru3]{} Ruelle, D. {\it Natural nonequilibrium states in quantum
statistical mechanics},  J. Stat. Phys. {\bf 98} , no. 1-2, 57-75
(2000)


\bibitem[Sa]{} Sakai, S., {\it $C^*$-Algebras and $W^*$-Algebras}, Berlin, Springer, 1971

\bibitem[Te]{}Teufel, S., {\em A note on the adiabatic theorem},
Lett. Math. Phys. {\bf 58}, 261-266 (2001)

\bibitem[Uh]{} Uhlmann, A., {\it Relative entropy and the Wigner-Yanase-Dyson-Lieb concavity in an interpolation
theory}, Commun. Math. Phys. {\bf 54}, 123 (1977)

\bibitem[Ya1]{} Yajima, K., {\it Scattering theory for Schr\"odinger equations with potentials periodic in
time}, J. Math. Soc. Japan {\bf 29}, 729-743 (1977)

\bibitem[Ya2]{} Yajima, K., {\it Resonances for AC-Stark effect},
Commun. Math. Phys. {\bf 87}, 331-352 (1982)

\bibitem[Yo]{} Yosida, K.: {\em Functional Analysis}, 6th ed.,
Springer-Verlag, Berlin, 1998


\end{thebibliography}
\end{document}